\documentclass[reprint,superscriptaddress,amsmath,amssymb,aps,pra,floatfix,nofootinbib]{revtex4-2}

\usepackage{graphicx}
\usepackage{physics}

\usepackage{amsthm}
\usepackage{mathtools}
\usepackage{dsfont}
\usepackage{tabularx}
\usepackage{units}
\usepackage{comment}
\usepackage{enumerate}
\usepackage{algorithm}
\usepackage{algpseudocode}
\usepackage{thmtools}
\usepackage{thm-restate}

\usepackage[dvipsnames]{xcolor}

\usepackage[colorlinks=true,citecolor=NavyBlue,linkcolor=NavyBlue,urlcolor=NavyBlue]{hyperref}
\usepackage[english,capitalise]{cleveref}

\newcommand{\R}{\mathbb{R}}
\newcommand{\C}{\mathbb{C}}
\newcommand{\N}{\mathbb{N}}

\newcommand{\eps}{\varepsilon}
\newcommand{\defvar}{\coloneqq} %Symbol to use for defining variables
\newcommand{\mbf}[1]{\mathbf{#1}} 
\newcommand{\bsym}[1]{\boldsymbol{#1}} 
\newcommand{\supp}{\operatorname{supp}} %Support
\newcommand{\idop}{\mathbb{I}} %Identity operator

\newcommand*\widebar[1]{%
  \hbox{%
    \vbox{%
      \hrule height 0.5pt % The actual bar
      \kern0.4ex%         % Distance between bar and symbol
      \hbox{%
        \kern-0.1em%      % Shortening on the left side
        \ensuremath{#1}%
        \kern-0.0em%      % Shortening on the right side
      }%
    }%
  }%
} 

\newcommand{\renyiEnt}     {\mathbb{H}}

\newcommand{\renyiSandUp}  {\widetilde{H}^\uparrow}
\newcommand{\renyiSandDown}{\widetilde{H}^\downarrow}
\newcommand{\renyiPetzUp}  {\widebar{H}^\uparrow}
\newcommand{\renyiPetzDown}{\widebar{H}^\downarrow}

\newcommand{\frenyiSandUp}  {\widetilde{H}^{\uparrow,f}}
\newcommand{\frenyiSandDown}{\widetilde{H}^{\downarrow,f}}

\newcommand{\renyiDiv}     {\mathbb{D}}
\newcommand{\renyiSandDiv} {\widetilde{D}}
\newcommand{\renyiPetzDiv} {\widebar{D}}

\newcommand{\Qbar}{\widebar{Q}}
\newcommand{\Qopt}{\widehat{Q}}

\newcommand{\ffull} {\hat{f}_\mathrm{full}}
\newcommand{\hQKD} {h_{\alpha}^{\mathrm{QKD}}}
\newcommand{\hUpQKD} {h_{\alpha}^{\uparrow,\mathrm{QKD}}}
\newcommand{\fprot} {\hat{f}_\mathrm{prot}}
\newcommand{\kapup}{\kappa^\uparrow} %Upparrow normalization constant
\newcommand{\kapupbnd}{\underline{\kappa}^\uparrow} %Lower bound on uparrow normalization constant

\newcommand{\rel}[2]{\!\left(#1 \middle \| #2 \right)}
\newcommand{\con}[2]{\!\left(#1 \middle | #2 \right)}

\DeclareMathOperator{\diag}{diag} %diagonal matrix
\DeclareMathOperator{\id}{\mathord{\rm id}} %identity map
\DeclareMathOperator{\pow}{Pow} %power function in words
\newcommand{\floor}[1]{\left\lfloor #1 \right\rfloor} %floor
\newcommand{\ceil}[1]{\left\lceil #1 \right\rceil} %ceil
\newcommand{\Pos}{\operatorname{Pos}} %Positive operators
\newcommand{\Herm}{\operatorname{Herm}} %Hermitian operators
\newcommand{\dop}[1]{\operatorname{S}_{#1}} %Set of density operators

\newcommand{\eigmin}{\operatorname{\lambda_{min}}}

\DeclareMathOperator{\EX}{\mathbb{E}}

\newcommand{\Isom}{\operatorname{U}}
\newcommand{\CPTP}{\operatorname{CPTP}} % completely positive trace preserving
\newcommand{\inner}[2]{\left\langle #1,#2 \right\rangle}

\newcommand{\leak}{\lambda_{\text{EC}}}
\newcommand{\eEV}{\eps_\mathrm{EV}}
\newcommand{\ePA}{\eps_\mathrm{PA}}
\newcommand{\eSec}{\eps_\mathrm{sec}}
\newcommand{\eSct}{\eps_\mathrm{sct}}
\newcommand{\eCor}{\eps_\mathrm{cor}}
\newcommand{\EATchannQKD}{\EATchann_{\text{QKD}}}
\newcommand{\OmegaAcc}{\Omega_{\mathrm{acc}}}
\newcommand{\OmegaLen}[1]{\Omega_{\mathrm{len} = #1}}
\newcommand{\OmegaEV}{\Omega_{\mathrm{EV}}}
\newcommand{\OmegaAT}{\Omega_{\mathrm{AT}}}
\newcommand{\Sacc}{S_{\mathrm{acc}}}
\newcommand{\fEC}{f_{\mathrm{EC}}}
\newcommand{\Ctest}{\mathcal{C}^{\mathrm{test}}_{AB}}

\newcommand{\Ncons}{N_{\mathrm{cons}}} %N_cons
\newcommand{\Nent}{N_{\mathrm{ent}}} %N_cons
\newcommand{\Nph}{N_{\mathrm{ph}}} %N_ph

\newcommand{\probst}{\mathbf{\Phi}}
\newcommand{\probstJ}{\mathbf{\Psi}}

\newcommand{\lvar}{l_{\mathrm{var}}}

\DeclareMathOperator{\GMap}{\mathcal{G}}
\DeclareMathOperator{\GMapDelta}{\GMap_\delta}
\DeclareMathOperator{\ZMap}{\mathcal{Z}}

\DeclareMathOperator{\pert}{\Delta}

\newcommand{\renObj}{\bar{g}}
\newcommand{\renObjNoG}{\hat{g}}

\newcommand{\gen}{\mathtt{gen}}
\newcommand{\test}{\mathtt{test}}
\newcommand{\Ct}{\widehat{\mathcal{C}}_{\setminus \mathrm{gen}}}

\theoremstyle{definition}
\newtheorem{definition}{Definition}

\theoremstyle{plain}
\newtheorem{theorem}[definition]{Theorem}

\theoremstyle{plain}
\newtheorem{lemma}[definition]{Lemma}

\theoremstyle{plain}
\newtheorem{cor}[definition]{Corollary}

\theoremstyle{plain}
\newtheorem{prop}[definition]{Proposition}

\theoremstyle{definition}
\newtheorem{prot}{Protocol}

\theoremstyle{remark}
\newtheorem{rem}{Remark}

\newcommand{\Renyi}{R\'{e}nyi }

\newcommand{\EATchann}{\mathcal{M}}
\newcommand{\CP}{\widehat{C}} 
\newcommand{\alphCP}{\widehat{\mathcal{C}}}
\newcommand{\cP}{\hat{c}} 
\newcommand{\Fobs}{\mbf{F}^\mathrm{obs}} %Observed frequency vector
\newcommand{\Fobsc}{F^\mathrm{obs}} %Observed frequency vector component

\begin{document}

%\title{Flexible finite-size security proofs against coherent attacks via Rényi entropies}
%\title{Finite-Size QKD Security Proofs via Renyi Entropies}
%\title{Source Imperfection resistant Variable-length security proofs against Coherent Attacks}
%\title{Security Framework for Coherent Attacks and Variable-length Key Rates via Rényi entropies}
%\title{Rényi QKD security framework for decoy-state protocols secure against coherent attacks}
\title{Rényi security framework against coherent attacks applied to decoy-state QKD}
%\title{Rényi security framework against coherent attacks for practical QKD with imperfect devices}

\author{Lars Kamin}
\email{lars.kamin@outlook.com}
\affiliation{Institute for Quantum Computing and Department of Physics and Astronomy, University of Waterloo, Waterloo, Ontario N2L 3G1, Canada}

\author{John Burniston}
\email{jburniston@uwaterloo.ca}
\affiliation{Institute for Quantum Computing and Department of Physics and Astronomy, University of Waterloo, Waterloo, Ontario N2L 3G1, Canada}

\author{Ernest Y.-Z.\ Tan}
\email{yzetan@uwaterloo.ca}
\affiliation{Institute for Quantum Computing and Department of Physics and Astronomy, University of Waterloo, Waterloo, Ontario N2L 3G1, Canada}

\date{\today}

\begin{abstract}
We develop a flexible and robust framework for finite-size security proofs of quantum key distribution (QKD) protocols under coherent attacks, applicable to both fixed- and variable-length protocols. Our approach achieves high finite-size key rates across a broad class of protocols while imposing minimal requirements. In particular, it eliminates the need for restrictive conditions such as limited repetition rates or the implementation of virtual tomography procedures. To achieve this goal, we introduce new numerical techniques for the evaluation of conditional sandwiched \Renyi entropies. In doing so, we also find an alternative formulation of the ``QKD cone'' studied in previous work. We illustrate the versatility of our framework by applying it to several practically relevant protocols, including decoy-state protocols. Furthermore, we extend the analysis to accommodate realistic device imperfections, such as independent intensity and phase imperfections. Overall, our framework provides both greater scope of applicability and better key rates than existing techniques, especially for small block sizes, 
hence offering a scalable path toward secure quantum communication under realistic conditions.
\end{abstract}

\maketitle

\section{Introduction}

Quantum key distribution (QKD) allows for the establishment of a shared secret key between two parties, Alice and Bob, through the use of an insecure quantum channel that can be accessed by an eavesdropper Eve. In order to establish the security of the key produced in a QKD protocol, it is important that the security analysis takes into account \emph{all} possible attacks Eve could perform in the channel. This is a rather challenging task, as it must address the most general forms of attacks that Eve could perform, often referred to as \emph{coherent attacks}. Some earlier works in QKD focused only on specific classes of attacks, for instance assuming that Eve attacks the transmitted states in an independent and identically distributed (IID) manner across the rounds, usually referred to as \emph{IID collective attacks}; however, such attacks would in general not capture the full scope of actions available to Eve. To safely deploy QKD, one should prove security against the entire class of coherent attacks, rather than restricting to IID collective attacks. Furthermore, it is important that the security proof accounts for the fact that any physical implementation of QKD can only run for a finite number of rounds, which introduces various ``finite-size effects'' as compared to the asymptotic limit. 

To achieve this goal, various proof methods have been developed, such as phase error correction \cite{koashi_simple_2005,koashi_efficient_2006,koashi_simple_2009,hayashi_concise_2012,hayashi_security_2014}, entropic uncertainty relations (EURs) \cite{tomamichel_uncertainty_2011} and their applications to QKD \cite{tomamichel_tight_2012,lim_concise_2014,rusca_finite-key_2018,tupkary_phase_2024,wiesemann_consolidated_2024}, the postselection technique \cite{christandl_postselection_2009,nahar_postselection_2024}, and the entropy accumulation theorem (EAT) and subsequent variants~\cite{dupuis_entropy_2020,dupuis_entropy_2019,metger_generalised_2022-1,metger_generalised_2024} with its applications to  QKD \cite{george_finite-key_2022,metger_security_2023,bauml_security_2024,kamin_finite-size_2025}. However, thus far there has often been a tradeoff between the finite-size key rates and the flexibility of the proof techniques. 

For instance, while the phase error correction and EUR techniques typically have good finite-size performance, they require that the analysis of the protocol must be reducible to a scenario that is close to the original qubit BB84 protocol \cite{bennett_quantum_2014} for QKD --- this often limits the scope of protocols where these techniques can be applied. On the other hand, while the postselection technique flexibly applies to a wide range of protocols, it usually has poor finite-size performance. As for entropy accumulation, the versions developed in~\cite{dupuis_entropy_2020,dupuis_entropy_2019,metger_generalised_2022-1,metger_generalised_2024} offer better finite-size key rates than the postselection technique~\cite{george_finite-key_2022,kamin_finite-size_2025}, while maintaining some flexibility. However, the rates are often still worse than those obtained from phase error correction or EURs. 
Furthermore, those versions of entropy accumulation imposed some technical restrictions on the protocol, which we briefly discuss later. 
Recent improvements to the entropy accumulation framework~\cite{inprep_HB24,arqand_generalized_2024,fawzi_additivity_2025,arqand_marginal-constrained_2025} indicate that it is possible to overcome the aforementioned drawbacks, but these versions have thus far only been applied to specialized, simple protocols. We use these versions as the technical foundation of our work, as we describe in \cref{sec:intro_foundation}.

\subsection{Contributions}

In this work, we establish a framework for QKD security proofs that performs well on both fronts, being flexibly applicable to a wide variety of protocols while still achieving high finite-size key rates. A core feature of this framework is that it is based entirely on (conditional) \emph{\Renyi entropies}, which are a generalization of the standard von Neumann entropy~\cite{nielsen_quantum_2010} considered in quantum information. We apply our framework to examples of various QKD protocols, including decoy-state protocols~\cite{hwang_quantum_2003,lo_decoy_2004,ma_practical_2005,wang_beating_2005}, which are highly relevant in practical QKD implementations. With these examples, we demonstrate notably better finite-size key rates as compared to all previous techniques, while still benefiting from the flexibility of this framework. 

We highlight that this framework is compatible with both \emph{fixed-length protocols}, which always output a key of some specific fixed length whenever they accept, and \emph{variable-length protocols}, where instead the key length may vary depending on the observations in the protocol. Security proofs for variable-length protocols can be fairly challenging, and various approaches have been developed for this purpose~\cite{hayashi_concise_2012,tupkary_security_2024,nahar_postselection_2024,inprep_HB24}. Our framework is most compatible with the proof developed in~\cite{inprep_HB24}, and hence we directly apply that approach in this work. As variable-length protocols are typically more convenient than fixed-length protocols in practical implementations, the explicit key rates we present in this work are focused on the former.

Additionally, this framework can easily accommodate various forms of device imperfections. 
In particular, we consider two classes of imperfections that have previously only been studied in the asymptotic regime against IID collective attacks or for BB84-like protocols. For these imperfections, our framework allows us to compute finite-size key rates against coherent attacks for \emph{generic protocols}, which is a task that appears challenging to handle using other proof techniques --- we explain this further in \cref{sec:Intensity imperfections,sec:Phase imperfections}. Hence, our work addresses an important area in security proofs for QKD with device imperfections.

At the technical level, our main contribution to achieve the above results is to build up a broad-ranging framework for bounding \Renyi entropies (or recently introduced~\cite{inprep_HB24} ``weighted'' versions of them, described in Refs.~\cite{arqand_generalized_2024,arqand_marginal-constrained_2025,fawzi_additivity_2025}) in a QKD protocol, accompanied by a detailed code implementation of algorithms to compute these bounds. We do so by developing a suitable reformulation of the task of bounding the \Renyi conditional entropy (which is generally not a convex function of the state) into a convex optimization problem. In particular, we construct a reformulation such that we can practically implement algorithms to \emph{reliably} lower-bound the optimal values, in the sense that the results we obtain from this framework will never be an over-estimate of the true secure key rate. As for the tightness of our results, this is demonstrated by the aforementioned improvements we obtain over all previous techniques. We incorporate these numerical methods into the \verb|openQKDSecurity| software suite of Ref.~\cite{burniston_open_2024}.
 
Some of our results regarding this reformulation were independently obtained in a separate work~\cite{chung_generalized_2025}, though that work was focused on the case of IID collective attacks, whereas we prove security against all coherent attacks. Also, their work analyzes a slightly different choice of \Renyi conditional entropy from ours, which we elaborate on in \cref{sec:Calculating halpha}.

Note that as a special case of our results, one can recover the entire framework developed in an earlier work~\cite{winick_reliable_2018} for analyzing von Neumann entropy, since von Neumann entropy is a special case of \Renyi entropies. In this work however, we address a variety of challenges regarding the general \Renyi entropies that did not arise in the special case of von Neumann entropy. For instance, the proof technique in~\cite{winick_reliable_2018} does not translate to \Renyi entropy in general, and thus we derived a new approach in order to prove a suitable analogous result. As a corollary of this approach, we obtain a novel formulation of the von Neumann entropy of states arising in QKD, which may be of independent interest for recent work~\cite{hu_robust_2022,lorente_quantum_2025,he_exploiting_2024} studying convex optimizations over a corresponding cone (sometimes referred to in those works as the \emph{QKD cone}). Similarly, for analyzing decoy-state protocols and device imperfections, we developed suitable techniques to address \Renyi entropies, accommodating the differences in properties as compared to von Neumann entropy. 

\subsection{Technical foundation}
\label{sec:intro_foundation}

We highlight two recent theoretical developments that were critical for us to establish this framework. The first is a \Renyi version of the \emph{leftover-hashing lemma}~\cite{dupuis_privacy_2023}, which is an important tool in studying the \emph{privacy amplification} step in QKD protocols. While previous versions of the leftover-hashing lemma were based on \emph{smooth min-entropy}, this version based on \Renyi entropies often provides better finite-size performance. Furthermore, with this version it is more straightforward to apply recent techniques developed to analyze \Renyi entropies, which we now describe.

The second critical development is a series of techniques to bound the overall \Renyi entropy of the raw data string generated in a QKD protocol (while accounting for finite-size effects and coherent attacks), by only analyzing \Renyi entropies in single rounds of the protocol. The original version of this technique was first proposed in Ref.~\cite{inprep_HB24}, though for the purposes of this work we rely on more recent work in Refs.~\cite{arqand_marginal-constrained_2025,fawzi_additivity_2025}, which provided self-contained proofs of generalized versions of the result in Ref.~\cite{inprep_HB24}. We shall broadly refer to this entire family of results as \emph{marginal-constrained entropy accumulation theorems} (MEAT), as the key difference between them and other entropy accumulation versions is that they allow imposing constraints on some particular marginal states (also often known as reduced states~\cite{nielsen_quantum_2010}). 
This makes them particularly suited for analyzing prepare-and-measure (P\&M) protocols, where most security proofs use a technique known as the source-replacement scheme \cite{bennett_quantum_1992, ferenczi_symmetries_2012} that involves such marginal constraints.
{In particular, we emphasize that security proofs for P\&M protocols using the MEAT are \emph{not} subject to restrictions that applied to previous entropy accumulation versions, such as requiring virtual tomography procedures~\cite{bauml_security_2024} or repetition-rate limitations~\cite{metger_security_2023}.}

However, thus far the aforementioned MEAT results have only been applied to a small selection of simple protocols. In particular, they have not been applied in the context of decoy-state protocols and device imperfections. As mentioned above, in this work we significantly extend the results in those works by addressing multiple novel aspects of analyzing such scenarios using \Renyi entropies, since some of the existing prior literature based on von Neumann entropy does not apply for \Renyi entropies.

\subsection{Structure}

The structure of this manuscript is as follows. In \cref{sec:Protocol description,sec:Notation} we describe a general version of a P\&M protocol covered by our framework and present important notation used throughout this work.

Next, in \cref{sec:Generic framework}, we lay the foundations of our framework. We formalize the QKD protocol described in \cref{Prot:PM Protocol} in terms of what we will define as the QKD channel \(\EATchannQKD\), which satisfies the assumptions of the MEAT frameworks~\cite{inprep_HB24,arqand_marginal-constrained_2025,fawzi_additivity_2025}. Using this setup, we then prove security against coherent attacks in \cref{thrm:Fixed-length security generic bnd}. This theorem requires a lower bound on the \(n\)-round conditional \Renyi entropy, which in turn requires minimizing a single round conditional \Renyi entropy.

Therefore, connecting to this in \cref{sec:Calculating halpha}, we then focus on the framework of minimizing \Renyi entropies in the QKD setting. Next, in \cref{sec:Variable-Length protocols} we prove security of variable-length QKD protocols, and also reformulate the associated minimization problem as a convex optimization tractable in our framework. Then we conclude this initial part in \cref{sec:Qubit BB84 and Comparisons with other Proof Techniques} with a comparison between our results and other proof techniques for a qubit BB84 protocol.

Afterwards, we focus on more sophisticated protocol classes. First, in \cref{sec:Block-Diag states}, we extend our methods to protocols that obey some block-diagonal structure, such as, but not limited to, decoy-state protocols~\cite{hwang_quantum_2003,lo_decoy_2004,ma_practical_2005,wang_beating_2005}. In \cref{sec:Decoy state} we then present the example of a decoy-state BB84 protocol with an active detection setup.

Next, we turn our attention to device imperfections, first discussing ``simple" device imperfections in \cref{sec:Simple Imperfections}. Then, in \cref{sec:Intensity imperfections} we analyze generic decoy-state protocols with intensity imperfections, i.e.~the sources do not send out the intensity specified in the protocol. We present variable-length key rates for a decoy-state BB84 protocol with a \emph{passive} detection setup and compare the results with the perfect case.

We conclude our work on imperfections with \cref{sec:Phase imperfections}, where we focus on decoy-state protocols with phase imperfections, i.e. the weak coherent pulses are not fully phase-randomized. Here we present a decoy-state version of the reference-frame-independent (RFI) 4-6 protocol \cite{laing_reference-frame-independent_2010} and compare our variable-length key rates for the perfect case with recent results using the postselection technique \cite{kamin_improved_2025}.

Finally, in \cref{sec:Conclusion} we draw concluding remarks on the performance of our techniques and possible future improvements.

\section{Protocol description}\label{sec:Protocol description}
We start by stating a description of a generic P\&M protocol. We note that this framework can easily be adapted to entanglement-based protocols, but for this work we focus on P\&M protocols.

\begin{prot}{Prepare-and-Measure Protocol.}\label{Prot:PM Protocol} \\
		\textbf{Parameters:} \\
	\begin{tabularx}{0.9\linewidth}{r c X}
			\(n \in \mathbb{N}_0\) 			    &:&     Total number of rounds \\
			\(l \in \mathbb{N}_0\)				&:&		Length of final key\\
			\(\{\sigma_i\}_{i=1 \dots d_A}\) 	&:& 	States sent by Alice \\
            \(d_A\)                             &:&     Number of different signal states \\
            \(\mathcal{S}\)				        &:&		alphabet of raw-key registers, usually \(\{0,1,\perp\}\) \\
			\(\{M^{B}_{i}\}_{i=1\dots d_B}\) 	&:& 	POVM elements acting on Bob's system describing his measurements outcomes\\
            \(d_B\)                             &:&     Number of Bob's POVM elements \\
            \(\gamma\)						    &:& 	Probability that Alice chooses a round to be a test round \\
		      \(\Sacc\)				  &:&     Convex acceptance set of accepted frequencies \\
            \(\lambda_{\mathrm{EC}}\)			&:& 	Number of bits exchanged during error correction step \\
			\(\ePA\)			                &:&		Security parameter contribution from privacy amplification \\
			\(\eEV\)			                &:&		Security parameter contribution from error verification \\
			\(\Omega_{\text{AT}}\)				&:&		Event of passing the acceptance test \\
			\(\Omega_{\text{EV}}\)				&:&		Event of passing the error verification \\
			\(\Omega_{\text{acc}} = \Omega_{\text{AT}} \land \Omega_{\text{EV}}\)				                                     &:&		Event of the protocol not aborting \\
	\end{tabularx}
	\vspace{20pt}
	
	\textbf{Protocol steps:}
    \begin{enumerate}
    \item For each round $i \in \{1,2,\dots,n\}$, Alice and Bob perform the following steps:\footnote{Note that in contrast to the analysis in e.g.~\cite{metger_security_2023}, here we do \emph{not} limit the repetition rate of the protocol such that Bob performs the $i^\text{th}$ measurement before Alice sends the $(i+1)^\text{th}$ state --- in our protocol, the states can be sent and measured in any time-ordering with respect to each other, apart from the trivial physical constraint that Bob can only perform the $i^\text{th}$ measurement after Alice sends the $i^\text{th}$ state. In fact, Alice can even send out all her signal states before Bob performs any measurements.}
	\begin{enumerate}
		\item \textbf{State preparation and transmission:} In each round, with probability $\gamma$ Alice independently chooses it to be a \emph{test round} or \emph{generation round}. Next, Alice prepares one of \(d_A\) possible signal states \(\{\sigma_k\}_{k=1\dots d_A}\), according to some distribution (which could depend on the choice of a test or generation round). She stores the label for her choice of the signal state in a classical register \(X_i\) with alphabet \(\mathcal{X}\), and computes a classical register \(C^A_i\) with alphabet \(\mathcal{C}^A\) for public announcement (including for instance the test/generation decision). Finally, Alice sends the signal state to Bob via a quantum channel.
  
		\item \textbf{Measurements:} Bob measures his received states described by a POVM with POVM elements \(\{M_k^B\}_{k=1 \dots d_B}\), and stores his results in a classical register \(Y_i\) with alphabet \(\mathcal{Y}\). Furthermore, he computes a classical register \(C^B_i\) with alphabet \(\mathcal{C}^B\) for public announcement.
	\end{enumerate}

        \item \label{step:announce} \textbf{Public announcement:} Alice and Bob announce their values $C^A_1 \dots C^A_n, C^B_1 \dots C^B_n$. For each $i$, Alice and Bob can perform some further public classical communication (which can be two-way) based only on the values $X_i Y_i C^A_i C^B_i$ and local randomness; let $I_i$ be a register denoting all such public communication with alphabet \(\mathcal{I}\).
        Then Alice and Bob both compute a value $\CP_i$ that is found by applying some deterministic function \(\phi\) on $I_i$ \footnote{The map \mbox{$\phi$} maps all announcements in a generation round to a special generation flag \mbox{$\ket{\gen}_{\CP}$}.}. These values $\CP_1^n$ will later be used in the acceptance test (to decide whether to abort) or the variable length decision (to decide the final key length). We require that it is set to a fixed symbol $\CP_i = \gen$ whenever Alice chose the round to be a generation round (informally, this corresponds to the fact that the acceptance test or variable length decision later will only depend on data from the test rounds). We also suppose that all the classical registers $\CP_i$ are isomorphic to each other, with a common alphabet $\alphCP$.

        For notational convenience in our later analysis, let $T_1^n$ denote registers that are set to $T_i = \test$ in test rounds and otherwise set to $T_i = I_i$. Let $\tilde{T}_1^n$ denote a copy of $T_1^n$ that Eve holds (which she can indeed compute since she has access to the public communication $I_i$).

		\item \label{step:sift} \textbf{Sifting and key map:} For each $i$,
        Alice applies a sifting step\footnote{For the purposes of this work, when a round is sifted out, we take this to mean Alice sets $S_i$ to a fixed value in the alphabet of $S_i$ (say, $0$). However, it should be possible to modify this to a version where such rounds are actually discarded (i.e.~not included in the privacy amplification step) by using the techniques in~\cite{tupkary_security_2024}.} and/or key map procedure based on her raw data $X_i$ and the public announcements $I_i$ (and local randomness, if needed), to produce a classical register $S_i$.
		
		\item \label{step:AT} \textbf{Acceptance test (parameter estimation) or variable length decision:} In a fixed length protocol, Alice and Bob compute the frequency distribution $\Fobs$ (see \cref{eq:freqdefn} below) of the observed string of values $\cP^n_1$ on the registers \(\CP_1^n\). Then they accept if \(\Fobs \in \Sacc \) where \(\Sacc\) is the predetermined acceptance set, and abort the protocol otherwise. We call the event of passing this stage \(\OmegaAT\).
        
        Alternatively, in a variable-length protocol, they compute the variable key length \(\lvar\)
        by computing a particular function of $\cP_1^n$
        (see \cref{thrm:Variable-length security} for further details).
        
		\item \label{step:ECandEV} \textbf{Error correction and verification:} In a fixed-length protocol, Alice and Bob publicly communicate $\lambda_{\mathrm{EC}}$ bits for error correction, whereas in a variable-length protocol they communicate $\lambda_{\mathrm{EC}}(\cP_1^n)$ bits. Next, Bob uses those bits together with his data $Y_1^n I_1^n$ to produce a guess $\widebar{S}_1^n$ for Alice's string $S_1^n$. This is followed by an error-verification step, where Alice sends a 2-universal hash of $S_1^n$ with length $\ceil{\log(1/\eEV)}$ to Bob, who compares it to the hash of his guess and accepts if the hashes match (and otherwise aborts). Let register \(L\) contain the full data communicated in this step and we call the event of passing the error correction procedure \(\OmegaEV\).
        
		\item \textbf{Privacy amplification:} In a fixed length protocol, Alice randomly chooses a 2-universal hash function from some family (with fixed output length $l$) announces it and applies it to her string $S_1^n$ producing her final key $K_A$ of \emph{fixed} length \(l\). Bob then also applies the hash function to his guess $\widebar{S}_1^n$, producing his final key $K_B$ of \emph{fixed} length \(l\).
        
        Alternatively, in a variable-length protocol, Alice and Bob still randomly choose a 2-universal hash function from some family. However, this hash function now maps \(n\) bits to \(\lvar\) bits. Then, they apply the hash function to their strings $S_1^n$ and $\widebar{S}_1^n$, producing the final keys $K_A$ and $K_B$ of length \(\lvar\).
	\end{enumerate}
\end{prot}

In the above description, we used the concept of the \emph{frequency vector} $\Fobs$ corresponding to the observed string of values $\cP_1^n$ on the registers $\CP_1^n$.
Formally, this is defined as the vector in $\mathbb{R}^{\abs{\alphCP}}$ with components given by
\begin{equation}\label{eq:freqdefn}
\Fobsc(\cP) \defvar \frac{\text{number of occurrences of $\cP$ in $\cP_1^n$}}{n}.
\end{equation}

\section{Notation and definitions}\label{sec:Notation}

\begin{table}[h!]
\def\arraystretch{1.5} %Spacing above/below text in cells
\setlength\tabcolsep{.28cm}
\begin{tabular}{c p{4.5cm}}
\toprule
\textit{Symbol} & \textit{Definition} \\
\toprule
$\log$ & Base-$2$ logarithm \\
\hline
$H$ & Base-$2$ von Neumann entropy \\
\hline
$\renyiPetzUp_{\alpha}, \renyiPetzDown_{\alpha}, \renyiSandUp_{\alpha}, \renyiSandDown_{\alpha}$ & Base-$2$ \Renyi entropies \\
\hline
$\floor{\cdot}$ (resp.~$\ceil{\cdot}$) & Floor (resp.~ceiling) function \\
\hline
$\norm{\cdot}_p$ & Schatten $p$-norm \\
\hline
$\supp$ & Support of an operator or a probability distribution \\
\hline
$\Pos(A)$ & Set of positive 
% semi-definite
semidefinite operators on register $A$\\
\hline
$\dop{=}(A)$ (resp.~$\dop{\leq}(A)$) & Set of normalized (resp.~subnormalized) states on register $A$ \\
\hline
$A_1^n$ & Abbreviated notation for registers $A_1 \dots A_n$ \\
\toprule
\end{tabular}
\def\arraystretch{1}
\caption{List of notation}\label{tab:notation}
\end{table}

We may denote the distribution on a classical register $C$ induced by a state $\rho$ as the tuple
\begin{align}\label{eq:stateprobvec}
\bsym{\rho}_C \defvar \left(\rho(1),\rho(2),\dots\right).
\end{align}

Furthermore, we will make use of the following definitions of \Renyi entropies as in \cite{tomamichel_quantum_2016}.

\begin{definition}\label{def:Renyi Divergence}
    Let $\alpha \in (0,1) \cup (1,\infty)$, and $\rho,\sigma \in \dop{\leq}(A)$ with \(\Tr[\rho] \neq 0\), the \emph{minimal quantum \Renyi divergence} (or sandwiched quantum \Renyi divergence) is defined as
    \begin{equation}
        \renyiSandDiv_\alpha \rel{\rho}{\sigma} \defvar \frac{1}{\alpha-1} \log \frac{\Tr\left[ \left( \sigma^\frac{1-\alpha}{2\alpha} \rho \sigma^\frac{1-\alpha}{2\alpha} \right)^{\alpha} \right]}{\Tr[\rho]},
    \end{equation}
    for \((\alpha < 1 \wedge \rho \not\perp \sigma ) \vee \supp(\rho) \subseteq \supp{\sigma}\), and \(\infty\) otherwise. 
    
    The \emph{Petz quantum \Renyi divergence} is defined as
    \begin{equation}
        \renyiPetzDiv_\alpha\rel{\rho}{\sigma}\defvar \frac{1}{\alpha-1} \log \frac{\Tr[\rho^\alpha \sigma^{1-\alpha}]}{\Tr[\rho]},
    \end{equation}
    for \((\alpha < 1 \wedge \rho \not\perp \sigma ) \vee \supp(\rho) \subseteq \supp{\sigma}\),
    and \(\infty\) otherwise. In any statement that applies to both divergences, we will write $\renyiDiv_\alpha$ to denote either divergence.
\end{definition}

\begin{definition}\label{def:Conditional Renyi entropy}
For $\alpha \geq 0$ and $\rho_{AB} \in \dop{=}(AB)$ the \emph{quantum conditional \Renyi entropies} are defined as:
\begin{align}
    \renyiPetzDown_\alpha \con{A}{B}_\rho &\defvar -\renyiPetzDiv_\alpha \rel{\rho_{AB}}{I_A \otimes \rho_B} \\
    \renyiPetzUp_\alpha \con{A}{B}_\rho &\defvar \sup_{\sigma_B\in\dop{=}(B)} -\renyiPetzDiv_\alpha \rel{\rho_{AB}}{I_A \otimes \sigma_B} \\
    \renyiSandDown_\alpha \con{A}{B}_\rho &\defvar -\renyiSandDiv_\alpha \rel{\rho_{AB}}{I_A \otimes \rho_B} \\
    \renyiSandUp_\alpha \con{A}{B}_\rho &\defvar \sup_{\sigma_B\in\dop{=}(B)} -\renyiSandDiv_\alpha \rel{\rho_{AB}}{I_A \otimes \sigma_B}
\end{align}
Again, in any statement that applies to all conditional \Renyi entropies, we use $\renyiEnt_\alpha$.
\end{definition}

Furthermore, we will make use of the following definitions of variable-length security as in \cite[Sec. VI.A]{portmann_security_2022}.

\begin{definition}[Variable-Length $\varepsilon$-security]\label{Def:Variable length eps security}
	Let \(\OmegaLen{m}\) be the event of a QKD protocol generating a final key of length \(m \in \N_0\). A variable-length QKD protocol is \emph{\(\eSec\)-secure} if for any input state \(\sigma_{A^n B^n}\) the resulting output state \(\sigma_{K_AK_BE}\) satisfies
	\begin{equation} \label{eq:security}
        \begin{split}
             \frac{1}{2} \sum_{m=0}^{\infty} \Pr[\OmegaLen{m}] \Bigg\lVert &\sigma_{K_AK_BE| \OmegaLen{m}} \\ & - \mathbb{K}_{K_AK_B}^m \otimes \sigma_{E| \OmegaLen{m}} \Bigg\rVert_1 \leq \eSec,
         \end{split}
	\end{equation}
    where \(\mathbb{K}_{K_AK_B}^m\) denotes the state of a perfect uniform shared key of length $m$:
    \begin{equation}
        \mathbb{K}_{K_AK_B}^m = \sum_{k \in \{0,1\}^m } \frac{\dyad{kk}_{K_AK_B}}{2^m}.
    \end{equation}
    Furthermore, a variable-length QKD protocol is \emph{\(\eSct\)-secret} if
	\begin{equation} \label{eq:secrecy}
        \begin{split}
             \frac{1}{2} \sum_{m=0}^{\infty} \Pr[\OmegaLen{m}] \Bigg\lVert &\sigma_{K_AE| \OmegaLen{m}} \\ & - \mathbb{U}_{K_A}^m \otimes \sigma_{E| \OmegaLen{m}} \Bigg\rVert_1 \leq \eSct,
         \end{split}
	\end{equation}
    where \(\mathbb{U}_{K_A}^m\) is a fully mixed state of dimension \(2^m\), and \emph{\(\eCor\)-correct} if
	\begin{equation}
		\Pr[K_A \neq K_B \wedge \Omega_{\text{acc}}] \leq \eCor.
	\end{equation}
    A variable-length QKD protocol that is \(\eCor\)-correct and \(\eSct\)-secret, is \(\eSec = \eSct + \eCor \) secure \cite{tupkary_security_2024}. 
\end{definition}

\begin{rem}
    For P\&M protocols one can restrict the input states to those which satisfy the marginal constraint \(\Tr_{B^n}[\sigma_{A^nB^n}] = \tau_A^{\otimes n}\) for some state \(\tau_A\) defined by Alice's choices. We can do so because we assume Eve has no access to Alice's lab and we will always implicitly assume that we have such a marginal constraint.
\end{rem}

\begin{rem}\label{rem:Fixed-Length security}
    Fixed-length security follows from variable-length security, by selecting only one length in the sum in \cref{eq:security} and identifying this with the fixed key length \(l\) of \cref{Prot:PM Protocol}; all other events are combined to one single abort event. Furthermore, secrecy and correctness imply security in the same way.
\end{rem}

\section{Generic Framework for Key Rates from the MEAT}\label{sec:Generic framework}
Traditionally, one uses the leftover-hashing lemma (LHL) \cite{renner_security_2006} for smooth min-entropies to prove secrecy of a QKD protocol. In this process, lower bounds on the \(n\)-round smooth min-entropy of an output state of the QKD protocol are derived. Several results proving security based on this approach exist, e.g. \cite{tomamichel_largely_2017,lim_concise_2014,george_finite-key_2022}. 

Recently, in Ref.~\cite{dupuis_privacy_2023} an alternative was presented, showing a similar leftover-hashing lemma based on \Renyi entropies, which we restate below.

\subsection{\texorpdfstring{\Renyi}{Renyi }Leftover-hashing Lemma \& Motivation}
\begin{theorem}[\Renyi Leftover-hashing Lemma (Theorem 9 restated from \cite{dupuis_privacy_2023})]
\label{thrm:Renyi LHL}
Let $\rho_{AE}$ be a classical-quantum state, and $\mathcal{F}_{\mathcal{A}\rightarrow\mathcal{Z}}$ be a set of two-universal hash functions with $\mathcal{Z}=\{0,1\}^l$, \(l\in \N_0\). Then if $g:A\rightarrow Z$ is a function drawn uniformly at random from $\mathcal{F}_{\mathcal{A}\rightarrow\mathcal{Z}}$, for $\alpha\in(1,2)$ we have
\begin{align}\label{eq:privacy_amp}
    \begin{split}
    &\frac{1}{2}\mathbb{E}_g\left\|\rho_{ZE|g}-\frac{1}{|\mathcal{Z}|}\idop_Z\otimes\rho_{E} \right\|_1 \\ 
    &= \frac{1}{2} \left\|\rho_{ZEG}-\frac{1}{|\mathcal{Z}|}\idop_Z\otimes\rho_{EG} \right\|_1 \\ 
    &\leq 2^{\frac{2(1-\alpha)}{\alpha}}2^{\frac{1-\alpha}{\alpha}\big(\renyiSandUp_{\alpha}(A|E)_\rho-l \big)},
    \end{split}
\end{align}
where $G$ is the register that stores the choice of hash function.
\end{theorem}

In contrast to the original smooth min-entropy version of the LHL, the \Renyi version requires bounds on the \(n\)-round \Renyi entropy and thus, enables one to prove security of QKD protocols fully based on \Renyi entropies. Before fully formalizing a QKD protocol or stating a security proof, let us build some intuition on how to use the \Renyi LHL.

Let us assume we have access to some lower bound \(n\underline{h}\) on the \(n\)-round \Renyi entropy of the secret data \(S_1^n\) conditioned on Eve's quantum side information including all publicly available data. Then, with the \Renyi LHL, one can show secrecy of a fixed-length protocol with key length
\begin{equation}
    l \leq n \underline{h} - \text{corrections}.
\end{equation}
Here we informally filled the error correction cost and hash length during error verification, etc. under the ``corrections'' term. 

The important point to note here is that with such a lower bound on the \(n\)-round \Renyi entropy, one does not require going through additional steps like for some smooth min-entropy based security proofs, which introduce a lot of looseness. 

A framework for bounding \Renyi entropies especially useful for security proofs of QKD protocols has been presented in Refs.~\cite{arqand_marginal-constrained_2025,fawzi_additivity_2025} (originally proposed in Ref.~\cite{inprep_HB24}). Their framework requires some special channel structure (satisfied by QKD protocols) and in \cref{sec:Formalizing QKD} we start by formalizing a QKD protocol under this framework. Afterwards, we will show a theorem leading to a lower bound like \(\underline{h}\).

\subsection{Formalizing QKD protocols}\label{sec:Formalizing QKD}
In this section we formalize a QKD protocol such that the methods of Refs.~\cite{inprep_HB24,arqand_marginal-constrained_2025,fawzi_additivity_2025} can be used. Intuitively, in these works the action of each round of a QKD protocol is specified by a channel \(\EATchann\) which creates the secret data and all public announcements.

\begin{figure}
    \centering
    \includegraphics[width=\linewidth]{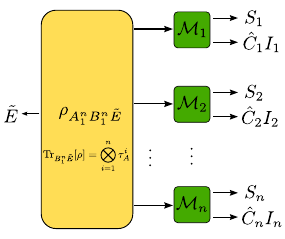}
    \caption{Representation of the tensor product structure of the MEAT theorem \cite[Corollary 4.2]{arqand_marginal-constrained_2025}.}
    \label{fig:MEAT Channels}
\end{figure}

Furthermore, the models of Refs.~\cite{inprep_HB24,arqand_marginal-constrained_2025,fawzi_additivity_2025} require that the channel \(\EATchann\) acts as a tensor-product of channels, i.e. \(\EATchann^{\otimes n}\), or as concatenation of channels \(\EATchann_n \circ \dots \circ \EATchann_1\) to create the final state of the \emph{full} $n$-round protocol. In \cref{fig:MEAT Channels} one can see a schematic representation of the tensor product structure which we will use for security proofs in this work.

We chose the tensor product model for simplicity, and note that by the same argument as in Ref.~\cite{metger_generalised_2022} the same key rates are valid under the model of concatenated channels. Therefore, in this section, using a similar approach as in Ref.~\cite{winick_reliable_2018} and Ref.~\cite[App A]{lin_asymptotic_2019} in the asymptotic regime, we define a channel \(\EATchannQKD\) suitable for describing QKD protocols in this tensor-product manner.

\subsubsection{Source-Replacement Scheme}
As a first step, let us briefly introduce the source-replacement scheme \cite{bennett_quantum_1992, ferenczi_symmetries_2012}, which we will use to analyze P\&M protocols. Under the source-replacement scheme, Alice's state preparation process in each round is recast as having her first prepare a \emph{pure} state, 
\begin{equation}
    \ket{\psi}_{AA'} = \sum_{k=1}^{d_A}  \sqrt{p(k)} \ket{k}_A \ket{s_k}_{A'},
\end{equation}
where \(p(k)\) is the total probability of Alice sending state \(\ket{s_k}_{A'}\) to Bob, and then performing a measurement on $A$ described by the following POVM elements:
\begin{equation}
    M_k^A = \ketbra{k}{k},
\end{equation}
for all \(k=1 \dots d_A\).

Importantly, this measurement on the $A$ registers commutes with any arbitrary coherent attack Eve performs on the signal registers $A'$, since these operations act on different systems. Therefore, the source-replacement scheme states that the state generated in the protocol can be equivalently described as follows. Alice first prepares the state $\ket{\psi}_{AA'}^{\otimes n}$, then Eve performs an arbitrary attack across all the signal-state registers $A_1^{\prime\, n}$ that maps them to some registers $B_1^n \widetilde{E}$, resulting in a new state \(\rho_{A_1^nB_1^n \widetilde{E}}\). 
Eve then forwards the $B_1^n$ registers to Bob, and Alice and Bob measure the registers $A_1^nB_1^n$ to generate the raw data strings $X_1^nY_1^n$, which are then processed further as described in the protocol. Note that for protocols of the form we described, the process of generating the registers $S_1^n I_1^n \CP_1^n$ from the registers $A_1^nB_1^n$ can be written as a channel of the form $\EATchannQKD^{\otimes n}$, for a channel $\EATchannQKD$ we describe in \cref{subsubsec:QKDchannel} below.

Furthermore, let us write \(\tau_A := \Tr_{A'}[\ketbra{\psi}{\psi}]\) to denote the marginal of the single-round state Alice prepares in the source-replaced picture. Since in this picture Eve has no access to the $A_1^n$ registers, the marginal state on these registers will be unchanged under Eve's attack. This will later be reflected by requiring that the state \(\rho_{A_1^nB_1^n \widetilde{E}}\) satisfies a marginal constraint \(\rho_{A_1^n} = \tau_A^{\otimes n} \).

Additionally, if Alice sends mixed states, for example in the case of weak coherent pulse (WCP) sources, one first has to purify the states with a shield system, which we will describe later in \cref{sec:Block-Diag states,sec:Decoy state} in more detail. 

Finally, we note that, for a numerical implementation, it might be preferred to incorporate a Schmidt decomposition to remove unnecessary dimensions. We will focus on this case in \cref{app:Conditional POVM construction} in the appendix. However, for the discussion presented here, we assume that each of Alice's POVM elements is a projector as given by the generic source-replacement scheme above.

\subsubsection{Announcements and Conditioning on Test and Generation rounds}
Next, we need to take care of Alice's and Bob's announcements, such as basis announcements. Given the distinction between test and generation rounds, we split Alice's set of announcements and measurement outcomes into test and generation subsets, i.e.
\begin{align}
    \mathcal{C}^A &= \mathcal{C}^{A,\test} \cup \mathcal{C}^{A,\gen},\\
    \mathcal{X} &= \mathcal{X}^{\test} \cup \mathcal{X}^{\gen}.
\end{align}
This split allows us to partition Alice's measurement outcomes similarly to \cite[App. A]{lin_asymptotic_2019},
\begin{align}
    \mathcal{X^{\test}} = \bigcup_{\alpha \in \mathcal{C}^{A,\test}} \mathcal{X}^{\test}_{\alpha}, \quad \mathcal{X^{\gen}} = \bigcup_{\alpha \in \mathcal{C}^{A,\gen}} \mathcal{X}^{\gen}_{\alpha},
\end{align}
and equivalently for Bob
\begin{equation}
    \mathcal{Y} = \bigcup_{\beta \in \mathcal{C}^B} \mathcal{Y}_{\beta}.
\end{equation}

Since we perform different operations conditioned on test and generation rounds, let us first define the partition operators for test and generation rounds respectively as
\begin{align}\label{eq:test and gen partition ops}
    \Pi^{\test} = \sum_{\substack{\alpha \in \mathcal{C}^{A,\test}, \\ x \in \mathcal{X}^{\test}_{\alpha}}} M_{\alpha,x}^A, \quad
    \Pi^{\gen} = \sum_{\substack{\alpha \in \mathcal{C}^{A,\gen}, \\ x \in \mathcal{X}^{\gen}_{\alpha}}} M_{\alpha,x}^A.
\end{align}

Next, in the same spirit to the announcements, we aim to partition Alice's POVM elements into ``\(\test\)" and ``\(\gen\)" subsets. Again, since we perform different operations conditioned on test and generation rounds, in our subsequent description it is useful to have a notion of POVM elements conditioned on these events.

Under the basic source-replacement scheme described above, we can set these to be equal to the original POVM elements, though the test subset will then form a POVM on the subspace $\supp(\Pi^{\test})$ rather the original space, and analogously for the generation subset. However, we note that when the analysis is simplified using a Schmidt decomposition of $\ket{\psi}_{AA'}$, this might not be the case, and we discuss this situation in more detail in \cref{app:Conditional POVM construction}. To clearly indicate the required conditioning, especially for the general case, we add \(|\test\) or \(|\gen\) where appropriate, and we will denote the resulting POVMs as \(\{M^{A| \test}_k\}\) and \(\{M^{A| \gen}_k\}\), respectively.

Therefore, after partitioning Alice's POVM elements appropriately, those sets can be written as
\begin{align}
    \{M^{A|\test}_{\alpha,x} \}_{\alpha \in \mathcal{C}^{A,\test}, x \in \mathcal{X}^{\test}_{\alpha} } &= \{M^{A| \test}_k\}_k ,\\
    \{M^{A|\gen}_{\alpha,x} \}_{\alpha \in \mathcal{C}^{A,\gen}, x \in \mathcal{X}^{\gen}_{\alpha} } &= \{M^{A| \gen}_k\}_k .
\end{align}

On Bob's side we also partition his POVM elements according to the announcements such that
\begin{align}
    \{M^B_{\beta,y} \}_{\beta \in \mathcal{C}^B, y \in \mathcal{Y}_{\beta}} = \{M^B_k\}_{k=1\dots d_B}.
\end{align}

Finally, for notational convenience, let us define
\begin{equation}
    \begin{split}
    \Ctest \defvar &\{(\alpha,x,\beta,y) | 
     \alpha \in \mathcal{C}^{A,\test}, x \in \mathcal{X}^{\test}_{\alpha} , \\ 
     &\qquad \beta \in \mathcal{C}^{B}, y \in \mathcal{Y}_{\beta} \},
    \end{split}
\end{equation}
as the set containing all announcements during test rounds. In addition, we assume that the alphabet \(\mathcal{I}\) of registers \(I_i\) contains \(\Ctest\).

\subsubsection{Defining the QKD channel and Proving Security}
\label{subsubsec:QKDchannel}

After these preliminary definitions, we have all tools available to define the channel \(\EATchannQKD\) characterizing a QKD protocol acting as a tensor product as shown in \cref{fig:MEAT Channels}.

Before we formally state the definition, let us discuss the constituents of the definition of \(\EATchannQKD\) on an intuitive level. We define \(\EATchannQKD\) based on four isometries. We will discuss them in the order in which they will be applied.

First, the isometry \(V_{\test}\) applies the split into test and generation rounds because different processing takes place based on each branch.
Next, are the isometries \(V_{\GMap}\) and \(V_{\mathrm{meas}}\). Informally, \(V_{\GMap}\) creates an intermediate version of the secret register, \(S_Q\), and Alice's and Bob's announcements and private data in generation rounds. The isometry \(V_{\mathrm{meas}}\) applies the measurements in test rounds and creates the corresponding announcements in register \(I\).

Afterwards, the isometry \(V_{\phi}\) applies a post-processing that applies the deterministic function \(\phi\) to register \(I\) and stores the outcomes in register \(\CP\). Thus, it effectively creates the announcements that are used for the acceptance test or the variable-length decision.

Finally, the isometry \(V_{\ZMap}\) effectively measures the intermediate secret register \(S_Q\) and creates the final classical secret register \(S\).

\begin{definition}\label{def:QKD channel}
    Let \(\hat{\GMap}: AB \rightarrow ABXY\hat{S}_QI\) be the CPTNI map as described in \cite[App A]{lin_asymptotic_2019} of a QKD protocol defined with Alice's POVM elements conditioned on a generation round, i.e. \(\{M_{\alpha,x}^{A|\gen}\}\)\footnote{Ref.~\cite{lin_asymptotic_2019} derives the \(\hat{\GMap}\) for reverse reconciliation, but the replacements for direct reconciliation are straightforward and already partly described in Ref.~\cite{lin_asymptotic_2019}}. We extend \(\hat{\GMap}\) to a CPTP map \(\GMap\) through the following construction. Let \(\hat{\mathcal{S}}\) be the alphabet of \(\hat{S}_Q\) and \(\mathcal{S}= \hat{\mathcal{S}}\cup\{\perp\}\) be the alphabet of the extended register \(S_Q\) and define \(\GMap\) as
    \begin{align}
        &\GMap: AB \rightarrow ABXYS_QI, \\
        &\GMap(\rho) \defvar \hat{\GMap}(\rho) + \left(1 - \Tr(\hat{\GMap}(\rho))\right) \ketbra{\perp}{\perp},
    \end{align}
    for all \(\rho \in \dop{=}(AB)\), and where \(\ket{\perp}\) is orthogonal to \(\hat{\GMap}(\rho)\) for any \(\rho \in \dop{=}(AB)\). Additionally, let \(V_{\GMap}:AB \rightarrow ABXYS_QI \tilde{T}\) be the Stinespring dilation of \(\GMap\)\footnote{Due to the properties of \(\hat{\GMap}\) its Kraus operators are indexed by the announcements contained in register \(I\), which in generation rounds is equal to \(T\). Therefore, the environment in a Stinespring dilation of \(\GMap\) only contains a copy of \(T\) which is exactly Eve's copy \(\tilde{T}\)}. Furthermore, we define the following isometries
    \begin{align}
        V_{\test} &\defvar \sqrt{\Pi^{\test}} \otimes \ket{t}_F + \sqrt{\Pi^{\gen}} \otimes \ket{g}_F, \\
        V_{\ZMap} &\defvar \sum_{j \in \mathcal{S}} \ket{j}_{S} \otimes Z_j, \\
        V_{\mathrm{meas}} &\defvar \sum_{\substack{(\alpha, x, \beta, y) \in \Ctest}} \sqrt{M_{\alpha,x}^{A|\test}} \otimes \sqrt{M_{\beta,y}^{B}} \\
        &\qquad \otimes \ket{\alpha,x,\beta,y}_{I} \otimes \ket{\perp}_{S_QXY}^{\otimes3} \otimes \ket{\test}_{\tilde{T}}, \\
        V_{\phi} &\defvar \sum_{j \in \mathcal{I}} \ketbra{j}{j}_I  \otimes \ket{\phi(j)}_{\CP}.
    \end{align}
    where the orthonormal projectors \(Z_j\) satisfy \(\sum_{j \in \mathcal{S}} Z_j = \mathbb{I}_{S_Q} \). Finally, let us define the concatenation of all isometries as
    \begin{equation}
        W \defvar \begin{aligned}[t]
            V_{\ZMap} V_{\phi} \big( &V_{\mathrm{meas}}\otimes\ketbra{t}{t}_F \\
            &\quad + V_{\GMap}\otimes\ketbra{g}{g}_F \big)V_{\test}.
        \end{aligned}
    \end{equation}
    Then, we define the channel \(\EATchannQKD\) of a QKD protocol for any \(\rho_{AB}\) as
    \begin{equation}
        \EATchannQKD[\rho_{AB}] \defvar \Tr_{\mathrm{all} \setminus SI \CP}\left[W \rho_{AB} W^{\dagger}\right],
    \end{equation}
    where with \(\Tr_{\mathrm{all} \setminus SI \CP}\) we indicate tracing out all systems apart from \(SI \CP\). Finally, we also define the map creating the statistics in test rounds as
    \begin{equation}
        \probst[\sigma] \defvar \sum_{\cP \in \Ct} \sum_{\substack{(\alpha,x,\beta,y) \\ \in \phi^{-1} (\cP)}} \Tr\left[\left(M_{\alpha,x}^{A|\test} \otimes M_{\beta,y}^{B}\right) \sigma \right] \hat{e}_{\cP},
    \end{equation}
    where \(\hat{e}_{\cP}\) is a unit vector with a one at the \(\cP\)-th position and \(\phi^{-1}\) stands for the preimage.
\end{definition}

\subsection{Fixed-Length Security}\label{subsec:Fixed-Length security}
Now, we will use this construction of the channel \(\EATchannQKD\) and apply the results of \cite{arqand_marginal-constrained_2025} to find a lower bound on the \Renyi entropy of a QKD protocol.

\begin{theorem}\label{thrm:halpha QKD}
    Let \(\EATchannQKD: AB \rightarrow SI\CP\) be the channel of a QKD protocol as in \cref{def:QKD channel}. Then, for \(\omega_{S_1^nI_1^n\CP_1^n \widetilde{E}} = \EATchannQKD^{\otimes n}[\rho_{A_1^nB_1^n \widetilde{E}}]\) with \(\Tr_{B_1^n \widetilde{E}}[\rho_{A_1^nB_1^n \widetilde{E}}] = \tau_{A}^{\otimes n}\) and \(\rho_{A_1^nB_1^n \widetilde{E}}\) a purification of \(\rho_{A_1^nB_1^n}\), it holds
    \begin{align}
        \renyiSandUp_{\alpha}(S_1^n|I_1^n \widetilde{E})_{\omega_{|\OmegaAT}} &\geq n \hUpQKD - \frac{\alpha}{\alpha-1} \log \frac{1}{\Pr[\OmegaAT]}.
    \end{align}
    Here the quantity \( \hUpQKD\) is given by
    \begin{equation}
        \hUpQKD =  \begin{aligned}[t]
        \inf_{\mbf{q} \in \Sacc} &\inf_{\substack{\rho \in \dop{=}(AB)\\ \text{s.t.} \Tr_B[\rho] = \tau_A}} \Bigg( \frac{\alpha}{\alpha - 1}D\left(\mbf{q} \middle\Vert \bsym{\nu}^{\rho}_{\CP}\right) \\
        &+ q(\gen) \renyiSandUp_{\alpha}(S|\tilde{T}E)_{\nu^{\rho}_{|\gen}} \Bigg),
        \end{aligned}
    \end{equation}
    where \(D\rel{.}{.}\) is the relative entropy and
    \begin{align}
        \nu^{\rho}_{S\tilde{T}\CP E} &= \EATchannQKD(\rho_{ABE}), \\
        \bsym{\nu}^{\rho}_{\CP} &= (\gamma \bsym{\nu}^{\rho}_{\CP|\test}, 1-\gamma)^T, \\
        \bsym{\nu}^{\rho}_{\CP|\test} &= \probst[\rho_{AB|\test}],
    \end{align}
    and \(E\) is a purifying system such that \(\rho_{ABE}\) is pure.
\end{theorem}
\begin{proof}
    First, note that since each \(\CP_i\) can be computed from \(I_i\) with the deterministic function \(\phi\), one finds
    \begin{equation}
        \renyiSandUp_{\alpha}(S_1^n|I_1^n \widetilde{E})_{\omega_{|\OmegaAT}} = \renyiSandUp_{\alpha}(S_1^n|I_1^n \CP_1^n \widetilde{E})_{\omega_{|\OmegaAT}}.
    \end{equation}
    Next, we will apply \cite[Corollary 4.2]{arqand_marginal-constrained_2025}, which gives us a lower bound on $\renyiSandUp_{\alpha}(S_1^n|I_1^n \CP_1^n \widetilde{E})_{\omega_{|\OmegaAT}}$ in terms of an expression equivalent to \(\hUpQKD\). Therefore, observe that the channel \(\EATchannQKD\) satisfies the assumptions of \cite[Corollary 4.2]{arqand_marginal-constrained_2025}. Hence, making the following identifications between the register and variable names in this work and in \cite[Corollary 4.2]{arqand_marginal-constrained_2025},
    \begin{equation}
    \begin{split}
        I &\mapsto I, \CP \mapsto \CP,  \bar{C} = \emptyset, A_j \mapsto A_{j-1}, \\
        B_j &\mapsto B_{j-1}, \widetilde{E} \mapsto \widehat{E}, E \mapsto \widetilde{E},\\         
        \omega &\mapsto \rho, \rho \mapsto \omega \:, \tau \mapsto \sigma, \\
        S_{\OmegaAT} &\mapsto S_{\Omega}.
    \end{split}
    \end{equation}
    we can apply \cite[Corollary 4.2]{arqand_marginal-constrained_2025}, and \(\hUpQKD\) has the following intermediate form
     \begin{equation}
        \hUpQKD =  \begin{aligned}[t]
        \inf_{\mbf{q} \in \Sacc} &\inf_{\substack{\rho \in \dop{=}(AB)\\ \text{s.t.} \Tr_B[\rho] = \tau_A}} \Bigg( \frac{\alpha}{\alpha - 1}D\left(\mbf{q} \middle\Vert \bsym{\nu}^{\rho}_{\CP}\right) \\
        &+ q(\gen) \renyiSandUp_{\alpha}(S|IE)_{\nu^{\rho}_{|\gen}} \Bigg).
        \end{aligned}
    \end{equation}
    However, this is equal to the formula claimed in the theorem statement, since conditioned on a generation round, we have \(T = I\). Thus, also Eve's copy satisfies \(\tilde{T} =I\) in generation rounds.

    Moreover, the equality between the probability distribution \(\bsym{\nu}^{\rho}_{\CP|\test}\) on \(\CP\) induced by \(\EATchannQKD[\rho_{ABE}]\) and \(\probst[\rho_{AB|\test}]\) follows from a direct calculation.
\end{proof}

We have not yet laid out how to calculate \(h^{\uparrow,\text{QKD}}_{\alpha}\) for a QKD protocol. One requires a significant amount of reformulations to reach an expression which is easily implementable with numerical methods. Therefore, we will focus on simplifying the expression in \cref{thrm:halpha QKD} in the next \cref{sec:Calculating halpha}, for now concluding this section with the security proof assuming that one has access to a lower bound \(\underline{h}^{\text{QKD}}\) on \(h^{\uparrow,\text{QKD}}_{\alpha}\).

\begin{theorem}\label{thrm:Fixed-length security generic bnd}
    For any $\alpha \in (1,2)$, $\ePA,\eEV \in(0,1]$, the fixed-length version of \cref{Prot:PM Protocol} is $\ePA$-secret, $\eEV$-correct, and hence $(\ePA+\eEV)$-secure, when the length $l$ of the final key satisfies
    \begin{align}\label{eq:generic fixed key length}
    l\leq n \underline{h}^{\text{QKD}} - \leak - \ceil{\log\frac{1}{\eEV}} - \frac{\alpha}{\alpha-1} \log\frac{1}{\ePA} + 2,
    \end{align}
    where \(\underline{h}^{\text{QKD}}\) is a lower bound on $\hUpQKD$, which is defined as in \cref{thrm:halpha QKD} and $\leak$ is the length of the error-correction string.
\end{theorem}
\begin{proof}
    As stated in \cref{rem:Fixed-Length security} below \cref{Def:Variable length eps security}, secrecy and correctness imply security. Therefore, we prove each part separately.

    Starting with correctness; the protocol is \(\eEV\)-correct since,
    \begin{align}
        &\Pr[K_A\neq K_B\land \OmegaAcc]\leq \Pr[K_A\neq K_B \wedge \Omega_{\mathrm{EV}}] \\ &\leq \Pr[S_1^n \neq \widebar{S}_1^n \wedge \OmegaEV]\leq\Pr[\OmegaEV|S_1^n \neq \widebar{S}_1^n] \\ &\leq 2^{-\ceil{\log \frac{1}{\eEV}}}\leq \eEV,
    \end{align}
    by the property of 2-universal hashing.
    
    For secrecy, let us start with the secrecy definition and apply the LHL for \Renyi entropies, \cref{thrm:Renyi LHL}, yielding 
    \begin{align}
        &\frac{1}{2}\Pr[\OmegaAcc]\left\|\sigma_{K_AI_1^nELG_{|_{\OmegaAcc}}} - \mathbb{U}_{K_A}^l \otimes \sigma_{I_1^nELG_{|_{\OmegaAcc}}}\right\|_1 \\
        &\leq \Pr[\OmegaAcc] 2^{\frac{2(1-\alpha)}{\alpha}} 2^{\frac{1-\alpha}{\alpha}\left(\renyiSandUp_{\alpha}(S_1^n|I_1^nEL)_{\omega_{|_{\OmegaAcc}}}-l\right)} \\
        &=\Pr[\OmegaEV|\OmegaAT] \Pr[\OmegaAT] 2^{\frac{1-\alpha}{\alpha}\left(\renyiSandUp_{\alpha}(S_1^n|I_1^nEL)_{\omega_{|_{\OmegaAcc}}} - l + 2\right)}. \label{eq:fixed length proof secrecy after LHL}
    \end{align}
    Continuing with the \Renyi entropy itself, we apply \cite[Lemma B.5]{dupuis_entropy_2020} together with \(\OmegaAcc = \OmegaAT \wedge \OmegaEV\) and find
    \begin{align}      
        &\renyiSandUp_{\alpha}(S_1^n|I_1^nEL)_{\omega_{|_{\OmegaAT \wedge \OmegaEV}}} \\
        &\geq \renyiSandUp_{\alpha}(S_1^n|I_1^nEL)_{\omega_{|_{\OmegaAT}}} - \frac{\alpha}{\alpha-1} \log \frac{1}{\Pr[\OmegaEV|\OmegaAT]}.
    \end{align}
    Next, we split off the error correction data contained in register \(L\) by applying the chain rule of \cite[Eqn. (5.94)]{tomamichel_quantum_2016}, which combines \cite[Lemmas 5.3 \& 5.4]{tomamichel_quantum_2016}, to find
    \begin{equation}
        \begin{split}
        &\renyiSandUp_{\alpha}(S_1^n|I_1^nEL)_{\omega_{|_{\OmegaAT \wedge \OmegaEV}}} \\
        &\geq \renyiSandUp_{\alpha}(S_1^n|I_1^nE)_{\omega_{|_{\OmegaAT}}} - \leak - \ceil{\log \frac{1}{\eEV}} \\
        & - \frac{\alpha}{\alpha-1} \log \frac{1}{\Pr[\OmegaEV|\OmegaAT]}.
        \end{split}
    \end{equation}
    Due to the source-replacement scheme, we only need to consider states with a marginal  \(\Tr_{B_1^n \widetilde{E}}[\rho_{A_1^nB_1^n \widetilde{E}}] = \tau_{A}^{\otimes n}\) defined by Alice's sending probabilities and signal states, which we can include in \cref{thrm:halpha QKD}. Then, by \cref{thrm:halpha QKD} the \Renyi entropy conditioned on the event \(\OmegaAT\) is bounded by
    \begin{equation}
        \renyiSandUp_{\alpha}(S_1^n|I_1^nE)_{\omega_{|_{\OmegaAT}}} \geq n \hUpQKD - \frac{\alpha}{\alpha-1} \log\frac{1}{\Pr[\OmegaAT]}.
    \end{equation}
    By assumption, \(\underline{h}^{\mathrm{QKD}}\) is a lower bound on \(\hUpQKD\), therefore
    \begin{equation}\label{eq:fixed length proof lower bound renyi entr}
        \begin{split}
        &\renyiSandUp_{\alpha}(S_1^n|I_1^nEL)_{\omega_{|_{\OmegaAT \wedge \OmegaEV}}} \\
        &\geq n \underline{h}^{\mathrm{QKD}} - \frac{\alpha}{\alpha-1} \log\frac{1}{\Pr[\OmegaAT]} - \leak - \ceil{\log \frac{1}{\eEV}} \\
        & - \frac{\alpha}{\alpha-1} \log \frac{1}{\Pr[\OmegaEV|\OmegaAT]}.
        \end{split}
    \end{equation}
    Now, if we insert \cref{eq:fixed length proof lower bound renyi entr} and the key length expression from \cref{eq:generic fixed key length} into \cref{eq:fixed length proof secrecy after LHL}, we find
    \begin{align}
        \frac{\Pr[\OmegaAcc]}{2} \left\|\sigma_{K_AI_1^nEL_{|_{\OmegaAcc}}} - \mathbb{U}_{K_A}^l \otimes \sigma_{I_1^nEL_{|_{\OmegaAcc}}}\right\|_1 \leq \ePA.
    \end{align}
    Hence, the protocol is \(\ePA\)-secret and \(\eEV\)-correct, and thus \(\ePA+\eEV\)-secure.
\end{proof}

\section{Calculating Versions of \texorpdfstring{\(\hUpQKD\)}{halpha}}\label{sec:Calculating halpha}

In general, numerical minimization of \Renyi entropies or divergences for generic values of \(\alpha\) can be a challenging task. For example, for \(\alpha = 1/2 \), the sandwiched divergence can be reformulated as a semidefinite program; however, for \(\alpha > 1/2 \) a similar reformulation remains unknown~\cite[Problem 1]{fawzi_liebs_2017}.\footnote{Very recently, an independent work~\cite{he_operator_2025} developed an interior-point solver for minimizing sandwiched \Renyi divergences. However, to use this to instead minimize \Renyi conditional entropies, one would still need to implement some of the reformulations we develop below. We aim to study in future work the task of integrating that solver into our framework.}

In this work, we address this by following a similar approach as in Ref.~\cite{winick_reliable_2018}, where numerical methods for calculating QKD key rates have been presented. Their numerical approach relied on the reformulation of the required conditional von Neumann entropy in terms of the relative entropy presented in \cite[Theorem 1]{coles_unification_2012}, which then was minimized with the Frank-Wolfe algorithm~\cite{frank_algorithm_1956}, an algorithm for constrained convex optimization problems.

The reformulation was required because the conditional von Neumann entropy \(H(S|E)\) is concave with respect to the state on $SE$, but when evaluated in terms of a purification \(\rho_{ABE}\) of the pre-measurement state \(\rho_{AB}\), it is convex with respect to \(\rho_{AB}\). This allows one to address it using convex optimization methods. 

Here, for \Renyi entropies we need a similar reformulation because of the same reasons. However, certain \Renyi entropies are harder to evaluate than others. Thus, in this section we will present a path to a computable lower bound on \(\hUpQKD\) presented in \cref{thrm:halpha QKD}.

We start by proving a similar duality relation as in \cite[Theorem 1]{coles_unification_2012}, but for \Renyi entropies. In \cite[Proposition 17]{gour_entropy_2021} a special case of the relations in the following lemma was found. Somewhat different formulations for $\renyiSandUp_\alpha$ and  $\renyiSandDown_\alpha$ were also obtained in~\cite[Lemma~A.2]{anco_how_2024}, but it appears unclear whether those formulas have the required convexity properties for our numerical methods.
Finally, recent and concurrent work~\cite{chung_generalized_2025} implemented Frank-Wolfe methods to minimize some forms of sandwiched \Renyi divergence; in particular, they used this to compute lower bounds on \(\renyiSandUp_{\alpha}\) by deriving a relation similar to those we present here.\footnote{Based on the more general \cref{lemma:RenyiQKDcones} we present in \cref{app:RenyiQKDcones}, we believe the computations they implemented are equivalent to evaluating  \(\renyiPetzUp_{\alpha}\) as a lower bound on \(\renyiSandUp_{\alpha}\). This would hence complement our computations, which are instead based on \(\renyiSandDown_{\alpha}\).}

\begin{theorem}\label{thrm:Petz Renyi simplifications}
    Let \(\rho_{QE} \in \dop{=}(QE)\) be pure and let \(\{Z_j\}_{1 \dots d_S}\) be a set of orthogonal projections on \(Q\) such that \(\sum_{j=1}^{d_S} Z_j = \idop_Q\). Furthermore, define the isometry
    \begin{equation}
        V_{\ZMap} \defvar \sum_j \ket{j}_S \otimes Z_j,
    \end{equation}
    and the state \(\sigma_{SQE}\) after applying the isometry \(V_{\ZMap}\) onto \(\rho_{QE}\) as
    \begin{equation}
        \sigma_{SQE} \defvar V_{\ZMap} \rho_{QE} V_{\ZMap}^{\dagger}.
    \end{equation}
    Then, it holds
    \begin{align}
        \renyiSandDown_{\alpha}(S|E)_{\sigma_{SE}} &= \frac{1}{1-\alpha} \log \Tr\left[ \left(\ZMap\left(\rho_Q^{\frac{1}{\alpha}}\right)\right)^{\alpha}\right], \label{eq:Simplified Sand Down}\\
        \renyiPetzDown_{\alpha}(S|E)_{\sigma_{SE}} &= \renyiPetzDiv_{2-\alpha}\left(\rho_Q||\ZMap\left(\rho_Q\right)\right) \label{eq:Simplified Petz Down}
    \end{align}
    where
    \begin{equation}
        \ZMap(\tau) \defvar \sum_{j=1}^{d_S} Z_j \tau Z_j^{\dagger},
    \end{equation}
    for all \(\tau \in \dop{=}(Q)\).
\end{theorem}
\begin{proof}
We first note that \(\sigma_Q = \ZMap\left(\rho_{Q}\right)\) since
    \begin{equation}
    \begin{split}\label{eq:sigmaQ = rhoQ}
        \sigma_{Q} &= \Tr_{SE}\left[\sigma_{SQE}\right] = \Tr_{SE}\left[V_{\ZMap} \rho_{QE} V_{\ZMap}^{\dagger}\right] \\
        &= \ZMap\left(\rho_{Q}\right).
    \end{split}
    \end{equation}
    Furthermore, it holds
    \begin{equation}\label{eq:trace sigma to Zmap rho}
        \Tr_S\left[\sigma_{SQ}^\beta\right] = \Tr_S\left[V_{\ZMap} \rho_{Q}^{\beta} V_{\ZMap}^{\dagger}\right] = \ZMap\left(\rho_Q^{\beta}\right)\; \forall \beta \geq 0.
    \end{equation}
    Using both properties, one can prove both theorem statements.
    Due to \cite[Theorem 2]{tomamichel_relating_2014}, it holds,
    \begin{align}
        \renyiSandDown_{\alpha}(S|E)_{\sigma_{SE}} = - \renyiPetzUp_{\frac{1}{\alpha}}(S|Q)_{\sigma_{SQ}}.
    \end{align}
    Furthermore, we can use~\cite[Lemma 1]{tomamichel_relating_2014}, to find
    \begin{align}
        \renyiSandDown_{\alpha}(S|E)_{\sigma_{SE}} = \frac{1}{1-\alpha} \log \Tr\left[ \left(\Tr_S\left(\sigma_{SQ}^{\frac{1}{\alpha}}\right)\right)^{\alpha}\right],
    \end{align}
    and inserting \cref{eq:trace sigma to Zmap rho} already yields \cref{eq:Simplified Sand Down}.
    
    Similarly, by applying the duality relation for $\renyiPetzDown_{\alpha}(S|E)_{\sigma_{SE}}$ presented in \cite[Lemma 6]{tomamichel_fully_2009}, we obtain 
    \begin{align}
    \renyiPetzDown_{\alpha}(S|E)_{\sigma_{SE}} 
    &= - \renyiPetzDown_{2-\alpha}(S|Q)_{\rho_{SQ}}
    \nonumber\\&= \renyiPetzDiv_{2-\alpha}\left(\rho_{SQ}||\idop_S \otimes \rho_Q \right) ,
    \end{align}
    after which \cref{eq:Simplified Petz Down} follows by further simplifying $\renyiPetzDiv_{2-\alpha}\left(\rho_{SQ}||\idop_S \otimes \rho_Q \right)$ using the definition of $\renyiPetzDiv$ and the properties in \cref{eq:sigmaQ = rhoQ,eq:trace sigma to Zmap rho}.
\end{proof}

The core idea here, of using duality relations to re-express the entropy in terms of only the initial state on $Q$, can be generalized to obtain formulas for all of the conditional \Renyi entropies in \cref{def:Conditional Renyi entropy}, though in a slightly different form from the above lemma. 
We defer these expressions to \cref{lemma:RenyiQKDcones} in the appendices, as we will not be making use of them in this work. 
However, we highlight  that in the case of von Neumann entropy, they yield an alternative expression for the QKD cone~\cite{hu_robust_2022,lorente_quantum_2025,he_exploiting_2024} that may be of independent interest.

For this work, we choose to focus only on the $\renyiSandDown_{\alpha}$ case presented in the above lemma, because it provides a reasonable balance between tightness of the bounds and ease of numerical work. Specifically, amongst the conditional entropies in \cref{def:Conditional Renyi entropy}, $\renyiPetzDown_{\alpha}$ is smaller than all the others~\cite[Fig.~5.1]{tomamichel_quantum_2016} and hence we choose not to use it. Out of the remaining options, the expressions we obtain for $\renyiSandUp_{\alpha}$ and $\renyiPetzUp_{\alpha}$ in \cref{lemma:RenyiQKDcones} involve the sandwiched divergence, while the expression we obtain for $\renyiSandDown_{\alpha}$ involves the Petz divergence. The sandwiched divergences have more complicated gradient expressions (required to implement the Frank-Wolfe algorithm) than the Petz divergences, and hence we choose to focus on $\renyiSandDown_{\alpha}$. (Note that as observed in~\cite[Fig.~5.1]{tomamichel_quantum_2016}, while we would theoretically always obtain the tightest bounds by using \mbox{$\renyiSandUp_{\alpha}$}, it is unknown in general which of \mbox{$\renyiSandDown_{\alpha}$} or \mbox{$\renyiPetzUp_{\alpha}$} provides tighter bounds.)

Additionally, for our later approaches involving variable-length key rates we require a formalism to find rates for either \(\renyiSandUp_{\alpha}\) or \(\renyiSandDown_{\alpha}\). Again, the above result suffices to lower bound both, simply because \(\renyiSandUp_{\alpha}\) is lower bounded by \(\renyiSandDown_{\alpha}\), which is covered by the above result.

In any case, for typical values of $\alpha$ used in this work (which approach $1$ as $n$ increases), all of the entropies in \cref{def:Conditional Renyi entropy} would only differ by small amounts, due to explicit converse bounds derived in Ref.~\cite[Corollary~4]{tomamichel_relating_2014} that constrain how much they can differ. 
With this in mind, we will lower bound the quantity \(h^{\uparrow,\text{QKD}}_{\alpha}\) by replacing the \Renyi entropy \(\renyiSandUp_{\alpha}\) with its counterpart \(\renyiSandDown_{\alpha}\) (expressed in the form in \cref{thrm:Petz Renyi simplifications}), with the understanding that this will make little difference whenever $\alpha$ is close to $1$. 
We note also that it has been empirically observed in Ref.~\cite{anco_how_2024} that in the context of randomness generation, this relaxation only makes a difference for very small sample sizes below \(n=10^4\) signals.

Therefore, we formulate \cref{cor:Fixed-length security lower bnd halpha}, which proves security of a generic P\&M QKD protocol against coherent attacks.

\begin{cor}[Fixed-Length security]\label{cor:Fixed-length security lower bnd halpha}
    For any $\alpha \in (1,2)$, $\ePA,\eEV \in(0,1]$, the fixed-length version of \cref{Prot:PM Protocol} is $\ePA$-secret, $\eEV$-correct, and hence $(\ePA+\eEV)$-secure, when the length $l$ of the final key satisfies
    \begin{align}\label{eq:Fixed key length}
    \begin{split}
    l\leq n \hQKD &- \leak - \ceil{\log\frac{1}{\eEV}} \\
    &- \frac{\alpha}{\alpha-1} \log\frac{1}{\ePA} + 2,
    \end{split}
    \end{align}
    where $\leak$ is the length of the error-correction string and \(\hQKD\) is defined by
    \begin{equation}\label{eq:defn_hQKD}
        \hQKD \defvar \begin{aligned}[t]
        \inf_{\mbf{q} \in \Sacc} &\inf_{\substack{\rho \in \dop{=}(AB), \\ \text{s.t.} \Tr_B[\rho] = \tau_A }} \Bigg( \frac{\alpha}{\alpha - 1}D\left(\mbf{q} \middle\Vert \bsym{\nu}^{\rho}_{\CP}\right) \\
        &+ q(\gen) \renyiSandDown_{\alpha}(S|\tilde{T}E)_{\nu_{|\gen}^{\rho}} \Bigg),
        \end{aligned}
    \end{equation}
    where
    \begin{align}
        \nu^{\rho}_{S\tilde{T}\CP E} &= \EATchannQKD(\rho_{ABE}), \\
        \bsym{\nu}^{\rho}_{\CP} &= (\gamma \bsym{\nu}^{\rho}_{\CP|\test}, 1-\gamma)^T, \\
        \bsym{\nu}^{\rho}_{\CP|\test} &= \probst[\rho_{AB|\test}].
    \end{align}
    The \Renyi entropy \(\renyiSandDown_{\alpha}(S|\tilde{T}E)_{\nu^{\rho}_{|\gen}}\) can further be simplified to
    \begin{equation}\label{eq:H rewritten in terms of rho cond gen}
        \begin{split} 
            \renyiSandDown_{\alpha}(S|\tilde{T}E)_{\nu_{|\gen}} = &\frac{1}{1-\alpha} \log \Tr\bigg[ \left(\hat{\ZMap}\left(\hat{\GMap}(\rho_{AB{|\gen}})^{\frac{1}{\alpha}}\right)\right)^{\alpha}\\ &+ 1- \Tr[\hat{\GMap}(\rho_{AB{|\gen}})]\bigg],
            \end{split}
    \end{equation}
    where \(\hat{\GMap}\) is the CPTNI map defined in \cref{def:QKD channel} omitting the discard symbol. The CPTNI map \(\hat{\ZMap}\) similarly omits the discard symbol.
\end{cor}
Before we state the proof, we note that in \cite[Lemma 5.1]{arqand_generalized_2024} it was shown that the optimization problem for \(\hQKD\) is convex in its arguments. Hence, it could be evaluated with convex optimization methods, as claimed earlier.
\begin{proof}
    Correctness follows immediately from \cref{thrm:Fixed-length security generic bnd} and secrecy also follows from \cref{thrm:Fixed-length security generic bnd} if \(\hQKD \leq \hUpQKD \). 
    This is true since \(\renyiSandDown_{\alpha}(S|IE)_{\nu^{\rho}_{|\gen}} \leq \renyiSandUp_{\alpha}(S|IE)_{\nu^{\rho}_{|\gen}}\) for any \(\nu\) and thus the infimum in \(\hQKD\) is also smaller.
    
    Hence, it only remains to show that \(\renyiSandDown_{\alpha}(S|\tilde{T}E)_{\nu^{\rho}_{|\gen}}\) can be simplified to the expression stated in the corollary. Applying the definition of \(\EATchannQKD\) yields that \(\nu_{|\gen}\) is given by
    \begin{equation}
        \nu_{S\tilde{T}ES_QIABXY|\gen} = V_{\ZMap} V_{\GMap} \rho_{AB|\gen} V_{\GMap}^{\dagger} V_{\ZMap}^{\dagger}.
    \end{equation}
    Now, by identifying \(\tilde{T}E \mapsto E\) and \(S_QIABXY \mapsto Q\) in \cref{eq:Simplified Sand Down} of \cref{thrm:Petz Renyi simplifications}, one finds
    \begin{equation}
    \begin{split}
        &\renyiSandDown_{\alpha}(S|\tilde{T}E)_{\nu_{S\tilde{T}E|\gen}}\\
        = &\frac{1}{1-\alpha} \log \Tr\left[ \left(\ZMap\left(\Tr_{\tilde{T}}\left[V_{\GMap}\rho_{AB|\gen}V_{\GMap}^{\dagger}\right]^{\frac{1}{\alpha}}\right)\right)^{\alpha}\right].
    \end{split}
    \end{equation}
    Finally, we can simplify the trace over \(\tilde{T}\) to find
    \begin{equation}
        \Tr_{\tilde{T}}\left[V_{\GMap}\rho_{AB|\gen}V_{\GMap}^{\dagger}\right] = \GMap(\rho_{AB|\gen}),
    \end{equation}
    which leaves us with
    \begin{equation}
    \begin{split}
        &\renyiSandDown_{\alpha}(S|\tilde{T}E)_{\nu^{\rho}_{S\tilde{T}E|\gen}}\\
        = &\frac{1}{1-\alpha} \log \Tr\left[ \left(\ZMap\left(\GMap(\rho_{AB|\gen})^{\frac{1}{\alpha}}\right)\right)^{\alpha}\right].
    \end{split}
    \end{equation}
    By inserting the definitions of \(\ZMap\) and \(\GMap\)
    \begin{align}
        \ZMap(\rho) &= \hat{\ZMap}(\rho) + \ketbra{\perp}{\perp} \rho \ketbra{\perp}{\perp}, \\
        \GMap(\rho) &= \hat{\GMap}(\rho) + \left(1 - \Tr(\hat{\GMap}(\rho))\right) \ketbra{\perp}{\perp},
    \end{align}
    and using the orthogonality of \(\ketbra{\perp}{\perp}\), the corollary statement follows.
\end{proof}

We will spend the remainder of this section to simplify \(\hQKD\) even further and bring it into a form that should be more numerically stable.

Hence, let us consider the following situation of small testing fractions. In general, the components of the state \(\rho_{AB}\) resulting in test constraints in \(\hQKD\) will be proportional to the testing fraction $\gamma$. However, due to this proportionality, when choosing a sufficiently small testing fraction, these constraints can quickly be obscured by numerical noise. 

To address this, we first make the trivial observation that we can always ``undo" the source-replacement argument on single rounds, to say that the set of all states $\EATchannQKD(\rho_{ABE})$ satisfying $\rho_A = \tau_A$ 
% (with pure $\rho_{ABE}$) 
can equivalently be generated by Alice preparing the classical-quantum state $\sigma_{XA'}$ in the original P\&M description, Eve applying some arbitrary attack channel $A'\to BE$, and Bob measuring $B$, followed by the relevant processing of Alice and Bob's classical values.
Furthermore, the simplification in \cref{eq:H rewritten in terms of rho cond gen} tells us that in fact we do not need to consider Eve's full attack channel $A'\to BE$, but rather only its restriction \(\mathcal{E}:A' \to B\) obtained by tracing out $E$ from the output.

Now, the state $\sigma_{XA'}$ is a mixture of the test and generation cases according to the protocol structure. However, the critical observation is that the channel \(\mathcal{E}\) must still act in the same way on both the state conditioned on a test round and conditioned on a generation round. 
Numerically, it would be much more desirable to optimize over the channel instead of the state \(\rho_{AB}\), as it avoids the components of order $\gamma$ in the latter.
Hence, similar to the approach in Ref.~\cite{kamin_finite-size_2025}, we will now rephrase the optimization problem of \(\hQKD\) in terms of the Choi state \(J\) of Eve's channel \(\mathcal{E}\) creating the states conditioned on test and generation rounds. 

To define the optimization problem in terms of Eve's channel, we now apply the source-replacement argument again, but \emph{separately} on the test and generation components. That is to say, we view Alice sending some signal state to Bob and storing her choice as instead preparing some pure states $\ket{\xi^t}_{AA'}$ and $\ket{\xi^g}_{AA'}$ with probabilities $\gamma$ and $1-\gamma$ respectively, and then measuring $A$ with some suitable measurement (which can differ in the two cases). To describe this, we define these pure states:
\begin{align}
    \ket{\xi^t}_{AA'} \defvar \sqrt{\Pi^{\test}} \frac{\ket{\psi}_{AA'}}{\sqrt{\gamma}}, \\  
    \ket{\xi^g}_{AA'} \defvar \sqrt{\Pi^{\gen}} \frac{\ket{\psi}_{AA'}}{\sqrt{1 -\gamma}}.
\end{align}
as the states Alice prepares conditioned on test and generation rounds, respectively. Then we can write
\begin{align}\label{eq:rhoAB in terms of channel}
    \begin{split}
        \rho_{AB|\gen} &=  \left( \id_A \otimes \; \mathcal{E} \right) \left( \ketbra{\xi^g}{\xi^g}_{AA'} \right) , \\
        \rho_{AB|\test} &=\left( \id_A \otimes \; \mathcal{E} \right) \left( \ketbra{\xi^t}{\xi^t}_{AA'} \right),
    \end{split}
\end{align}
where again \(\mathcal{E}\) defines Eve's channel. Thus, we could immediately write \(\rho_{AB|\gen}\) required for \cref{eq:H rewritten in terms of rho cond gen} in terms of Eve's channel. Moreover, \(\bsym{\nu}^{\rho}_{\CP|\test} = \probst[\rho_{AB|\test}]\) could also straightforwardly be written as Eve's channel acting on the state \(\ketbra{\xi^t}{\xi^t}_{AA'}\).

Next, we rephrase Eve's channel in terms of its Choi state. This allows for a treatment with convex optimization methods because Choi states of CPTP maps need to be positive semidefinite and satisfy a partial trace constraint. 

To simplify this formulation, we define the CPTNI maps \(\chi_g: A' \rightarrow A\) and \(\chi_t : A' \rightarrow A\) as
\begin{equation}\label{eq:xi test and gen}
    \begin{split}
        \chi_g[X] &\defvar \Tr_{A'}\left[\left( \idop_A \otimes X^{T_{A'}} \right) \ketbra{\xi^g}{\xi^g} \right], \\
        \chi_t[X] &\defvar \Tr_{A'}\left[ \left( \idop_A \otimes X^{T_{A'}} \right) \ketbra{\xi^t}{\xi^t} \right],
    \end{split}
\end{equation}
which let us write
\begin{align}
    \rho_{AB|\gen/\test} &= \chi_{g/t}(J),
\end{align}
i.e. an linear map acting directly on the Choi state. Additionally, to achieve the same for the statistics in test rounds, let us define the map \(\probstJ\) generating the statistics in test rounds directly from a Choi state by
\begin{equation}\label{eq:def probstJ}
    \probstJ[J] \defvar \probst\circ \chi_t[J].
\end{equation}

Finally, for a condensed notation of the \Renyi entropy evaluated on the state \(\nu_{|\gen}\), let us define the function \(g_{\alpha}: \mathrm{Pos}(A'B) \rightarrow \R_{\geq 0}\) as
\begin{equation}\label{eq:defn galpha}
    g_{\alpha}(J) \defvar \renyiSandDown_{\alpha}(S|\tilde{T}E)_{\EATchannQKD\circ \chi_g(J)},
\end{equation}
for all Choi states \(J \in \mathrm{Pos}(A'B)\), such that \(\Tr_B[J] = \idop_{A'}\).

Combining all these definitions we can recast the optimization problem of \(\hQKD\), originally defined in terms of the state \(\rho_{AB}\), as an optimization problem over the Choi state \(J\) of Eve's channel. The resulting optimization problem is given by
\begin{equation}\label{eq:halpha final}
    \begin{aligned}[t]
        \hQKD = &\inf_{\substack{\mbf{q} \in \Sacc, \\ J \in \mathrm{Pos}(A'B)}} \Bigg( \frac{\alpha D\rel{\mbf{q}}{\bsym{\nu}_{\CP}}}{\alpha - 1} + q(\gen) g_{\alpha}(J) \Bigg)\\
        \textrm{s.t. } &\Tr_B[J] = \idop_{A'}, \\
        &\bsym{\nu}_{\CP} = (\gamma \bsym{\nu}_{\CP|\test}, 1-\gamma)^T, \\
        &\bsym{\nu}_{\CP|\test} = \probstJ[J].
        \end{aligned}
\end{equation}
In \cref{cor:Joint convexity halpha channel}, in the appendix, we show that this reformulation of \(\hQKD\) is indeed convex. Therefore, similar to the algorithm of Ref.~\cite{winick_reliable_2018}, we employ a Frank-Wolfe method \cite{frank_algorithm_1956} to find the minimum. Hence, we require the gradient of \(\hQKD\) with respect to \(\mbf{q}\) and \(J\). Again similar to the case of the relative entropy in \cite{winick_reliable_2018}, we need some form of perturbation for the gradient to exist everywhere on its domain. In \cref{app:Details numerics} we present the details of our algorithm evaluating \(\hQKD\) including the gradient and theorems for the perturbation procedure. However, we leave the details to the appendix and conclude the section for now.

\section{Variable-Length Protocols}\label{sec:Variable-Length protocols}
So far we only considered fixed-length QKD protocols, and in this section we will prove the security of variable-length protocols. In preparation for this, we need to establish a few more definitions as presented in Ref.~\cite{arqand_generalized_2024}.

\begin{definition}[$f$-weighted \Renyi entropies (partially restated from \cite{arqand_generalized_2024,inprep_HB24})]\label{def:f-weighted entropies}
Let $\rho \in \dop{=}(\CP Q Q')$ be a state where $\CP$ is classical with alphabet $\alphCP$. A \emph{tradeoff function\footnote{Ref.~\cite{arqand_generalized_2024} instead referred to this as a ``quantum estimation score-system'' (QES).} on $\CP$} is a function $f:\alphCP \to \mathbb{R}$; equivalently, we may denote it as a real-valued tuple $\mbf{f} \in \mathbb{R}^{|\alphCP|}$ where each term in the tuple specifies the value $f(\cP)$. Given a tradeoff function $f$ and a value $\alpha\in(0,1)\cup (1,\infty)$, we define two versions of an \emph{$f$-weighted \Renyi entropy}  of order $\alpha$ for $\rho$, as follows:
\begin{equation}\label{eq:upfweighted entropy}
\begin{split}
&\frenyiSandUp_\alpha(Q|\CP Q')_{\rho} \defvar \\
&\frac{\alpha}{1-\alpha} \log \left( \sum_{\cP} \rho(\cP) \, 2^{\frac{1-\alpha}{\alpha} \left(-f(\cP) + \renyiSandUp_{\alpha}(Q|Q')_{\rho_{|\cP}} \right) } \right) ,
\end{split}
\end{equation}
\begin{equation}\label{eq:downfweighted entropy}
    \begin{split}
    &\frenyiSandDown_\alpha(Q|\CP Q')_{\rho} \defvar \\
    &\frac{1}{1-\alpha} \log \left( \sum_{\cP} \rho(\cP) \, 2^{(1-\alpha) \left(-f(\cP) + \renyiSandDown_{\alpha}(Q|Q')_{\rho_{|\cP}} \right) } \right),
    \end{split}
\end{equation}
where the sums run over all $\cP$ values such that $\rho(\cP)>0$. We extend both definitions to $\alpha=\infty$ by taking the $\alpha\to\infty$ limit.
\end{definition}

\newcommand{\normset}{\mathcal{S}}
\begin{definition}[Normalized tradeoff functions]\label{def:Full f-weighting}
Let \(f\) be a tradeoff function on a register \(\CP\) as in \cref{def:f-weighted entropies}. Given any set of states $\normset \subseteq \dop{=}(\CP Q Q')$, we define the \emph{$\frenyiSandUp_\alpha$-normalization constant} and \emph{$\frenyiSandDown_\alpha$-normalization constant} for that set to be, respectively,
\begin{gather} 
\kapup \defvar \inf_{\rho\in\normset} \frenyiSandUp_\alpha(Q|\CP Q')_{\rho} ,\\
\kappa \defvar \inf_{\rho\in\normset} \frenyiSandDown_\alpha(Q|\CP Q')_{\rho} .
\end{gather}
Given either of the above values, we then define a corresponding \emph{$\frenyiSandUp_\alpha$-normalized} or \emph{$\frenyiSandDown_\alpha$-normalized} tradeoff function \(f\) respectively, via
\begin{gather}
\hat{f}(\cP) \defvar f(\cP) + \kapup, \\
\hat{f}(\cP) \defvar f(\cP) + \kappa.
\end{gather}
\end{definition}

With these definitions, we are already able to prove security of variable-length QKD protocols.

\begin{theorem}[Variable-Length Security]\label{thrm:Variable-length security}
    Consider \cref{Prot:PM Protocol}, and take any $\alpha \in (1,2)$, $\ePA,\eEV \in(0,1]$,
    and any tradeoff function $f:\alphCP \rightarrow \R$. Let $\kapup$ be the $\frenyiSandUp_\alpha$-normalization  constant 
    (\cref{def:Full f-weighting}) corresponding to the set of all states that can be produced in a single round, i.e.
    \begin{align}\label{eq:singleroundHf}
    \begin{aligned}
        \kapup \defvar &\inf_{\substack{\rho_{AB} \in \dop{=}(AB)\\ \text{s.t.} \Tr_B[\rho_{AB}] = \tau_A}} \frenyiSandUp_\alpha(S|\CP I E)_{\nu} , 
    \end{aligned}
    \end{align}
    where the set of possible output states $\nu_{S\CP I E}$ is parametrized via the input state $\rho_{AB}$ on Alice and Bob's registers as described in the previous sections. We then define the following tradeoff function on $\CP_1^n$:
    \begin{align}\label{eq:ffull}
        \ffull(\cP_1^n) \defvar \sum_i (f(\cP_i) + \kapup).
    \end{align} 
    
    Then, the variable-length version of \cref{Prot:PM Protocol} is $\ePA$-secret, $\eEV$-correct, and hence $(\ePA+\eEV)$-secure, whenever the length $\lvar$ of the final key (when error verification does not abort) is chosen as some function of $\cP_1^n$ satisfying
    \begin{equation}\label{eq:Variable key length bound}
    \begin{split}
        \lvar \leq \max \Bigg\{0, \ffull(\cP_1^n) &- \leak(\cP_1^n) - \ceil{\log\frac{1}{\eEV}} \\
        &- \frac{\alpha}{\alpha-1} \log\frac{1}{\ePA} + 2 \Bigg\},
    \end{split}
    \end{equation}
    where \(\leak(\cP_1^n)\) is the length of error correction data used. In particular, this holds if we define the function 
    \begin{align}\label{eq:protcol tradeoff function}
        \fprot(\cP_1^n) \defvar \sum_i (f(\cP_i) + \kapupbnd),
    \end{align} 
    where $\kapupbnd$ is any lower bound on $\kapup$, and choose
    \begin{equation}\label{eq:Variable key length exact}
    \begin{split}
        \lvar = \max \Bigg\{0, \Bigg\lfloor \fprot(\cP_1^n) &- \leak(\cP_1^n) - \ceil{\log\frac{1}{\eEV}} \\
        &- \frac{\alpha}{\alpha-1} \log\frac{1}{\ePA} + 2 \Bigg\rfloor \Bigg\}.
    \end{split}
    \end{equation} 
\end{theorem}

\begin{rem}\label{rem:kappup lower bounded by kappa}
For the purposes of this work, we will consider the variable-length protocol to be implemented by setting $\lvar$ as specified in \cref{eq:Variable key length exact}. Observe that to do so, we need to compute a lower bound $\kapupbnd$ on the $\frenyiSandUp_\alpha$-normalization constant $\kapup$, but we are free to use any method that securely computes such a lower bound. 
As shown in \cite[Lemma~6.2]{arqand_generalized_2024}, the two versions of $f$-weighted \Renyi entropy are related by \(\frenyiSandUp_\alpha(Q|\CP Q')_{\rho} \geq \frenyiSandDown_\alpha(Q|\CP Q')_{\rho}\), which means the $\frenyiSandUp_\alpha$-normalization constant $\kapup$ is always lower bounded by the $\frenyiSandDown_\alpha$-normalization constant $\kappa$. Hence, throughout the rest of this work, we will often focus only on evaluating or lower bounding $\kappa$ rather than $\kapup$ (as $\kappa$ is more straightforwardly compatible with our $\renyiSandDown_\alpha$-based analysis in the preceding sections) without further elaboration, understanding that this indeed provides a valid lower bound on $\kapup$.
\end{rem}

The main claim in \cref{thrm:Variable-length security} was proven in Ref.~\cite{inprep_HB24}. However, at the current time of writing, the manuscript for that work is still in preparation. Thus for the sake of completeness, we re-state here the essential steps of their analysis, but we emphasize that this is simply an exposition of their proof, and that work should be cited as the source whenever possible.

\begin{proof}
As stated in \cref{Def:Variable length eps security}, secrecy and correctness imply security. Therefore, we prove each part separately.
As a reminder, in line with the protocol description, we use the following register names, \(K_A\) for Alice's final key, \(L\) for error correction and verification data, \(G\) storing the choice of hash function, and \(I,\CP\) to indicate the different public announcements. 

Starting with correctness; the protocol is \(\eEV\)-correct since,
\begin{align}
    &\Pr[K_A\neq K_B\land \OmegaAcc]\leq \Pr[K_A\neq K_B \wedge \Omega_{\mathrm{EV}}] \\ &\leq \Pr[S_1^n \neq \widebar{S}_1^n \wedge \OmegaEV]\leq\Pr[\OmegaEV|S_1^n \neq \widebar{S}_1^n] \\ &\leq 2^{-\ceil{\log \frac{1}{\eEV}}}\leq \eEV.
\end{align}

Next, to show secrecy, we observe the following critical bound: by \cite[Theorem 4.1a and Corollary 4.2]{arqand_marginal-constrained_2025} (also proven in~\cite{inprep_HB24,fawzi_additivity_2025}), the state $\omega$ at the end of the public announcement step satisfies
\begin{align}\label{eq:chainMEAT}
\widetilde{H}^{\uparrow,\ffull}_{\alpha}(S_1^n| \CP_1^n I_1^n \tilde{E} )_\omega \geq 0.
\end{align}
With this, we can prove that the variable-length definition of secrecy in \cref{Def:Variable length eps security} holds. We again highlight that the main steps shown here were first presented in \cite{inprep_HB24}. We find:
\begin{widetext}
\begin{align}
    &\frac{1}{2} \sum_{m=0}^{\infty} \Pr[\OmegaLen{m}] \Bigg\lVert \sigma_{K_A I_1^n L E G| \OmegaLen{m}}  - \mathbb{U}_{K_A}^m \otimes \sigma_{I_1^n L E G| \OmegaLen{m}} \Bigg\rVert_1 \\
    &\leq \frac{1}{2} \sum_{\cP_1^n} \Pr[\CP_1^n = \cP_1^n \wedge \OmegaEV ] \Bigg\lVert \sigma_{K_A I_1^n L E G| \CP_1^n = \cP_1^n \wedge \OmegaEV}  - \mathbb{U}_{K_A}^m \otimes \sigma_{I_1^n L E G| \CP_1^n = \cP_1^n \wedge \OmegaEV} \Bigg\rVert_1 \label{eq:Proof var length line 2} \\
    &\leq \sum_{\cP_1^n \text{ s.t. } \lvar>0} \Pr[\CP_1^n = \cP_1^n \wedge \OmegaEV ] 2^{\frac{1-\alpha}{\alpha}\big(\renyiSandUp_{\alpha}(S_1^n|I_1^n L E)_{\omega|\CP_1^n = \cP_1^n \wedge \OmegaEV }-\lvar + 2 \big)} \label{eq:Proof var length line 3} \\
    &\leq \sum_{\cP_1^n \text{ s.t. } \lvar>0} \Pr[\CP_1^n = \cP_1^n \wedge \OmegaEV ] 2^{\frac{1-\alpha}{\alpha}\big(\renyiSandUp_{\alpha}(S_1^n|I_1^n L E)_{\omega|\CP_1^n = \cP_1^n} - \frac{\alpha}{\alpha-1} \log \frac{1}{\Pr[\OmegaEV|\CP_1^n = \cP_1^n]} - \lvar + 2 \big)} \label{eq:Proof var length line 4} \\
    &= \sum_{\cP_1^n \text{ s.t. } \lvar>0} \Pr[\CP_1^n = \cP_1^n ] 2^{\frac{1-\alpha}{\alpha}\big(\renyiSandUp_{\alpha}(S_1^n|I_1^n L E)_{\omega|\CP_1^n = \cP_1^n} - \lvar + 2 \big)} \label{eq:Proof var length line 5} \\
    &\leq \sum_{\cP_1^n \text{ s.t. } \lvar>0} \Pr[\CP_1^n = \cP_1^n ] 2^{\frac{1-\alpha}{\alpha}\big(\renyiSandUp_{\alpha}(S_1^n|I_1^n L E)_{\omega|\CP_1^n = \cP_1^n} - \ffull(\cP_1^n) + \leak(\cP_1^n) + \ceil{\log\frac{1}{\eEV}} + \frac{\alpha}{\alpha-1} \log\frac{1}{\ePA} \big)} \label{eq:Proof var length line 6} \\
    &\leq \sum_{\cP_1^n \text{ s.t. } \lvar>0} \Pr[\CP_1^n = \cP_1^n ] 2^{\frac{1-\alpha}{\alpha}\big(\renyiSandUp_{\alpha}(S_1^n|I_1^n E)_{\omega|\CP_1^n = \cP_1^n} - \ffull(\cP_1^n) + \frac{\alpha}{\alpha-1} \log\frac{1}{\ePA} \big)} \label{eq:Proof var length line 7} \\
    &= \ePA \sum_{\cP_1^n \text{ s.t. } \lvar>0} \Pr[\CP_1^n = \cP_1^n ] 2^{\frac{1-\alpha}{\alpha}\big(\renyiSandUp_{\alpha}(S_1^n|I_1^n E)_{\omega|\CP_1^n = \cP_1^n} - \ffull(\cP_1^n) \big)} \label{eq:Proof var length line 8} \\
    &\leq \ePA \sum_{\cP_1^n} \Pr[\CP_1^n = \cP_1^n ] 2^{\frac{1-\alpha}{\alpha}\big(\renyiSandUp_{\alpha}(S_1^n|I_1^n E)_{\omega|\CP_1^n = \cP_1^n} - \ffull(\cP_1^n) \big)} \label{eq:Proof var length line 9} \\
    &= \ePA 2^{\frac{1-\alpha}{\alpha}\big(\widetilde{H}^{\uparrow,\ffull}_{\alpha}(S_1^n|I_1^n\CP_1^nE)_{\omega} \big)} \leq \ePA \label{eq:Proof var length line 10}.
    %\leq \ePA 2^{\frac{1-\alpha}{\alpha}\big(\widetilde{H}^{\downarrow,\ffull}_{\alpha}(S_1^n|\tilde{T}_1^n\CP_1^nE)_{\omega} \big)}
\end{align}
\end{widetext}
The second line (\cref{eq:Proof var length line 2}) is due to the key length \(m\) depending only on the announcements \(\cP_1^n\) and error verification succeeding. Furthermore, for any \(\cP_1^n\) leading to the same key length \(m\), we used strong convexity to bound the trace distance. Thus, one can rewrite the sum in terms of \(\cP_1^n\) as in the second line. The third line (\cref{eq:Proof var length line 3}) follows by applying the Renyi LHL (\cref{thrm:Renyi LHL}) together with the fact that the trace-distance term is zero whenever $\lvar=0$ (note that here we view $\lvar$ as the function of $\cP_1^n$ satisfying \cref{eq:Variable key length bound}, observing that in all the trace-distance terms in the preceding expression, we have conditioned on error verification accepting and thus the key length would indeed be given by that function). The fourth line (\cref{eq:Proof var length line 4}) is due to \cite[Lemma B.5]{dupuis_entropy_2020}. Line six (\cref{eq:Proof var length line 6}) simply inserts \cref{eq:Variable key length bound} (recalling the sum is restricted only to terms with $\lvar>0$) and line seven (\cref{eq:Proof var length line 7}) follows by using the chain rule of \cite[Eqn. (5.94)]{tomamichel_quantum_2016}, which combines \cite[Lemmas 5.3 \& 5.4]{tomamichel_quantum_2016}. The ninth line (\cref{eq:Proof var length line 9}) holds simply because we have extended the sum with only non-negative terms, and the tenth line (\cref{eq:Proof var length line 10}) follows by the definition of $f$-weighted \Renyi entropies, \cref{def:f-weighted entropies}. Finally, the last inequality follows from the fundamental bound noted in \cref{eq:chainMEAT} above. 
\end{proof}

\subsection{Properties of variable-length key rates}
We now discuss an important connection between our variable-length key lengths and the expected key length.
First, recall the definition of the observed frequency vector $\Fobs$ from \cref{eq:freqdefn}. Observe that given any $\kapupbnd$ as described in \cref{thrm:Variable-length security}, the resulting function $\fprot$ applied to any \(\cP_1^n\) can be rewritten in terms of \(\Fobs\) as follows:
\begin{equation}
    \fprot(\cP_1^n) = n \left(\mbf{f}\cdot \Fobs
    + \kapupbnd\right),
\end{equation}
where $\mbf{f}$ denotes the tradeoff function $f:\alphCP \to \mathbb{R}$ viewed as a vector $\mbf{f} \in \mathbb{R}^{|\alphCP|}$.
It thus follows that the expected value of this quantity is (since expectation commutes with affine functions):
\begin{align}
    \EX[\fprot(\cP_1^n)] = n\left(\mbf{f}\cdot \EX[\Fobs] + \kapupbnd\right).
\end{align}
If we write \(\mbf{q}^\mathrm{hon}\) to denote the probability vector produced on a single-round test register $\CP$ (that is to say, the probability of obtaining outcome \(\cP\) is given by the component $\mathrm{q}^\mathrm{hon}(\cP)$ of the vector \(\mbf{q}^\mathrm{hon}\)) by an honest IID channel behavior, then the expected value of the frequency vector given by this behavior is connected to it by
\begin{equation}
    \EX[\Fobs] = \mbf{q}^\mathrm{hon}.
\end{equation}
Therefore, the expectation value of the function \(\fprot\) satisfies
\begin{align}
    \EX[\fprot(\cP_1^n)] = n\left(\mbf{f}\cdot \mbf{q}^\mathrm{hon} + \kapupbnd\right).
\end{align}

With this, we see that if the error correction procedure is chosen such that it satisfies \(\EX[\leak(\cP_1^n)] \leq \leak( \mbf{q}^\mathrm{hon} ) \), 
and has negligible probability of failing for the honest behaviour,
then when the key length is chosen according to \cref{eq:Variable key length exact}, its expected value would satisfy 
\begin{equation}
    \begin{split}
    \EX[\lvar] \geq &n \left(\mbf{f}\cdot \mbf{q}^\mathrm{hon} + \kapupbnd \right) - \leak( \mbf{q}^\mathrm{hon} ) - \ceil{\log\frac{1}{\eEV}} \\
    &- \frac{\alpha}{\alpha-1} \log\frac{1}{\ePA} + 2,
    \end{split} \label{eq:expectedvarlengthrate}
\end{equation}
since the maximum on the right-hand-side of \cref{eq:Variable key length exact} is lower bounded by its second argument (here we ignored the minor effect of taking the floor function). Note that the above expression is simply equal to the key length in \cref{eq:Variable key length exact} evaluated at the expected frequency distribution of the honest channel. In the subsequent sections, when plotting the expected key rates in the variable-length case, we will be taking the expression in \cref{eq:expectedvarlengthrate} as the formula for the expected key length. (While this implicitly assumes that the error correction procedure satisfies the stated conditions, without these conditions it would typically not be possible to find a closed-form expression for the expected key rates of variable-length protocols, see e.g. \cite{kamin_improved_2025}.\footnote{In any case, the \(\EX[\leak(\cP_1^n)] \leq \leak( \mbf{q}^\mathrm{hon} ) \) condition holds whenever the error correction procedure satisfies $\leak(\cP_1^n) = g(\Fobs)$ for some concave function $g$ of the observed frequency distribution, since in that case we have $\EX[\leak(\cP_1^n)] = \EX[g(\Fobs)] \leq g(\EX[\Fobs]) = \leak( \mbf{q}^\mathrm{hon} )$.} Hence this simplification has also been implicitly used in many previous works computing expected key rates of such protocols.)

\subsection{Optimality of the tradeoff function}
Note that while \cref{thrm:Variable-length security} is valid for {any} choice of the (single-round) tradeoff function \(f\), it does not by itself indicate how to choose $f$ in a manner that produces good finite-size performance. 
However, in \cite[Lemma~4.12]{arqand_marginal-constrained_2025}, a method was presented to address this, by giving a numerical technique to find the optimal $f$ in
\begin{equation}\label{eq:dualf}
\sup_{f} \left(\mbf{f}\cdot \mbf{q}^\mathrm{hon} + \kapup\right).
\end{equation}
Critically, observe that this is equivalent to finding the optimal choice of $f$ that maximizes our expected key length formula in \cref{eq:expectedvarlengthrate} for $\kapupbnd=\kapup$, i.e.~the tightest possible lower bound $\kapupbnd$. Hence we shall use this method to choose $f$, as it would maximize that formula for the $\kapupbnd=\kapup$ case, and any suboptimality in the expected key rate caused by using a smaller $\kapupbnd$ would then be at most the difference $\kapup-\kapupbnd$. 

Specifically, this method simply consists of modifying the optimization problem for \(\hQKD\) by introducing an auxiliary variable \(\bsym{\lambda} \in \mathbb{P}(\alphCP) \) that is a probability vector on register \(\CP\), leading to a modified optimization problem
\begin{align}\label{eq:halpha tradeoff}
        % \hQKD = 
        &\inf_{\substack{
        % \mbf{q} \in \Sacc, \\ 
        J \in \mathrm{Pos}(A'B), \\
        \varphi \geq 0, \; \bsym{\lambda} \in \mathbb{P}(\alphCP) }} \Bigg( \frac{\alpha}{\alpha - 1}\varphi  + q^\mathrm{hon}(\gen) g_{\alpha}(J) \Bigg) \\
        \textrm{s.t. } &\Tr_B[J] = \idop_{A'}, \nonumber \\
        &\varphi \geq D\rel{\bsym{\lambda}}{\bsym{\nu}_{\CP}}, \nonumber \\
        &\bsym{\nu}_{\CP} = (\gamma \bsym{\nu}_{\CP|\test}, 1-\gamma)^T, \nonumber \\
        &\bsym{\nu}_{\CP|\test} = \probstJ[J], \nonumber \\
        &\mbf{q}^\mathrm{hon} - \bsym{\lambda} = 0 . \tag{solve for dual} \label{eq:dualizedconstr}
\end{align}
This optimization problem is written as the epigraph form of \cref{eq:halpha final} with a fixed $\mbf{q}^\mathrm{hon}$. Thus, since the optimization of $\hQKD$ is jointly convex, this optimization problem is jointly convex by \cite[Section 4.1.3]{boyd_convex_2004}.

With regards to this optimization problem,~\cite[Lemma~4.12]{arqand_marginal-constrained_2025} states\footnote{In fact, that lemma handles a more general family of optimizations where rather than the single value $\mbf{q}^\mathrm{hon}$, one has a minimization over some convex set, similar to the minimization over $\mbf{q}\in\Sacc$ in \cref{eq:defn_hQKD}. Also, it furthermore states that the optimal values in \cref{eq:dualf} and \cref{eq:halpha tradeoff} are in fact equal to each other. However, we do not use that property in this work, since the actual protocol implementation using \cref{eq:Variable key length exact} would be based on the actual $f$ we choose rather than the theoretically optimal choice, and numerical methods would in general not find \emph{exactly} the latter. (We emphasize however that this does not affect the security of our results; see \cref{rem:reliable}.)} that the optimal $f$ in \cref{eq:dualf} is given by the optimal dual variables to the constraint \(\mbf{q}^\mathrm{hon} - \bsym{\lambda} = 0\) labeled by \eqref{eq:dualizedconstr} in the above expression, interpreting those dual variables as a tuple \(\mbf{f} \in \R^{\abs{\alphCP}}\) defining \(f(\cP)\) for all \(\cP\in\alphCP\). 
We can apply Frank-Wolfe methods to this optimization problem in a similar fashion to solving for $\hQKD$, which in the process also returns a dual solution to that constraint --- this hence gives us a choice of $f$ which should be close to optimal, up to the convergence of our numerical methods.

\begin{rem}\label{rem:reliable}
We stress that while numerical methods for finding the dual solution might not find the exact optimal choice of $f$, this does not affect the security of our results. This is because \cref{thrm:Variable-length security} holds for \emph{any} choice of $f$, and hence using a suboptimal $f$ would only mean that the resulting key rates we obtain might not be optimal. The main requirement for security is to ensure that we compute a rigorous lower bound $\kapupbnd$ on $\kapup$, which we now discuss.
\end{rem}

\subsection{Computing the normalization constant \texorpdfstring{$\kappa$}{k}}
Given a choice of $f$, to compute key lengths according to \cref{eq:Variable key length exact}, we now need to find a lower bound $\kapupbnd$ on the $\frenyiSandUp_\alpha$-normalization constant \(\kapup\). (There exists independent ongoing work on developing interior-point methods for this computation~\cite{inprep_HB24}; however, in this work we apply the Frank-Wolfe algorithm due to its simplicity.) 

As already mentioned in \cref{rem:kappup lower bounded by kappa}, the \(\frenyiSandDown_{\alpha}\)-normalization constant is a lower bound on the \(\frenyiSandUp_{\alpha}\)-normalization constant. Thus, we will find $\kapupbnd$ by minimizing \(\frenyiSandDown_{\alpha}\), because it is directly compatible with our methods shown in \cref{sec:Calculating halpha}. 

Therefore, similar to \cref{eq:halpha final} we now simplify the optimization problem for 
\begin{equation}
    \kappa = \inf_{\substack{\rho_{AB} \\ \text{s.t.} \Tr_B[\rho_{AB}] = \tau_A}} \frenyiSandDown_\alpha(S|\CP I E)_{\nu},
\end{equation}
and rewrite it in terms of the Choi state \(J\) of Eve's channel. First, as in the proof of \cref{thrm:Fixed-length security generic bnd}, we note that 
\begin{equation}
    \renyiSandDown_{\alpha}(S|IE)_{\nu^{\rho}_{|\gen}} = \renyiSandDown_{\alpha}(S|\tilde{T}E)_{\nu^{\rho}_{|\gen}} = g_{\alpha}(J).
\end{equation}
Additionally, we note that the entropy in test rounds is zero, and applying our definition of \(\probstJ\) of \cref{eq:def probstJ} we can write
\begin{equation}
    \Pr(\cP|\test) = \probstJ[J].
\end{equation}
Thus, we can rewrite the optimization problem of the $\frenyiSandDown_\alpha$-normalization constant \(\kappa\) as
\begin{align}\label{eq:kappa const}
\begin{aligned}
    \kappa = &\inf_{\substack{J \in \mathrm{Pos}(A'B), \\ \bsym{\nu} \in \mathbb{P}(\Ct) }} \Bigg( \frac{1}{1-\alpha} \log \Big( \gamma \sum_{\cP\in \Ct} \bsym{\nu}(\cP)2^{(\alpha - 1) f(\cP)} \\
    &\qquad + \left(1-\gamma\right)2^{-(\alpha-1)\left(g_{\alpha}(J) - f(\gen)  \right)} \Big) \Bigg) \\
    \quad \textrm{s.t. } &\Tr_B[J] = \idop_{A'}, \\
    &\bsym{\nu} = \probstJ[J],
    \end{aligned}
\end{align}
where we defined \(\bsym{\nu}\defvar \Pr(\cP|\test) \) and assigned \(f(\cP) = \mathrm{f}_{\cP}\). This optimization problem is again a convex problem that can be solved by iterative methods such as the Frank-Wolfe algorithm. In the appendix, in \cref{thrm:kappa convex}, we prove the joint convexity in \(J\) and \(\bsym{\nu}\) of the optimization problem for computing \(\kappa\).

\begin{rem}
    We emphasize that for an experimental implementation of a QKD protocol utilizing our \Renyi framework, one is only required to find the gradient of the tradeoff function \(\fprot\) \emph{once} before running the protocol. Moreover, since \(\fprot\) is an affine function determined by its gradient \(\mbf{f}\) and the $\frenyiSandDown_\alpha$-normalization constant \(\kappa\), it can be implemented very efficiently. This is a huge advantage compared to other proof techniques such as the postselection technique~\cite{nahar_postselection_2024} or EUR- and phase error correction correction-based techniques~\cite{tupkary_phase_2024,curras-lorenzo_finite-key_2021}.
\end{rem}

We have now established the necessary framework for calculating both fixed- and variable-length key rates within the context of finite-dimensional protocols, specifically qubit-based implementations. Fixed-length key rates are derivable from \cref{eq:halpha final}, which in turn facilitates the direct determination of the tradeoff function via \cref{eq:halpha tradeoff}. Variable-length key rates can then be obtained by solving \cref{eq:kappa const}.

Therefore, in the subsequent section, we will present the qubit BB84 protocol as an example and provide a comparative analysis with existing proof techniques.

\section{Qubit BB84 and Comparisons with other Proof Techniques}\label{sec:Qubit BB84 and Comparisons with other Proof Techniques}
\newcommand{\lossparam}{\zeta_\mathrm{hon}}
\newcommand{\depol}{p^\mathrm{hon}_\mathrm{depol}}

Before we turn our focus to more advanced protocols, in this section we present a comparison between our variable-length key rates for the qubit BB84 protocol and variable-length key rates obtained in the EUR framework. For the comparison, we present EUR-based variable-length key rates from Ref.~\cite{tupkary_phase_2024} incorporating the improvements from \cite{mannalath_sharp_2025}, which at the time of this writing should amount to the best known variable-length key rates for this protocol. Other works using the phase error correction approach, such as e.g. \cite{curras-lorenzo_finite-key_2021}, would give similar results.

Our protocol choices are the following. The \(Z\)-basis is always used for key generation rounds and the \(X\)-basis is always used for test rounds. Hence, with probability \(1-\gamma = p_z\) Alice sends a state in the \(Z\)-basis and with \(\gamma =p_x\) she sends a state in the \(X\) basis. Bob in turn measures his incoming states with the same probabilities in the \(Z\)- and \(X\)-basis, respectively. 

For the error correction cost we assume
\begin{equation}
	\begin{split}
		\leak = &n p_{\text{sift}}^{\mathrm{hon}} \fEC H(X| Y; \text{sift} )_{\mathrm{hon}},
	\end{split}
\end{equation}
with \(\fEC =1.1\) (see e.g. \cite{elkouss_information_2011}) and \(p_{\text{sift}}^{\mathrm{hon}} = (1-\eta)(1-\gamma)^2 \) being the probability of a signal surviving the sifting procedure. Furthermore, we chose the security parameters \(\ePA = \eEV = \frac{1}{2}10^{-80}\) for a total security of \(\eSec = 10^{-80}\).

This choice is motivated by the Advanced Encryption Standard with a 256-bit key, in short AES256. The probability of randomly guessing such a key used for AES256 is \(2^{-256}\leq 10^{-80}\).

We model channel loss with \(\eta = 10^{-\frac{\lossparam}{ \unit[10]{dB}}}\) and depolarization with the channel
\begin{align}
\mathcal{E}_\mathrm{depol}[\rho] \defvar \left(1-\depol \right) \rho + \depol \frac{\idop}{2},
\end{align}
where for this example we chose \(\depol = 0.03\). Finally, to calculate the key rates of our work, we use the Kraus operators for the maps \(\GMap\), \(\ZMap\) and states \(\ket{\xi^t}\), \(\ket{\xi^g}\) as stated in \cref{app:Kraus ops qubit BB84}.

\begin{figure}[th]
	\centering
	\includegraphics[width=\linewidth]{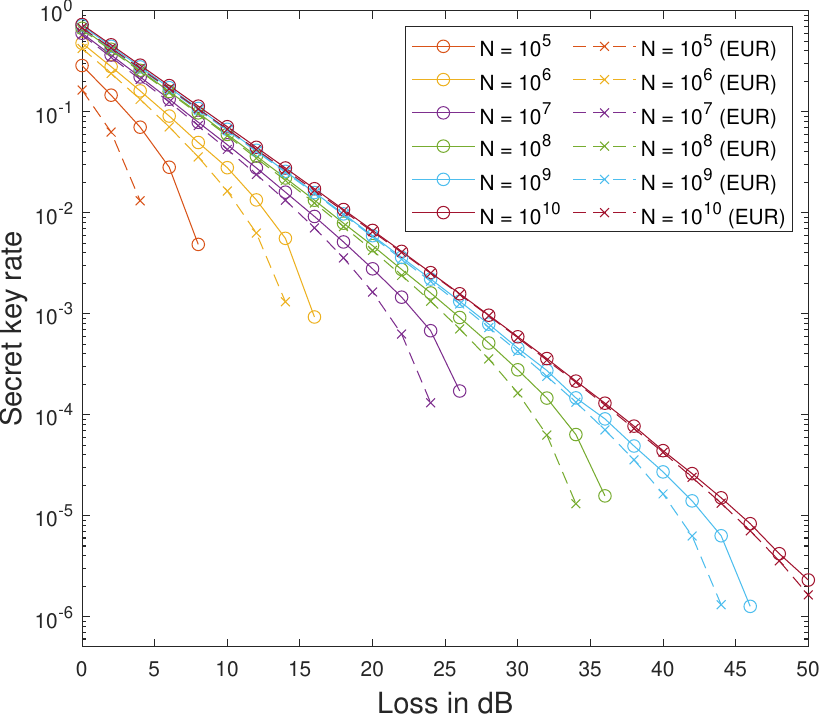}
	\caption{Results for variable-length secret key rates of the qubit BB84 protocol plotted against the channel loss in \(\unit{dB}\) and a constant depolarization with \(\depol = 0.03\) with a varying number of total signals sent \(N=10^5, \dots, 10^{10} \). The security parameters, and error correction efficiency were chosen as \(\ePA = \eEV = \frac{1}{2}10^{-80}\) and \(\fEC=1.1\). As a comparison, we showcase results using the EUR approach of \cite{tupkary_phase_2024} incorporating the improvements from \cite{mannalath_sharp_2025}. For both approaches, we optimize the 
    probability $\gamma$ of using the
    \(Z\)-basis choice in each round, 
    and for our work we additionally optimize the \Renyi parameter \(\alpha\).}
	\label{fig:QubitBB84Depol}
\end{figure}

The resulting key rates for both this work and the comparison with EUR-based security proofs can be seen in \cref{fig:QubitBB84Depol}. For both approaches we optimized the testing probability \(\gamma\) and for our work additionally the \Renyi parameter \(\alpha \in (1,2)\).

Especially for small block sizes our approach significantly outperforms the EUR key rates. Although not shown explicitly, we are even able to achieve a positive key rate with \(n=10^4\) signals, but only at zero loss, hence we omitted this data point for clarity of the plot. With larger block sizes, the differences become smaller and smaller and both techniques converge to more or less the same key rates. 

At least for qubit protocols, this example justifies our claim of achieving the best key rates to date. Moreover, in contrast to EUR- or complementarity-based security proofs, we can incorporate any detection setup.

\section{Extension to Block-Diagonal States}\label{sec:Block-Diag states}
In this section, we derive the application of \cref{cor:Fixed-length security lower bnd halpha,thrm:Variable-length security} to e.g. decoy-state protocols. \Cref{thrm:halpha QKD} as it is stated can already be validly applied to decoy-state protocols; however, the system \(A'\) at least would have an infinite-dimensional Hilbert space, albeit a separable one. 
WCP sources exhibit a special block-diagonal structure in the photon number. Hence, to solve the issue of infinite dimensions and simplify calculations, we exploit this block-diagonal structure exhibited in the signal states. The approach chosen here is similar to other works; see e.g. \cite{kamin_finite-size_2025} in the GEAT framework, and \cite{li_improving_2020} in the asymptotic limit.

Therefore, let us assume:
\begin{enumerate}[(1)]
    \item \label{Decoy assump block-diag} The signal states \(\{\sigma_i\}_{i=1\dots d_A}\) are simultaneously block-diagonal in some basis with block numbers \(m=0,1\dots\), such that we can write \(\ket{\xi^g}\) and \(\ket{\xi^t}\) as
    \begin{align}\label{eq:blocksum states}
        \ket{\xi^g}_{A_S AA'} &= \sum_{m=0}^{\infty} \sqrt{p(m|\gen)} \ket{m}_{A_S} \ket{\xi^g_{|m}}_{AA'}, \\
        \ket{\xi^t}_{A_S AA'} &= \sum_{m=0}^{\infty} \sqrt{p(m|\test)} \ket{m}_{A_S} \ket{\xi^t_{|m}}_{AA'}.
    \end{align} 
    \item In addition to the system $A$, the system \(A_S\) (often called a shield system) also remains with Alice and is inaccessible to Eve. \label{Decoy assump shield}
    \item Eve can perform a quantum non-demolishing (QND) measurement of system \(A'\) and learn the block-diagonal structure without disturbing the state. \label{Decoy assump QND Eve}
\end{enumerate}

Note that assumption~\ref{Decoy assump QND Eve} is made without loss of generality. This can be seen with the following construction based on a decomposition for WCP sources in \cite[App. B]{li_improving_2020}, which is also valid for any block-diagonal state where Eve could perform a QND measurement.

Following Ref.~\cite{li_improving_2020}, for a single round, Eve's attack in Stinespring form, including a QND measurement, can be written as
\begin{equation}
    V_{A'\rightarrow BE \tilde{E}} = \sum_m V_{A'\rightarrow BE}^m \Pi^m_{A'}\otimes \ket{m}_{\tilde{E}},
\end{equation}
where \(\Pi^m\) is a projector for the QND measurement and register \(\tilde{E}\) stores the block number \(m\). On the other hand, if Eve does not attempt to measure the block number, her attack acts the same on each block, i.e. \(V_{A'\rightarrow BE}^m = V_{A'\rightarrow BE}\) for all \(m\).

Hence, for states satisfying assumption~\ref{Decoy assump block-diag}, any feasible point in \cref{eq:halpha final}, characterizing Eve's channel applied to system \(A'\), is also a feasible point in an optimization over direct-sum channels. Thus, we only further reduce the infimum when the optimization is performed over direct-sum channels, and assumption~\ref{Decoy assump QND Eve} does not restrict Eve's attack.

In summary, the assumptions \ref{Decoy assump block-diag}--\ref{Decoy assump QND Eve} are naturally satisfied for WCP sources, and have been exploited in many works \cite{hwang_quantum_2003,lo_decoy_2004,ma_practical_2005} to formulate the first analytical decoy-state methods and have also been extended to numerical methods \cite{wang_numerical_2022,Rice2009}. We will discuss how these assumptions are met in more detail in \cref{sec:Decoy state}.

Additionally, let us also quickly discuss how one can satisfy assumption~\ref{Decoy assump block-diag} if Eve cannot perform a QND measurement. In this case, one can apply a source-map \cite{nahar_imperfect_2023} that announces the block number \(m\) to Eve, which intuitively only reduces the infimum. In \cref{app:Source maps}, we formally show that such source maps can be incorporated and we make use of source-maps in \cref{sec:Phase imperfections}.

When there are shield systems, we slightly modify the definition of the single-round channel $\EATchannQKD$ so that it acts on $A_S A B$ rather than just $A B$. In most scenarios, we take $\EATchannQKD$ to just apply the operations already described in \cref{def:QKD channel} on $A B$, while simply tracing out $A_S$. However, for the sake of generality we note that it can also be allowed to perform joint operations across all the registers $A_S A B$ if necessary. Then when applying the MEAT, we take the marginal constraints to be imposed on all the $A_S A$ registers, rather than just the $A$ registers --- this is valid due to assumption~\ref{Decoy assump shield}, i.e.~Eve cannot act on the $A_S A$ registers. Finally, when performing the single-round analysis, we again make the observation that we can ``undo'' the source-replacement argument, viewing a minimization over states $\rho_{A_S ABE}$ satisfying a marginal constraint on $A_S A$ as being equivalent to minimizing over Eve's channel acting on $A'$.

Now, we will incorporate the block-diagonal structure into the optimization problem for \(\hQKD\) and the $\frenyiSandDown_\alpha$-normalization constant \(\kappa\). As a consequence of assumption~\ref{Decoy assump QND Eve} and the discussion above, Eve's channel \(\mathcal{E}\) can be written as a direct sum acting on each block \(m\) separately,
\begin{equation}
    \mathcal{E}= \bigoplus_{m=0}^{\infty} \mathcal{E}_m,
\end{equation}
or equivalently, in terms of Choi states
\begin{equation}
    J = \bigoplus_{m=0}^{\infty} J_m,
\end{equation}
see for example \cite[Sec. 7.2]{kamin_finite-size_2025}. Moreover, assumptions \ref{Decoy assump block-diag}--\ref{Decoy assump QND Eve} imply that the state \(\rho_{A_SAB|\test}\) has the form
\begin{equation}\label{eq:Blocksum rho}
    \rho_{A_SAB|\test} = \sum_{m=0}^{\infty} p(m|\test) \ketbra{m}{m} \otimes \rho_{AB|\test,m},
\end{equation}
and similarly for \(\rho_{A_SAB|\gen}\). An equivalent decomposition was also found in \cite[App. B]{li_improving_2020}. 

Next, we will simplify the expression for \(\hQKD\) in \cref{eq:halpha final} such that only finite-dimensional spaces and finite sums are involved. We will start with the entropy and later simplify the constraints.

\subsection{Bounding the Entropy}\label{subsec:Bounding the Entropy}
Since, by assumption~\ref{Decoy assump QND Eve}, Eve can always perform a QND measurement, we can assume her quantum side-information to include a classical register \(M\) determining the block-number \(m\). Thus, by exploiting the properties of \Renyi entropies when conditioned on classical registers \cite[Sec. III.B.2 and Prop. 9]{muller-lennert_quantum_2013} (or \cite[Prop. 5.1]{tomamichel_quantum_2016}), we find for the entropy in a generation round
\begin{equation}
    \begin{split}
        &\renyiSandDown_{\alpha}(S|\tilde{T}EM)_{\nu^{\rho}_{|\gen}} \\ 
        &= \frac{1}{1-\alpha} \log\left( \sum_{m=0}^{\infty} p(m|\gen) 2^{(1-\alpha)\renyiSandDown_{\alpha}(S|\tilde{T}E)_{\nu^{\rho}_{|m,\gen}}} \right).
    \end{split}
\end{equation}
Because \(\nu_{S\tilde{T}E|m,\gen}\) is classical on \(S\)  (and thus is a separable state across $S$ and $\tilde{T}E$), we have \(\renyiSandDown_{\alpha}(S|\tilde{T}E)_{\nu^{\rho}_{|m,\gen}} \geq 0\) by~\cite[Lemma 5.2]{tomamichel_quantum_2016}. Hence, picking a cut-off \(\Nent \in \N_0\) allows us to bound the entropy by
\begin{equation}
    \begin{split}
        &\renyiSandDown_{\alpha}(S|\tilde{T}EM)_{\nu^{\rho}_{|\gen}} \\ 
        &\geq \frac{1}{1-\alpha} \log\Bigg\{ 1- \sum_{m \leq \Nent} p(m|\gen) \\
        &\quad +  \sum_{m \leq \Nent} p(m|\gen) 2^{(1-\alpha)\renyiSandDown_{\alpha}(S|\tilde{T}E)_{\nu^{\rho}_{|m,\gen}}} \Bigg\},
    \end{split}
\end{equation}
where the inequality can be made arbitrarily tight by increasing \(\Nent\).

Finally, in line with \cref{eq:defn galpha}, we aim to define \(g_{\alpha}^{\Nent}\), a version of \(g_{\alpha}\) incorporating the cutoff. Therefore, similarly to \cref{eq:xi test and gen}, let us define the CPTNI maps \(\chi_{g/t}^m: A' \rightarrow A\),
\begin{equation}
    \begin{split}
        \chi_g^m[X] &\defvar \Tr_{A'}\left[\left( \idop_A \otimes X^{T_{A'}} \right)  \ketbra{\xi^g_{|m}}{\xi^g_{|m}}  \right], \\
         \chi_t^m[X] &\defvar \Tr_{A'}\left[ \left( \idop_A \otimes X^{T_{A'}} \right) \ketbra{\xi^t_{|m}}{\xi^t_{|m}} \right],
    \end{split}
\end{equation}
which are now based on the states conditioned on block \(m\).

This lets us write
\begin{equation}\label{eq:nu_m of J}
    \nu_{S\tilde{T}\CP E|m,\gen} = \EATchannQKD \circ \chi^m_{g}[J_m],
\end{equation}
for all \(m\in \N_0\). Thus, we define \(g_{\alpha}^{\Nent}\) incorporating the cutoff \(\Nent\) by
\begin{equation}
    \begin{split}
        &g_{\alpha}^{\Nent}\left(J_0,\dots, J_{\Nent} \right) \\ 
        &\defvar \frac{1}{1-\alpha} \log\Bigg\{ 1- \sum_{m \leq \Nent} p(m|\gen) \\
        &\quad +  \sum_{m \leq \Nent} p(m|\gen) 2^{(1-\alpha)\renyiSandDown_{\alpha}(S|\tilde{T}E)_{\nu(J_m)_{|m,\gen}}} \Bigg\},
    \end{split}
\end{equation}
where \(\nu\) depends on \(J_m\) as stated above in \cref{eq:nu_m of J}.

\subsection{Bounding the Constraints}\label{subsec:Bounding the Constraints}
Next, we turn our attention to the constraints. Here, the only constraint in \cref{eq:halpha final} depending on Eve's channel is \(\bsym{\nu}_{\CP|\test} = \probstJ[J]\). In other words, we need to reformulate \(\probstJ\) using the direct-sum structure in Eve's channel.

Similarly to the discussion preceding the definition of the QKD channel \(\EATchannQKD\), see \cref{eq:test and gen partition ops} and below, we require the POVM elements conditioned on \(m\) and \(\test\) (or \(\gen\)). Again, under the generic source-replacement scheme these conditional POVM elements remain the same as the unconditional ones, as they are projectors. However, under a Schmidt decomposition the POVM elements not conditioned on block \(m\) may contain factors proportional to the probability of sending block \(m\). Then, one needs to appropriately incorporate the conditioning as described in \cref{app:Conditional POVM construction}.

For the discussion presented here, we again assume Alice's POVM elements are projectors, however, for clarity we indicate the conditioning on \(m\) by 
\begin{align}
    &\{M^{A|\test,m}_{\alpha,x} \}_{\alpha \in \mathcal{C}^{A,\test}, x \in \mathcal{X}^{\test}_{\alpha} } \\
    &\{M^{A|\gen,m}_{\alpha,x} \}_{\alpha \in \mathcal{C}^{A,\gen}, x \in \mathcal{X}^{\gen}_{\alpha} }.
\end{align}

Therefore, let us define the maps \(\probst_m\) generating the statistics conditioned on block \(m\),
\begin{align}
    \probst_m[\sigma] \defvar \sum_{\cP \in \Ct} \sum_{\substack{(\alpha,x,\beta,y) \\ \in \phi^{-1} (\cP)}} \Tr\left[\left(M_{\alpha,x}^{A|\test,m} \otimes M_{\beta,y}^{B}\right) \sigma \right] \hat{e}_{\cP},
\end{align}
and their concatenation with the maps \(\chi^m_t\) as
\begin{equation}\label{eq:def probstJm}
    \probstJ_m[J_m] \defvar \probst_m\circ \chi^m_t[J_m].
\end{equation}

Then, a straightforward calculation using \cref{eq:Blocksum rho} yields
\begin{equation}
    \bsym{\nu}_{\CP|\test} = \probstJ[J] = \sum_{m=0}^{\infty} p(m|\test) \probstJ_m[J_m].
\end{equation}
To consider only finitely many parts of the sum in the constraints, we pick another cutoff \(\Ncons \in \N_0\). Each entry of the vector \(\probstJ_m[J_m]\) must be greater than zero, thus
\begin{equation}
    \bsym{\nu}_{\CP|\test} \geq \sum_{m\leq \Ncons} p(m|\test) \probstJ_m[J_m].
\end{equation}
On the other hand, each entry in \(\probstJ_m[J_m]\) must be smaller than the probability of Alice sending a particular state determined by \((\alpha,x)\) (bit value and announcement) conditioned on block \(m\) and \(\test\). Hence, the outcomes in test rounds \(\bsym{\nu}_{\CP|\test}\) can be upper bounded by
\begin{equation}
    \begin{split}
        \bsym{\nu}_{\CP|\test} \leq 
        &\sum_{m\leq \Ncons} p(m|\test) \probstJ_m[J_m] +1 \\ 
        &- \sum_{\cP \in \Ct} \sum_{\substack{(\alpha,x) \\ \in \phi^{-1} (\cP)}} \sum_{m \leq \Ncons} p(\alpha,x,m|\test) \hat{e}_{\cP},
    \end{split} 
\end{equation}
and to simplify the notation, we define the vector
\begin{equation}\label{eq:ptot block diag}
    \mbf{p}_{\mathrm{tot}} \defvar \sum_{\cP \in \Ct} \sum_{\substack{(\alpha,x) \\ \in \phi^{-1} (\cP)}} \sum_{m \leq \Ncons} p(\alpha,x,m|\test) \; \hat{e}_{\cP}.
\end{equation}
In summary, we can bound \(\bsym{\nu}_{\CP|\test}\) by
\begin{align}\label{eq:nuc constr blocksum}
    \bsym{\nu}_{\CP|\test} &\leq 
    \sum_{m\leq \Ncons} p(m|\test) \probstJ_m[J_m]
    + 1-\mbf{p}_{\mathrm{tot}}, \\
    \bsym{\nu}_{\CP|\test} &\geq \sum_{m\leq \Ncons} p(m|\test) \probstJ_m[J_m],
\end{align}
and also these bounds can be made arbitrarily tight by now increasing \(\Ncons\).

\subsection{Reformulated Optimization problems}
When we combine both \cref{subsec:Bounding the Entropy} and \cref{subsec:Bounding the Constraints}, we can bound \(\hQKD\) by
\begin{equation}\label{eq:halpha blocksum}
    \begin{aligned}[t]
        \hQKD \geq 
        &\begin{aligned}[t]
            \inf_{\substack{\mbf{q} \in \Sacc, \bsym{\nu}_{\CP} \in \mathbb{P}(\alphCP) \\ J_m \in \mathrm{Pos}(A'B)}} \Bigg( &\frac{\alpha D\rel{\mbf{q}}{\bsym{\nu}_{\CP}}}{\alpha - 1} \\ 
            &+ q(\gen) g_{\alpha}^{\Nent}(J_0 \dots J_{\Nent}) \Bigg)
        \end{aligned}\\
        \textrm{s.t. } &\Tr_B[J_m] = \idop_{A'}, \\
        &\bsym{\nu}_{\CP} = (\gamma \bsym{\nu}_{\CP|\test}, 1-\gamma)^T, \\
        &\bsym{\nu}_{\CP|\test} \leq \sum_{m\leq \Ncons} p(m|\test) \probstJ_m[J_m]
        + 1-\mbf{p}_{\mathrm{tot}}, \\
        &\bsym{\nu}_{\CP|\test} \geq \sum_{m\leq \Ncons} p(m|\test) \probstJ_m[J_m].
        \end{aligned}
\end{equation}

Computing variable-length key rates requires a tradeoff function \(f\) and the $\frenyiSandDown_\alpha$-normalization constant \(\kappa\). Regarding the tradeoff function, similar to \cref{eq:halpha tradeoff}, one can add a constraint and an auxiliary variable \(\bsym{\lambda}\) and solve for its dual solution to find the tuple \(\mbf{f}\) defining the best choice of tradeoff function $f$. For the $\frenyiSandDown_\alpha$-normalization constant \(\kappa\), one can follow exactly the same steps as laid out above for \(\hQKD\) and finds
\begin{align}\label{eq:kappa const blocksum}
\begin{aligned}
    \kappa \geq &\inf_{\substack{J_m \in \mathrm{Pos}(A'B), \\ \bsym{\nu} \in \mathbb{P}(\Ct) }} \Bigg( \frac{1}{1-\alpha} \log \Big( \gamma \sum_{c\in \Ct} \bsym{\nu}(\cP)2^{(\alpha - 1) f(\cP)} \\
    &\qquad + \left(1-\gamma\right)2^{-(\alpha-1)\left(g_{\alpha}^{\Nent}(J_0,\dots J_{\Nent}) - f(\gen)  \right)} \Big) \Bigg) \\
    \quad \textrm{s.t. } &\Tr_B[J_m] = \idop_{A'}, \\
    &\bsym{\nu} \leq \sum_{m\leq \Ncons} p(m|\test) \probstJ_m[J_m]
    + 1-\mbf{p}_{\mathrm{tot}}, \\
    &\bsym{\nu} \geq \sum_{m\leq \Ncons} p(m|\test) \probstJ_m[J_m].
    \end{aligned}
\end{align}

Again, we emphasize that the inequalities in both \cref{eq:halpha blocksum} and \cref{eq:kappa const blocksum} can be made arbitrarily tight, by increasing the cutoffs \(\Nent\) and \(\Ncons\). Thus, with enough such terms, the fixed-length and variable-length key rates resulting from these lower bounds will converge to the best possible values available under this general proof approach.

\section{Decoy-state Protocols}\label{sec:Decoy state}
In decoy-state protocols, the state preparation and announcements fit the general framework laid out in \cref{Prot:PM Protocol}, but here we state slightly more details. In terms of the protocol, we replace step 1. (a) of \cref{Prot:PM Protocol} with (a') stated below. 

\begin{enumerate}[(a')]
    \item \textbf{State preparation and transmission:} In each round, with probability $\gamma$ Alice independently chooses it to be a \emph{test round} or \emph{generation round}. In the case of a generation round Alice prepares one of \(d_A\) states \(\{\ket{s_j}\}_{j=1\dots d_A}\) with intensity \(\mu_s\). For test rounds Alice selects the intensity \(\mu_j\) (from a finite predefined list) with probability \(p(\mu_j | \text{test})\). She stores the label for her choice of the signal state in a classical register \(X_i\), and computes a classical register \(C^A_i\) with alphabet \(\mathcal{C}^A\) for public announcement (including for instance the test/generation decision). Finally, Alice sends the signal state to Bob via a quantum channel.
\end{enumerate}	
\begin{rem}
    Our framework allows Alice to send all intensities in key generation rounds. In particular, all of the equations in this section are still valid in this situation.
    
    However, assuming that the signal intensity has the largest probability of sending a single photon, for optimal key rates only the signal intensity should be used in generation rounds.
\end{rem}

Decoy-state protocols possess even more structure than the generic format required to reach \cref{eq:halpha blocksum}. One can exploit this to further simplify the formulations of \cref{eq:halpha blocksum} and \cref{eq:kappa const blocksum}. Thus, in this section, we show how decoy-state protocols satisfy the assumptions \ref{Decoy assump block-diag}--\ref{Decoy assump QND Eve} and simplify the optimization problems for decoy-state protocols.

We start with the assumptions \ref{Decoy assump block-diag}--\ref{Decoy assump QND Eve}; in particular the block-diagonal structure of the signal states in \cref{eq:blocksum states}. As before, we include a shield system \cite{horodecki_general_2009}, and apply the source-replacement scheme \cite{bennett_quantum_1992, ferenczi_symmetries_2012}. Furthermore, we note that for weak coherent pulses the states in system \(A'\) sent to Bob with \(m\) photons do not depend on the intensity. Therefore, the states Alice prepares in generation rounds can be written as
\begin{equation}\label{eq:Decoy source replacement}
    \begin{split}
    \ket{\phi^g}_{A\bar{A}A_SA'} = \sum_{m=0}^{\infty} \sum_{a,\mu} &\sqrt{p(a,\mu,m|\gen)} \ket{a}_A \ket{\mu}_{\bar{A}} \\
    &\otimes \ket{m}_{A_S} \ket{s_m^{(a)}}_{A'},
    \end{split}
\end{equation}
where we abbreviated Alice's choice \((\alpha,x)\) as \(a\). A similar expression upon appropriate replacements holds for \(\test\) rounds. Again, we note that for weak coherent pulses the signal states \(\ket{s_m^{(a)}}_{A'}\) only depend on \(a\), and therefore, we stated no explicit dependence on the intensity \(\mu\). Hence, assumption~\ref{Decoy assump block-diag} is already satisfied.

This is one crucial difference from the generic block-diagonal framework presented above and we will exploit this fact in the following simplifications. For this purpose, we also kept the intensity choice in a separate register \(\bar{A}\). In the context of defining the channels $\EATchannQKD$ and applying the MEAT, we define $\EATchannQKD$ to act on $A \bar{A} A_S B$, and apply the marginal constraint to the $A \bar{A} A_S$ registers.

Next, assumption~\ref{Decoy assump shield} is trivially satisfied through the construction of the shield system. 

Finally, due to the photon number splitting attack \cite{bennett_experimental_1992,brassard_limitations_2000} we can assume without loss of generality that for WCP sources Eve always performs a QND measurement of the photon number first and then applies an attack based on the photon number. Thus, assumption~\ref{Decoy assump QND Eve} is satisfied as well, and by \cite[App. B]{li_improving_2020} the channel can immediately be decomposed in the block-diagonal form.

\subsection{Simplifications of the Optimization Problems}\label{subsec:Decoy Simplifications}
Now, we turn our attention to the simplifications specific to decoy-state protocols. The goal of the reformulations here is to find an equivalent form of the constraints in \cref{eq:halpha blocksum} that instead uses the statistics conditioned on each intensity choice \(\mu_j\) and has a similar form to commonly used decoy-state methods, e.g. \cite{wang_numerical_2022}. We again highlight that this reformulation heavily relies on phase randomized weak coherent pulses and is in general not possible.

First, we note that the vector \(\probstJ_m[J_m]\), is ordered by the announcements \(\mu\) and \(a =(\alpha,x)\) for Alice and \(b=(\beta,y)\) for Bob. Hence, we can write for each component of \(\probstJ_m[J_m]\)
\begin{align}
    &\left(\Psi_m[J_m]\right)_{\mu,a,b} \nonumber \\ 
    &= \Tr\left[\left(M_a^{A|\test,m}\otimes \dyad{\mu} \otimes M_b^B\right) \chi_m^t[J_m] \right] \\
    &= \Tr\left[\left(M_a^{A|\test,m} \otimes  M_b^B\right) \bra{\mu}\chi_m^t[J_m]\ket{\mu}\right].
\end{align}
To simplify the inner product \(\bra{\mu}\chi_m^t[J_m]\ket{\mu}\) further, let us define the states conditioned on Alice sending intensity \(\mu\),
\begin{equation}
    \ket{\xi_{|m,\mu}^t}_{AA'} \defvar \sum_{a} \sqrt{p(a|m,\mu,\test)}\ket{a}_A\ket{s^{(a)}_m}_{A'},
\end{equation}
where we again omitted any dependence on the intensity \(\mu\) of the states sent to Bob due to the properties of WCP sources. With these states we define the CPTNI map \(\chi_t^{m,\mu}\) as
\begin{equation}
     \chi_t^{m,\mu}[X] \defvar \Tr_{A'}\left[ \left( \idop_A \otimes X^{T_{A'}} \right) \ketbra{\xi^t_{|m,\mu}}{\xi^t_{|m,\mu}} \right],
\end{equation}
which allows us to write
\begin{equation}
    \bra{\mu}\chi_m^t[J_m]\ket{\mu} = p(\mu|m,\test) \chi_t^{m,\mu}[J_m].
\end{equation}
Furthermore, we define the map generating the statistics conditioned on an intensity \(\mu\) and photon number \(m\) equivalently to before, see \cref{eq:def probstJ},
\begin{equation}
    \probstJ_{m,\mu} [J_m] \defvar \Phi_m \circ \chi_{t}^{m,\mu}[J_m].
\end{equation}
Then, we find for the \(\mu\) component of each \(\probstJ_m[J_m]\)
\begin{align}
    &p(m|\test)(\probstJ_m[J_m])_{\mu} \nonumber \\
    &= p(m|\test) p(\mu|m,\test) \probstJ_{m,\mu}[J_m] \\
    &= p(\mu|\test) p(m|\mu,\test) \probstJ_{m,\mu}[J_m].
\end{align}
Thus, we can rewrite the constraints of \cref{eq:nuc constr blocksum} additionally conditioned on the intensity \(\mu_j\) as
\begin{align}\label{eq:nuc constr blocksum decoy}
    \bsym{\nu}_{\CP|\test,\mu_j} &\leq 
    \sum_{m\leq \Ncons} p(m|\test,\mu_j) \probstJ_{m,\mu_j}[J_m]
    \\ &\quad + 1-\mbf{p}_{\mathrm{tot}|\mu_j}, \nonumber\\
    \bsym{\nu}_{\CP|\test,\mu_j} &\geq \sum_{m\leq \Ncons} p(m|\test,\mu_j) \probstJ_{m,\mu_j}[J_m],
\end{align}
where \(\mbf{p}_{\mathrm{tot}|\mu_j}\) is defined as in \cref{eq:ptot block diag} with additional conditioning on the intensity \(\mu_j\).

Finally, for typical decoy-state protocols, signals with the same encoding \(a = (\alpha,x)\) but different intensities are usually mapped to the same key. This is especially true if only the signal intensity is used in key generation rounds. In this case, we define
\begin{equation}
    \begin{split}
    \ket{\xi^g}_{A_S AA'} &= \sum_{m=0}^{\infty} \sqrt{p(m|\gen)} \ket{m}_{A_S} \ket{\xi^g_{|m}}_{AA'},
    \end{split}
\end{equation}
as Alice's signal state in generation rounds, where
\begin{equation}
    \ket{\xi^g_{|m}} = \sum_{a} \sqrt{p(a|m,\gen)} \ket{a}_A \ket{s_m^{(a)}}_{A'}.
\end{equation}
Crucially, here \(p(m|\gen) = \sum_{\mu_j} p(m,\mu_j|\gen) \) is the total probability of sending \(m\) photons in a key generation round. 
Then, the objective function \(g_{\alpha}^{\Nent}\) is defined in terms of the states \(\ket{\xi^g}\). If this assumption is not satisfied, we simply use \(\ket{\phi^g}\).

Incorporating these reformulations into \cref{eq:halpha blocksum} we find for \(\hQKD\)
\begin{equation}\label{eq:halpha blocksum decoy}
    \begin{aligned}[t]
        \hQKD \geq 
        &\begin{aligned}[t]
            \inf_{\substack{\mbf{q} \in \Sacc, \bsym{\nu}_{\CP} \in \mathbb{P}(\alphCP) \\ J_m \in \mathrm{Pos}(A'B) }} \Bigg( &\frac{\alpha D\rel{\mbf{q}}{\bsym{\nu}_{\CP}}}{\alpha - 1} \\
            &+ q(\gen) g_{\alpha}^{\Nent}(J_0 \dots J_{\Nent}) \Bigg)
        \end{aligned}\\
        \textrm{s.t. } &\Tr_B[J_m] = \idop_{A'}, \\
        &\bsym{\nu}_{\CP} = (\gamma p(\mu_s|\test) \bsym{\nu}_{\CP|\test,\mu_s},\dots, 1-\gamma)^T, \\
        &\bsym{\nu}_{\CP|\test,\mu_j} \leq \sum_{m\leq \Ncons} p(m|\test,\mu_j) \probstJ_{m,\mu_j}[J_m] \\
        &\qquad \qquad + 1-\mbf{p}_{\mathrm{tot}|\mu_j}, \\
        &\bsym{\nu}_{\CP|\test,\mu_j} \geq \sum_{m\leq \Ncons} p(m|\test,\mu_j) \probstJ_{m,\mu_j}[J_m].
        \end{aligned}
\end{equation}
This formulation of \(\hQKD\) already improves the numerical precision because the probability vector \(\bsym{\nu}_{\CP|\test,\mu_j}\) is conditioned on the intensity \(\mu_j\). However, the optimization problem would still require solving for \(\Ncons\) Choi states that typically only have a very small influence on the constraints for higher photon numbers. 

Alternatively, one could replace the optimization over the Choi state with an optimization over the probabilities or so-called yields, similar to \cite[Sec. 7]{kamin_finite-size_2025}. This version is also much closer to common numerical decoy-sate methods such as \cite{wang_numerical_2022,nahar_imperfect_2023,kamin_improved_2024}.

Hence, to reduce the overhead from optimizing many Choi states, we introduce the yields
\begin{equation}
    Y_m^{a,b} \defvar \Pr(b|a,\mu,m,\test).
\end{equation}
These yields are actually independent of the intensity \(\mu\), because the state sent to Bob does not depend on the intensity. Therefore, we can relate the components \((a,b)\) of each vector \(\probstJ_{m,\mu_j}[J_m]\) to the yields by
\begin{equation}
    \begin{split}
        &(\Psi_{m,\mu}[J_m])_{a,b} = p(a,b|m,\mu,\test) \\
        &= p(a|m,\mu,\test) p(b|a,\mu,m,\test) \\
        &= p(a|m,\mu,\test) Y_m^{a,b}.
    \end{split}
\end{equation}
Moreover, to simplify the notation, we appropriately stack all yields \(Y_m^{a,b}\) of the same photon number \(m\) into a vector \(\mbf{Y}_m \), such that it holds component-wise
\begin{equation}
    \probstJ_{m,\mu_j}[J_m] = p(a|m,\mu_j,\test) \mbf{Y}_m,
\end{equation}
or equivalently
\begin{equation}
    p(m|\test,\mu_j) \probstJ_{m,\mu_j}[J_m] = p(a,m|\mu_j,\test) \mbf{Y}_m.
\end{equation}

Using this relation and inserting it into the previous version of the constraints \cref{eq:nuc constr blocksum decoy}, one finds the fully simplified version of \(\hQKD\) for decoy-state protocols
\begin{equation}\label{eq:halpha decoy}
    \begin{aligned}[t]
        \hQKD \geq &\begin{aligned}[t]
            \inf_{\substack{\mbf{q} \in \Sacc, \bsym{\nu}_{\CP} \in \mathbb{P}(\alphCP) \\ J_m \in \mathrm{Pos}(A'B), \\ \mbf{Y}_m}} \Bigg( &\frac{\alpha D\rel{\mbf{q}}{\bsym{\nu}_{\CP}}}{\alpha - 1} \\
            &+ q(\gen) g_{\alpha}^{\Nent}(J_0 \dots J_{\Nent}) \Bigg)
        \end{aligned}\\
        \textrm{s.t. } &\Tr_B[J_m] = \idop_{A'}, \\
        &\bsym{\nu}_{\CP} = (\gamma p(\mu_s|\test) \bsym{\nu}_{\CP|\test,\mu_s},\dots, 1-\gamma)^T, \\
        &\bsym{\nu}_{\CP|\test,\mu_j} \leq \sum_{m\leq \Nph} p(a,m|\test,\mu_j) \mbf{Y}_m \\
        &\qquad \qquad + 1-\mbf{p}_{\mathrm{tot}|\mu_j}, \\
        &\bsym{\nu}_{\CP|\test,\mu_j} \geq \sum_{m\leq \Nph} p(a,m|\test,\mu_j) \mbf{Y}_m, \\
        &p(a|m,\mu_j,\test) \mbf{Y}_m = \probstJ_{m,\mu_j}[J_m] \; \forall m\leq \Nent.
        \end{aligned}
\end{equation}
Repeating the same arguments as before for the $\frenyiSandDown_\alpha$-normalization constant \(\kappa\), one finds the decoy reformulation to be
\begin{align}\label{eq:kappa const decoy}
\begin{aligned}
    \kappa \geq &\inf_{\substack{J_m \in \mathrm{Pos}(A'B), \\ \bsym{\nu} \in \mathbb{P}(\Ct), \\ \mbf{Y}_m }} \Bigg( \frac{1}{1-\alpha} \log \Big( \gamma \sum_{c\in \Ct} \bsym{\nu}(\cP)2^{(\alpha - 1) f(\cP)} \\
    &\qquad + \left(1-\gamma\right)2^{-(\alpha-1)\left(g_{\alpha}^{\Nent}(J_0,\dots J_{\Nent}) - f(\gen)  \right)} \Big) \Bigg) \\
    \quad \textrm{s.t. } &\Tr_B[J_m] = \idop_{A'}, \\
    &\bsym{\nu} = (p(\mu_s|\test) \bsym{\nu}_{|\mu_s},\dots)^T, \\
    &\bsym{\nu}_{|\mu_j} \leq \sum_{m\leq \Nph} p(a,m|\test,\mu_j) \mbf{Y}_m
    + 1-\mbf{p}_{\mathrm{tot}|\mu_j}, \\
    &\bsym{\nu}_{|\mu_j} \geq \sum_{m\leq \Nph} p(a,m|\test,\mu_j) \mbf{Y}_m, \\
    &p(a|m,\mu_j,\test) \mbf{Y}_m = \probstJ_{m,\mu_j}[J_m] \; \forall m\leq \Nent.        
    \end{aligned}
\end{align}

\subsection{Example: Active Decoy BB84}\label{subsec:Ex Active BB84}
\newcommand{\misalign}{\theta^{\mathrm{misalign}}_\mathrm{hon}}
In this section, we present results for variable-length secret key rates of a decoy version of BB84 protocol with WCP sources and an active detection setup. This setup allows us to use the squashing map from Refs.~\cite{beaudry_squashing_2008,gittsovich_squashing_2014}. 

In each round, Alice decides with probability \(\gamma\) if it is a test round and with probability \(1-\gamma\) if it is a generation round and always selects the signal intensity \(\mu_s\) and the \(Z\) basis if it was a key generation round. Therefore, as in the qubit example of \cref{sec:Qubit BB84 and Comparisons with other Proof Techniques}, Alice's basis choices are given by \(p_z =1-\gamma\) and \(p_x = \gamma\). In a test round, Alice always selects the \(X\) basis and all possible intensities with equal probability, i.e. \(p(\mu|\test) = 1/3\). For the intensity choices we assume
\begin{equation}
    \mu_2 = 0.02, \quad \mu_3 = 10^{-3},
\end{equation}
and always optimize the signal intensity \(\mu_s\). For the one-decoy protocol we show later, we simply omit the third intensity \(\mu_3\).

On the receiver side, Bob, actively chooses the \(Z\) and \(X\) basis, also with probabilities \(p_z =1-\gamma\) and \(p_x = \gamma\), respectively.

We assume the same security parameters \(\ePA = \eEV = \frac{1}{2}10^{-80}\) for a total security of \(\eSec = 10^{-80}\) and error correction model with efficiency \(\fEC=1.1\) as in \cref{sec:Qubit BB84 and Comparisons with other Proof Techniques}. We model channel loss similarly to the qubit protocol and include a misalignment with an angle of \(\misalign = 0.03\). The Kraus operators of the maps \(\GMap\) and \(\ZMap\) and the states can be found in \cref{app:Kraus ops active decoy BB84}.

For comparison, we present the decoy-state key rates using the EUR approach first presented by Lim et. al. \cite{lim_concise_2014}. However, we use the results of Ref.~\cite{tupkary_phase_2024} again incorporating the improvements from Ref.~\cite{mannalath_sharp_2025} because it removes the detection efficiency mismatch assumption and corrected other issues \cite{tupkary_qkd_2025}.

Furthermore, this combination gave, at the time of writing, the highest variable-length secret key rates for decoy-state protocols\footnote{In Ref.~\cite{attema_optimizing_2021} EUR-based key rates were presented with the additional improvement of performing the decoy-state analysis with a linear program. However, the authors' claim of using the Chernoff bound requires the knowledge of the moment generating function of dependent variables which is not derived. Therefore, we left it out of this comparison.}.

In both approaches, we optimize all free parameters, such as the testing probability \(\gamma\), the signal intensity \(\mu_s\) and the \Renyi parameter \(\alpha\) (only for this work's results).

\begin{figure}[tb]
	\centering
	\includegraphics[width=\linewidth]{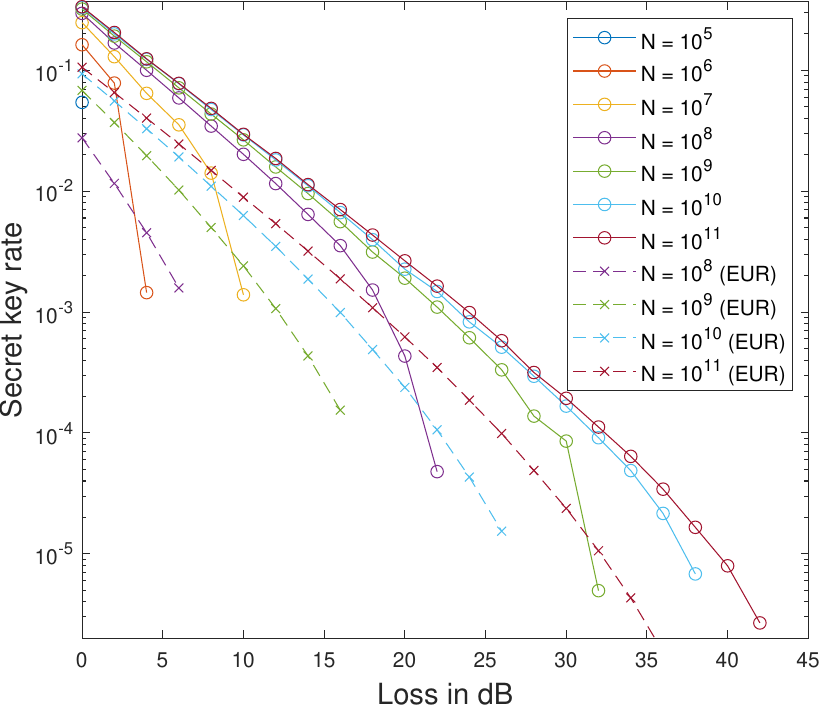}
	\caption{Results for variable-length secret key rates of the active decoy BB84 protocol with \emph{two} decoy intensities plotted against the channel loss in \(\unit{dB}\) with a varying number of total signals sent \(N=10^5, \dots, 10^{11} \). The security parameters, and error correction efficiency were chosen as \(\ePA = \eEV = \frac{1}{2}10^{-80}\) and \(\fEC=1.1\). As a comparison, we showcase results using the EUR approach of \cite{tupkary_phase_2024} incorporating the improvements from \cite{mannalath_sharp_2025}, omitting any curve leading to zero key rate. For both approaches, we optimize the probability $\gamma$ of using the
    \(Z\)-basis choice in each round and the signal intensity \(\mu_s\). Additionally, for our work we optimize the \Renyi parameter \(\alpha\).}
	\label{fig:ActiveDecoyBB84}
\end{figure}

The resulting variable-length key rates can be seen in \cref{fig:ActiveDecoyBB84}. Our key rates clearly outperform those using the EUR-based security proof. Compared to the qubit key rates of \cref{sec:Qubit BB84 and Comparisons with other Proof Techniques}, the improvements are much more significant.

Those improvements in key rate are due to multiple reasons in conjunction. First, the security analysis itself is tighter for our \Renyi entropy formulation compared to the EUR approach, especially for small block sizes as already shown by the qubit BB84 example in \cref{sec:Qubit BB84 and Comparisons with other Proof Techniques}. 

On an intuitive level, for decoy-state protocols, the number of signals sent that are useful for key generation is roughly only those sent with a single photon, effectively reducing the block size. Thus, one would expect that decoy-state protocols 
magnify the gap in key rate between our work and the EUR approach, which is exactly what we observe. In \cref{fig:QubitBB84Depol} although key rates were different, one required the same amount of signals sent to generate positive key rates. Now, the EUR approach requires three orders of magnitude more signals to be sent to generate non-zero key rates.

Second, our proof technique allows one to send \emph{only} the signal intensity in key generation rounds. The EUR approach of \cite{tupkary_phase_2024} cannot accommodate such a protocol, yet. For the example we present here this roughly corresponds to a factor of three improvement in key rates.

Third, we employed a decoy-state analysis similar to \cite{kamin_finite-size_2025}, where it was already discussed that this approach already yields better key rates even in the asymptotic limit. The main difference of our technique is that we perform the decoy analysis and the key rate optimization in one step in contrast to the ``traditional" two-step process \cite{wang_numerical_2022,nahar_imperfect_2023,kamin_improved_2024}. For \Renyi entropies, as previously emphasized in \cref{sec:Block-Diag states,sec:Decoy state}, the key rates presented here can be made arbitrarily close to the ``true" decoy-state key rate by simply including more terms.

Finally, we do not distinguish the influence of each individual improvement mentioned in the three points above --- disentangling them may not be possible, as certain steps are necessary to be able to calculate variable-length key rates at all under our approach. However, we conclude that the results presented here pose a significant step forward in QKD performance even purely in terms of achievable key rates.

Similarly to the results shown in Refs.~\cite{rusca_finite-key_2018,wiesemann_consolidated_2024}, we can also employ only a single decoy intensity. 

\begin{figure}[t]
	\centering
	\includegraphics[width=\linewidth]{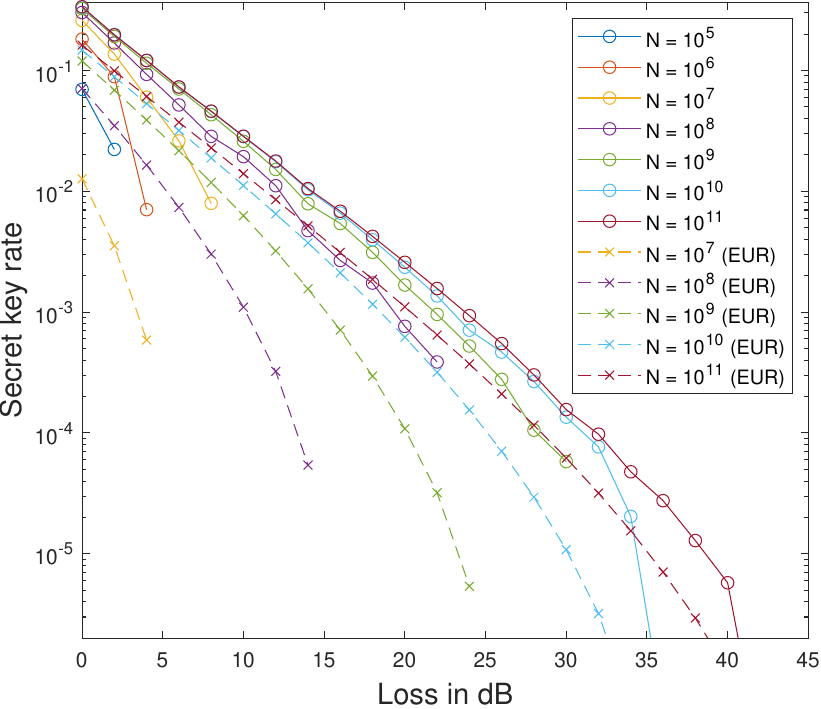}
	\caption{Results for variable-length secret key rates of the active decoy BB84 protocol with \emph{one} decoy intensity plotted against the channel loss in \(\unit{dB}\) with a varying number of total signals sent \(N=10^5, \dots, 10^{11} \). The security parameters, and error correction efficiency were chosen as \(\ePA = \eEV = \frac{1}{2}10^{-80}\) and \(\fEC=1.1\). As a comparison, we showcase results using the EUR approach of Refs.~\cite{rusca_finite-key_2018,wiesemann_consolidated_2024}, omitting any curve leading to zero key rate. For both approaches, we optimize probability $\gamma$ of using the
    \(Z\)-basis choice in each round and the signal intensity \(\mu_s\). Additionally, for our work we optimize the \Renyi parameter \(\alpha\).}
	\label{fig:ActiveDecoyBB84SingleDecoy}
\end{figure}

To showcase this protocol variant, in \cref{fig:ActiveDecoyBB84SingleDecoy}, we present the same protocol as above with one decoy intensity less, i.e. we use the same decoy intensity \(\mu_2=0.02\) and drop \(\mu_3\). Additionally, we adjust the probability of sending an intensity in a test round to \(p(\mu|\test) = 1/2\). The signal intensity \(\mu_s\) is still optimized for each data point.

To compare our results to the literature, we present the fixed-length key rates from \cite{rusca_finite-key_2018,wiesemann_consolidated_2024}. Ref.~\cite{wiesemann_consolidated_2024} improved upon the results from Ref.~\cite{rusca_finite-key_2018} and corrected some gaps in the analysis. For a survey of the gaps in those methods see Ref.~\cite{tupkary_qkd_2025}. We will interpret these key rates from Ref.~\cite{wiesemann_consolidated_2024} as an approximation of variable-length key rates for the same protocol, since we expect the key rates to stay roughly the same if the variable-length results from Ref.~\cite{tupkary_phase_2024} are incorporated.

Again, we observe that our methods significantly outperform the EUR approach for small block sizes, while the difference reduces as the number of signals grows. Similarly to before, Refs.~\cite{rusca_finite-key_2018,wiesemann_consolidated_2024} also require that all intensities are sent in both test and key generation rounds. Hence, we now observe that our key rates at zero loss are roughly a factor of two better (previously it was a factor of three).

As already observed in \cite{rusca_finite-key_2018} and later in \cite{kamin_improved_2025} (for an IID analysis that is then lifted to coherent attacks via the postselection technique), one single decoy intensity can perform better. Here we observe that especially for small block sizes, the performance of the one-decoy protocol is indeed better, for example see the \(10^5\) key rates which now extend to \(\unit[2]{dB}\).

This advantage vanishes for larger block sizes where again two decoy intensities perform better; see for example the last positive key rate for \(10^{11}\) signals. This observation is in line with the expectation that asymptotically three decoy intensities are better than two; see \cite{kamin_improved_2024}, which applied the old two-step process for the decoy analysis. 

However, we were unable to determine the point of transition with much precision, as the global optimization routine we used turned out to be very unstable. With better tuning of the protocol parameters, one could potentially achieve even higher key rates for both the one- and two-decoy protocol. Crucially, however, we emphasize that the key rates we report here are valid lower bounds.

At this point we also highlight that in contrast to our methods, the global optimization routine does converge well for all EUR key rates we presented here. Thus, those key rates do represent the optimal ones under the respective approaches.

\section{Device Imperfections and Implications}\label{sec:Simple Imperfections}
The previous section essentially concludes our work on perfect devices, and we will next focus on two examples of imperfections of particular interest. However, before we present those, let us quickly discuss some implications of imperfections.

First, there are different types of imperfections. For example, the source device could prepare an erroneous state as compared to the perfect case, but simply does so in an IID manner in each round (supposing that the signal state choice in each round is IID). Alternatively, the states prepared across rounds could be independent but not identical. The worst case would be that there are correlations between rounds and the devices involved even possess a memory. 

Thus far, in the absence of additional assumptions, the MEAT framework \cite{arqand_marginal-constrained_2025,fawzi_additivity_2025} can only accommodate imperfections without a memory, i.e.~that last scenario cannot currently be handled. However, we note that at the time of writing, no other framework has demonstrated a full security analysis for that scenario either.
For source correlations, one could assume a specific correlation length as in Ref.~\cite{pereira_quantum_2020}. Then, we conjecture that by analyzing the protocol in terms of several ``interleaved virtual protocols'' in the manner proposed in that work, our MEAT framework would also address such correlations. However, that work did not present a formal proof of the claim that the ``interleaved virtual protocols'' analysis indeed yields security of the real protocol, so we leave it at this conjecture for the present work.

Next, let us see how IID or independent imperfections without memories could be included. An example of an independent imperfection (which we will discuss in the upcoming section) is a source sending out a signal with a slightly different intensity in each round without any memory or correlations.

To handle independent imperfections, we consider a scenario where the channel \(\EATchannQKD\) can be different in each round, with the overall channel describing the mapping from $A_1^nB_1^n$ to $S_1^n I_1^n \CP_1^n$ being a tensor product channel of the form
\begin{equation}
    \bigotimes_{i=1}^{n} \EATchannQKD^i.
\end{equation}
IID imperfections would simply correspond to having the same channel and marginal state for each round \(i\), so we would be able to immediately apply our previous analysis from \cref{sec:Formalizing QKD}.
For general non-IID independent imperfections, we note that \cite[Corollary 4.2]{arqand_marginal-constrained_2025} (and also~\cite[Corollary~5.1]{fawzi_additivity_2025}) still apply even if the channels and marginal states are different in each round, and hence one can still construct both fixed- and variable-length security proofs by following similar arguments as in the preceding sections, with minor changes. 

Specifically, for variable-length proofs,~\cite[Corollary 4.2]{arqand_marginal-constrained_2025} states that \cref{eq:chainMEAT} (the critical step in the proof) still holds if we simply instead define $\ffull$ in \cref{eq:ffull} with a different \(\frenyiSandUp_\alpha\)-normalization constant  \(\kapup_i\) for each round \(i\), defined with respect to the channel \(\EATchannQKD^i\).
If we lower bound each of these constants \(\kapup_i\) with the worst case over all $i$, we can then obtain a secure protocol by choosing $\fprot$ as in \cref{eq:protcol tradeoff function}, with a single lower bound $\kapupbnd$ for all the rounds. We will make use of this in  \cref{sec:Intensity imperfections,sec:Phase imperfections} below.

For fixed-length proofs, a technicality arises because when the channels $\EATchannQKD^i$ for each round are different, the formula for $\hQKD$ according to \cite[Corollary 4.2]{arqand_marginal-constrained_2025} requires considering a particular convex hull, which may not be simple to describe for some imperfections. Since in this work we focus only on presenting the variable-length key rates, this technicality does not affect the computations we performed (i.e.~the key rates in our subsequent plots are all valid for independent non-IID imperfections), though for the sake of accuracy we restrict our subsequent theorem statements involving $\hQKD$ to the case of IID imperfections only.\footnote{This does imply that when we modify these $\hQKD$ optimizations to choose the tradeoff function $\mbf{f}$ according to the dual solution, we might in principle not be finding the exact optimal $\mbf{f}$ to use for variable-length key rates against independent imperfections. However, we once again emphasize that our framework is still computing secure lower bounds on the key rates; this suboptimality merely makes the bounds potentially less tight.}

Finally, let us discuss other ``simple" imperfections. Detector imperfections under the model of Ref.~\cite{nahar_imperfect_2025} can be readily included. This model only reduces the weight of the subspace inside the subspace after applying the flag-state squasher \cite{zhang_security_2021}. Here in \cref{app:Flag-state squasher}, we explicitly show that the flag-state can be included in our framework as well.

Furthermore, imperfections that effectively only change Alice's marginal state \(\tau_A\) or require a ball around it, are also easily implemented in the above framework by adjusting the optimization problem. Therefore, in the next two \cref{sec:Intensity imperfections,sec:Phase imperfections}, we focus on the examples of intensity and phase imperfections, respectively, which are not immediate to include.

\section{Intensity Imperfections}\label{sec:Intensity imperfections}
Our security framework using \Renyi entropies allows for a simple treatment of imperfections. Again, one caveat is that the imperfections must occur in an independent fashion and must not possess a memory. We will show two examples, intensity imperfections in this section and phase imperfections in the following \cref{sec:Phase imperfections}. 

In \cite{wang_tight_2016, zhou_finite-key_2022,zapatero_security_2021,sixto_security_2022} similar intensity imperfections were already considered for a very specific active BB84 protocol (\cite{wang_tight_2016, zhou_finite-key_2022}) or in the asymptotic limit \cite{zapatero_security_2021,sixto_security_2022}. In contrast, our methods are valid for \emph{generic} decoy-state protocols and do not require any additional assumptions on the receiver, such as an active detection setup.

The analysis is simplified a lot by the fact that one only needs to find a lower bound on the single-round quantity \(\hQKD\) and the $\frenyiSandDown_\alpha$-normalization constant \(\kappa\). The only catch is to ensure that the channel \(\EATchannQKD\) with imperfections still satisfies \cref{thrm:Fixed-length security generic bnd,thrm:Variable-length security}. Then, one can employ techniques very similar to the asymptotic limit. 

In this section, we focus on independent intensity imperfections of decoy-state protocols. The model we assume is as follows. Alice's source still produces weak coherent pulses, but the exact intensity is unknown. More concretely, we assume that in each round Alice sends a signal with some intensity \(\mu_j^{\mathrm{true}}\) in an interval \([\mu_j,\mu_j+\delta_j]\) in an independent manner. Here \(\mu_j\) is the intensity specified in the protocol and \(\delta_j\) is a parameter that specifies the deviation.

There are two immediate examples under which such a behavior could occur. For the first example, assume that a WCP source works correctly, but due to uncertainties in the characterization process, the true intensity \(\mu_j^{\mathrm{true}}\) is only known with an interval of size \(\delta_j\). Then Alice's source still emits the same intensity every round, namely \(\mu_j^{\mathrm{true}}\), but it is unknown.

The second example is a source that emits an intensity \(\mu_j^{\mathrm{true}} \in [\mu_j,\mu_j+\delta_j]\) sampled independently in each round. Thus, in each round the intensity \(\mu_j^{\mathrm{true}} \) is chosen according to some unknown (potentially different) probability distribution. In particular, since the probability distribution is unknown, we cannot make use of any properties of it. Coincidentally, both cases can be analyzed in the same way if one assumes some form of worst case for each round.

Therefore, under this model, we will show how to bound the decoy-state version of \(\hQKD\) (and the $\frenyiSandDown_\alpha$-normalization constant \(\kappa\)) as presented in \cref{eq:halpha blocksum decoy} (and \cref{eq:kappa const decoy}). In principle, minimizing \(\hQKD\) of \cref{eq:halpha decoy} with respect to each intensity in its respective intervals would result in a valid lower bound. However, this is a non-convex problem and general optimization routines do not yield guaranteed lower bounds.

Therefore, we bound both the constraints and the objective function separately, such that we still reach a convex optimization problem summarized in \cref{prop:Bound galpha for intensity imp}.

Starting with the constraints, for each intensity choice \(\mu_j\), we can bound the probability of sending \(m\) photons and signal \(a=(\alpha,x)\) by
\begin{align}\label{eq:Bounds probs intensity imp}
    p(a,m|\test,\mu_j^{\mathrm{true}}) &\geq \min_{\substack{\mu \in \\ [\mu_j,\mu_j+\delta_j]}} p(a,m|\test,\mu) 
    \\ & =: p^L(a,m|\test,\mu_j), \\
    p(a,m|\test,\mu_j^{\mathrm{true}}) &\leq \max_{\substack{\mu \in \\ [\mu_j,\mu_j+\delta_j]}} p(a,m|\test,\mu)
    \\ & =: p^U(a,m|\test,\mu_j).
\end{align}
In \cref{subsec:Decoy Simplifications}, we introduced the so-called yields to bound the statistics \(\bsym{\nu}_{\CP|\test,\mu_j}\) conditioned on \(\test\) and the intensity choice \(\mu_j\). These yields are defined as
\begin{equation}
    Y_m^{a,b}\defvar\Pr(b|a,\mu_j,m,\test),
\end{equation}
and are independent of the intensity choice \(\mu_j\) for fully phase randomized weak coherent pulses. Furthermore, for the true intensities \(\mu_j^{\mathrm{true}}\), the yields satisfy the following relation to \(\bsym{\nu}_{\CP|\test,\mu_j}\),
\begin{align}
    \bsym{\nu}_{\CP|\test,\mu_j} &\leq \sum_{m\leq \Nph} p(a,m|\test,\mu_j^{\mathrm{true}}) \mbf{Y}_m \\
        &+ 1-\mbf{p}_{\mathrm{tot}|\mu_j^{\mathrm{true}}}, \nonumber \\
    \bsym{\nu}_{\CP|\test,\mu_j} &\geq \sum_{m\leq \Nph} p(a,m|\test,\mu_j^{\mathrm{true}}) \mbf{Y}_m.
\end{align}

Then, inserting \cref{eq:Bounds probs intensity imp}, we can immediately bound the constraints on the statistics involving the yields by
\begin{align}\label{eq:pL and pU yields}
    \bsym{\nu}_{\CP|\test,\mu_j} &\leq \sum_{m\leq \Nph} p^U(a,m|\test,\mu_j) \mbf{Y}_m, \\
        &+ 1-\mbf{p}_{\mathrm{tot}|\mu_j}^L, \nonumber \\
    \bsym{\nu}_{\CP|\test,\mu_j} &\geq \sum_{m\leq \Nph}p^L(a,m|\test,\mu_j) \mbf{Y}_m,
\end{align}
where
\begin{equation}
    \mbf{p}^L_{\mathrm{tot}|\mu_j} \defvar  \min_{\substack{\mu \in \\ [\mu_j,\mu_j+\delta_j]}} \sum_{\substack{\cP \in \Ct \\ a \in \phi^{-1}(\cP)}} \sum_{m \leq \Ncons} p(a,m|\test,\mu) \hat{e}_{\cP}.
\end{equation}

Next, let us see how the objective function depends on the deviations in the intensities. If Alice only sends the signal intensity \(\mu_s\) in generation rounds and we set \(\Nent =1\), \(g_{\alpha}^{\Nent}\) simplifies to
\begin{equation}
    \begin{split}
        &g_{\alpha}^{1}\left(J_0,J_1; \mu_s \right) \\ 
        &= \frac{1}{1-\alpha} \log\Bigg\{ 1- \sum_{m=0,1} p(m|\mu_s) \\
        &\quad +  \sum_{m =0,1} p(m|\mu_s) 2^{(1-\alpha)\renyiSandDown_{\alpha}(S|\tilde{T}E)_{\nu(J_m)_{|m,\gen}}} \Bigg\},
    \end{split}
\end{equation}
where we made the dependence on \(\mu_s\) explicit. Furthermore, for each \(m\), the state \(\nu(J_m)_{|m,\gen}\), is independent of \(\mu_s\) since
\begin{equation}\label{eq:xi independent of mus}
    \ket{\xi^g_{|m}} = \sum_{a} \sqrt{p(a|m,\gen)} \ket{a}_A \ket{s_m^{(a)}}_{A'},
\end{equation}
is independent of \(\mu_s\). Hence, finding \(\mu\) in \([\mu_s,\mu_s+\delta_s]\) such that the resulting probabilities minimize \(g_{\alpha}^{\Nent}\) will lead to a valid lower bound. We state this result in the following proposition.

\begin{prop}\label{prop:Bound galpha for intensity imp}
    Let all intensities \(\mu_j^{\mathrm{true}}\) vary in intervals \([\mu_j,\mu_j+\delta_j]\), but the photon number distribution follows a Poisson distribution for any possible \(\mu_j^{\mathrm{true}} \in [\mu_j,\mu_j+\delta_j]\). Let us also assume that, in each round \(i = 1,\dots,n\), the intensity choice of this round is sampled independently from \([\mu_j,\mu_j+\delta_j]\), and that Alice only sends the signal intensity in generation rounds. Additionally, assume it holds
    \begin{align}
        p(a|m,\mu_j,\test) = p(a|m,\test) \; \forall j.
    \end{align} 
    Then, all $\frenyiSandDown_\alpha$-normalization constants \(\kapup_i\) can be bounded by
    \begin{equation}\label{eq:kappa const decoy int imp}
    \begin{aligned}
        \kapup_i \geq &\inf_{\substack{J_m \in \mathrm{Pos}(A'B), \\ \bsym{\nu} \in \mathbb{P}(\Ct), \\ \mu \in \{\mu_s,\mu_s+\delta_s\} }} \Bigg( \frac{1}{1-\alpha} \log \Big( \gamma \sum_{c\in \Ct} \bsym{\nu}(\cP)2^{(\alpha - 1) f(\cP)} \\
        &\qquad + \left(1-\gamma\right)2^{-(\alpha-1)\left(g_{\alpha}^{1}(J_0, J_1;\mu) - f(\gen)  \right)} \Big) \Bigg) \\
        \quad \textrm{s.t. } &\Tr_B[J_m] = \idop_{A'}, \\
        &\bsym{\nu} = (p(\mu_s|\test) \bsym{\nu}_{|\mu_s},\dots)^T, \\
        &\bsym{\nu}_{|\mu_j} \leq \sum_{m\leq \Nph} p^U(a,m|\test,\mu_j) \mbf{Y}_m \\
        &\qquad \qquad + 1-\mbf{p}^L_{\mathrm{tot}|\mu_j}, \\
        &\bsym{\nu}_{|\mu_j} \geq \sum_{m\leq \Nph} p^L(a,m|\test,\mu_j) \mbf{Y}_m, \\
        &p(a|m,\mu_j,\test) \mbf{Y}_m = \probstJ_{m,\mu_j}[J_m], \; m=0,1.
        \end{aligned}
    \end{equation}
    Furthermore, if the sampling from \([\mu_j,\mu_j+\delta_j]\) is IID, then \(\hQKD\) is bounded from below by 
    \begin{equation}\label{eq:halpha decoy int imp}
    \begin{aligned}[t]
        \hQKD \geq &\begin{aligned}[t]
            \inf_{\substack{\mbf{q} \in \Sacc, \bsym{\nu}_{\CP} \in \mathbb{P}(\alphCP) \\ J_m \in \mathrm{Pos}(A'B), \\ \mu \in \{\mu_s,\mu_s+\delta_s\} }} \Bigg( &\frac{\alpha D\rel{\mbf{q}}{\bsym{\nu}_{\CP}}}{\alpha - 1} \\
            &+ q(\gen) g_{\alpha}^{\Nent}(J_0,J_1;\mu) \Bigg)
        \end{aligned}\\
        \textrm{s.t. } &\Tr_B[J_m] = \idop_{A'}, \\
        &\bsym{\nu}_{\CP} = (\gamma p(\mu_s|\test) \bsym{\nu}_{\CP|\test,\mu_s},\dots, 1-\gamma)^T, \\
        &\bsym{\nu}_{\CP|\test,\mu_j} \leq \sum_{m\leq \Nph} p^U(a,m|\test,\mu_j) \mbf{Y}_m \\
        &\qquad \qquad + 1-\mbf{p}^L_{\mathrm{tot}|\mu_j}, \\
        &\bsym{\nu}_{\CP|\test,\mu_j} \geq \sum_{m\leq \Nph} p^L(a,m|\test,\mu_j) \mbf{Y}_m, \\
        &p(a|m,\test) \mbf{Y}_m = \probstJ_{m,\mu_j}[J_m], \; m=0,1.
        \end{aligned}
    \end{equation}
    In both equations, we stated the dependence of \(g_{\alpha}^{1}\) on \(\mu_s\) explicitly.
\end{prop}
\begin{proof}
    For simplicity, we only explicitly prove the statement involving \(\hQKD\). The proof for the $\frenyiSandUp_\alpha$-normalization constants \(\kapup_i\) follows similar arguments. 
    
    For the statement involving \(\hQKD\), Alice's source prepares the states in an IID manner. Thus, the channel \(\EATchannQKD\) in \cref{thrm:Fixed-length security generic bnd} and \cref{thrm:Variable-length security} would in principle be defined in terms of the true intensities \(\mu_j^{\mathrm{true}}\). Most importantly, although these true intensities are unknown, the channel still satisfies the assumptions of \cref{thrm:halpha QKD,thrm:Fixed-length security generic bnd}. Then, one follows the same steps as in \cref{subsec:Decoy Simplifications} to reach the expression of \cref{eq:halpha decoy} with the intensities \(\mu_j^{\mathrm{true}}\) in place of \(\mu_j\).
    
    However, since the true intensities \(\mu_j^{\mathrm{true}}\) are unknown, we need to replace the optimization problem for \(\hQKD\) with the ``worst possible value'' of \(\mu_j^{\mathrm{true}}\) over the entire interval \([\mu_j,\mu_j+\delta_j]\). Our goal is now to show that this is indeed lower bounded by the expression given in the claim.
    
    First, the weakened constraints follow immediately from \cref{eq:pL and pU yields} together with the assumption that \(p(a|m,\mu,\test) = p(a|m,\test)\). Hence, since \( D\rel{\mbf{q}}{\bsym{\nu}_{\CP}}\) is independent of the true signal intensity, we are left with showing that
    \begin{equation}
        \begin{split}
        &\min_{\mu \in [\mu_s,\mu_s+\delta_s]} g_{\alpha}^{1}\left(J_0,J_1; \mu\right) \\ &=\min_{\mu \in \{\mu_s,\mu_s+\delta_s\}} g_{\alpha}^{1}\left(J_0,J_1; \mu \right).
        \end{split}
    \end{equation}
    To simplify the notation, let us define 
    \begin{equation}
        \beta_m \defvar 1-2^{(1-\alpha)\renyiSandDown_{\alpha}(S|\tilde{T}E)_{\nu(J_m)_{|m,\gen}}},
    \end{equation}
    which is always nonnegative and lets us rewrite \(g_{\alpha}^{1}\) as
    \begin{equation}
    \begin{split}
        &g_{\alpha}^{1}\left(J_0, J_{1}; \mu\right) = \frac{1}{1-\alpha} \log\Bigg\{ 1 - \sum_{m =0,1} p(m|\mu) \beta_m \Bigg\}.
    \end{split}
    \end{equation}
    Next we note that minimizing \(-\log(v(x))\) over \(x\), for some function \(v\), is equivalent to maximizing \(v\) over \(x\). Therefore, minimizing \(g_{\alpha}^{1}\) is equivalent to
    \begin{align}
        &\max_{\mu \in [\mu_s,\mu_s+\delta_s]} 1- \sum_{m=0,1} p(m|\mu) \beta_m \\
        &= 1- \min_{\mu \in [\mu_s,\mu_s+\delta_s]} \sum_{m=0,1} p(m|\mu) \beta_m \\
        &= 1- \min_{x\in [0,\delta_s]} e^{-(\mu_s+x)}\left[ \beta_0 + (\mu_s +x) \beta_1 \right],
    \end{align}
    where we introduced the variable \(x \in [0,\delta_s]\) to represent the deviation and used that \(p(m|\mu+x)\) is distributed according to a Poisson distribution for all \(x \in [0,\delta_s]\). By the assumption that Alice sends only the signal intensity in generation rounds, \(\beta_m\) is independent of \(x\), as shown in \cref{eq:xi independent of mus}.
    
    Next, if we define
    \begin{equation}
        u(x) \defvar  e^{-(\mu_s+x)}\left[ \beta_0 + (\mu_s +x) \beta_1 \right],
    \end{equation}
    one can find the local extrema by taking the first derivative, which leads to an optimal value \(x_{\text{opt}} = 1- \beta_0/\beta_1 -\mu_s\). Taking the second derivative of \(f\) yields
    \begin{equation}
        u''(x_{\text{opt}}) = -\beta_1 \exp\left(\frac{\beta_0}{\beta_1} - 1\right) < 0,
    \end{equation}
    since for \(\beta_1 =0\), \(x_{\text{opt}} = -\infty \notin [0,\delta_s]\) and \(\beta_0,\beta_1 \geq 0\) Hence, the local extremum is a maximum and by Fermat's theorem the minimum is achieved on the boundary \(x \in \{0, \delta_s\}\).

    With regards to the $\frenyiSandUp_\alpha$-normalization constants \(\kapup_i\) one can follow a very similar line of argument. Here, one needs to consider different channels \(\EATchannQKD^i\) for each \(\kapup_i\) as described in \cref{sec:Simple Imperfections}. For each \(\kapup_i\), we can still apply the simplifications of \cref{subsec:Decoy Simplifications} with the intensities \(\mu_j^{\mathrm{true}}\) in place of \(\mu_j\). Then, the loosened constraints again follow from \cref{eq:pL and pU yields} for each round \(i\).
    
    Next, one can follow the same argument of minimizing the negative logarithm of a function, and relate the minimum of each \(\kapup_i\) to the same function \(u\) from above. This yields that for all rounds \(i=1,\dots, n\), the constants \(\kapup_i\) are bounded by the same worst case and the statement follows.
\end{proof}

There are several ways in which one could extend the above proposition. For example, one could straightforwardly extend it to allow all intensities to be sent in generation rounds. Another possible extension might be to allow the signal choice \(a=(\alpha,x)\) to depend on the intensity, which violates the assumption \(p(a|m,\mu,\test) = p(a|m,\test)\), but we will leave these extensions for future work. For now, in the next section we present an example of a biased passive BB84 protocol using WCP sources with intensity imperfections under the model described above.

\subsection{Example: Biased passive BB84 with Intensity Imperfections}\label{subsec:Ex passive BB84 with Intensity Imp}
\newcommand{\epsInt}{\epsilon_{\text{int}}}
In this section, we present results for variable-length secret key rates of a decoy-state version of a BB84 protocol with WCP sources and a \emph{biased} passive detection setup. This setup does not allow for simple squashing methods to reduce it back to a qubit space, such as \cite{beaudry_squashing_2008,gittsovich_squashing_2014}. However, we can apply the flag-state squasher \cite{zhang_security_2021} together with the bounds from \cite{kamin_improved_2024}.

In each round, Alice decides with probability \(\gamma\) if it is a test or a generation round, and if it is a generation round, she always selects the signal intensity \(\mu_s\) and the \(Z\) basis. Therefore, as before, Alice's basis choice probabilities are given by \(p_z =1-\gamma\) and \(p_x = \gamma\). In a test round, Alice chooses uniformly at random from all the intensities considered, and always uses the \(X\) basis. We again present the one- and two-decoy variant of the protocol. In contrast to the active decoy BB84 protocol, we choose a \emph{fixed testing probability} of \(\gamma=0.2\).

Additionally, we assume that Alice's setup has intensity imperfections satisfying the assumptions of the previous section. Hence, we assume that in each round Alice independently sends out a signal with an intensity in the ranges
\begin{equation}
\begin{aligned}
    \mu_s^{\mathrm{true}} &\in [\mu_s,\mu_s +\delta_s], \quad \mu_2^{\mathrm{true}} \in [\mu_2,\mu_2 +\delta_2], \\
    \mu_3^{\mathrm{true}} &\in [\mu_3,\mu_3 +\delta_3].
\end{aligned}
\end{equation}
We set the intensities specified by the protocol to be
\begin{equation}
    \mu_s = 0.8, \quad \mu_2 = 0.2, \quad \mu_3 = 0.01,
\end{equation}
and omit the second decoy intensity for the one-decoy version of this protocol.

For a better comparison with the literature, we chose \(\delta_j\) in a similar way as in Refs.~\cite{zapatero_security_2021,sixto_security_2022}, such that
\begin{equation}
    \abs{1 - \frac{\mu_j}{\mu_j +\delta_j}} \leq \epsInt,
\end{equation}
where for \(\epsInt\) we will chose \(\epsInt = 10\%,\; 25\%\) and \(\epsInt=0\), which recovers the perfect case.

On the receiver side (Bob), we assume a passive four-state detection setup that chooses the \(Z\) and \(X\) basis with splitting ratios \(p_z =1-\gamma\) and \(p_x = \gamma\), respectively. To incorporate this into our security proof, we apply the flag-state squasher from \cite{zhang_security_2021} with the subspace bound from \cite[Theorem 1]{kamin_improved_2024}. Additionally, in \cref{app:Flag-state squasher}, we prove that it is applicable in this \Renyi framework.

Finally, we assume the same security parameters \(\ePA = \eEV = \frac{1}{2}10^{-80}\) and error correction model with efficiency \(\fEC=1.1\) as in \cref{sec:Qubit BB84 and Comparisons with other Proof Techniques}. We again model channel loss similarly to the qubit protocol and include a misalignment with an angle of \(\misalign = 0.03\). The Kraus operators of the maps \(\GMap\) and \(\ZMap\) and the states can be found in \cref{app:Kraus ops passive decoy BB84}.

\begin{figure}[tbh]
	\centering
	\includegraphics[width=\linewidth]{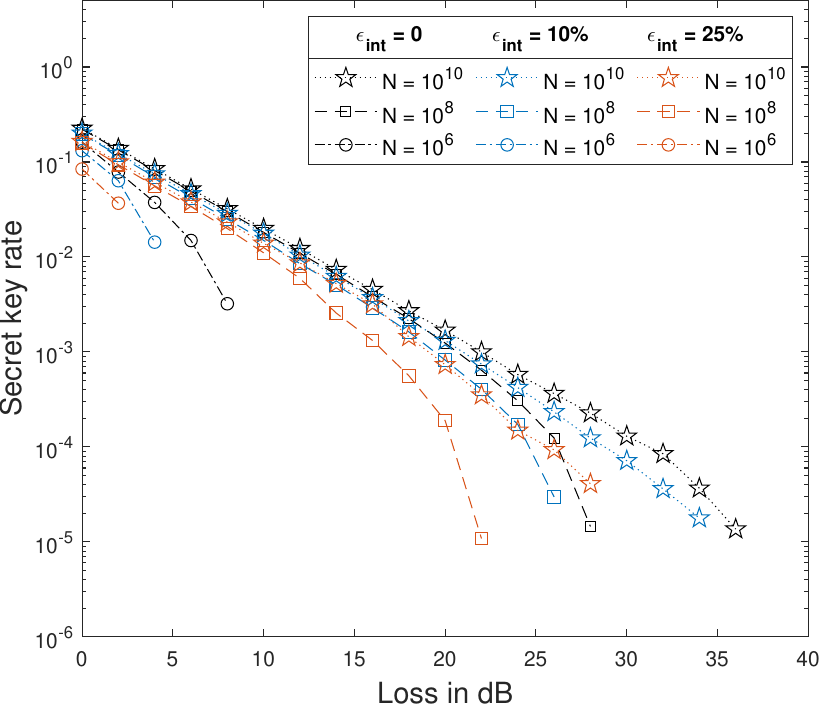}
	\caption{Results for variable-length secret key rates of the passive decoy BB84 protocol with \emph{two} decoy intensities and intensity imperfection plotted against the channel loss in \(\unit{dB}\) with a varying number of total signals sent \(N=10^6, 10^8, 10^{10} \). The security parameters, and error correction efficiency were chosen as \(\ePA = \eEV = \frac{1}{2}10^{-80}\) and \(\fEC=1.1\).}
    \label{fig:PassiveDecoyBB84}
\end{figure}

\begin{figure}[bth]
	\centering
	\includegraphics[width=\linewidth]{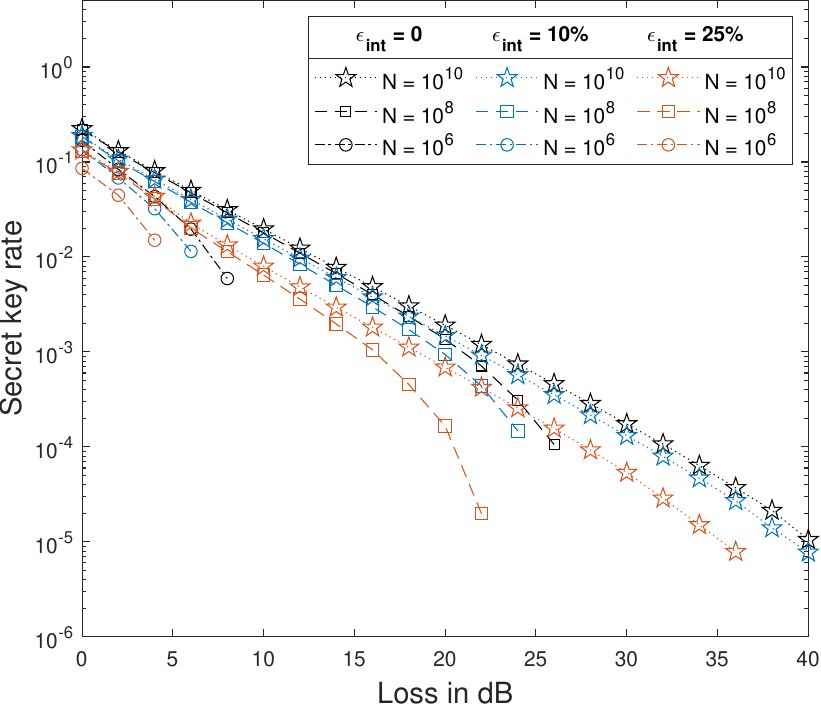}
	\caption{Results for variable-length secret key rates of the passive decoy BB84 protocol with \emph{one} decoy intensity and intensity imperfection plotted against the channel loss in \(\unit{dB}\) with a varying number of total signals sent \(N=10^6, 10^8, 10^{10} \). The security parameters, and error correction efficiency were chosen as \(\ePA = \eEV = \frac{1}{2}10^{-80}\) and \(\fEC=1.1\).}
	\label{fig:PassiveDecoyBB84_1Decoy}
\end{figure}

The resulting variable-length key rates for the one- and two-decoy protocols are shown in \cref{fig:PassiveDecoyBB84_1Decoy} and \cref{fig:PassiveDecoyBB84}, respectively, versus channel loss in \(\unit{dB}\).

Both the one- and two-decoy protocol react very similarly to intensity imperfections and, for small imperfections, \(\epsInt < 10\%\), the difference from a perfect setup is negligible. Moreover, in the perfect case \(\epsInt = 0\), both variants achieve key rates comparable to the active decoy BB84 protocol as shown in \cref{subsec:Ex Active BB84}. Both observations are true across the range of number of signals sent that we present here.

Additionally, the performance of the one-decoy variant of the protocol is better than that of the two-decoy variant. This is especially true if very few signals are sent or if in the high-loss regime. In both scenarios, the largest finite-size penalties would be expected due to small sample sizes. Our observations here are therefore in line with similar results previously shown in \cite{rusca_finite-key_2018,kamin_improved_2025}.

Since the behavior of the one- and two-decoy protocol with respect to the intensity imperfections is very similar, we discuss their results together. Compared to the perfect case (black curves), for $\epsInt = 10\%$ (blue curves), one can see a small but noticeable difference in key rates, which are shifted downward along the $y$-axis but still roughly follow the same trend as the perfect setup. The reduction in the key rate is approximately $15\%$, which is again valid across the full range of the numbers of signals sent presented here.

The influence of intensity imperfections with $\epsInt = 25\%$ is much more pronounced. For any number of total signals sent, the overall achievable key rate is reduced by roughly $40\%$ as well as the tolerable loss. The highest tolerable loss will be roughly $4 - \unit[6]{dB}$ smaller than with a perfect setup.

Therefore, our results show that we can effectively tolerate experimentally viable amounts of intensity imperfections without drastically reducing the key rates.

Finally, we briefly comment on the capabilities of other proof techniques in accounting for independent intensity imperfections. In principle, the postselection technique could incorporate IID intensity imperfections by loosening the constraints similarly to our results in \cref{sec:Intensity imperfections}. However, independent imperfections are currently not possible to include, since the postselection technique relies on IID sources (or sources representable as IID sources through source maps).

On the other hand, phase error correction based approaches cannot treat these forms of imperfections in the finite-size regime yet. For the EUR approach, these imperfections have only been treated for active one-decoy \cite{zhou_finite-key_2022} and two-decoy \cite{wang_tight_2016} BB84 protocols. Most importantly, their techniques would not be applicable for the passive detection setup presented here. Furthermore, it is not immediately clear how well their techniques would translate to other protocols with active detection setups e.g. an active 6-state protocol. In contrast, our methods here are tighter, since we solved parts of the optimization problem exactly, and can be readily applied to generic decoy-state protocols.

In terms of possible future improvements, we conclude the following. The lower bound on the objective function is already as tight as possible using our \Renyi framework, since we found the true minimum of that quantity over the possible intensity imperfections. However, future research could focus on tightening the constraints in \cref{eq:pL and pU yields}, as the looseness introduced here appears to be the remaining reason for the reduction in key rates.

\section{Phase Imperfections}\label{sec:Phase imperfections}
\newcommand{\PiNA}{\Pi_{\leq N_A}}
\newcommand{\PibarNA}{\Pi_{>N_A}}
\newcommand{\XiPhase}{\Xi_{\mathrm{phase}}}
\newcommand{\epsProb}{\epsilon_{\mathrm{prob}}}
\newcommand{\epsProj}{\epsilon_{\mathrm{proj}}}
\newcommand{\epsFull}{\epsilon_{\mathrm{full}}}
Another common imperfection which occurs especially under high repetition rates of decoy-state protocols are so-called phase imperfections. Typically, decoy-state protocols use a fully phase-randomized weak coherent pulse. However, if the repetition rate becomes too high, the phase-randomization can become imperfect. Then, the sources do not satisfy the assumptions of typical decoy-state protocols anymore. Previously, in Refs.~\cite{curras-lorenzo_security_2023,nahar_imperfect_2023} asymptotic key rates of decoy-state protocols with phase imperfections have been derived. Here, we present a finite-size security proof against coherent attacks.

Similarly to intensity imperfections presented in \cref{sec:Intensity imperfections}, one can also include phase imperfections as long as there is no memory process involved and states are sent out in an independent manner. Hence, we assume that in each round Alice's source emits a phase-imperfect state that is independent of all other rounds. However, the imperfection might not be identical in all rounds. To simplify notation, as done before, we will abbreviate \(a = (\alpha,x)\) in all of this section.

We use the same model of phase imperfections as described in \cite{curras-lorenzo_security_2023,nahar_imperfect_2023}, which connects phase imperfect states to a model state via a source map. In particular, this model assumed that in each round \(i =1,\dots,n\), instead of a fully phase-randomized weak coherent pulse, Alice sends out the following mixture
\begin{equation}\label{eq:phase imperfect signal state}
    \rho_{A'}^{(a,\mu)} =  q \sum_{m=0}^{\infty} p(m|\mu) \ketbra{m}{m}_{a}  + (1-q) \ketbra{\sqrt{\mu}}{\sqrt{\mu}}_a,
\end{equation}
where \(\ket{m}_{a}\) indicates \(m\) photons sent with \(a=(\alpha,x)\) in partition \(\alpha\) and bit value \(x\). This could be for example horizontal polarized light with \(m\) photons. Similarly, \(\ket{\sqrt{\mu}}_a\) represents a coherent state sent with the choice \(a=(\alpha,x)\) and intensity \(\mu\). The quality factor \(q \in [0,1] \) quantifies how close the state is to a fully phase-randomized state. Then the connection to independent phase imperfections is made by a source map \(\XiPhase^i\), which is described in \cite[App, A]{nahar_imperfect_2023}, and can be different in each round \(i= 1,\dots,n\).

In order to prove the security of such a protocol, we follow a similar approach as in \cite{nahar_imperfect_2023} and translate it into our framework. We will mainly focus on the state \(\rho_{A'}^{(a,\mu)}\) because the additional source map will be included in Eve's attack by \cref{app:Source maps}.

First, let us note that each state \(\rho_{A'}^{(a,\mu)}\) is block-diagonal in some basis, which is only for \(q=1\) the photon-number basis. We will exploit this block-diagonal structure and derive a formulation which can be incorporated in our \Renyi framework presented for block-diagonal states \cref{sec:Block-Diag states}. 

However, there is the caveat that this block-diagonal structure can only be found approximately \cite{nahar_imperfect_2023} for arbitrary values of \(q\). If numerical methods are used to find this decomposition, one can only perform the block-diagonalization on a finite-dimensional subspace. Furthermore, states with phase imperfections as in \cref{eq:phase imperfect signal state} do not satisfy the assumptions \ref{Decoy assump block-diag}--\ref{Decoy assump QND Eve}. Hence, the goal of this section is to find a formulation of phase imperfections that can be immediately incorporated into our framework.

\subsection{Definition of block-tagged states}\label{subsec:Block-tagged states}
We start by constructing states that satisfy the assumptions \ref{Decoy assump block-diag}--\ref{Decoy assump QND Eve}, incorporating a cutoff required for a numerical block-diagonalization. Later we will show how to rigorously incorporate the cutoff.

For this purpose, let us define the state projected onto the space with \(\leq N_A\) photons
\begin{align}\label{eq:phase imperfect states with projection}
    \sigma_{A'}^{(a,\mu)} &\defvar \PiNA \rho_{A'}^{(a,\mu)} \PiNA + \PibarNA \rho_{A'}^{(a,\mu)} \PibarNA \\
    &= \sum_{m=0}^{\infty} \omega_m(a,\mu) \ketbra{w_m(a,\mu)}{w_m(a,\mu)}_{A'},
\end{align}
where the projectors onto the subspaces are
\begin{align}
    \Pi_{\leq N_A} \defvar \sum_{m \leq N_A} \ketbra{m}{m}, \quad \Pi_{> N_A} \defvar \sum_{m > N_A} \ketbra{m}{m}
\end{align}

Crucially, for different signal and intensity settings, (\(a,\mu\)), the states \(\ket{w_m(a,\mu)}_{A'}\) are not orthogonal for different values of \(m\). That means that the states \(\sigma^{(a,\mu)}\) violate the assumption~\ref{Decoy assump block-diag} because the block-diagonal structure over \(m\) is not the same for all signal choices (omitting the trivial case of only a single block of the full space). Furthermore, Eve could not apply a QND measurement without disturbing the state.

Thus, one cannot without loss of generality assume a block-diagonal channel as in \cref{sec:Block-Diag states}.

Therefore, we introduce block-tagged states that allow us to recover the block structure. We can incorporate block-tagged states by making use of source maps, see \cref{app:Source maps}. Hence, let us define the block-tagged version of the state \(\sigma_{A'}\) from above as
\begin{align}
    \tau_{A'\bar{A}}^{(a,\mu)} = \sum_{m=0}^{\infty} \omega_m(a,\mu) \ketbra{w_m(a,\mu)}{w_m(a,\mu)}_{A'} \otimes \ketbra{m}{m}_{\bar{A}}.
\end{align}
The states \(\tau_{A'\bar{A}}^{(a,\mu)}\) are now simultaneously block-diagonal in the same basis for all \(a,\mu\). Then, Alice's pure state after the source-replacement scheme including a shield system can be written as
\begin{equation}
    \ket{\psi}_{AA_sA'\bar{A}} = \sum_{m=0}^{\infty}\sqrt{\omega_m} \ket{m}_{A_s} \ket{\psi_{|m}}_{AA'\bar{A}},
\end{equation}
where
\begin{equation}
    \ket{\psi_{|m}}_{AA'\bar{A}} = \sum_{a,\mu} \sqrt{p(a,\mu|m)} \ket{a,\mu}_{A} \ket{w_m(a,\mu)}_{A'} \ket{m}_{\bar{A}}.
\end{equation}
Now, this state satisfies assumptions \ref{Decoy assump block-diag}--\ref{Decoy assump QND Eve} and intuitively by using a source map one can find a lower bound on the key rate.

Hence, based on the state \(\ket{\psi}\) we can then define the states
\begin{align}\label{eq:phase imperfect key rate states}
    \ket{\xi^g}_{A_S AA'\bar{A}} &= \sum_{m=0}^{\infty} \sqrt{\omega_{m|\gen}} \ket{m}_{A_S} \ket{\xi^g_{|m}}_{AA'\bar{A}}, \\
    \ket{\xi^t}_{A_S AA'\bar{A}} &= \sum_{m=0}^{\infty} \sqrt{\omega_{m|\test}} \ket{m}_{A_S} \ket{\xi^t_{|m}}_{AA'\bar{A}},
\end{align}
by conditioning on ``test" and ``gen" appropriately. This concludes the construction of the block-tagged states. However, we need to show that including the cutoff \(N_A\) indeed results in valid lower bounds on the key rates, which we will show in the next subsection.

\subsection{Approximate Diagonalization with Cutoff}\label{subsec:Approx Diag}
Now, we show how to rigorously incorporate the cutoff \(N_A\) required for the approximate diagonalization. For both \(\hQKD\) and the $\frenyiSandDown_\alpha$-normalization constant \(\kappa\) we cannot \textit{a priori} assume a cutoff, but intuitively, if we choose \(N_A\) large enough, the difference should be insignificant. We will formalize this intuition in this subsection.

Therefore, let us first define the states \(\ket{s_m(a,\mu)}_{A'}\) and probabilities \(\zeta_m(a,\mu)\) such that
\begin{equation}
    \rho_{A'}^{(a,\mu)} = \sum_{m=0}^{\infty} \zeta_m(a,\mu) \ketbra{s_m(a,\mu)}{s_m(a,\mu)}_{A'},
\end{equation}
i.e. they are found by diagonalizing \cref{eq:phase imperfect signal state} directly.

Then, similarly to \cref{eq:phase imperfect key rate states} let us define the source-replaced states without a cutoff (or equivalently \(N_A \rightarrow \infty\)) as
\begin{align}\label{eq:phase imperfect key rate state no cutoff}
    \ket{\widetilde{\xi}^g}_{A_S AA'\bar{A}} &= \sum_{m=0}^{\infty} \sqrt{\zeta_{m|\gen}} \ket{m}_{A_S} \ket{\widetilde{\xi}^g_{|m}}_{AA'\bar{A}}, \\
    \ket{\widetilde{\xi}^t}_{A_S AA'\bar{A}} &= \sum_{m=0}^{\infty} \sqrt{\zeta_{m|\test}} \ket{m}_{A_S} \ket{\widetilde{\xi}^t_{|m}}_{AA'\bar{A}},
\end{align}
where
\begin{equation}
    \begin{split}
        \ket{\widetilde{\xi}^g_{|m}}_{AA'\bar{A}} = \sum_{a,\mu} &\sqrt{p(a,\mu|m,\gen)} \\
        &\cdot\ket{a,\mu}_{A} \ket{s_m(a,\mu)}_{A'} \ket{m}_{\bar{A}},
    \end{split}
\end{equation}
and similarly for \(\ket{\widetilde{\xi}^t_{|m}}\). Here again the probabilities \(\zeta_{m|\test}\) are found by conditioning \(\zeta_m(a,\mu)\) appropriately on \(\test\) and \(\gen\) rounds.

Now, as a first step towards rigorously accounting for the cutoff, we aim to relate the conditional entropies with and without a cutoff \(N_A\) by
\begin{equation}
    \abs{\renyiSandUp_{\alpha}(S|\tilde{T}E)_{\widetilde{\nu}_{|\gen}} - \renyiSandUp_{\alpha}(S|\tilde{T}E)_{\nu_{|\gen}}} \leq \Delta,
\end{equation}
where \(\widetilde{\nu}\) and \(\nu\) are generated from \(\ket{\widetilde{\xi}^g}\) and \(\ket{\xi^g}\), respectively. This form of inequality can be understood as the worst case in which the cut-off could influence the final entropy.

One can derive a bound of this form immediately from the continuity bounds for conditional \Renyi entropies found in \cite[Corollary 5.1]{bluhm_unified_2024}, if a bound on the trace distance,
\begin{equation}
    \frac{1}{2} \norm{\nu_{|\gen} - \tilde{\nu}_{|\gen}},
\end{equation}
between the states \(\nu\) and \(\widetilde{\nu}\) is available.

In order to achieve such a bound on the trace distance, for now, let us assume that we have
\begin{equation}
    \frac{1}{2} \norm{\rho_{A'}^{(a,\mu)} - \sigma_{A'}^{(a,\mu)}}_1 \leq \epsProj^{a,\mu},
\end{equation}
for all \((a,\mu)\), which was proven in App. B of Ref.~\cite{nahar_imperfect_2023}. We will show the exact definition of \(\epsProj^{a,\mu}\) in \cref{subsec:Phase Imperfect Key Rate}, but for now, let us see how we could apply it to rigorously incorporate the cutoff.

Next, \cref{Lem:trace distance bounds source-replacement} in the appendix, together with the data-processing inequality, see e.g. \cite[Theorem 9.2]{nielsen_quantum_2010}, allows us to conclude
\begin{align}
    &\frac{1}{2} \norm{\nu_{|\gen} - \tilde{\nu}_{|\gen}} \\ 
    &\leq \frac{1}{2} \norm{ \Tr_{A_s} \circ \GMap \circ \left( \id_A \otimes \; \mathcal{E} \right) \left( \ketbra{\xi^g}{\xi^g} - \ketbra{\widetilde{\xi}^g}{\widetilde{\xi}^g} \right) }_1 \\
    &\leq \sum_{a,\mu} p(a,\mu|\gen) \epsProj^{a,\mu} \eqqcolon \epsFull,
\end{align}
where we defined \(\epsFull\) as the probabilistic mixture of the bounds \(\epsProj^{a,\mu}\) on the signal states.

Using this bound on the trace distance, one could directly apply \cite[Eq. (9) in Corollary 5.1]{bluhm_unified_2024} to find a version of \(\Delta\). However, we can slightly improve the bound in \cite[Corollary 5.1]{bluhm_unified_2024} because the register \(S\) is classical, i.e. \(\renyiSandUp_{\alpha}(S|\tilde{T}E)_\nu \geq 0\) for all \(\nu\). Hence, appropriately identifying the registers between our notation and the one in \cite{bluhm_unified_2024}, one can replace \(d_S^2\) with \(d_S\) in Eq. (9) of Ref.~\cite[Corollary 5.1]{bluhm_unified_2024} and use this as the definition of \(\Delta\). This will reduce the cost due to the continuity bound slightly and result in
\begin{equation}\label{eq:bounds-renyi-conditional-entropy-bluhm-replaced}
\begin{split}
    &\Delta(\epsilon) = \\
    &\min \begin{cases}
        \begin{aligned}
            &\log(1 + \varepsilon) \\ 
            &+ \frac{1}{\alpha - 1}\log(1 + \varepsilon d_S^{\alpha - 1} - \frac{\varepsilon^\alpha}{(1 + \varepsilon)^{\alpha - 1}})
        \end{aligned},\\
        \frac{\alpha}{\alpha - 1}\log(1 + \varepsilon d_S^{\frac{\alpha - 1}{\alpha}}) ,\\
        \begin{aligned}
             &\log(1 + \varepsilon) \\ 
             &+ \frac{\alpha}{\alpha - 1}\log(1 + \varepsilon d_S^{ \frac{\alpha - 1}{\alpha}} - \frac{\varepsilon^{2 - \frac{1}{\alpha}}}{(1 + \varepsilon)^{\frac{\alpha - 1}{\alpha}}}).
        \end{aligned}
     \end{cases}
\end{split}
\end{equation}

Thus, we find the following lower bound on the entropy excluding a cutoff in terms of the state including a cutoff \(N_A\)
\begin{align}\label{eq:Continuity lower bnd key rate states}
    \renyiSandUp_{\alpha}(S|\tilde{T}E)_{\widetilde{\nu}_{|\gen}} &\geq \renyiSandUp_{\alpha}(S|\tilde{T}E)_{\nu_{|\gen}} - \Delta(\epsFull) \\ 
    &\geq \renyiSandDown_{\alpha}(S|\tilde{T}E)_{\nu_{|\gen}} - \Delta(\epsFull),
\end{align}
where the correction term \(\Delta\) is given by the tightened version of \cref{eq:bounds-renyi-conditional-entropy-bluhm-replaced} from above.

Next, we need to incorporate the difference in the states due to the cutoff \(N_A\) into the map \(\probst\) or \(\probstJ\). Therefore, let us assume the signal states in system \(A'\) and the probabilities satisfy
\begin{align}
    \begin{split}
        &\Big\lVert \ketbra{s_{m}(a,\mu)}{s_{m}(a,\mu)} \\
        &\qquad - \ketbra{w_{m}(a,\mu)}{w_{m}(a,\mu)} \Big \rVert_1 \leq \epsilon_{a,\mu},
    \end{split} \\
    &\abs{\zeta_{m}(a,\mu) - \omega_{m}(a,\mu)} \leq \epsProb,
\end{align}
which was proven in \cite[Theorem 6]{nahar_imperfect_2023}. Again in \cref{subsec:Phase Imperfect Key Rate} we will present the exact definitions, but for now, let us see how we could apply it to rigorously incorporate the cutoff in the map \(\probst\). We summarize the bounds in the following lemma.

\begin{restatable}[Bounds on Statistics under Approximate Diagonalization]{lemma}{LemBoundsOnStatisticsApprox}\label{lem:Bounds on Statistics under Approx Diag}
    Let \(\ket{\xi^t}\) and \(\ket{\widetilde{\xi}^t}\) be defined as in \cref{eq:phase imperfect key rate states} and \cref{eq:phase imperfect key rate state no cutoff}, respectively. Furthermore, let the states and probabilities satisfy
    \begin{align}
        \begin{split}
            &\Big\lVert \ketbra{s_{m}(a,\mu)}{s_{m}(a,\mu)} \\
            &\qquad - \ketbra{w_{m}(a,\mu)}{w_{m}(a,\mu)} \Big \rVert_1 \leq \epsilon_{a,\mu},
        \end{split} \\
        &\abs{\zeta_{m}(a,\mu) - \omega_{m}(a,\mu)} \leq \epsProb.
    \end{align}
    Then, it holds
    \begin{align}
    \begin{split}
        &\sum_{m} \zeta_{m|\test} \Phi_{m}\circ\mathcal{E}_m\left[\ketbra{\widetilde{\xi}^t_{|m}}{\widetilde{\xi}^t_{|m}}\right] \\
        &\geq \sum_{m} \omega_{m|\test}^L \left(\Phi_{m}\circ\mathcal{E}_m\left[\ketbra{\xi^t_{|m}}{\xi^t_{|m}}\right] - \epsilon_m \right),
    \end{split}
    \end{align}
    and
    \begin{align}
    \begin{split}
        &\sum_{m} \zeta_{m|\test} \Phi_{m}\circ\mathcal{E}_m\left[\ketbra{\widetilde{\xi}^t_{|m}}{\widetilde{\xi}^t_{|m}}\right] + 1-\mbf{p}_{\mathrm{tot}}(\widetilde{\xi}) \\
        &\leq \sum_{m} \omega_{m|\test}^U \left(\Phi_{m}\circ\mathcal{E}_m\left[\ketbra{\xi^t_{|m}}{\xi^t_{|m}}\right] + \epsilon_m \right) \\
        &\quad + 1-\mbf{p}_{\mathrm{tot}}^L ,
    \end{split}
    \end{align}
    where 
    \begin{align}
        &\omega_{m|\test}^U \defvar  \min\{\omega_{m|\test} + \epsProb,1\}, \\
        &\omega_{m|\test}^L \defvar  \max\{\omega_{m|\test} - \epsProb,0\}, \\
        &\mbf{p}_{\mathrm{tot}}^L:= \sum_{\substack{\cP \in \Ct \\ a \in \phi^{-1}(\cP)}} \sum_{m \leq \Ncons} \omega_{m|\test}^L p(a|m,\test) \hat{e}_{\cP},
    \end{align}
    and
    \begin{align}
        \epsilon_m &\defvar \sum_{a,\mu} p(a,\mu|\test,m) \epsilon_{a,\mu}, \\
        \epsProb &\defvar \sum_{a,\mu} p(a,\mu|\test) \epsProj^{a,\mu}.
    \end{align}
\end{restatable}
\begin{proof}
    See the proof of the restated \cref{lem:Bounds on Statistics under Approx Diag} in \cref{Proof:LemBoundsOnStatisticsApprox}.
\end{proof}

Hence, we have all the tools to state the final expression for \(\hQKD\) and the $\frenyiSandDown_\alpha$-normalization constant \(\kappa\) for phase imperfections. We will summarize the result in the next subsection.

\subsection{Key rate construction}\label{subsec:Phase Imperfect Key Rate}
In the previous subsections, we found relations that allow us to rigorously incorporate the cutoff \(N_A\), requiring the parameters \(\epsilon_m, \epsProb\) and \(\epsProj\). First, we will show how these can be calculated and then state \cref{thrm:hQKD and kappa phase imperfections} which gives the final expressions required for fixed- and variable-length key rates.

Hence, let us start with the bound \(\epsFull\) on the trace distance of the final states in generation rounds. As already mentioned, in App. B of Ref.~\cite{nahar_imperfect_2023} it was shown that
\begin{equation}
    \frac{1}{2} \norm{\rho_{A'}^{(a,\mu)} - \sigma_{A'}^{(a,\mu)}}_1 \leq \epsProj^{a,\mu},
\end{equation}
where
\begin{equation}
    \epsProj^{a,\mu} = (1-q) \sqrt{W_{a,\mu}},
\end{equation}
and \(W_{a,\mu} \defvar \Tr[\PibarNA \rho_{A'}^{(a,\mu)}]\) is the weight contained outside the subspace. Furthermore, note that the weights \(W_{a,\mu}\) satisfy \(W_{a,\mu} = \Tr[\PibarNA \sigma_{A'}^{(a,\mu)}]\). Thus,
\begin{align}
    \epsFull &\defvar \sum_{a,\mu} p(a,\mu|\gen) \epsProj^{a,\mu} \\
    &= (1-q) \sum_{a,\mu} p(a,\mu|\gen) \sqrt{W_{a,\mu}}.
\end{align}

Next, we continue with \(\epsilon_m\) and \(\epsProb\). In Ref.~\cite[Theorem 6]{nahar_imperfect_2023}, it was also found that
\begin{align}
    &\abs{\zeta_{m}(a,\mu) - \omega_{m}(a,\mu)} \leq \epsProj^{a,\mu} \\
    \begin{split}
        &\Big\lVert \ketbra{s_{m}(a,\mu)}{s_{m}(a,\mu)} \\
        &\qquad - \ketbra{w_{m}(a,\mu)}{w_{m}(a,\mu)} \Big \rVert_1 \leq \frac{2\epsProj^{a,\mu}}{\delta_m(a,\mu)},
    \end{split}
\end{align}
where
\begin{equation}
\begin{split}
    \delta_m(a,\mu) &= \min\Big\{\omega_{m}(a,\mu) - \omega_{m-1}(a,\mu) - \epsProj^{a,\mu},  \\ &\quad \omega_{m+1}(a,\mu) -\omega_{m}(a,\mu) - \epsProj^{a,\mu} \Big\} \; \forall m\geq 1, \\
    \delta_0(a,\mu) &= \omega_{1}(a,\mu) - \omega_{0}(a,\mu) -\epsProj^{a,\mu}.
    \end{split}
\end{equation}
Thus, we find
\begin{align}
    \epsProb &= \sum_{a,\mu} p(a,\mu|\test) \epsProj^{a,\mu}\\
    &= (1-q) \sum_{a,\mu} p(a,\mu|\test) \sqrt{W_{a,\mu}}, \\
    \epsilon_m &= \sum_{a,\mu} p(a,\mu|\test,m) \epsilon_{a,\mu} \\
    &= 2 (1-q) \sum_{a,\mu} p(a,\mu|\test,m) \frac{ \sqrt{W_{a,\mu}}}{\delta_m(a,\mu)}.
\end{align}

Finally, in the theorem below, we summarize the results for \(\hUpQKD\) and the bound on the $\frenyiSandUp_\alpha$-normalization constants \(\kapup_i\) assuming the availability of the parameters \(\epsilon_m,\epsProb\) and \(\epsProj\).

\begin{restatable}[Fixed and Variable-Length key rates with Phase Imperfections]{theorem}{ThrmPhaseImperfectRates}\label{thrm:hQKD and kappa phase imperfections}
    Let Alice's source prepare states with independent phase imperfections, and let
    \begin{equation}
        \XiPhase = \bigotimes_{i=1}^n \XiPhase^i,
    \end{equation}
    be the source map from \cite[App A.]{nahar_imperfect_2023}. Furthermore, define the model state
    \begin{equation}
        \rho_{A'}^{(a,\mu)} =  q \sum_{m=0}^{\infty} p(m|\mu) \ketbra{m}{m}_{a}  + (1-q) \ketbra{\sqrt{\mu}}{\sqrt{\mu}},
    \end{equation}
    with \(q \in [0,1]\) and assume that \(\Nent,\Ncons\leq N_A\).

    Then, all $\frenyiSandUp_\alpha$-normalization constants \(\kapup_i\) can be bounded by
    \begin{align}
    \begin{aligned}
        \kapup_i \geq &\inf_{\substack{J_m \in \mathrm{Pos}(A'\bar{A}B), \\ \bsym{\nu} \in \mathbb{P}(\Ct) }} \Bigg( \frac{1}{1-\alpha} \log \Big( \gamma \sum_{c\in \Ct} \bsym{\nu}(\cP)2^{(\alpha - 1) f(\cP)} \\
        &\qquad + \left(1-\gamma\right)2^{(\alpha-1)\left(\Delta(\epsFull)+ f(\gen)\right)} \\ 
        &\qquad \cdot 2^{-(\alpha-1)g_{\alpha}^{\Nent}(J_0,\dots J_{\Nent}) } \Big) \Bigg) \\
        \quad \textrm{s.t. } &\Tr_B[J_m] = \idop_{A'\bar{A}}, \\
        &\bsym{\nu} \leq \sum_{m\leq \Ncons} \omega_{m|\test}^U \left(\probstJ_m[J_m] + \epsilon_m \right) \\
        &\quad + 1-\mbf{p}_{\mathrm{tot}}^L, \\
        &\bsym{\nu} \geq \sum_{m\leq \Ncons} \omega_{m|\test}^L \left(\probstJ_m[J_m] - \epsilon_m \right).
        \end{aligned}
    \end{align}    
    Furthermore, if the phase imperfections are IID, i.e. all source maps \(\XiPhase^i\) are equal, then \(\hUpQKD\) is bounded from below by 
    \begin{equation}\label{eq:halpha phase imperfect}
    \begin{aligned}[t]
        \hUpQKD \geq &\inf_{\substack{\mbf{q} \in \Sacc, \bsym{\nu}_{\CP} \in \mathbb{P}(\alphCP), \\ J_m \in \mathrm{Pos}(A'\bar{A}B)}} \Bigg( \frac{\alpha D\rel{\mbf{q}}{\bsym{\nu}_{\CP}}}{\alpha - 1} - \Delta(\epsFull) \\ 
        &\quad + q(\gen) g_{\alpha}^{\Nent}\left(J_0,\dots, J_{\Nent} \right) \Bigg)\\
        \textrm{s.t. } &\Tr_B[J_m] = \idop_{A'\bar{A}}, \\
        &\bsym{\nu}_{\CP} = (\gamma \bsym{\nu}_{\CP|\test}, 1-\gamma)^T, \\
        &\bsym{\nu}_{\CP|\test} \leq \sum_{m\leq \Ncons} \omega_{m|\test}^U \left(\probstJ_m[J_m]  + \epsilon_m \right) \\
        &\quad + 1-\mbf{p}_{\mathrm{tot}}^L, \\
        &\bsym{\nu}_{\CP|\test} \geq \sum_{m\leq \Ncons} \omega_{m|\test}^L \left(\probstJ_m[J_m] - \epsilon_m \right).
    \end{aligned}
    \end{equation}
    For both quantities, we defined
    \begin{equation}
        \begin{split}
            &\omega_{m|\test}^U \defvar  \min\{\omega_{m|\test} + \epsProb,1\}, \\
            &\omega_{m|\test}^L \defvar  \max\{\omega_{m|\test} - \epsProb,0\}, \\
            &\mbf{p}_{\mathrm{tot}}^L \defvar \sum_{\substack{\cP \in \Ct \\ a \in \phi^{-1}(\cP)}} \sum_{m \leq \Ncons} \omega_{m|\test}^L p(a|m,\test) \hat{e}_{\cP},
        \end{split}
    \end{equation}
    and both \(g_{\alpha}^{\Nent}(J_0,\dots J_{\Nent})\) and \(\probstJ_m\) are defined in terms of the states \(\ket{\xi^g_{|m}}\) and \(\ket{\xi^t_{|m}}\), respectively.
\end{restatable}
\begin{proof}
    See the proof of the restated \cref{thrm:hQKD and kappa phase imperfections} in \cref{Proof:ThrmPhaseImperfectRates}.
\end{proof}

\subsection{Example: 4-6 Protocol with Phase Imperfections}\label{subsec:Ex 4-6 Protocol with Phase Imp}

In this section, we present results for variable-length secret key rates for a decoy-state 4-6 protocol with phase imperfections. Again, we model phase imperfections as in \cref{eq:phase imperfect signal state} derived from earlier works \cite{nahar_imperfect_2023,curras-lorenzo_security_2023}. Under this model, the quality of the phase randomization is characterized by the quality factor \(q \in[0,1]\). If \(q=1\) the states are perfectly phase-randomized, i.e. no phase imperfections, and any value of \(q<1\) states that Alice's source has phase imperfections.

As in the previous sections, in each round, Alice decides with probability \(\gamma\) if it is a test or a generation round and always selects the signal intensity \(\mu_s\) and the \(Z\) basis if it was a key generation round. Therefore, as before, Alice's basis choice probabilities are given by \(p_z =1-\gamma\) and \(p_x = \gamma\). In a test round, Alice always selects the \(X\) basis and all possible intensities with uniform probability. Now, we present only the one-decoy version of the protocol and choose a \emph{fixed testing probability} of \(\gamma=1/2\). Furthermore, we chose the intensities to be 
\begin{equation}
    \mu_s = 0.9 \quad \text{and} \quad \mu_2 = 0.2.
\end{equation}

On the receiver side (Bob), we now assume a passive six-state detection setup that chooses the \(Z\), \(X\), and \(Y\) basis with splitting ratios \(p_z =1-\gamma\), \(p_x = \gamma/2\) and \(p_y = \gamma/2\) respectively. To incorporate this into our security proof, we apply the flag-state squasher from \cite{zhang_security_2021} with the subspace bound from \cite[Theorem 1]{kamin_improved_2024}. Again, in \cref{app:Flag-state squasher} we show that the flag-state squasher can be incorporated.

Finally, we assume the same security parameters \(\ePA = \eEV = \frac{1}{2}10^{-80}\) and error correction model with efficiency \(\fEC=1.1\) as in \cref{sec:Qubit BB84 and Comparisons with other Proof Techniques}. We again model channel loss similarly to the qubit protocol and include a misalignment with an angle of \(\misalign = 0.03\). 

For this protocol we calculate the Kraus operators, states and POVM elements numerically and we describe the procedure in \cref{app:Kraus ops passive decoy 46}.

\begin{figure}[tb]
	\centering
	\includegraphics[width=\linewidth]{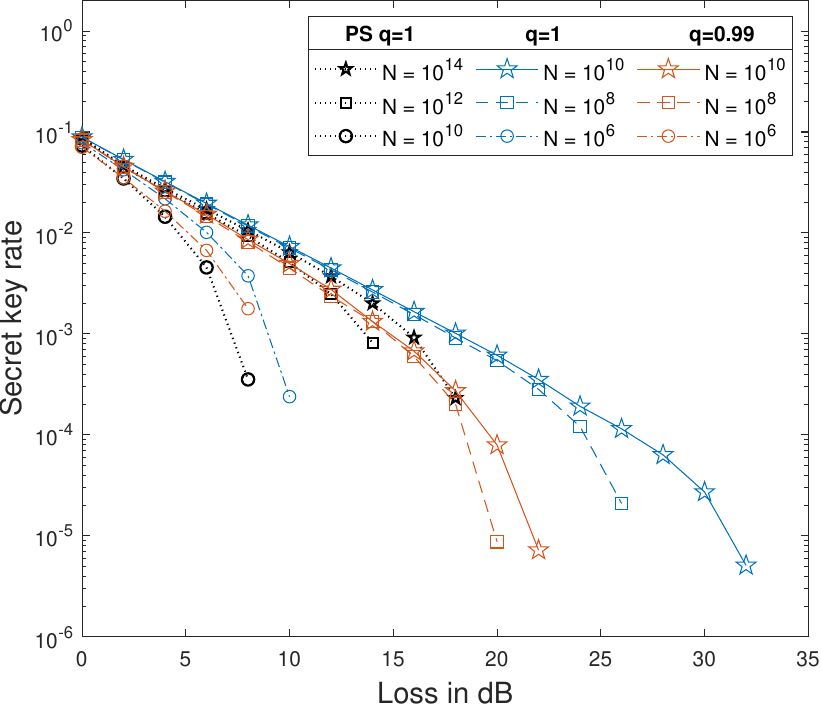}
	\caption{Results for variable-length secret key rates of the passive decoy-state 4-6 protocol with \emph{one} decoy intensity and phase imperfections plotted against the channel loss in \(\unit{dB}\) with a varying number of total signals sent \(N=10^6, 10^{8}, 10^{10} \) and different quality factors \(q=1\) and \(q=0.99\). The security parameters, and error correction efficiency were chosen as \(\ePA = \eEV = \frac{1}{2}10^{-80}\) and \(\fEC=1.1\). As a comparison, we showcase results from \cite{kamin_improved_2025}, which applies the postselection technique \cite{nahar_postselection_2024}, for \(N=10^{10}, 10^{12}, 10^{14}\) marked with (PS). For our work we optimize the \Renyi parameter \(\alpha\).}
	\label{fig:Decoy46}
\end{figure}

In \cref{fig:Decoy46} we show our resulting variable-length key rates for several choices of number of signals sent (\(N= 10^{6}, 10^{8}\) and \(10^{10}\)) and the quality factor (\(q =1\) and \(q=0.99\)),  plotted against the loss in \(\unit{dB}\). Moreover, we compare our key rates with the findings in Ref.~\cite{kamin_improved_2025} for \(q=1\). Since the key rates from Ref.~\cite{kamin_improved_2025} perform significantly worse, we show its results for \(N=10^{10},10^{12}\) and \(10^{14}\). 

First, let us discuss the different variable-length key rates for the perfect case, i.e. \(q=1\). The improvements of our methods presented in this work compared to Ref.~\cite{kamin_improved_2025} (which makes use of the postselection technique from \cite{nahar_postselection_2024}) are so dramatic that even a difference of four orders of magnitude in $N$ does not suffice to achieve comparable rates. For example, our rates for \(N=10^6\) signals (blue circles) still outperform \(N=10^{10}\) signals with the postselection technique.

Furthermore, in \cref{fig:Decoy46} we present the imperfect case with \(q=0.99\) in orange curves. The reduction in key rates is relatively significant, especially for higher channel losses. For instance, for \(n=10^6\) signals sent, the difference between the perfect case (blue curves) and the imperfect case (orange curves) is rather small, whereas for \(n=10^{10}\) signals sent this difference becomes much bigger.

Here we also mention that similar to all previous examples, at high channel loss,
our numerical methods encounter difficulty in converging to a tight enough bound to certify positive key rates. For the imperfect case, this is a particular issue since we cannot use the same simplifications of the optimization problem as for perfect decoy-state protocols. We are required to use \(\Ncons\)-many Choi states as optimization variables. That not only slows down the optimization routine, but also from our experience affects the precision of
the underlying solver. Therefore, we again highlight that for future research one should focus on improving the performance of the numerical methods. Then, the key rates in the high-loss regime might actually improve even further.

Finally, let us discuss the prospects of including phase imperfections into other proof techniques. First, the postselection technique could also incorporate independent phase imperfections in the same way as presented by us. However, the key rates would drop even more drastically due to the dimensionality scaling of the correction factor. That would make the key rates even worse than the perfect case presented in \cref{fig:Decoy46} (black curves).

Next, as of now, EUR and phase error correction based proofs cannot incorporate this form of imperfection in the finite-size regime. Hence, so far our framework presented here is the only available option for obtaining reasonable finite-size key rates in this scenario.

%\newpage
\section{Conclusion}\label{sec:Conclusion}
In this work, we developed a flexible framework for security proofs against coherent attacks for both fixed- and variable-length QKD protocols. 
Our approach introduces new numerical methods for tightly bounding the sandwiched conditional entropy \(\renyiSandDown_{\alpha}\), by relating it to a particular form of Petz divergence and then applying a Frank-Wolfe optimization method. This complements independent work~\cite{chung_generalized_2025} that implemented Frank-Wolfe methods to evaluate some forms of sandwiched divergences, which they used to lower bound \(\renyiSandUp_{\alpha}\) (we believe that the computations in their work are equivalent to evaluating a tight bound on the Petz conditional entropy \(\renyiPetzUp_{\alpha}\)). Along the way, we derived \Renyi versions of the dual relations found in \cite{coles_unification_2012}, which later formed the basis for the numerical framework for asymptotic key rates in Ref.~\cite{winick_reliable_2018}.

To demonstrate the versatility of our framework, we applied it to various protocol examples, including a qubit BB84 and an active decoy-state BB84 protocol. In addition, we extended our framework to protocols that possess experimental imperfections, further showcasing its adaptability. Our results demonstrate significant improvements over existing proof techniques, particularly for small block sizes, where our approach consistently outperforms reference methods in achievable secret key rates.

Furthermore, our framework is capable of handling generic block-diagonal protocols, including active and passive decoy-state protocols. Moreover, our work accommodates arbitrary passive detection setups via the flag-state squasher \cite{zhang_security_2021,kamin_improved_2024}, and also yields performance surpassing all existing proof techniques capable of including those setups, at the time of writing.

In addition, in \cref{subsec:Ex passive BB84 with Intensity Imp} and \cref{subsec:Ex 4-6 Protocol with Phase Imp}, we presented the first finite-size security analysis for generic protocols with intensity and phase imperfections. In the case of phase imperfections, we extend beyond prior work that only considered the asymptotic limit, Refs.~\cite{nahar_imperfect_2023,curras-lorenzo_security_2023}. For intensity imperfections, Refs.~\cite{wang_tight_2016,zhou_finite-key_2022} specialized to an active BB84 protocol, and Refs.~\cite{zapatero_security_2021,sixto_security_2022} proved security in the asymptotic limit, whereas our methods provide finite-size security against coherent attacks for generic decoy-state protocols.

For experimental implementations, our framework offers significant practical advantages. Specifically, in variable-length protocols, the computational burden of solving optimization problems is a one-time process. Once solved, variable-length rates can be efficiently calculated using a simple affine function dramatically reducing the overheads and computational demands on QKD devices.

Despite these advancements, some challenges remain. Our current method evaluates an ``unoptimized" version of sandwiched \Renyi entropy, \(\renyiSandDown_{\alpha}\), instead of its ``optimized'' counterpart, \(\renyiSandUp_{\alpha}\), leading to minor suboptimalities in high-loss regimes and small block sizes. Additionally, the numerical methods exhibit some instability that requires tuning for reliable results. Future work should focus on improving numerical stability (particularly in high-loss scenarios), better global optimization routines, and exploring alternative optimization techniques (for instance generalizing the interior-point solvers in~\cite{hu_robust_2022,lorente_quantum_2025,he_exploiting_2024,he_operator_2025} to handle $\frenyiSandUp_\alpha$ or $\frenyiSandDown_\alpha$) to enhance performance.

Overall, our work establishes a robust and scalable security proof framework for QKD protocols, imposing minimal restrictions while achieving the highest reported secret key rates at the time of writing. Moreover, our framework allows one to include many protocols and imperfections that are relevant for practical implementations. As such, our work can serve as a foundation for future advancements in security proofs against coherent attacks in practical QKD implementations.

\section*{Author Contributions}
LK was the principal investigator and lead author, responsible for the theoretical formulation, numerical implementation, data analysis, figure preparation, and manuscript writing. JB implemented parts of the Frank–Wolfe algorithm, derived the required gradient expressions, and developed the initial perturbation methods later extended by LK. ET provided conceptual guidance and critical feedback on multiple drafts of the manuscript, and developed the methods in \cref{app:RenyiQKDcones,app:Monotonicity of Log-Mean-Exp}.

\section*{Code Availability}
The code used to prepare the results in this paper will be available through \href{https://openqkdsecurity.wordpress.com/repositories-for-publications/}{https://openqkdsecurity.wordpress.com/repositories-for-publications/} or directly at \cite{kamin_renyi_2025}.
\newline

\section*{Acknowledgments}
We would like to thank Norbert Lütkenhaus for feedback and discussions. Furthermore, we thank Amir Arqand for many insightful discussions explaining the GREAT and MEAT framework. Additionally, we thank Devashish Tupkary for explanations of EUR-based security proofs and Shlok Nahar for discussions on device imperfections. Finally, we thank Laura Gracie for contributing her new optimization module to the openQKDsecurity software suite.

The research has been conducted at the Institute for Quantum Computing, at the University of Waterloo, which is supported by Innovation, Science, and Economic Development Canada and the NSERC Alliance. NSERC provided support under the Discovery Grants Program, Grant No. 341495, Alliance Grant QUINT, and Alliance Grant ReFQ.

%References
\bibliography{InPrepRefs,references,SpecialCharRefs}

\begin{thebibliography}{94}%
\makeatletter
\providecommand \@ifxundefined [1]{%
 \@ifx{#1\undefined}
}%
\providecommand \@ifnum [1]{%
 \ifnum #1\expandafter \@firstoftwo
 \else \expandafter \@secondoftwo
 \fi
}%
\providecommand \@ifx [1]{%
 \ifx #1\expandafter \@firstoftwo
 \else \expandafter \@secondoftwo
 \fi
}%
\providecommand \natexlab [1]{#1}%
\providecommand \enquote  [1]{``#1''}%
\providecommand \bibnamefont  [1]{#1}%
\providecommand \bibfnamefont [1]{#1}%
\providecommand \citenamefont [1]{#1}%
\providecommand \href@noop [0]{\@secondoftwo}%
\providecommand \href [0]{\begingroup \@sanitize@url \@href}%
\providecommand \@href[1]{\@@startlink{#1}\@@href}%
\providecommand \@@href[1]{\endgroup#1\@@endlink}%
\providecommand \@sanitize@url [0]{\catcode `\\12\catcode `\$12\catcode
  `\&12\catcode `\#12\catcode `\^12\catcode `\_12\catcode `\%12\relax}%
\providecommand \@@startlink[1]{}%
\providecommand \@@endlink[0]{}%
\providecommand \url  [0]{\begingroup\@sanitize@url \@url }%
\providecommand \@url [1]{\endgroup\@href {#1}{\urlprefix }}%
\providecommand \urlprefix  [0]{URL }%
\providecommand \Eprint [0]{\href }%
\providecommand \doibase [0]{https://doi.org/}%
\providecommand \selectlanguage [0]{\@gobble}%
\providecommand \bibinfo  [0]{\@secondoftwo}%
\providecommand \bibfield  [0]{\@secondoftwo}%
\providecommand \translation [1]{[#1]}%
\providecommand \BibitemOpen [0]{}%
\providecommand \bibitemStop [0]{}%
\providecommand \bibitemNoStop [0]{.\EOS\space}%
\providecommand \EOS [0]{\spacefactor3000\relax}%
\providecommand \BibitemShut  [1]{\csname bibitem#1\endcsname}%
\let\auto@bib@innerbib\@empty
%</preamble>
\bibitem [{\citenamefont {Koashi}(2005)}]{koashi_simple_2005}%
  \BibitemOpen
  \bibfield  {author} {\bibinfo {author} {\bibfnamefont {M.}~\bibnamefont
  {Koashi}},\ }\href {https://doi.org/10.48550/arXiv.quant-ph/0505108}
  {\bibinfo {title} {Simple security proof of quantum key distribution via
  uncertainty principle}} (\bibinfo {year} {2005}),\ \bibinfo {note}
  {arXiv:quant-ph/0505108}\BibitemShut {NoStop}%
\bibitem [{\citenamefont {Koashi}(2006)}]{koashi_efficient_2006}%
  \BibitemOpen
  \bibfield  {author} {\bibinfo {author} {\bibfnamefont {M.}~\bibnamefont
  {Koashi}},\ }\href {https://doi.org/10.48550/arXiv.quant-ph/0609180}
  {\bibinfo {title} {Efficient quantum key distribution with practical sources
  and detectors}} (\bibinfo {year} {2006}),\ \bibinfo {note}
  {arXiv:quant-ph/0609180}\BibitemShut {NoStop}%
\bibitem [{\citenamefont {Koashi}(2009)}]{koashi_simple_2009}%
  \BibitemOpen
  \bibfield  {author} {\bibinfo {author} {\bibfnamefont {M.}~\bibnamefont
  {Koashi}},\ }\bibfield  {title} {\bibinfo {title} {Simple security proof of
  quantum key distribution based on complementarity},\ }\href
  {https://doi.org/10.1088/1367-2630/11/4/045018} {\bibfield  {journal}
  {\bibinfo  {journal} {New Journal of Physics}\ }\textbf {\bibinfo {volume}
  {11}},\ \bibinfo {pages} {045018} (\bibinfo {year} {2009})}\BibitemShut
  {NoStop}%
\bibitem [{\citenamefont {Hayashi}\ and\ \citenamefont
  {Tsurumaru}(2012)}]{hayashi_concise_2012}%
  \BibitemOpen
  \bibfield  {author} {\bibinfo {author} {\bibfnamefont {M.}~\bibnamefont
  {Hayashi}}\ and\ \bibinfo {author} {\bibfnamefont {T.}~\bibnamefont
  {Tsurumaru}},\ }\bibfield  {title} {\bibinfo {title} {Concise and tight
  security analysis of the {Bennett}–{Brassard} 1984 protocol with finite key
  lengths},\ }\href {https://doi.org/10.1088/1367-2630/14/9/093014} {\bibfield
  {journal} {\bibinfo  {journal} {New Journal of Physics}\ }\textbf {\bibinfo
  {volume} {14}},\ \bibinfo {pages} {093014} (\bibinfo {year} {2012})},\
  \bibinfo {note} {publisher: IOP Publishing}\BibitemShut {NoStop}%
\bibitem [{\citenamefont {Hayashi}\ and\ \citenamefont
  {Nakayama}(2014)}]{hayashi_security_2014}%
  \BibitemOpen
  \bibfield  {author} {\bibinfo {author} {\bibfnamefont {M.}~\bibnamefont
  {Hayashi}}\ and\ \bibinfo {author} {\bibfnamefont {R.}~\bibnamefont
  {Nakayama}},\ }\bibfield  {title} {\bibinfo {title} {Security analysis of the
  decoy method with the {Bennett}–{Brassard} 1984 protocol for finite key
  lengths},\ }\href {https://doi.org/10.1088/1367-2630/16/6/063009} {\bibfield
  {journal} {\bibinfo  {journal} {New Journal of Physics}\ }\textbf {\bibinfo
  {volume} {16}},\ \bibinfo {pages} {063009} (\bibinfo {year} {2014})},\
  \bibinfo {note} {publisher: IOP Publishing}\BibitemShut {NoStop}%
\bibitem [{\citenamefont {Tomamichel}\ and\ \citenamefont
  {Renner}(2011)}]{tomamichel_uncertainty_2011}%
  \BibitemOpen
  \bibfield  {author} {\bibinfo {author} {\bibfnamefont {M.}~\bibnamefont
  {Tomamichel}}\ and\ \bibinfo {author} {\bibfnamefont {R.}~\bibnamefont
  {Renner}},\ }\bibfield  {title} {\bibinfo {title} {Uncertainty {Relation} for
  {Smooth} {Entropies}},\ }\href
  {https://doi.org/10.1103/PhysRevLett.106.110506} {\bibfield  {journal}
  {\bibinfo  {journal} {Physical Review Letters}\ }\textbf {\bibinfo {volume}
  {106}},\ \bibinfo {pages} {110506} (\bibinfo {year} {2011})},\ \bibinfo
  {note} {publisher: American Physical Society}\BibitemShut {NoStop}%
\bibitem [{\citenamefont {Tomamichel}\ \emph {et~al.}(2012)\citenamefont
  {Tomamichel}, \citenamefont {Lim}, \citenamefont {Gisin},\ and\ \citenamefont
  {Renner}}]{tomamichel_tight_2012}%
  \BibitemOpen
  \bibfield  {author} {\bibinfo {author} {\bibfnamefont {M.}~\bibnamefont
  {Tomamichel}}, \bibinfo {author} {\bibfnamefont {C.~C.~W.}\ \bibnamefont
  {Lim}}, \bibinfo {author} {\bibfnamefont {N.}~\bibnamefont {Gisin}},\ and\
  \bibinfo {author} {\bibfnamefont {R.}~\bibnamefont {Renner}},\ }\bibfield
  {title} {\bibinfo {title} {Tight finite-key analysis for quantum
  cryptography},\ }\href {https://doi.org/10.1038/ncomms1631} {\bibfield
  {journal} {\bibinfo  {journal} {Nature Communications}\ }\textbf {\bibinfo
  {volume} {3}},\ \bibinfo {pages} {634} (\bibinfo {year} {2012})},\ \bibinfo
  {note} {number: 1 Publisher: Nature Publishing Group}\BibitemShut {NoStop}%
\bibitem [{\citenamefont {Lim}\ \emph {et~al.}(2014)\citenamefont {Lim},
  \citenamefont {Curty}, \citenamefont {Walenta}, \citenamefont {Xu},\ and\
  \citenamefont {Zbinden}}]{lim_concise_2014}%
  \BibitemOpen
  \bibfield  {author} {\bibinfo {author} {\bibfnamefont {C.~C.~W.}\
  \bibnamefont {Lim}}, \bibinfo {author} {\bibfnamefont {M.}~\bibnamefont
  {Curty}}, \bibinfo {author} {\bibfnamefont {N.}~\bibnamefont {Walenta}},
  \bibinfo {author} {\bibfnamefont {F.}~\bibnamefont {Xu}},\ and\ \bibinfo
  {author} {\bibfnamefont {H.}~\bibnamefont {Zbinden}},\ }\bibfield  {title}
  {\bibinfo {title} {Concise security bounds for practical decoy-state quantum
  key distribution},\ }\href {https://doi.org/10.1103/PhysRevA.89.022307}
  {\bibfield  {journal} {\bibinfo  {journal} {Physical Review A}\ }\textbf
  {\bibinfo {volume} {89}},\ \bibinfo {pages} {022307} (\bibinfo {year}
  {2014})}\BibitemShut {NoStop}%
\bibitem [{\citenamefont {Rusca}\ \emph {et~al.}(2018)\citenamefont {Rusca},
  \citenamefont {Boaron}, \citenamefont {Grünenfelder}, \citenamefont
  {Martin},\ and\ \citenamefont {Zbinden}}]{rusca_finite-key_2018}%
  \BibitemOpen
  \bibfield  {author} {\bibinfo {author} {\bibfnamefont {D.}~\bibnamefont
  {Rusca}}, \bibinfo {author} {\bibfnamefont {A.}~\bibnamefont {Boaron}},
  \bibinfo {author} {\bibfnamefont {F.}~\bibnamefont {Grünenfelder}}, \bibinfo
  {author} {\bibfnamefont {A.}~\bibnamefont {Martin}},\ and\ \bibinfo {author}
  {\bibfnamefont {H.}~\bibnamefont {Zbinden}},\ }\bibfield  {title} {\bibinfo
  {title} {Finite-key analysis for the 1-decoy state {QKD} protocol},\ }\href
  {https://doi.org/10.1063/1.5023340} {\bibfield  {journal} {\bibinfo
  {journal} {Applied Physics Letters}\ }\textbf {\bibinfo {volume} {112}},\
  \bibinfo {pages} {171104} (\bibinfo {year} {2018})},\ \bibinfo {note}
  {publisher: American Institute of Physics Inc.}\BibitemShut {Stop}%
\bibitem [{\citenamefont {Tupkary}\ \emph
  {et~al.}(2024{\natexlab{a}})\citenamefont {Tupkary}, \citenamefont {Nahar},
  \citenamefont {Sinha},\ and\ \citenamefont
  {Lütkenhaus}}]{tupkary_phase_2024}%
  \BibitemOpen
  \bibfield  {author} {\bibinfo {author} {\bibfnamefont {D.}~\bibnamefont
  {Tupkary}}, \bibinfo {author} {\bibfnamefont {S.}~\bibnamefont {Nahar}},
  \bibinfo {author} {\bibfnamefont {P.}~\bibnamefont {Sinha}},\ and\ \bibinfo
  {author} {\bibfnamefont {N.}~\bibnamefont {Lütkenhaus}},\ }\href
  {https://doi.org/10.48550/arXiv.2408.17349} {\bibinfo {title} {Phase error
  rate estimation in {QKD} with imperfect detectors}} (\bibinfo {year}
  {2024}{\natexlab{a}}),\ \bibinfo {note} {arXiv:2408.17349
  [quant-ph]}\BibitemShut {NoStop}%
\bibitem [{\citenamefont {Wiesemann}\ \emph {et~al.}(2024)\citenamefont
  {Wiesemann}, \citenamefont {Krause}, \citenamefont {Tupkary}, \citenamefont
  {Lütkenhaus}, \citenamefont {Rusca},\ and\ \citenamefont
  {Walenta}}]{wiesemann_consolidated_2024}%
  \BibitemOpen
  \bibfield  {author} {\bibinfo {author} {\bibfnamefont {J.}~\bibnamefont
  {Wiesemann}}, \bibinfo {author} {\bibfnamefont {J.}~\bibnamefont {Krause}},
  \bibinfo {author} {\bibfnamefont {D.}~\bibnamefont {Tupkary}}, \bibinfo
  {author} {\bibfnamefont {N.}~\bibnamefont {Lütkenhaus}}, \bibinfo {author}
  {\bibfnamefont {D.}~\bibnamefont {Rusca}},\ and\ \bibinfo {author}
  {\bibfnamefont {N.}~\bibnamefont {Walenta}},\ }\href
  {https://doi.org/10.48550/arXiv.2405.16578} {\bibinfo {title} {A consolidated
  and accessible security proof for finite-size decoy-state quantum key
  distribution}} (\bibinfo {year} {2024}),\ \bibinfo {note} {arXiv:2405.16578
  [quant-ph]}\BibitemShut {NoStop}%
\bibitem [{\citenamefont {Christandl}\ \emph {et~al.}(2009)\citenamefont
  {Christandl}, \citenamefont {König},\ and\ \citenamefont
  {Renner}}]{christandl_postselection_2009}%
  \BibitemOpen
  \bibfield  {author} {\bibinfo {author} {\bibfnamefont {M.}~\bibnamefont
  {Christandl}}, \bibinfo {author} {\bibfnamefont {R.}~\bibnamefont {König}},\
  and\ \bibinfo {author} {\bibfnamefont {R.}~\bibnamefont {Renner}},\
  }\bibfield  {title} {\bibinfo {title} {Postselection {Technique} for
  {Quantum} {Channels} with {Applications} to {Quantum} {Cryptography}},\
  }\href {https://doi.org/10.1103/PhysRevLett.102.020504} {\bibfield  {journal}
  {\bibinfo  {journal} {Physical Review Letters}\ }\textbf {\bibinfo {volume}
  {102}},\ \bibinfo {pages} {020504} (\bibinfo {year} {2009})},\ \bibinfo
  {note} {publisher: American Physical Society}\BibitemShut {NoStop}%
\bibitem [{\citenamefont {Nahar}\ \emph {et~al.}(2024)\citenamefont {Nahar},
  \citenamefont {Tupkary}, \citenamefont {Zhao}, \citenamefont {Lütkenhaus},\
  and\ \citenamefont {Tan}}]{nahar_postselection_2024}%
  \BibitemOpen
  \bibfield  {author} {\bibinfo {author} {\bibfnamefont {S.}~\bibnamefont
  {Nahar}}, \bibinfo {author} {\bibfnamefont {D.}~\bibnamefont {Tupkary}},
  \bibinfo {author} {\bibfnamefont {Y.}~\bibnamefont {Zhao}}, \bibinfo {author}
  {\bibfnamefont {N.}~\bibnamefont {Lütkenhaus}},\ and\ \bibinfo {author}
  {\bibfnamefont {E.~Y.-Z.}\ \bibnamefont {Tan}},\ }\bibfield  {title}
  {\bibinfo {title} {Postselection {Technique} for {Optical} {Quantum} {Key}
  {Distribution} with {Improved} de {Finetti} {Reductions}},\ }\href
  {https://doi.org/10.1103/PRXQuantum.5.040315} {\bibfield  {journal} {\bibinfo
   {journal} {PRX Quantum}\ }\textbf {\bibinfo {volume} {5}},\ \bibinfo {pages}
  {040315} (\bibinfo {year} {2024})},\ \bibinfo {note} {publisher: American
  Physical Society}\BibitemShut {NoStop}%
\bibitem [{\citenamefont {Dupuis}\ \emph {et~al.}(2020)\citenamefont {Dupuis},
  \citenamefont {Fawzi},\ and\ \citenamefont {Renner}}]{dupuis_entropy_2020}%
  \BibitemOpen
  \bibfield  {author} {\bibinfo {author} {\bibfnamefont {F.}~\bibnamefont
  {Dupuis}}, \bibinfo {author} {\bibfnamefont {O.}~\bibnamefont {Fawzi}},\ and\
  \bibinfo {author} {\bibfnamefont {R.}~\bibnamefont {Renner}},\ }\bibfield
  {title} {\bibinfo {title} {Entropy {Accumulation}},\ }\href
  {https://doi.org/10.1007/s00220-020-03839-5} {\bibfield  {journal} {\bibinfo
  {journal} {Communications in Mathematical Physics}\ }\textbf {\bibinfo
  {volume} {379}},\ \bibinfo {pages} {867} (\bibinfo {year}
  {2020})}\BibitemShut {NoStop}%
\bibitem [{\citenamefont {Dupuis}\ and\ \citenamefont
  {Fawzi}(2019)}]{dupuis_entropy_2019}%
  \BibitemOpen
  \bibfield  {author} {\bibinfo {author} {\bibfnamefont {F.}~\bibnamefont
  {Dupuis}}\ and\ \bibinfo {author} {\bibfnamefont {O.}~\bibnamefont {Fawzi}},\
  }\bibfield  {title} {\bibinfo {title} {Entropy {Accumulation} {With}
  {Improved} {Second}-{Order} {Term}},\ }\href
  {https://doi.org/10.1109/TIT.2019.2929564} {\bibfield  {journal} {\bibinfo
  {journal} {IEEE Transactions on Information Theory}\ }\textbf {\bibinfo
  {volume} {65}},\ \bibinfo {pages} {7596} (\bibinfo {year} {2019})},\ \bibinfo
  {note} {conference Name: IEEE Transactions on Information Theory}\BibitemShut
  {NoStop}%
\bibitem [{\citenamefont {Metger}\ \emph
  {et~al.}(2022{\natexlab{a}})\citenamefont {Metger}, \citenamefont {Fawzi},
  \citenamefont {Sutter},\ and\ \citenamefont
  {Renner}}]{metger_generalised_2022-1}%
  \BibitemOpen
  \bibfield  {author} {\bibinfo {author} {\bibfnamefont {T.}~\bibnamefont
  {Metger}}, \bibinfo {author} {\bibfnamefont {O.}~\bibnamefont {Fawzi}},
  \bibinfo {author} {\bibfnamefont {D.}~\bibnamefont {Sutter}},\ and\ \bibinfo
  {author} {\bibfnamefont {R.}~\bibnamefont {Renner}},\ }\bibfield  {title}
  {\bibinfo {title} {Generalised entropy accumulation},\ }in\ \href
  {https://doi.org/10.1109/FOCS54457.2022.00085} {\emph {\bibinfo {booktitle}
  {2022 {IEEE} 63rd {Annual} {Symposium} on {Foundations} of {Computer}
  {Science} ({FOCS})}}}\ (\bibinfo {year} {2022})\ pp.\ \bibinfo {pages}
  {844--850},\ \bibinfo {note} {iSSN: 2575-8454}\BibitemShut {NoStop}%
\bibitem [{\citenamefont {Metger}\ \emph {et~al.}(2024)\citenamefont {Metger},
  \citenamefont {Fawzi}, \citenamefont {Sutter},\ and\ \citenamefont
  {Renner}}]{metger_generalised_2024}%
  \BibitemOpen
  \bibfield  {author} {\bibinfo {author} {\bibfnamefont {T.}~\bibnamefont
  {Metger}}, \bibinfo {author} {\bibfnamefont {O.}~\bibnamefont {Fawzi}},
  \bibinfo {author} {\bibfnamefont {D.}~\bibnamefont {Sutter}},\ and\ \bibinfo
  {author} {\bibfnamefont {R.}~\bibnamefont {Renner}},\ }\bibfield  {title}
  {\bibinfo {title} {Generalised {Entropy} {Accumulation}},\ }\href
  {https://doi.org/10.1007/s00220-024-05121-4} {\bibfield  {journal} {\bibinfo
  {journal} {Communications in Mathematical Physics}\ }\textbf {\bibinfo
  {volume} {405}},\ \bibinfo {pages} {261} (\bibinfo {year}
  {2024})}\BibitemShut {NoStop}%
\bibitem [{\citenamefont {George}\ \emph {et~al.}(2022)\citenamefont {George},
  \citenamefont {Lin}, \citenamefont {van Himbeeck}, \citenamefont {Fang},\
  and\ \citenamefont {Lütkenhaus}}]{george_finite-key_2022}%
  \BibitemOpen
  \bibfield  {author} {\bibinfo {author} {\bibfnamefont {I.}~\bibnamefont
  {George}}, \bibinfo {author} {\bibfnamefont {J.}~\bibnamefont {Lin}},
  \bibinfo {author} {\bibfnamefont {T.}~\bibnamefont {van Himbeeck}}, \bibinfo
  {author} {\bibfnamefont {K.}~\bibnamefont {Fang}},\ and\ \bibinfo {author}
  {\bibfnamefont {N.}~\bibnamefont {Lütkenhaus}},\ }\href
  {http://arxiv.org/abs/2203.06554} {\bibinfo {title} {Finite-{Key} {Analysis}
  of {Quantum} {Key} {Distribution} with {Characterized} {Devices} {Using}
  {Entropy} {Accumulation}}} (\bibinfo {year} {2022}),\ \bibinfo {note}
  {arXiv:2203.06554 [quant-ph]}\BibitemShut {NoStop}%
\bibitem [{\citenamefont {Metger}\ and\ \citenamefont
  {Renner}(2023)}]{metger_security_2023}%
  \BibitemOpen
  \bibfield  {author} {\bibinfo {author} {\bibfnamefont {T.}~\bibnamefont
  {Metger}}\ and\ \bibinfo {author} {\bibfnamefont {R.}~\bibnamefont
  {Renner}},\ }\bibfield  {title} {\bibinfo {title} {Security of quantum key
  distribution from generalised entropy accumulation},\ }\href
  {https://doi.org/10.1038/s41467-023-40920-8} {\bibfield  {journal} {\bibinfo
  {journal} {Nature Communications}\ }\textbf {\bibinfo {volume} {14}},\
  \bibinfo {pages} {5272} (\bibinfo {year} {2023})},\ \bibinfo {note}
  {publisher: Nature Publishing Group}\BibitemShut {NoStop}%
\bibitem [{\citenamefont {Bäuml}\ \emph {et~al.}(2024)\citenamefont {Bäuml},
  \citenamefont {Pascual-García}, \citenamefont {Wright}, \citenamefont
  {Fawzi},\ and\ \citenamefont {Acín}}]{bauml_security_2024}%
  \BibitemOpen
  \bibfield  {author} {\bibinfo {author} {\bibfnamefont {S.}~\bibnamefont
  {Bäuml}}, \bibinfo {author} {\bibfnamefont {C.}~\bibnamefont
  {Pascual-García}}, \bibinfo {author} {\bibfnamefont {V.}~\bibnamefont
  {Wright}}, \bibinfo {author} {\bibfnamefont {O.}~\bibnamefont {Fawzi}},\ and\
  \bibinfo {author} {\bibfnamefont {A.}~\bibnamefont {Acín}},\ }\bibfield
  {title} {\bibinfo {title} {Security of discrete-modulated continuous-variable
  quantum key distribution},\ }\href
  {https://doi.org/10.22331/q-2024-07-18-1418} {\bibfield  {journal} {\bibinfo
  {journal} {Quantum}\ }\textbf {\bibinfo {volume} {8}},\ \bibinfo {pages}
  {1418} (\bibinfo {year} {2024})},\ \bibinfo {note} {publisher: Verein zur
  Förderung des Open Access Publizierens in den
  Quantenwissenschaften}\BibitemShut {NoStop}%
\bibitem [{\citenamefont {Kamin}\ \emph
  {et~al.}(2025{\natexlab{a}})\citenamefont {Kamin}, \citenamefont {Arqand},
  \citenamefont {George}, \citenamefont {Lütkenhaus},\ and\ \citenamefont
  {Tan}}]{kamin_finite-size_2025}%
  \BibitemOpen
  \bibfield  {author} {\bibinfo {author} {\bibfnamefont {L.}~\bibnamefont
  {Kamin}}, \bibinfo {author} {\bibfnamefont {A.}~\bibnamefont {Arqand}},
  \bibinfo {author} {\bibfnamefont {I.}~\bibnamefont {George}}, \bibinfo
  {author} {\bibfnamefont {N.}~\bibnamefont {Lütkenhaus}},\ and\ \bibinfo
  {author} {\bibfnamefont {E.~Y.-Z.}\ \bibnamefont {Tan}},\ }\bibfield  {title}
  {\bibinfo {title} {Finite-{Size} {Analysis} of {Prepare}-and-{Measure} and
  {Decoy}-{State} {Quantum} {Key} {Distribution} via {Entropy}
  {Accumulation}},\ }\href {https://doi.org/10.1103/PRXQuantum.6.020342}
  {\bibfield  {journal} {\bibinfo  {journal} {PRX Quantum}\ }\textbf {\bibinfo
  {volume} {6}},\ \bibinfo {pages} {020342} (\bibinfo {year}
  {2025}{\natexlab{a}})},\ \bibinfo {note} {publisher: American Physical
  Society}\BibitemShut {NoStop}%
\bibitem [{\citenamefont {Bennett}\ and\ \citenamefont
  {Brassard}(2014)}]{bennett_quantum_2014}%
  \BibitemOpen
  \bibfield  {author} {\bibinfo {author} {\bibfnamefont {C.~H.}\ \bibnamefont
  {Bennett}}\ and\ \bibinfo {author} {\bibfnamefont {G.}~\bibnamefont
  {Brassard}},\ }\bibfield  {title} {\bibinfo {title} {Quantum cryptography:
  {Public} key distribution and coin tossing},\ }\href
  {https://doi.org/https://doi.org/10.1016/j.tcs.2014.05.025} {\bibfield
  {journal} {\bibinfo  {journal} {Theoretical Computer Science}\ }\textbf
  {\bibinfo {volume} {560}},\ \bibinfo {pages} {7} (\bibinfo {year}
  {2014})}\BibitemShut {NoStop}%
\bibitem [{\citenamefont {van Himbeeck}\ and\ \citenamefont
  {Brown}(2025)}]{inprep_HB24}%
  \BibitemOpen
  \bibfield  {author} {\bibinfo {author} {\bibfnamefont {T.}~\bibnamefont {van
  Himbeeck}}\ and\ \bibinfo {author} {\bibfnamefont {P.}~\bibnamefont
  {Brown}},\ }\href@noop {} {\bibinfo {title} {{Tight and general finite-size
  security of quantum key distribution}}} (\bibinfo {year} {2025}),\ \bibinfo
  {note} {in preparation.}\BibitemShut {Stop}%
\bibitem [{\citenamefont {Arqand}\ \emph {et~al.}(2024)\citenamefont {Arqand},
  \citenamefont {Hahn},\ and\ \citenamefont {Tan}}]{arqand_generalized_2024}%
  \BibitemOpen
  \bibfield  {author} {\bibinfo {author} {\bibfnamefont {A.}~\bibnamefont
  {Arqand}}, \bibinfo {author} {\bibfnamefont {T.~A.}\ \bibnamefont {Hahn}},\
  and\ \bibinfo {author} {\bibfnamefont {E.~Y.-Z.}\ \bibnamefont {Tan}},\
  }\href {https://doi.org/10.48550/arXiv.2405.05912} {\bibinfo {title}
  {Generalized {Rényi} entropy accumulation theorem and generalized quantum
  probability estimation}} (\bibinfo {year} {2024}),\ \bibinfo {note}
  {arXiv:2405.05912v4 [quant-ph] version: 4}\BibitemShut {NoStop}%
\bibitem [{\citenamefont {Fawzi}\ \emph {et~al.}(2025)\citenamefont {Fawzi},
  \citenamefont {Kochanowski}, \citenamefont {Rouzé},\ and\ \citenamefont
  {Himbeeck}}]{fawzi_additivity_2025}%
  \BibitemOpen
  \bibfield  {author} {\bibinfo {author} {\bibfnamefont {O.}~\bibnamefont
  {Fawzi}}, \bibinfo {author} {\bibfnamefont {J.}~\bibnamefont {Kochanowski}},
  \bibinfo {author} {\bibfnamefont {C.}~\bibnamefont {Rouzé}},\ and\ \bibinfo
  {author} {\bibfnamefont {T.~V.}\ \bibnamefont {Himbeeck}},\ }\href
  {https://doi.org/10.48550/arXiv.2502.01611} {\bibinfo {title} {Additivity and
  chain rules for quantum entropies via multi-index {Schatten} norms}}
  (\bibinfo {year} {2025}),\ \bibinfo {note} {arXiv:2502.01611
  [quant-ph]}\BibitemShut {NoStop}%
\bibitem [{\citenamefont {Arqand}\ and\ \citenamefont
  {Tan}(2025)}]{arqand_marginal-constrained_2025}%
  \BibitemOpen
  \bibfield  {author} {\bibinfo {author} {\bibfnamefont {A.}~\bibnamefont
  {Arqand}}\ and\ \bibinfo {author} {\bibfnamefont {E.~Y.-Z.}\ \bibnamefont
  {Tan}},\ }\href {https://doi.org/10.48550/arXiv.2502.02563} {\bibinfo {title}
  {Marginal-constrained entropy accumulation theorem}} (\bibinfo {year}
  {2025}),\ \bibinfo {note} {arXiv:2502.02563 [quant-ph] version:
  3}\BibitemShut {NoStop}%
\bibitem [{\citenamefont {Nielsen}\ and\ \citenamefont
  {Chuang}(2010)}]{nielsen_quantum_2010}%
  \BibitemOpen
  \bibfield  {author} {\bibinfo {author} {\bibfnamefont {M.~A.}\ \bibnamefont
  {Nielsen}}\ and\ \bibinfo {author} {\bibfnamefont {I.~L.}\ \bibnamefont
  {Chuang}},\ }\href@noop {} {\emph {\bibinfo {title} {Quantum computation and
  quantum information}}},\ \bibinfo {edition} {10th}\ ed.\ (\bibinfo
  {publisher} {Cambridge university press},\ \bibinfo {address} {Cambridge},\
  \bibinfo {year} {2010})\BibitemShut {NoStop}%
\bibitem [{\citenamefont {Hwang}(2003)}]{hwang_quantum_2003}%
  \BibitemOpen
  \bibfield  {author} {\bibinfo {author} {\bibfnamefont {W.-Y.}\ \bibnamefont
  {Hwang}},\ }\bibfield  {title} {\bibinfo {title} {Quantum {Key}
  {Distribution} with {High} {Loss}: {Toward} {Global} {Secure}
  {Communication}},\ }\href {https://doi.org/10.1103/PhysRevLett.91.057901}
  {\bibfield  {journal} {\bibinfo  {journal} {Physical Review Letters}\
  }\textbf {\bibinfo {volume} {91}},\ \bibinfo {pages} {057901} (\bibinfo
  {year} {2003})},\ \bibinfo {note} {publisher: American Physical
  Society}\BibitemShut {NoStop}%
\bibitem [{\citenamefont {Lo}\ \emph {et~al.}(2004)\citenamefont {Lo},
  \citenamefont {Ma},\ and\ \citenamefont {Chen}}]{lo_decoy_2004}%
  \BibitemOpen
  \bibfield  {author} {\bibinfo {author} {\bibfnamefont {H.-K.}\ \bibnamefont
  {Lo}}, \bibinfo {author} {\bibfnamefont {X.}~\bibnamefont {Ma}},\ and\
  \bibinfo {author} {\bibfnamefont {K.}~\bibnamefont {Chen}},\ }\bibfield
  {title} {\bibinfo {title} {Decoy {State} {Quantum} {Key} {Distribution}},\
  }\href {https://doi.org/10.1103/PhysRevLett.94.230504} {\bibfield  {journal}
  {\bibinfo  {journal} {Physical Review Letters}\ }\textbf {\bibinfo {volume}
  {94}},\ \bibinfo {pages} {230504} (\bibinfo {year} {2004})},\ \bibinfo {note}
  {arXiv: quant-ph/0411004 Publisher: American Physical Society}\BibitemShut
  {NoStop}%
\bibitem [{\citenamefont {Ma}\ \emph {et~al.}(2005)\citenamefont {Ma},
  \citenamefont {Qi}, \citenamefont {Zhao},\ and\ \citenamefont
  {Lo}}]{ma_practical_2005}%
  \BibitemOpen
  \bibfield  {author} {\bibinfo {author} {\bibfnamefont {X.}~\bibnamefont
  {Ma}}, \bibinfo {author} {\bibfnamefont {B.}~\bibnamefont {Qi}}, \bibinfo
  {author} {\bibfnamefont {Y.}~\bibnamefont {Zhao}},\ and\ \bibinfo {author}
  {\bibfnamefont {H.-K.}\ \bibnamefont {Lo}},\ }\bibfield  {title} {\bibinfo
  {title} {Practical decoy state for quantum key distribution},\ }\href
  {https://doi.org/10.1103/PhysRevA.72.012326} {\bibfield  {journal} {\bibinfo
  {journal} {Physical Review A}\ }\textbf {\bibinfo {volume} {72}},\ \bibinfo
  {pages} {012326} (\bibinfo {year} {2005})},\ \bibinfo {note} {publisher:
  American Physical Society}\BibitemShut {NoStop}%
\bibitem [{\citenamefont {Wang}(2005)}]{wang_beating_2005}%
  \BibitemOpen
  \bibfield  {author} {\bibinfo {author} {\bibfnamefont {X.-B.}\ \bibnamefont
  {Wang}},\ }\bibfield  {title} {\bibinfo {title} {Beating the
  {Photon}-{Number}-{Splitting} {Attack} in {Practical} {Quantum}
  {Cryptography}},\ }\href {https://doi.org/10.1103/PhysRevLett.94.230503}
  {\bibfield  {journal} {\bibinfo  {journal} {Physical Review Letters}\
  }\textbf {\bibinfo {volume} {94}},\ \bibinfo {pages} {230503} (\bibinfo
  {year} {2005})},\ \bibinfo {note} {publisher: American Physical
  Society}\BibitemShut {NoStop}%
\bibitem [{\citenamefont {Tupkary}\ \emph
  {et~al.}(2024{\natexlab{b}})\citenamefont {Tupkary}, \citenamefont {Tan},\
  and\ \citenamefont {Lütkenhaus}}]{tupkary_security_2024}%
  \BibitemOpen
  \bibfield  {author} {\bibinfo {author} {\bibfnamefont {D.}~\bibnamefont
  {Tupkary}}, \bibinfo {author} {\bibfnamefont {E.~Y.-Z.}\ \bibnamefont
  {Tan}},\ and\ \bibinfo {author} {\bibfnamefont {N.}~\bibnamefont
  {Lütkenhaus}},\ }\bibfield  {title} {\bibinfo {title} {Security proof for
  variable-length quantum key distribution},\ }\href
  {https://doi.org/10.1103/PhysRevResearch.6.023002} {\bibfield  {journal}
  {\bibinfo  {journal} {Physical Review Research}\ }\textbf {\bibinfo {volume}
  {6}},\ \bibinfo {pages} {023002} (\bibinfo {year} {2024}{\natexlab{b}})},\
  \bibinfo {note} {publisher: American Physical Society}\BibitemShut {NoStop}%
\bibitem [{\citenamefont {Burniston}\ \emph {et~al.}(2024)\citenamefont
  {Burniston}, \citenamefont {Wang}, \citenamefont {Kamin}, \citenamefont
  {Lin}, \citenamefont {Coles}, \citenamefont {Metodiev}, \citenamefont
  {George}, \citenamefont {Li}, \citenamefont {Fang}, \citenamefont {Chemtov},
  \citenamefont {Zhang}, \citenamefont {Böhm}, \citenamefont {Winick},
  \citenamefont {van Himbeeck}, \citenamefont {Johnstun}, \citenamefont
  {Nahar}, \citenamefont {Tupkary}, \citenamefont {Pan}, \citenamefont {Wang},
  \citenamefont {Corrigan}, \citenamefont {Kanitschar}, \citenamefont {Gracie},
  \citenamefont {Gu}, \citenamefont {Mathur}, \citenamefont {Upadhyaya},\ and\
  \citenamefont {Lutkenhaus}}]{burniston_open_2024}%
  \BibitemOpen
  \bibfield  {author} {\bibinfo {author} {\bibfnamefont {J.}~\bibnamefont
  {Burniston}}, \bibinfo {author} {\bibfnamefont {W.}~\bibnamefont {Wang}},
  \bibinfo {author} {\bibfnamefont {L.}~\bibnamefont {Kamin}}, \bibinfo
  {author} {\bibfnamefont {J.}~\bibnamefont {Lin}}, \bibinfo {author}
  {\bibfnamefont {P.}~\bibnamefont {Coles}}, \bibinfo {author} {\bibfnamefont
  {E.}~\bibnamefont {Metodiev}}, \bibinfo {author} {\bibfnamefont
  {I.}~\bibnamefont {George}}, \bibinfo {author} {\bibfnamefont {N.~K.~H.}\
  \bibnamefont {Li}}, \bibinfo {author} {\bibfnamefont {K.}~\bibnamefont
  {Fang}}, \bibinfo {author} {\bibfnamefont {M.}~\bibnamefont {Chemtov}},
  \bibinfo {author} {\bibfnamefont {Y.}~\bibnamefont {Zhang}}, \bibinfo
  {author} {\bibfnamefont {C.}~\bibnamefont {Böhm}}, \bibinfo {author}
  {\bibfnamefont {A.}~\bibnamefont {Winick}}, \bibinfo {author} {\bibfnamefont
  {T.}~\bibnamefont {van Himbeeck}}, \bibinfo {author} {\bibfnamefont
  {S.}~\bibnamefont {Johnstun}}, \bibinfo {author} {\bibfnamefont
  {S.}~\bibnamefont {Nahar}}, \bibinfo {author} {\bibfnamefont
  {D.}~\bibnamefont {Tupkary}}, \bibinfo {author} {\bibfnamefont
  {S.}~\bibnamefont {Pan}}, \bibinfo {author} {\bibfnamefont {Z.}~\bibnamefont
  {Wang}}, \bibinfo {author} {\bibfnamefont {A.}~\bibnamefont {Corrigan}},
  \bibinfo {author} {\bibfnamefont {F.}~\bibnamefont {Kanitschar}}, \bibinfo
  {author} {\bibfnamefont {L.}~\bibnamefont {Gracie}}, \bibinfo {author}
  {\bibfnamefont {S.}~\bibnamefont {Gu}}, \bibinfo {author} {\bibfnamefont
  {N.}~\bibnamefont {Mathur}}, \bibinfo {author} {\bibfnamefont
  {T.}~\bibnamefont {Upadhyaya}},\ and\ \bibinfo {author} {\bibfnamefont
  {N.}~\bibnamefont {Lutkenhaus}},\ }\href
  {https://doi.org/10.5281/zenodo.14262569} {\bibinfo {title} {Open {QKD}
  {Security}: {Version} 2.0.2}} (\bibinfo {year} {2024})\BibitemShut {NoStop}%
\bibitem [{\citenamefont {Chung}\ \emph {et~al.}(2025)\citenamefont {Chung},
  \citenamefont {Ng},\ and\ \citenamefont {Cai}}]{chung_generalized_2025}%
  \BibitemOpen
  \bibfield  {author} {\bibinfo {author} {\bibfnamefont {R.~R.~B.}\
  \bibnamefont {Chung}}, \bibinfo {author} {\bibfnamefont {N.~H.~Y.}\
  \bibnamefont {Ng}},\ and\ \bibinfo {author} {\bibfnamefont {Y.}~\bibnamefont
  {Cai}},\ }\bibfield  {title} {\bibinfo {title} {Generalized numerical
  framework for improved finite-sized key rates with {R}{\textbackslash}'enyi
  entropy},\ }\href {https://doi.org/10.1103/tyts-8v8j} {\bibfield  {journal}
  {\bibinfo  {journal} {Physical Review A}\ }\textbf {\bibinfo {volume}
  {112}},\ \bibinfo {pages} {012612} (\bibinfo {year} {2025})},\ \bibinfo
  {note} {publisher: American Physical Society}\BibitemShut {NoStop}%
\bibitem [{\citenamefont {Winick}\ \emph {et~al.}(2018)\citenamefont {Winick},
  \citenamefont {Lütkenhaus},\ and\ \citenamefont
  {Coles}}]{winick_reliable_2018}%
  \BibitemOpen
  \bibfield  {author} {\bibinfo {author} {\bibfnamefont {A.}~\bibnamefont
  {Winick}}, \bibinfo {author} {\bibfnamefont {N.}~\bibnamefont
  {Lütkenhaus}},\ and\ \bibinfo {author} {\bibfnamefont {P.~J.}\ \bibnamefont
  {Coles}},\ }\bibfield  {title} {\bibinfo {title} {Reliable numerical key
  rates for quantum key distribution},\ }\href
  {https://doi.org/10.22331/q-2018-07-26-77} {\bibfield  {journal} {\bibinfo
  {journal} {Quantum}\ }\textbf {\bibinfo {volume} {2}},\ \bibinfo {pages} {77}
  (\bibinfo {year} {2018})}\BibitemShut {NoStop}%
\bibitem [{\citenamefont {Hu}\ \emph {et~al.}(2022)\citenamefont {Hu},
  \citenamefont {Im}, \citenamefont {Lin}, \citenamefont {Lütkenhaus},\ and\
  \citenamefont {Wolkowicz}}]{hu_robust_2022}%
  \BibitemOpen
  \bibfield  {author} {\bibinfo {author} {\bibfnamefont {H.}~\bibnamefont
  {Hu}}, \bibinfo {author} {\bibfnamefont {J.}~\bibnamefont {Im}}, \bibinfo
  {author} {\bibfnamefont {J.}~\bibnamefont {Lin}}, \bibinfo {author}
  {\bibfnamefont {N.}~\bibnamefont {Lütkenhaus}},\ and\ \bibinfo {author}
  {\bibfnamefont {H.}~\bibnamefont {Wolkowicz}},\ }\bibfield  {title} {\bibinfo
  {title} {Robust {Interior} {Point} {Method} for {Quantum} {Key}
  {Distribution} {Rate} {Computation}},\ }\href
  {https://doi.org/10.22331/q-2022-09-08-792} {\bibfield  {journal} {\bibinfo
  {journal} {Quantum}\ }\textbf {\bibinfo {volume} {6}},\ \bibinfo {pages}
  {792} (\bibinfo {year} {2022})},\ \bibinfo {note} {publisher: Verein zur
  Förderung des Open Access Publizierens in den
  Quantenwissenschaften}\BibitemShut {NoStop}%
\bibitem [{\citenamefont {Lorente}\ \emph {et~al.}(2025)\citenamefont
  {Lorente}, \citenamefont {Parellada}, \citenamefont {Castillo-Celeita},\ and\
  \citenamefont {Araújo}}]{lorente_quantum_2025}%
  \BibitemOpen
  \bibfield  {author} {\bibinfo {author} {\bibfnamefont {A.~G.}\ \bibnamefont
  {Lorente}}, \bibinfo {author} {\bibfnamefont {P.~V.}\ \bibnamefont
  {Parellada}}, \bibinfo {author} {\bibfnamefont {M.}~\bibnamefont
  {Castillo-Celeita}},\ and\ \bibinfo {author} {\bibfnamefont {M.}~\bibnamefont
  {Araújo}},\ }\bibfield  {title} {\bibinfo {title} {Quantum key distribution
  rates from non-symmetric conic optimization},\ }\href
  {https://doi.org/10.22331/q-2025-03-10-1657} {\bibfield  {journal} {\bibinfo
  {journal} {Quantum}\ }\textbf {\bibinfo {volume} {9}},\ \bibinfo {pages}
  {1657} (\bibinfo {year} {2025})},\ \bibinfo {note} {publisher: Verein zur
  Förderung des Open Access Publizierens in den
  Quantenwissenschaften}\BibitemShut {NoStop}%
\bibitem [{\citenamefont {He}\ \emph {et~al.}(2024)\citenamefont {He},
  \citenamefont {Saunderson},\ and\ \citenamefont
  {Fawzi}}]{he_exploiting_2024}%
  \BibitemOpen
  \bibfield  {author} {\bibinfo {author} {\bibfnamefont {K.}~\bibnamefont
  {He}}, \bibinfo {author} {\bibfnamefont {J.}~\bibnamefont {Saunderson}},\
  and\ \bibinfo {author} {\bibfnamefont {H.}~\bibnamefont {Fawzi}},\ }\href
  {https://doi.org/10.48550/arXiv.2407.00241} {\bibinfo {title} {Exploiting
  {Structure} in {Quantum} {Relative} {Entropy} {Programs}}} (\bibinfo {year}
  {2024}),\ \bibinfo {note} {arXiv:2407.00241 [quant-ph]}\BibitemShut {NoStop}%
\bibitem [{\citenamefont {Dupuis}(2023)}]{dupuis_privacy_2023}%
  \BibitemOpen
  \bibfield  {author} {\bibinfo {author} {\bibfnamefont {F.}~\bibnamefont
  {Dupuis}},\ }\bibfield  {title} {\bibinfo {title} {Privacy {Amplification}
  and {Decoupling} {Without} {Smoothing}},\ }\href
  {https://doi.org/10.1109/TIT.2023.3301812} {\bibfield  {journal} {\bibinfo
  {journal} {IEEE Transactions on Information Theory}\ }\textbf {\bibinfo
  {volume} {69}},\ \bibinfo {pages} {7784} (\bibinfo {year} {2023})},\ \bibinfo
  {note} {conference Name: IEEE Transactions on Information Theory}\BibitemShut
  {NoStop}%
\bibitem [{\citenamefont {Bennett}\ \emph
  {et~al.}(1992{\natexlab{a}})\citenamefont {Bennett}, \citenamefont
  {Brassard},\ and\ \citenamefont {Mermin}}]{bennett_quantum_1992}%
  \BibitemOpen
  \bibfield  {author} {\bibinfo {author} {\bibfnamefont {C.~H.}\ \bibnamefont
  {Bennett}}, \bibinfo {author} {\bibfnamefont {G.}~\bibnamefont {Brassard}},\
  and\ \bibinfo {author} {\bibfnamefont {N.~D.}\ \bibnamefont {Mermin}},\
  }\bibfield  {title} {\bibinfo {title} {Quantum cryptography without {Bell}'s
  theorem},\ }\href {https://doi.org/10.1103/PhysRevLett.68.557} {\bibfield
  {journal} {\bibinfo  {journal} {Physical Review Letters}\ }\textbf {\bibinfo
  {volume} {68}},\ \bibinfo {pages} {557} (\bibinfo {year}
  {1992}{\natexlab{a}})},\ \bibinfo {note} {publisher: American Physical
  Society}\BibitemShut {NoStop}%
\bibitem [{\citenamefont {Ferenczi}\ and\ \citenamefont
  {Lütkenhaus}(2012)}]{ferenczi_symmetries_2012}%
  \BibitemOpen
  \bibfield  {author} {\bibinfo {author} {\bibfnamefont {A.}~\bibnamefont
  {Ferenczi}}\ and\ \bibinfo {author} {\bibfnamefont {N.}~\bibnamefont
  {Lütkenhaus}},\ }\bibfield  {title} {\bibinfo {title} {Symmetries in quantum
  key distribution and the connection between optimal attacks and optimal
  cloning},\ }\href {https://doi.org/10.1103/PhysRevA.85.052310} {\bibfield
  {journal} {\bibinfo  {journal} {Physical Review A}\ }\textbf {\bibinfo
  {volume} {85}},\ \bibinfo {pages} {052310} (\bibinfo {year} {2012})},\
  \bibinfo {note} {publisher: American Physical Society}\BibitemShut {NoStop}%
\bibitem [{\citenamefont {Laing}\ \emph {et~al.}(2010)\citenamefont {Laing},
  \citenamefont {Scarani}, \citenamefont {Rarity},\ and\ \citenamefont
  {O’Brien}}]{laing_reference-frame-independent_2010}%
  \BibitemOpen
  \bibfield  {author} {\bibinfo {author} {\bibfnamefont {A.}~\bibnamefont
  {Laing}}, \bibinfo {author} {\bibfnamefont {V.}~\bibnamefont {Scarani}},
  \bibinfo {author} {\bibfnamefont {J.~G.}\ \bibnamefont {Rarity}},\ and\
  \bibinfo {author} {\bibfnamefont {J.~L.}\ \bibnamefont {O’Brien}},\
  }\bibfield  {title} {\bibinfo {title} {Reference-frame-independent quantum
  key distribution},\ }\href {https://doi.org/10.1103/PhysRevA.82.012304}
  {\bibfield  {journal} {\bibinfo  {journal} {Physical Review A}\ }\textbf
  {\bibinfo {volume} {82}},\ \bibinfo {pages} {012304} (\bibinfo {year}
  {2010})},\ \bibinfo {note} {publisher: American Physical Society}\BibitemShut
  {NoStop}%
\bibitem [{\citenamefont {Kamin}\ \emph
  {et~al.}(2025{\natexlab{b}})\citenamefont {Kamin}, \citenamefont {Tupkary},\
  and\ \citenamefont {Lütkenhaus}}]{kamin_improved_2025}%
  \BibitemOpen
  \bibfield  {author} {\bibinfo {author} {\bibfnamefont {L.}~\bibnamefont
  {Kamin}}, \bibinfo {author} {\bibfnamefont {D.}~\bibnamefont {Tupkary}},\
  and\ \bibinfo {author} {\bibfnamefont {N.}~\bibnamefont {Lütkenhaus}},\
  }\href {https://doi.org/10.48550/arXiv.2502.05382} {\bibinfo {title}
  {Improved finite-size effects in {QKD} protocols with applications to
  decoy-state {QKD}}} (\bibinfo {year} {2025}{\natexlab{b}}),\ \bibinfo {note}
  {arXiv:2502.05382 [quant-ph]}\BibitemShut {NoStop}%
\bibitem [{\citenamefont {Tomamichel}(2016)}]{tomamichel_quantum_2016}%
  \BibitemOpen
  \bibfield  {author} {\bibinfo {author} {\bibfnamefont {M.}~\bibnamefont
  {Tomamichel}},\ }\href {https://doi.org/10.1007/978-3-319-21891-5} {\emph
  {\bibinfo {title} {Quantum {Information} {Processing} with {Finite}
  {Resources}}}},\ \bibinfo {series} {{SpringerBriefs} in {Mathematical}
  {Physics}}, Vol.~\bibinfo {volume} {5}\ (\bibinfo  {publisher} {Springer
  International Publishing},\ \bibinfo {address} {Cham},\ \bibinfo {year}
  {2016})\BibitemShut {NoStop}%
\bibitem [{\citenamefont {Portmann}\ and\ \citenamefont
  {Renner}(2022)}]{portmann_security_2022}%
  \BibitemOpen
  \bibfield  {author} {\bibinfo {author} {\bibfnamefont {C.}~\bibnamefont
  {Portmann}}\ and\ \bibinfo {author} {\bibfnamefont {R.}~\bibnamefont
  {Renner}},\ }\bibfield  {title} {\bibinfo {title} {Security in quantum
  cryptography},\ }\href {https://doi.org/10.1103/RevModPhys.94.025008}
  {\bibfield  {journal} {\bibinfo  {journal} {Reviews of Modern Physics}\
  }\textbf {\bibinfo {volume} {94}},\ \bibinfo {pages} {025008} (\bibinfo
  {year} {2022})},\ \bibinfo {note} {publisher: American Physical
  Society}\BibitemShut {NoStop}%
\bibitem [{\citenamefont {Renner}(2006)}]{renner_security_2006}%
  \BibitemOpen
  \bibfield  {author} {\bibinfo {author} {\bibfnamefont {R.}~\bibnamefont
  {Renner}},\ }\href {https://doi.org/10.48550/arXiv.quant-ph/0512258}
  {\bibinfo {title} {Security of {Quantum} {Key} {Distribution}}} (\bibinfo
  {year} {2006}),\ \bibinfo {note} {arXiv:quant-ph/0512258}\BibitemShut
  {NoStop}%
\bibitem [{\citenamefont {Tomamichel}\ and\ \citenamefont
  {Leverrier}(2017)}]{tomamichel_largely_2017}%
  \BibitemOpen
  \bibfield  {author} {\bibinfo {author} {\bibfnamefont {M.}~\bibnamefont
  {Tomamichel}}\ and\ \bibinfo {author} {\bibfnamefont {A.}~\bibnamefont
  {Leverrier}},\ }\bibfield  {title} {\bibinfo {title} {A largely
  self-contained and complete security proof for quantum key distribution},\
  }\href {https://doi.org/10.22331/q-2017-07-14-14} {\bibfield  {journal}
  {\bibinfo  {journal} {Quantum}\ }\textbf {\bibinfo {volume} {1}},\ \bibinfo
  {pages} {14} (\bibinfo {year} {2017})}\BibitemShut {NoStop}%
\bibitem [{\citenamefont {Metger}\ \emph
  {et~al.}(2022{\natexlab{b}})\citenamefont {Metger}, \citenamefont {Fawzi},
  \citenamefont {Sutter},\ and\ \citenamefont
  {Renner}}]{metger_generalised_2022}%
  \BibitemOpen
  \bibfield  {author} {\bibinfo {author} {\bibfnamefont {T.}~\bibnamefont
  {Metger}}, \bibinfo {author} {\bibfnamefont {O.}~\bibnamefont {Fawzi}},
  \bibinfo {author} {\bibfnamefont {D.}~\bibnamefont {Sutter}},\ and\ \bibinfo
  {author} {\bibfnamefont {R.}~\bibnamefont {Renner}},\ }\href
  {http://arxiv.org/abs/2203.04989} {\bibinfo {title} {Generalised entropy
  accumulation}} (\bibinfo {year} {2022}{\natexlab{b}}),\ \bibinfo {note}
  {arXiv:2203.04989 [quant-ph]}\BibitemShut {NoStop}%
\bibitem [{\citenamefont {Lin}\ \emph {et~al.}(2019)\citenamefont {Lin},
  \citenamefont {Upadhyaya},\ and\ \citenamefont
  {Lütkenhaus}}]{lin_asymptotic_2019}%
  \BibitemOpen
  \bibfield  {author} {\bibinfo {author} {\bibfnamefont {J.}~\bibnamefont
  {Lin}}, \bibinfo {author} {\bibfnamefont {T.}~\bibnamefont {Upadhyaya}},\
  and\ \bibinfo {author} {\bibfnamefont {N.}~\bibnamefont {Lütkenhaus}},\
  }\bibfield  {title} {\bibinfo {title} {Asymptotic {Security} {Analysis} of
  {Discrete}-{Modulated} {Continuous}-{Variable} {Quantum} {Key}
  {Distribution}},\ }\href {https://doi.org/10.1103/PhysRevX.9.041064}
  {\bibfield  {journal} {\bibinfo  {journal} {Physical Review X}\ }\textbf
  {\bibinfo {volume} {9}},\ \bibinfo {pages} {041064} (\bibinfo {year}
  {2019})},\ \bibinfo {note} {publisher: American Physical Society}\BibitemShut
  {NoStop}%
\bibitem [{\citenamefont {Fawzi}\ and\ \citenamefont
  {Saunderson}(2017)}]{fawzi_liebs_2017}%
  \BibitemOpen
  \bibfield  {author} {\bibinfo {author} {\bibfnamefont {H.}~\bibnamefont
  {Fawzi}}\ and\ \bibinfo {author} {\bibfnamefont {J.}~\bibnamefont
  {Saunderson}},\ }\bibfield  {title} {\bibinfo {title} {Lieb's concavity
  theorem, matrix geometric means, and semidefinite optimization},\ }\href
  {https://doi.org/10.1016/j.laa.2016.10.012} {\bibfield  {journal} {\bibinfo
  {journal} {Linear Algebra and its Applications}\ }\textbf {\bibinfo {volume}
  {513}},\ \bibinfo {pages} {240} (\bibinfo {year} {2017})}\BibitemShut
  {NoStop}%
\bibitem [{\citenamefont {He}\ \emph {et~al.}(2025)\citenamefont {He},
  \citenamefont {Saunderson},\ and\ \citenamefont {Fawzi}}]{he_operator_2025}%
  \BibitemOpen
  \bibfield  {author} {\bibinfo {author} {\bibfnamefont {K.}~\bibnamefont
  {He}}, \bibinfo {author} {\bibfnamefont {J.}~\bibnamefont {Saunderson}},\
  and\ \bibinfo {author} {\bibfnamefont {H.}~\bibnamefont {Fawzi}},\ }\href
  {https://doi.org/10.48550/arXiv.2502.05627} {\bibinfo {title} {Operator
  convexity along lines, self-concordance, and sandwiched {Rényi} entropies}}
  (\bibinfo {year} {2025}),\ \bibinfo {note} {arXiv:2502.05627
  [math]}\BibitemShut {NoStop}%
\bibitem [{\citenamefont {Coles}(2012)}]{coles_unification_2012}%
  \BibitemOpen
  \bibfield  {author} {\bibinfo {author} {\bibfnamefont {P.~J.}\ \bibnamefont
  {Coles}},\ }\bibfield  {title} {\bibinfo {title} {Unification of different
  views of decoherence and discord},\ }\href
  {https://doi.org/10.1103/PhysRevA.85.042103} {\bibfield  {journal} {\bibinfo
  {journal} {Physical Review A}\ }\textbf {\bibinfo {volume} {85}},\ \bibinfo
  {pages} {042103} (\bibinfo {year} {2012})},\ \bibinfo {note} {publisher:
  American Physical Society}\BibitemShut {NoStop}%
\bibitem [{\citenamefont {Frank}\ and\ \citenamefont
  {Wolfe}(1956)}]{frank_algorithm_1956}%
  \BibitemOpen
  \bibfield  {author} {\bibinfo {author} {\bibfnamefont {M.}~\bibnamefont
  {Frank}}\ and\ \bibinfo {author} {\bibfnamefont {P.}~\bibnamefont {Wolfe}},\
  }\bibfield  {title} {\bibinfo {title} {An algorithm for quadratic
  programming},\ }\href {https://doi.org/10.1002/nav.3800030109} {\bibfield
  {journal} {\bibinfo  {journal} {Naval Research Logistics Quarterly}\ }\textbf
  {\bibinfo {volume} {3}},\ \bibinfo {pages} {95} (\bibinfo {year}
  {1956})}\BibitemShut {NoStop}%
\bibitem [{\citenamefont {Gour}\ and\ \citenamefont
  {Wilde}(2021)}]{gour_entropy_2021}%
  \BibitemOpen
  \bibfield  {author} {\bibinfo {author} {\bibfnamefont {G.}~\bibnamefont
  {Gour}}\ and\ \bibinfo {author} {\bibfnamefont {M.~M.}\ \bibnamefont
  {Wilde}},\ }\bibfield  {title} {\bibinfo {title} {Entropy of a quantum
  channel},\ }\href {https://doi.org/10.1103/PhysRevResearch.3.023096}
  {\bibfield  {journal} {\bibinfo  {journal} {Physical Review Research}\
  }\textbf {\bibinfo {volume} {3}},\ \bibinfo {pages} {023096} (\bibinfo {year}
  {2021})},\ \bibinfo {note} {publisher: American Physical Society}\BibitemShut
  {NoStop}%
\bibitem [{\citenamefont {Anco}\ \emph {et~al.}(2024)\citenamefont {Anco},
  \citenamefont {Nemoz},\ and\ \citenamefont {Brown}}]{anco_how_2024}%
  \BibitemOpen
  \bibfield  {author} {\bibinfo {author} {\bibfnamefont {K.~G.}\ \bibnamefont
  {Anco}}, \bibinfo {author} {\bibfnamefont {T.}~\bibnamefont {Nemoz}},\ and\
  \bibinfo {author} {\bibfnamefont {P.}~\bibnamefont {Brown}},\ }\href
  {https://doi.org/10.48550/arXiv.2410.16447} {\bibinfo {title} {How much
  secure randomness is in a quantum state?}} (\bibinfo {year} {2024}),\
  \bibinfo {note} {arXiv:2410.16447 [quant-ph]}\BibitemShut {NoStop}%
\bibitem [{\citenamefont {Tomamichel}\ \emph {et~al.}(2014)\citenamefont
  {Tomamichel}, \citenamefont {Berta},\ and\ \citenamefont
  {Hayashi}}]{tomamichel_relating_2014}%
  \BibitemOpen
  \bibfield  {author} {\bibinfo {author} {\bibfnamefont {M.}~\bibnamefont
  {Tomamichel}}, \bibinfo {author} {\bibfnamefont {M.}~\bibnamefont {Berta}},\
  and\ \bibinfo {author} {\bibfnamefont {M.}~\bibnamefont {Hayashi}},\
  }\bibfield  {title} {\bibinfo {title} {Relating different quantum
  generalizations of the conditional {Rényi} entropy},\ }\href
  {https://doi.org/10.1063/1.4892761} {\bibfield  {journal} {\bibinfo
  {journal} {Journal of Mathematical Physics}\ }\textbf {\bibinfo {volume}
  {55}},\ \bibinfo {pages} {082206} (\bibinfo {year} {2014})}\BibitemShut
  {NoStop}%
\bibitem [{\citenamefont {Tomamichel}\ \emph {et~al.}(2009)\citenamefont
  {Tomamichel}, \citenamefont {Colbeck},\ and\ \citenamefont
  {Renner}}]{tomamichel_fully_2009}%
  \BibitemOpen
  \bibfield  {author} {\bibinfo {author} {\bibfnamefont {M.}~\bibnamefont
  {Tomamichel}}, \bibinfo {author} {\bibfnamefont {R.}~\bibnamefont
  {Colbeck}},\ and\ \bibinfo {author} {\bibfnamefont {R.}~\bibnamefont
  {Renner}},\ }\bibfield  {title} {\bibinfo {title} {A {Fully} {Quantum}
  {Asymptotic} {Equipartition} {Property}},\ }\href
  {https://doi.org/10.1109/TIT.2009.2032797} {\bibfield  {journal} {\bibinfo
  {journal} {IEEE Transactions on Information Theory}\ }\textbf {\bibinfo
  {volume} {55}},\ \bibinfo {pages} {5840} (\bibinfo {year} {2009})},\ \bibinfo
  {note} {conference Name: IEEE Transactions on Information Theory}\BibitemShut
  {NoStop}%
\bibitem [{\citenamefont {Boyd}\ and\ \citenamefont
  {Vandenberghe}(2004)}]{boyd_convex_2004}%
  \BibitemOpen
  \bibfield  {author} {\bibinfo {author} {\bibfnamefont {S.~P.}\ \bibnamefont
  {Boyd}}\ and\ \bibinfo {author} {\bibfnamefont {L.}~\bibnamefont
  {Vandenberghe}},\ }\href@noop {} {\emph {\bibinfo {title} {Convex
  optimization}}}\ (\bibinfo  {publisher} {Cambridge University Press},\
  \bibinfo {address} {Cambridge, UK ; New York},\ \bibinfo {year}
  {2004})\BibitemShut {NoStop}%
\bibitem [{\citenamefont {{Curr{\'a}s-Lorenzo}}\ \emph
  {et~al.}(2021)\citenamefont {{Curr{\'a}s-Lorenzo}}, \citenamefont
  {Navarrete}, \citenamefont {Pereira},\ and\ \citenamefont
  {Tamaki}}]{curras-lorenzo_finite-key_2021}%
  \BibitemOpen
  \bibfield  {author} {\bibinfo {author} {\bibfnamefont {G.}~\bibnamefont
  {{Curr{\'a}s-Lorenzo}}}, \bibinfo {author} {\bibfnamefont
  {{\'A}.}~\bibnamefont {Navarrete}}, \bibinfo {author} {\bibfnamefont
  {M.}~\bibnamefont {Pereira}},\ and\ \bibinfo {author} {\bibfnamefont
  {K.}~\bibnamefont {Tamaki}},\ }\bibfield  {title} {\bibinfo {title}
  {Finite-key analysis of loss-tolerant quantum key distribution based on
  random sampling theory},\ }\href
  {https://doi.org/10.1103/PhysRevA.104.012406} {\bibfield  {journal} {\bibinfo
   {journal} {Physical Review A}\ }\textbf {\bibinfo {volume} {104}},\ \bibinfo
  {pages} {012406} (\bibinfo {year} {2021})}\BibitemShut {NoStop}%
\bibitem [{\citenamefont {Mannalath}\ \emph {et~al.}(2025)\citenamefont
  {Mannalath}, \citenamefont {Zapatero},\ and\ \citenamefont
  {Curty}}]{mannalath_sharp_2025}%
  \BibitemOpen
  \bibfield  {author} {\bibinfo {author} {\bibfnamefont {V.}~\bibnamefont
  {Mannalath}}, \bibinfo {author} {\bibfnamefont {V.}~\bibnamefont
  {Zapatero}},\ and\ \bibinfo {author} {\bibfnamefont {M.}~\bibnamefont
  {Curty}},\ }\bibfield  {title} {\bibinfo {title} {Sharp {Finite} {Statistics}
  for {Quantum} {Key} {Distribution}},\ }\href
  {https://doi.org/10.1103/l735-x48g} {\bibfield  {journal} {\bibinfo
  {journal} {Physical Review Letters}\ }\textbf {\bibinfo {volume} {135}},\
  \bibinfo {pages} {020803} (\bibinfo {year} {2025})},\ \bibinfo {note}
  {publisher: American Physical Society}\BibitemShut {NoStop}%
\bibitem [{\citenamefont {Elkouss}\ \emph {et~al.}(2011)\citenamefont
  {Elkouss}, \citenamefont {Martinez-Mateo},\ and\ \citenamefont
  {Martin}}]{elkouss_information_2011}%
  \BibitemOpen
  \bibfield  {author} {\bibinfo {author} {\bibfnamefont {D.}~\bibnamefont
  {Elkouss}}, \bibinfo {author} {\bibfnamefont {J.}~\bibnamefont
  {Martinez-Mateo}},\ and\ \bibinfo {author} {\bibfnamefont {V.}~\bibnamefont
  {Martin}},\ }\href {https://doi.org/10.48550/arXiv.1007.1616} {\bibinfo
  {title} {Information {Reconciliation} for {Quantum} {Key} {Distribution}}}
  (\bibinfo {year} {2011}),\ \bibinfo {note} {arXiv:1007.1616
  [quant-ph]}\BibitemShut {NoStop}%
\bibitem [{\citenamefont {Li}\ and\ \citenamefont
  {Lütkenhaus}(2020)}]{li_improving_2020}%
  \BibitemOpen
  \bibfield  {author} {\bibinfo {author} {\bibfnamefont {N.~K.~H.}\
  \bibnamefont {Li}}\ and\ \bibinfo {author} {\bibfnamefont {N.}~\bibnamefont
  {Lütkenhaus}},\ }\bibfield  {title} {\bibinfo {title} {Improving key rates
  of the unbalanced phase-encoded {BB84} protocol using the flag-state
  squashing model},\ }\href {https://doi.org/10.1103/PhysRevResearch.2.043172}
  {\bibfield  {journal} {\bibinfo  {journal} {Physical Review Research}\
  }\textbf {\bibinfo {volume} {2}},\ \bibinfo {pages} {043172} (\bibinfo {year}
  {2020})},\ \bibinfo {note} {publisher: American Physical Society}\BibitemShut
  {NoStop}%
\bibitem [{\citenamefont {Wang}\ and\ \citenamefont
  {Lütkenhaus}(2022)}]{wang_numerical_2022}%
  \BibitemOpen
  \bibfield  {author} {\bibinfo {author} {\bibfnamefont {W.}~\bibnamefont
  {Wang}}\ and\ \bibinfo {author} {\bibfnamefont {N.}~\bibnamefont
  {Lütkenhaus}},\ }\bibfield  {title} {\bibinfo {title} {Numerical security
  proof for the decoy-state {BB84} protocol and measurement-device-independent
  quantum key distribution resistant against large basis misalignment},\ }\href
  {https://doi.org/10.1103/PhysRevResearch.4.043097} {\bibfield  {journal}
  {\bibinfo  {journal} {Physical Review Research}\ }\textbf {\bibinfo {volume}
  {4}},\ \bibinfo {pages} {043097} (\bibinfo {year} {2022})},\ \bibinfo {note}
  {publisher: American Physical Society}\BibitemShut {NoStop}%
\bibitem [{\citenamefont {Rice}\ and\ \citenamefont
  {Harrington}(2009)}]{Rice2009}%
  \BibitemOpen
  \bibfield  {author} {\bibinfo {author} {\bibfnamefont {P.}~\bibnamefont
  {Rice}}\ and\ \bibinfo {author} {\bibfnamefont {J.}~\bibnamefont
  {Harrington}},\ }\href {http://arxiv.org/abs/0901.0013} {\bibinfo {title}
  {Numerical analysis of decoy state quantum key distribution protocols}}
  (\bibinfo {year} {2009}),\ \bibinfo {note} {arXiv:0901.0013
  [quant-ph]}\BibitemShut {NoStop}%
\bibitem [{\citenamefont {Nahar}\ \emph {et~al.}(2023)\citenamefont {Nahar},
  \citenamefont {Upadhyaya},\ and\ \citenamefont
  {Lütkenhaus}}]{nahar_imperfect_2023}%
  \BibitemOpen
  \bibfield  {author} {\bibinfo {author} {\bibfnamefont {S.}~\bibnamefont
  {Nahar}}, \bibinfo {author} {\bibfnamefont {T.}~\bibnamefont {Upadhyaya}},\
  and\ \bibinfo {author} {\bibfnamefont {N.}~\bibnamefont {Lütkenhaus}},\
  }\bibfield  {title} {\bibinfo {title} {Imperfect phase randomization and
  generalized decoy-state quantum key distribution},\ }\href
  {https://doi.org/10.1103/PhysRevApplied.20.064031} {\bibfield  {journal}
  {\bibinfo  {journal} {Physical Review Applied}\ }\textbf {\bibinfo {volume}
  {20}},\ \bibinfo {pages} {064031} (\bibinfo {year} {2023})},\ \bibinfo {note}
  {publisher: American Physical Society}\BibitemShut {NoStop}%
\bibitem [{\citenamefont {Müller-Lennert}\ \emph {et~al.}(2013)\citenamefont
  {Müller-Lennert}, \citenamefont {Dupuis}, \citenamefont {Szehr},
  \citenamefont {Fehr},\ and\ \citenamefont
  {Tomamichel}}]{muller-lennert_quantum_2013}%
  \BibitemOpen
  \bibfield  {author} {\bibinfo {author} {\bibfnamefont {M.}~\bibnamefont
  {Müller-Lennert}}, \bibinfo {author} {\bibfnamefont {F.}~\bibnamefont
  {Dupuis}}, \bibinfo {author} {\bibfnamefont {O.}~\bibnamefont {Szehr}},
  \bibinfo {author} {\bibfnamefont {S.}~\bibnamefont {Fehr}},\ and\ \bibinfo
  {author} {\bibfnamefont {M.}~\bibnamefont {Tomamichel}},\ }\bibfield  {title}
  {\bibinfo {title} {On quantum {Rényi} entropies: {A} new generalization and
  some properties},\ }\href {https://doi.org/10.1063/1.4838856} {\bibfield
  {journal} {\bibinfo  {journal} {Journal of Mathematical Physics}\ }\textbf
  {\bibinfo {volume} {54}},\ \bibinfo {pages} {122203} (\bibinfo {year}
  {2013})}\BibitemShut {NoStop}%
\bibitem [{\citenamefont {Horodecki}\ \emph {et~al.}(2009)\citenamefont
  {Horodecki}, \citenamefont {Horodecki}, \citenamefont {Horodecki},\ and\
  \citenamefont {Oppenheim}}]{horodecki_general_2009}%
  \BibitemOpen
  \bibfield  {author} {\bibinfo {author} {\bibfnamefont {K.}~\bibnamefont
  {Horodecki}}, \bibinfo {author} {\bibfnamefont {M.}~\bibnamefont
  {Horodecki}}, \bibinfo {author} {\bibfnamefont {P.}~\bibnamefont
  {Horodecki}},\ and\ \bibinfo {author} {\bibfnamefont {J.}~\bibnamefont
  {Oppenheim}},\ }\bibfield  {title} {\bibinfo {title} {General {Paradigm} for
  {Distilling} {Classical} {Key} {From} {Quantum} {States}},\ }\href
  {https://doi.org/10.1109/TIT.2008.2009798} {\bibfield  {journal} {\bibinfo
  {journal} {IEEE Transactions on Information Theory}\ }\textbf {\bibinfo
  {volume} {55}},\ \bibinfo {pages} {1898} (\bibinfo {year} {2009})},\ \bibinfo
  {note} {conference Name: IEEE Transactions on Information Theory}\BibitemShut
  {NoStop}%
\bibitem [{\citenamefont {Bennett}\ \emph
  {et~al.}(1992{\natexlab{b}})\citenamefont {Bennett}, \citenamefont
  {Bessette}, \citenamefont {Brassard}, \citenamefont {Salvail},\ and\
  \citenamefont {Smolin}}]{bennett_experimental_1992}%
  \BibitemOpen
  \bibfield  {author} {\bibinfo {author} {\bibfnamefont {C.~H.}\ \bibnamefont
  {Bennett}}, \bibinfo {author} {\bibfnamefont {F.}~\bibnamefont {Bessette}},
  \bibinfo {author} {\bibfnamefont {G.}~\bibnamefont {Brassard}}, \bibinfo
  {author} {\bibfnamefont {L.}~\bibnamefont {Salvail}},\ and\ \bibinfo {author}
  {\bibfnamefont {J.}~\bibnamefont {Smolin}},\ }\bibfield  {title} {\bibinfo
  {title} {Experimental quantum cryptography},\ }\href
  {https://doi.org/10.1007/BF00191318} {\bibfield  {journal} {\bibinfo
  {journal} {Journal of Cryptology}\ }\textbf {\bibinfo {volume} {5}},\
  \bibinfo {pages} {3} (\bibinfo {year} {1992}{\natexlab{b}})}\BibitemShut
  {NoStop}%
\bibitem [{\citenamefont {Brassard}\ \emph {et~al.}(2000)\citenamefont
  {Brassard}, \citenamefont {Lütkenhaus}, \citenamefont {Mor},\ and\
  \citenamefont {Sanders}}]{brassard_limitations_2000}%
  \BibitemOpen
  \bibfield  {author} {\bibinfo {author} {\bibfnamefont {G.}~\bibnamefont
  {Brassard}}, \bibinfo {author} {\bibfnamefont {N.}~\bibnamefont
  {Lütkenhaus}}, \bibinfo {author} {\bibfnamefont {T.}~\bibnamefont {Mor}},\
  and\ \bibinfo {author} {\bibfnamefont {B.~C.}\ \bibnamefont {Sanders}},\
  }\bibfield  {title} {\bibinfo {title} {Limitations on {Practical} {Quantum}
  {Cryptography}},\ }\href {https://doi.org/10.1103/PhysRevLett.85.1330}
  {\bibfield  {journal} {\bibinfo  {journal} {Physical Review Letters}\
  }\textbf {\bibinfo {volume} {85}},\ \bibinfo {pages} {1330} (\bibinfo {year}
  {2000})},\ \bibinfo {note} {publisher: American Physical Society}\BibitemShut
  {NoStop}%
\bibitem [{\citenamefont {Kamin}\ and\ \citenamefont
  {Lütkenhaus}(2024)}]{kamin_improved_2024}%
  \BibitemOpen
  \bibfield  {author} {\bibinfo {author} {\bibfnamefont {L.}~\bibnamefont
  {Kamin}}\ and\ \bibinfo {author} {\bibfnamefont {N.}~\bibnamefont
  {Lütkenhaus}},\ }\bibfield  {title} {\bibinfo {title} {Improved decoy-state
  and flag-state squashing methods},\ }\href
  {https://doi.org/10.1103/PhysRevResearch.6.043223} {\bibfield  {journal}
  {\bibinfo  {journal} {Physical Review Research}\ }\textbf {\bibinfo {volume}
  {6}},\ \bibinfo {pages} {043223} (\bibinfo {year} {2024})},\ \bibinfo {note}
  {publisher: American Physical Society}\BibitemShut {NoStop}%
\bibitem [{\citenamefont {Beaudry}\ \emph {et~al.}(2008)\citenamefont
  {Beaudry}, \citenamefont {Moroder},\ and\ \citenamefont
  {Lütkenhaus}}]{beaudry_squashing_2008}%
  \BibitemOpen
  \bibfield  {author} {\bibinfo {author} {\bibfnamefont {N.~J.}\ \bibnamefont
  {Beaudry}}, \bibinfo {author} {\bibfnamefont {T.}~\bibnamefont {Moroder}},\
  and\ \bibinfo {author} {\bibfnamefont {N.}~\bibnamefont {Lütkenhaus}},\
  }\bibfield  {title} {\bibinfo {title} {Squashing {Models} for {Optical}
  {Measurements} in {Quantum} {Communication}},\ }\href
  {https://doi.org/10.1103/PhysRevLett.101.093601} {\bibfield  {journal}
  {\bibinfo  {journal} {Physical Review Letters}\ }\textbf {\bibinfo {volume}
  {101}},\ \bibinfo {pages} {093601} (\bibinfo {year} {2008})},\ \bibinfo
  {note} {publisher: American Physical Society}\BibitemShut {NoStop}%
\bibitem [{\citenamefont {Gittsovich}\ \emph {et~al.}(2014)\citenamefont
  {Gittsovich}, \citenamefont {Beaudry}, \citenamefont {Narasimhachar},
  \citenamefont {Alvarez}, \citenamefont {Moroder},\ and\ \citenamefont
  {Lütkenhaus}}]{gittsovich_squashing_2014}%
  \BibitemOpen
  \bibfield  {author} {\bibinfo {author} {\bibfnamefont {O.}~\bibnamefont
  {Gittsovich}}, \bibinfo {author} {\bibfnamefont {N.~J.}\ \bibnamefont
  {Beaudry}}, \bibinfo {author} {\bibfnamefont {V.}~\bibnamefont
  {Narasimhachar}}, \bibinfo {author} {\bibfnamefont {R.~R.}\ \bibnamefont
  {Alvarez}}, \bibinfo {author} {\bibfnamefont {T.}~\bibnamefont {Moroder}},\
  and\ \bibinfo {author} {\bibfnamefont {N.}~\bibnamefont {Lütkenhaus}},\
  }\bibfield  {title} {\bibinfo {title} {Squashing model for detectors and
  applications to quantum-key-distribution protocols},\ }\href
  {https://doi.org/10.1103/PhysRevA.89.012325} {\bibfield  {journal} {\bibinfo
  {journal} {Physical Review A}\ }\textbf {\bibinfo {volume} {89}},\ \bibinfo
  {pages} {012325} (\bibinfo {year} {2014})}\BibitemShut {NoStop}%
\bibitem [{\citenamefont {Tupkary}\ \emph {et~al.}(2025)\citenamefont
  {Tupkary}, \citenamefont {Tan}, \citenamefont {Nahar}, \citenamefont
  {Kamin},\ and\ \citenamefont {Lütkenhaus}}]{tupkary_qkd_2025}%
  \BibitemOpen
  \bibfield  {author} {\bibinfo {author} {\bibfnamefont {D.}~\bibnamefont
  {Tupkary}}, \bibinfo {author} {\bibfnamefont {E.~Y.-Z.}\ \bibnamefont {Tan}},
  \bibinfo {author} {\bibfnamefont {S.}~\bibnamefont {Nahar}}, \bibinfo
  {author} {\bibfnamefont {L.}~\bibnamefont {Kamin}},\ and\ \bibinfo {author}
  {\bibfnamefont {N.}~\bibnamefont {Lütkenhaus}},\ }\href
  {https://doi.org/10.48550/arXiv.2502.10340} {\bibinfo {title} {{QKD} security
  proofs for decoy-state {BB84}: protocol variations, proof techniques, gaps
  and limitations}} (\bibinfo {year} {2025}),\ \bibinfo {note}
  {arXiv:2502.10340 [quant-ph]}\BibitemShut {NoStop}%
\bibitem [{\citenamefont {Attema}\ \emph {et~al.}(2021)\citenamefont {Attema},
  \citenamefont {Bosman},\ and\ \citenamefont
  {Neumann}}]{attema_optimizing_2021}%
  \BibitemOpen
  \bibfield  {author} {\bibinfo {author} {\bibfnamefont {T.}~\bibnamefont
  {Attema}}, \bibinfo {author} {\bibfnamefont {J.~W.}\ \bibnamefont {Bosman}},\
  and\ \bibinfo {author} {\bibfnamefont {N.~M.~P.}\ \bibnamefont {Neumann}},\
  }\bibfield  {title} {\bibinfo {title} {Optimizing the decoy-state {BB84}
  {QKD} protocol parameters},\ }\href
  {https://doi.org/10.1007/s11128-021-03078-0} {\bibfield  {journal} {\bibinfo
  {journal} {Quantum Information Processing}\ }\textbf {\bibinfo {volume}
  {20}},\ \bibinfo {pages} {154} (\bibinfo {year} {2021})}\BibitemShut
  {NoStop}%
\bibitem [{\citenamefont {Pereira}\ \emph {et~al.}(2020)\citenamefont
  {Pereira}, \citenamefont {Kato}, \citenamefont {Mizutani}, \citenamefont
  {Curty},\ and\ \citenamefont {Tamaki}}]{pereira_quantum_2020}%
  \BibitemOpen
  \bibfield  {author} {\bibinfo {author} {\bibfnamefont {M.}~\bibnamefont
  {Pereira}}, \bibinfo {author} {\bibfnamefont {G.}~\bibnamefont {Kato}},
  \bibinfo {author} {\bibfnamefont {A.}~\bibnamefont {Mizutani}}, \bibinfo
  {author} {\bibfnamefont {M.}~\bibnamefont {Curty}},\ and\ \bibinfo {author}
  {\bibfnamefont {K.}~\bibnamefont {Tamaki}},\ }\bibfield  {title} {\bibinfo
  {title} {Quantum key distribution with correlated sources},\ }\href
  {https://doi.org/10.1126/sciadv.aaz4487} {\bibfield  {journal} {\bibinfo
  {journal} {Science Advances}\ }\textbf {\bibinfo {volume} {6}},\ \bibinfo
  {pages} {eaaz4487} (\bibinfo {year} {2020})},\ \bibinfo {note} {publisher:
  American Association for the Advancement of Science}\BibitemShut {NoStop}%
\bibitem [{\citenamefont {Nahar}\ and\ \citenamefont
  {Lütkenhaus}(2025)}]{nahar_imperfect_2025}%
  \BibitemOpen
  \bibfield  {author} {\bibinfo {author} {\bibfnamefont {S.}~\bibnamefont
  {Nahar}}\ and\ \bibinfo {author} {\bibfnamefont {N.}~\bibnamefont
  {Lütkenhaus}},\ }\href {https://doi.org/10.48550/arXiv.2503.06328} {\bibinfo
  {title} {Imperfect detectors for adversarial tasks with applications to
  quantum key distribution}} (\bibinfo {year} {2025}),\ \bibinfo {note}
  {arXiv:2503.06328 [quant-ph]}\BibitemShut {NoStop}%
\bibitem [{\citenamefont {Zhang}\ \emph {et~al.}(2021)\citenamefont {Zhang},
  \citenamefont {Coles}, \citenamefont {Winick}, \citenamefont {Lin},\ and\
  \citenamefont {Lütkenhaus}}]{zhang_security_2021}%
  \BibitemOpen
  \bibfield  {author} {\bibinfo {author} {\bibfnamefont {Y.}~\bibnamefont
  {Zhang}}, \bibinfo {author} {\bibfnamefont {P.~J.}\ \bibnamefont {Coles}},
  \bibinfo {author} {\bibfnamefont {A.}~\bibnamefont {Winick}}, \bibinfo
  {author} {\bibfnamefont {J.}~\bibnamefont {Lin}},\ and\ \bibinfo {author}
  {\bibfnamefont {N.}~\bibnamefont {Lütkenhaus}},\ }\bibfield  {title}
  {\bibinfo {title} {Security proof of practical quantum key distribution with
  detection-efficiency mismatch},\ }\href
  {https://doi.org/10.1103/PhysRevResearch.3.013076} {\bibfield  {journal}
  {\bibinfo  {journal} {Physical Review Research}\ }\textbf {\bibinfo {volume}
  {3}},\ \bibinfo {pages} {013076} (\bibinfo {year} {2021})}\BibitemShut
  {NoStop}%
\bibitem [{\citenamefont {Wang}\ \emph {et~al.}(2016)\citenamefont {Wang},
  \citenamefont {Bao}, \citenamefont {Zhou}, \citenamefont {Jiang},\ and\
  \citenamefont {Li}}]{wang_tight_2016}%
  \BibitemOpen
  \bibfield  {author} {\bibinfo {author} {\bibfnamefont {Y.}~\bibnamefont
  {Wang}}, \bibinfo {author} {\bibfnamefont {W.-S.}\ \bibnamefont {Bao}},
  \bibinfo {author} {\bibfnamefont {C.}~\bibnamefont {Zhou}}, \bibinfo {author}
  {\bibfnamefont {M.-S.}\ \bibnamefont {Jiang}},\ and\ \bibinfo {author}
  {\bibfnamefont {H.-W.}\ \bibnamefont {Li}},\ }\bibfield  {title} {\bibinfo
  {title} {Tight finite-key analysis of a practical decoy-state quantum key
  distribution with unstable sources},\ }\href
  {https://doi.org/10.1103/PhysRevA.94.032335} {\bibfield  {journal} {\bibinfo
  {journal} {Physical Review A}\ }\textbf {\bibinfo {volume} {94}},\ \bibinfo
  {pages} {032335} (\bibinfo {year} {2016})},\ \bibinfo {note} {publisher:
  American Physical Society}\BibitemShut {NoStop}%
\bibitem [{\citenamefont {Zhou}\ \emph {et~al.}(2022)\citenamefont {Zhou},
  \citenamefont {Zhou}, \citenamefont {Xu}, \citenamefont {Wang}, \citenamefont
  {Lu}, \citenamefont {Jiang}, \citenamefont {Zhang},\ and\ \citenamefont
  {Bao}}]{zhou_finite-key_2022}%
  \BibitemOpen
  \bibfield  {author} {\bibinfo {author} {\bibfnamefont {C.}~\bibnamefont
  {Zhou}}, \bibinfo {author} {\bibfnamefont {Y.}~\bibnamefont {Zhou}}, \bibinfo
  {author} {\bibfnamefont {Y.}~\bibnamefont {Xu}}, \bibinfo {author}
  {\bibfnamefont {Y.}~\bibnamefont {Wang}}, \bibinfo {author} {\bibfnamefont
  {Y.}~\bibnamefont {Lu}}, \bibinfo {author} {\bibfnamefont {M.}~\bibnamefont
  {Jiang}}, \bibinfo {author} {\bibfnamefont {X.}~\bibnamefont {Zhang}},\ and\
  \bibinfo {author} {\bibfnamefont {W.}~\bibnamefont {Bao}},\ }\bibfield
  {title} {\bibinfo {title} {Finite-{Key} {Analysis} of 1-{Decoy} {Method}
  {Quantum} {Key} {Distribution} with {Intensity} {Fluctuation}},\ }\href
  {https://doi.org/10.3390/app12094709} {\bibfield  {journal} {\bibinfo
  {journal} {Applied Sciences}\ }\textbf {\bibinfo {volume} {12}},\ \bibinfo
  {pages} {4709} (\bibinfo {year} {2022})},\ \bibinfo {note} {number: 9
  Publisher: Multidisciplinary Digital Publishing Institute}\BibitemShut
  {NoStop}%
\bibitem [{\citenamefont {Zapatero}\ \emph {et~al.}(2021)\citenamefont
  {Zapatero}, \citenamefont {Navarrete}, \citenamefont {Tamaki},\ and\
  \citenamefont {Curty}}]{zapatero_security_2021}%
  \BibitemOpen
  \bibfield  {author} {\bibinfo {author} {\bibfnamefont {V.}~\bibnamefont
  {Zapatero}}, \bibinfo {author} {\bibfnamefont {A.}~\bibnamefont {Navarrete}},
  \bibinfo {author} {\bibfnamefont {K.}~\bibnamefont {Tamaki}},\ and\ \bibinfo
  {author} {\bibfnamefont {M.}~\bibnamefont {Curty}},\ }\bibfield  {title}
  {\bibinfo {title} {Security of quantum key distribution with intensity
  correlations},\ }\href {https://doi.org/10.22331/q-2021-12-07-602} {\bibfield
   {journal} {\bibinfo  {journal} {Quantum}\ }\textbf {\bibinfo {volume} {5}},\
  \bibinfo {pages} {602} (\bibinfo {year} {2021})}\BibitemShut {NoStop}%
\bibitem [{\citenamefont {Sixto}\ \emph {et~al.}(2022)\citenamefont {Sixto},
  \citenamefont {Zapatero},\ and\ \citenamefont {Curty}}]{sixto_security_2022}%
  \BibitemOpen
  \bibfield  {author} {\bibinfo {author} {\bibfnamefont {X.}~\bibnamefont
  {Sixto}}, \bibinfo {author} {\bibfnamefont {V.}~\bibnamefont {Zapatero}},\
  and\ \bibinfo {author} {\bibfnamefont {M.}~\bibnamefont {Curty}},\ }\bibfield
   {title} {\bibinfo {title} {Security of {Decoy}-{State} {Quantum} {Key}
  {Distribution} with {Correlated} {Intensity} {Fluctuations}},\ }\href
  {https://doi.org/10.1103/PhysRevApplied.18.044069} {\bibfield  {journal}
  {\bibinfo  {journal} {Physical Review Applied}\ }\textbf {\bibinfo {volume}
  {18}},\ \bibinfo {pages} {044069} (\bibinfo {year} {2022})},\ \bibinfo {note}
  {publisher: American Physical Society}\BibitemShut {NoStop}%
\bibitem [{\citenamefont {Currás-Lorenzo}\ \emph {et~al.}(2023)\citenamefont
  {Currás-Lorenzo}, \citenamefont {Nahar}, \citenamefont {Lütkenhaus},
  \citenamefont {Tamaki},\ and\ \citenamefont
  {Curty}}]{curras-lorenzo_security_2023}%
  \BibitemOpen
  \bibfield  {author} {\bibinfo {author} {\bibfnamefont {G.}~\bibnamefont
  {Currás-Lorenzo}}, \bibinfo {author} {\bibfnamefont {S.}~\bibnamefont
  {Nahar}}, \bibinfo {author} {\bibfnamefont {N.}~\bibnamefont {Lütkenhaus}},
  \bibinfo {author} {\bibfnamefont {K.}~\bibnamefont {Tamaki}},\ and\ \bibinfo
  {author} {\bibfnamefont {M.}~\bibnamefont {Curty}},\ }\bibfield  {title}
  {\bibinfo {title} {Security of quantum key distribution with imperfect phase
  randomisation},\ }\href {https://doi.org/10.1088/2058-9565/ad141c} {\bibfield
   {journal} {\bibinfo  {journal} {Quantum Science and Technology}\ }\textbf
  {\bibinfo {volume} {9}},\ \bibinfo {pages} {015025} (\bibinfo {year}
  {2023})},\ \bibinfo {note} {publisher: IOP Publishing}\BibitemShut {NoStop}%
\bibitem [{\citenamefont {Bluhm}\ \emph {et~al.}(2024)\citenamefont {Bluhm},
  \citenamefont {Capel}, \citenamefont {Gondolf},\ and\ \citenamefont
  {M{\"o}bus}}]{bluhm_unified_2024}%
  \BibitemOpen
  \bibfield  {author} {\bibinfo {author} {\bibfnamefont {A.}~\bibnamefont
  {Bluhm}}, \bibinfo {author} {\bibfnamefont {{\'A}.}~\bibnamefont {Capel}},
  \bibinfo {author} {\bibfnamefont {P.}~\bibnamefont {Gondolf}},\ and\ \bibinfo
  {author} {\bibfnamefont {T.}~\bibnamefont {M{\"o}bus}},\ }\bibfield  {title}
  {\bibinfo {title} {Unified {{Framework}} for {{Continuity}} of {{Sandwiched
  R{\'e}nyi Divergences}}},\ }\bibfield  {journal} {\bibinfo  {journal}
  {Annales Henri Poincar{\'e}}\ }\href
  {https://doi.org/10.1007/s00023-024-01519-x} {10.1007/s00023-024-01519-x}
  (\bibinfo {year} {2024})\BibitemShut {NoStop}%
\bibitem [{\citenamefont {Kamin}\ \emph
  {et~al.}(2025{\natexlab{c}})\citenamefont {Kamin}, \citenamefont
  {Burniston},\ and\ \citenamefont {Tan}}]{kamin_renyi_2025}%
  \BibitemOpen
  \bibfield  {author} {\bibinfo {author} {\bibfnamefont {L.}~\bibnamefont
  {Kamin}}, \bibinfo {author} {\bibfnamefont {J.}~\bibnamefont {Burniston}},\
  and\ \bibinfo {author} {\bibfnamefont {E.~Y.-Z.}\ \bibnamefont {Tan}},\
  }\href
  {https://github.com/Optical-Quantum-Communication-Theory/Renyi-security-framework}
  {\bibinfo {title} {Rényi security framework against coherent attacks applied
  to decoy-state {QKD} - {GitHub} {Repository}}} (\bibinfo {year}
  {2025}{\natexlab{c}}),\ \bibinfo {note} {original-date:
  2025-04-21T19:42:04Z}\BibitemShut {NoStop}%
\bibitem [{\citenamefont {Beigi}(2013)}]{beigi_sandwiched_2013}%
  \BibitemOpen
  \bibfield  {author} {\bibinfo {author} {\bibfnamefont {S.}~\bibnamefont
  {Beigi}},\ }\bibfield  {title} {\bibinfo {title} {Sandwiched {Rényi}
  divergence satisfies data processing inequality},\ }\href
  {https://doi.org/10.1063/1.4838855} {\bibfield  {journal} {\bibinfo
  {journal} {Journal of Mathematical Physics}\ }\textbf {\bibinfo {volume}
  {54}},\ \bibinfo {pages} {122202} (\bibinfo {year} {2013})}\BibitemShut
  {NoStop}%
\bibitem [{\citenamefont {Frank}\ and\ \citenamefont
  {Lieb}(2013)}]{frank_monotonicity_2013}%
  \BibitemOpen
  \bibfield  {author} {\bibinfo {author} {\bibfnamefont {R.~L.}\ \bibnamefont
  {Frank}}\ and\ \bibinfo {author} {\bibfnamefont {E.~H.}\ \bibnamefont
  {Lieb}},\ }\bibfield  {title} {\bibinfo {title} {Monotonicity of a relative
  {Rényi} entropy},\ }\href {https://doi.org/10.1063/1.4838835} {\bibfield
  {journal} {\bibinfo  {journal} {Journal of Mathematical Physics}\ }\textbf
  {\bibinfo {volume} {54}},\ \bibinfo {pages} {122201} (\bibinfo {year}
  {2013})}\BibitemShut {NoStop}%
\bibitem [{\citenamefont {Mosonyi}\ and\ \citenamefont
  {Ogawa}(2015)}]{mosonyi_quantum_2015}%
  \BibitemOpen
  \bibfield  {author} {\bibinfo {author} {\bibfnamefont {M.}~\bibnamefont
  {Mosonyi}}\ and\ \bibinfo {author} {\bibfnamefont {T.}~\bibnamefont
  {Ogawa}},\ }\bibfield  {title} {\bibinfo {title} {Quantum {Hypothesis}
  {Testing} and the {Operational} {Interpretation} of the {Quantum} {Rényi}
  {Relative} {Entropies}},\ }\href {https://doi.org/10.1007/s00220-014-2248-x}
  {\bibfield  {journal} {\bibinfo  {journal} {Communications in Mathematical
  Physics}\ }\textbf {\bibinfo {volume} {334}},\ \bibinfo {pages} {1617}
  (\bibinfo {year} {2015})}\BibitemShut {NoStop}%
\bibitem [{\citenamefont {Epstein}(1973)}]{epstein_remarks_1973}%
  \BibitemOpen
  \bibfield  {author} {\bibinfo {author} {\bibfnamefont {H.}~\bibnamefont
  {Epstein}},\ }\bibfield  {title} {\bibinfo {title} {Remarks on two theorems
  of {E}. {Lieb}},\ }\href {https://doi.org/10.1007/BF01646492} {\bibfield
  {journal} {\bibinfo  {journal} {Communications in Mathematical Physics}\
  }\textbf {\bibinfo {volume} {31}},\ \bibinfo {pages} {317} (\bibinfo {year}
  {1973})}\BibitemShut {NoStop}%
\bibitem [{\citenamefont {Grant}\ \emph {et~al.}(2006)\citenamefont {Grant},
  \citenamefont {Boyd},\ and\ \citenamefont {Ye}}]{grant_disciplined_2006}%
  \BibitemOpen
  \bibfield  {author} {\bibinfo {author} {\bibfnamefont {M.}~\bibnamefont
  {Grant}}, \bibinfo {author} {\bibfnamefont {S.}~\bibnamefont {Boyd}},\ and\
  \bibinfo {author} {\bibfnamefont {Y.}~\bibnamefont {Ye}},\ }\bibfield
  {title} {\bibinfo {title} {Disciplined {Convex} {Programming}},\ }in\ \href
  {https://doi.org/10.1007/0-387-30528-9_7} {\emph {\bibinfo {booktitle}
  {Global {Optimization}: {From} {Theory} to {Implementation}}}},\ \bibinfo
  {editor} {edited by\ \bibinfo {editor} {\bibfnamefont {L.}~\bibnamefont
  {Liberti}}\ and\ \bibinfo {editor} {\bibfnamefont {N.}~\bibnamefont
  {Maculan}}}\ (\bibinfo  {publisher} {Springer US},\ \bibinfo {address}
  {Boston, MA},\ \bibinfo {year} {2006})\ pp.\ \bibinfo {pages}
  {155--210}\BibitemShut {NoStop}%
\bibitem [{\citenamefont {Fawzi}\ \emph {et~al.}(2019)\citenamefont {Fawzi},
  \citenamefont {Saunderson},\ and\ \citenamefont
  {Parrilo}}]{fawzi_semidefinite_2019}%
  \BibitemOpen
  \bibfield  {author} {\bibinfo {author} {\bibfnamefont {H.}~\bibnamefont
  {Fawzi}}, \bibinfo {author} {\bibfnamefont {J.}~\bibnamefont {Saunderson}},\
  and\ \bibinfo {author} {\bibfnamefont {P.~A.}\ \bibnamefont {Parrilo}},\
  }\bibfield  {title} {\bibinfo {title} {Semidefinite approximations of the
  matrix logarithm},\ }\href {https://doi.org/10.1007/s10208-018-9385-0}
  {\bibfield  {journal} {\bibinfo  {journal} {Foundations of Computational
  Mathematics}\ }\textbf {\bibinfo {volume} {19}},\ \bibinfo {pages} {259}
  (\bibinfo {year} {2019})},\ \bibinfo {note} {arXiv:1705.00812
  [math]}\BibitemShut {NoStop}%
\bibitem [{\citenamefont {Harville}(1997)}]{harville_matrix_1997}%
  \BibitemOpen
  \bibfield  {author} {\bibinfo {author} {\bibfnamefont {D.~A.}\ \bibnamefont
  {Harville}},\ }\href {https://doi.org/10.1007/b98818} {\emph {\bibinfo
  {title} {Matrix {Algebra} {From} a {Statistician}’s {Perspective}}}}\
  (\bibinfo  {publisher} {Springer},\ \bibinfo {address} {New York, NY},\
  \bibinfo {year} {1997})\BibitemShut {NoStop}%
\bibitem [{\citenamefont {Hiai}\ and\ \citenamefont
  {Petz}(2014)}]{hiai_introduction_2014}%
  \BibitemOpen
  \bibfield  {author} {\bibinfo {author} {\bibfnamefont {F.}~\bibnamefont
  {Hiai}}\ and\ \bibinfo {author} {\bibfnamefont {D.}~\bibnamefont {Petz}},\
  }\href {https://doi.org/10.1007/978-3-319-04150-6} {\emph {\bibinfo {title}
  {Introduction to {Matrix} {Analysis} and {Applications}}}},\ Universitext\
  (\bibinfo  {publisher} {Springer International Publishing},\ \bibinfo
  {address} {Cham},\ \bibinfo {year} {2014})\BibitemShut {NoStop}%
\bibitem [{\citenamefont {Magnus}\ and\ \citenamefont
  {Neudecker}(2002)}]{magnus_matrix_2002}%
  \BibitemOpen
  \bibfield  {author} {\bibinfo {author} {\bibfnamefont {J.~R.}\ \bibnamefont
  {Magnus}}\ and\ \bibinfo {author} {\bibfnamefont {H.}~\bibnamefont
  {Neudecker}},\ }\href@noop {} {\emph {\bibinfo {title} {Matrix differential
  calculus with applications in statistics and econometrics}}},\ \bibinfo
  {edition} {rev. ed., reprinted}\ ed.,\ Wiley series in probability and
  statistics\ (\bibinfo  {publisher} {John Wiley},\ \bibinfo {address}
  {Chichester},\ \bibinfo {year} {2002})\BibitemShut {NoStop}%
\bibitem [{\citenamefont {MATLAB}(2024)}]{MATLAB}%
  \BibitemOpen
  \bibfield  {author} {\bibinfo {author} {\bibnamefont {MATLAB}},\ }\href
  {https://www.mathworks.com} {\emph {\bibinfo {title} {MATLAB version:
  24.2.0.2806996 (R2024b)}}}\ (\bibinfo  {publisher} {The MathWorks Inc.},\
  \bibinfo {address} {Natick, Massachusetts},\ \bibinfo {year}
  {2024})\BibitemShut {NoStop}%
\end{thebibliography}%

%Appendix
\appendix

\section{Alternative formulations of a \Renyi QKD cone}
\label{app:RenyiQKDcones}

We present here some alternative ways to formulate all of the \Renyi conditional entropies described in \cref{def:Conditional Renyi entropy}, in a manner suitable for convex optimization. Compared to the expressions in the main text, these formulations may be more ``generic'' in that they do not make use of any specific structure of the channel $\EATchann$ (in fact, the output registers of $\EATchann$ can even be quantum rather than classical). Certainly, when focusing on the cases considered in the main text (i.e.~the channel has the form $\EATchannQKD$ and we consider either $\renyiSandUp_{\alpha}$ or $\renyiPetzDown_{\alpha}$),  these expressions would in principle be just a different way of writing exactly the same mathematical \emph{function} of the input state $\rho_Q$; however, it seems worth investigating in future work whether the expressions presented in the following lemma might provide any advantages in terms of simplicity or numerical stability.

\begin{widetext}
\begin{lemma}\label{lemma:RenyiQKDcones}
    Consider any CPTP map $\EATchann: Q \to SI$, take any Stinespring dilation $V: Q \to SIR$ of it, and define a new channel $\mathcal{N}:Q \to SR$ given by $\mathcal{N} \defvar \Tr_{I} \circ V$. Then for any pure \(\rho_{QE} \in \dop{=}(QE)\), we have
    \begin{align}
        \forall \alpha \in \left[\frac{1}{2},\infty\right], \quad &\renyiSandUp_{\alpha}(S|IE)_{\EATchann(\rho_{QE})} = 
            \inf_{\tau_R} \renyiSandDiv_{\frac{\alpha}{2\alpha -1}} \rel{\mathcal{N}(\rho_{Q})}{\idop_S \otimes \tau_R} , \label{eq:dual_sand_up}\\
        \forall \alpha \in \left[0,\infty\right],\quad &\renyiSandDown_{\alpha}(S|IE)_{\EATchann(\rho_{QE})} = \inf_{\tau_R} \renyiPetzDiv_{\frac{1}{\alpha}}\rel{\mathcal{N}(\rho_{Q})}{\idop_S \otimes \tau_R} = \frac{1}{1-\alpha} \log \Tr \left[ \left( \Tr_S \left[\mathcal{N}(\rho_{Q})^{\frac{1}{\alpha}}\right] \right)^{\alpha}\right] , \label{eq:dual_sand_down}\\
        \forall \alpha \in \left[0,\infty\right], \quad &\renyiPetzUp_{\alpha}(S|IE)_{\EATchann(\rho_{QE})} = \renyiSandDiv_{\frac{1}{\alpha}}\rel{\mathcal{N}(\rho_{Q})}{\idop_S \otimes \Tr_{S} \circ \mathcal{N}(\rho_{Q})}, \\
        \forall \alpha \in \left[0,2\right], \quad &\renyiPetzDown_{\alpha}(S|IE)_{\EATchann(\rho_{QE})} = \renyiPetzDiv_{2-\alpha}\rel{\mathcal{N}(\rho_{Q})}{\idop_S \otimes \Tr_{S} \circ \mathcal{N}(\rho_{Q})} \label{eq:dual_petz_down}.  
    \end{align}
\end{lemma}
\end{widetext}

\begin{proof}
    The relations again follow straightforwardly from the duality relations of \Renyi entropies as presented in \cite[Sec 5.3]{tomamichel_quantum_2016}. We will show the proof for \(\renyiSandUp_{\alpha}(S|IE)\) as an example and the remaining relations follow similarly. 
    Writing $\sigma_{SIRE} = V(\rho_{QE})$, note that $\sigma$ is a purification of $\EATchann(\rho_{QE})$. Therefore we can write 
    \begin{align}\renyiSandUp_{\alpha}(S|IE)_{\EATchann(\rho_{QE})} &= 
        \renyiSandUp_{\alpha}(S|IE)_{\sigma_{SIE}} \nonumber\\&= - \renyiSandUp_{\beta}(S|R)_{\sigma_{SR}},
    \end{align}
    where the second line is \cite[Theorem 9]{beigi_sandwiched_2013}, for \(\beta \) such that \(\frac{1}{\alpha} + \frac{1}{\beta} = 2\). Inserting the definition of the conditional sandwiched \Renyi entropy, see \cref{def:Conditional Renyi entropy}, we find
    \begin{equation}
        \renyiSandUp_{\beta}(S|R)_{\sigma_{SR}} = \sup_{\tau_R} - \renyiSandDiv_{\beta}\rel{\sigma_{SR}}{\idop_S \otimes \tau_R}.
    \end{equation}
    Finally, rearranging \(\frac{1}{\alpha} + \frac{1}{\beta} = 2\) for \(\beta\) and invoking \(\inf(A) = - \sup(-A)\), the claimed expression for $\renyiSandUp_{\alpha}(S|IE)_{\EATchann(\rho_{QE})}$ follows.

    For the case of $\renyiSandDown_{\alpha}$, we obtain an additional simplification since the optimization in the definition of $\renyiPetzUp_{\beta}$ has a closed-form solution~\cite[Lemma 1]{tomamichel_relating_2014}, yielding 
    \begin{align}
    \renyiPetzUp_{\beta}(S|R)_{\sigma_{SR}} = \frac{\beta}{1-\beta} \log \Tr \left[ \left( \Tr_S \left[\sigma_{SR}^\beta\right] \right)^{\frac{1}{\beta}}\right],
    \end{align}
    then substituting $\beta=\frac{1}{\alpha}$.
\end{proof}

Note that for $\alpha\geq 1$, all the divergence-based expressions in 
\crefrange{eq:dual_sand_up}{eq:dual_petz_down} have $\beta \leq 1$ and are thus convex in $\rho$ (and, jointly, $\tau$, when it appears). The rightmost expression in \cref{eq:dual_sand_down} is also convex in $\rho$, since taking the infimum of a jointly convex function over a convex set preserves convexity~\cite[Section~3.2.5]{boyd_convex_2004}, as long as the function does not take value $-\infty$.
Therefore, these expressions again allow us to formulate our analysis as a convex optimization problem --- though for the expressions involving $\tau$, one would need to introduce $\tau$ as another optimization variable when solving, which may be computationally intensive. 

As a special case of the above results (i.e.~setting $\alpha=1$, or simply a direct proof from the duality relation for von Neumann entropy), we obtain 
\begin{align}
H(S|IE)_{\EATchann(\rho_{QE})} &= 
\inf_{\tau_R} D\rel{\mathcal{N}(\rho_{Q})}{\idop_S \otimes \tau_R} \\
&= D\rel{\mathcal{N}(\rho_{Q})}{\idop_S \otimes \Tr_{S} \circ \mathcal{N}(\rho_{Q})},
\end{align}
which is a formulation of the von Neumann entropy of $\EATchann(\rho_{QE})$ in a manner somewhat different from previous numerical work in QKD~\cite{winick_reliable_2018,hu_robust_2022,lorente_quantum_2025,he_exploiting_2024}. Specifically, such works have often focused on expressing this entropy in terms of $\GMap$ and $\ZMap$ maps as defined in the main text, which defines a convex cone sometimes referred to in those works as the QKD cone. As mentioned above, while theoretically the above expression would simply be a different way of writing exactly the same mathematical function, it seems worth exploring in future work whether writing it in this form might improve numerical stability or ease of analysis.

\section{Flag-state squasher}\label{app:Flag-state squasher}
Here we show that the flag-state squasher of \cite{zhang_security_2021} can be incorporated into the optimization problems for fixed-length \cref{eq:halpha final} and variable-length \cref{eq:kappa const} key rates. The following theorem is written in a very general form; however, the CPTP map \(\Lambda\) should be thought of as a squashing map, which naturally satisfies all assumptions.

\begin{theorem}
Let \(\Lambda: B \rightarrow B' \) be a CPTP map, and for any CPTP map $\mathcal{E}:A' \to B$, let \(J'\) denote the Choi state of the combined CPTP map \(\mathcal{E}' \defvar \Lambda \circ \mathcal{E}: A' \rightarrow B' \). Let us define states \(\sigma_{AB'|\gen}\) and \(\sigma_{AB'|\test}\) in terms of $\mathcal{E}'$ similar to \cref{eq:rhoAB in terms of channel}, except with $\mathcal{E}'$ in place of $\mathcal{E}$, and similarly the CPTNI maps \(\chi_{t/g}\), such that
\begin{align}
    \sigma_{AB|\gen/\test} &= \chi_{g/t}[J'].
\end{align}
Furthermore, suppose there exists a quantum-to-classical channel \(\probst': AB'\rightarrow \CP\) such that 
\begin{align}
    \probst'[\sigma_{AB'}] = \probst'\circ \Lambda[\rho_{AB}] = \probst[\rho_{AB}],
\end{align}
for all attack channels \(\mathcal{E}\). Additionally, suppose there exist constants \(\theta_1,\theta_2\) and a subset \(\mathcal{F}_t \subseteq \alphCP\) such that 
\begin{equation}\label{eq:Squashing Constraint}
    \sum_{\cP\in\mathcal{F}_t} \Phi_{\cP}[\rho] \geq \theta_1\left( \theta_2 - \Tr[\left(M^A \otimes \Pi^B \right) \rho] \right) \; \forall \rho \in \dop{=}(AB),
\end{equation} 
where \(\Pi^B\) is a projector onto a subspace in \(B\) which is invariant under \(\Lambda\), and \(M^A\) is one of Alice's POVM elements. Finally, assume that
\begin{equation}
    \renyiSandDown_{\alpha}(S|\tilde{T}E)_{\EATchannQKD \circ \chi_g(J)} \geq \renyiSandDown_{\alpha}(S|\tilde{T}E)_{\EATchannQKD \circ \chi_g(J')},
\end{equation}
for all attack channels. Then, it holds for \(\hQKD\)
\begin{equation}\label{eq:halpha squashed}
    \begin{aligned}[t]
        \hQKD \geq &\inf_{\substack{\mbf{q} \in \Sacc, \\ J' \in \mathrm{Pos}(A'B')}} \Bigg( \frac{\alpha D\rel{\mbf{q}}{\bsym{\nu}_{\CP}}}{\alpha - 1} + q(\gen) g_{\alpha}(J') \Bigg)\\
        \textrm{s.t. } &\Tr_{B'}[J'] = \idop_{A'}, \\
        &\bsym{\nu}_{\CP} = (\gamma \bsym{\nu}_{\CP|\test}, 1-\gamma)^T, \\
        &\bsym{\nu}_{\CP|\test} = \probstJ'[J'], \\
        &\sum_{\cP\in\mathcal{F}_t} \Psi'_{\cP}[J'] \geq \theta_1\left( \theta_2 - \Tr[\left(M^A \otimes \Pi^{B'} \right) \chi_t(J')] \right), \\
        %&\sum_{\cP\in\mathcal{F}_g} \Psi'_{\cP}[J'] \geq \theta_1\left( \theta_2 - \Tr[\left(M^A \otimes \Pi^B \right) \chi_g(J')] \right),
        \end{aligned}
\end{equation}
where \(\Pi^{B'} = \Lambda[\Pi^B]\), i.e. it projects onto the image of \(\Pi^B\) under \(\Lambda\), and we defined
\begin{equation}
    \probstJ'[J'] \defvar \probst' \circ \chi_t[J'].
\end{equation}
Moreover, for the $\frenyiSandDown_\alpha$-normalization constant \(\kappa\) it holds
\begin{align}\label{eq:kappa const squashed}
\begin{aligned}
    \kappa = &\inf_{\substack{J \in \mathrm{Pos}(A'B'), \\ \bsym{\nu} \in \mathbb{P}(\Ct) }} \Bigg( \frac{1}{1-\alpha} \log \Big( \gamma \sum_{c\in \Ct} \bsym{\nu}(\cP)2^{(\alpha - 1) f(\cP)} \\
    &\qquad + \left(1-\gamma\right)2^{-(\alpha-1)\left(g_{\alpha}(J') - f(\gen)  \right)} \Big) \Bigg) \\
    \quad \textrm{s.t. } &\Tr_{B'}[J'] = \idop_{A'}, \\
    &\bsym{\nu} = \probstJ'[J], \\
    &\sum_{\cP\in\mathcal{F}_t} \Psi'_{\cP}[J'] \geq \theta_1\left( \theta_2 - \Tr[\left(M^A \otimes \Pi^B \right) \chi_t(J')] \right). \\
    %&\sum_{\cP\in\mathcal{F}_g} \Psi'_{\cP}[J'] \geq \theta_1\left( \theta_2 - \Tr[\left(M^A \otimes \Pi^B \right) \chi_g(J')] \right),
    \end{aligned}
\end{align}
\end{theorem}
\begin{proof}
    The proof follows the same idea as in \cite[App. B]{kamin_finite-size_2025}, but involves slightly different quantities. 
    First, we can bound \(\hQKD\) by
    \begin{equation}
    \begin{aligned}[t]
        \hQKD \geq &\inf_{\substack{\mbf{q} \in \Sacc, \\ J \in \mathrm{Pos}(A'B)}} \Bigg( \frac{\alpha D\rel{\mbf{q}}{\bsym{\nu}_{\CP}}}{\alpha - 1} + q(\gen) g_{\alpha}(J') \Bigg)\\
        \textrm{s.t. } &\Tr_{B}[J] = \idop_{A'}, \\
        &\bsym{\nu}_{\CP} = (\gamma \bsym{\nu}_{\CP|\test}, 1-\gamma)^T, \\
        &\bsym{\nu}_{\CP|\test} = \probstJ[J], \\
        &\sum_{\cP\in\mathcal{F}_t} \Psi_{\cP}[J] \geq \theta_1\left( \theta_2 - \Tr[\left(M^A \otimes \Pi^B \right) \chi_t(J)] \right), \\
        %&\sum_{\cP\in\mathcal{F}_g} \Psi_{\cP}[J] \geq \theta_1\left( \theta_2 - \Tr[\left(M^A \otimes \Pi^B \right) \chi_g(J)] \right),
        \end{aligned}
    \end{equation}
    since by assumption, every feasible point of \cref{eq:halpha final} is a feasible point of the equation above and we used that
    \begin{align}
        &\renyiSandDown_{\alpha}(S|\tilde{T}E)_{\EATchannQKD \circ \chi_g(J)} \\ 
        &\geq \renyiSandDown_{\alpha}(S|\tilde{T}E)_{\EATchannQKD \circ \chi_g(J')} = g_{\alpha}(J').
    \end{align}
    Next, we note that \(\probstJ'[J'] = \probstJ[J]\) because
    \begin{align}
        &\probstJ'[J'] = \probst' \circ \chi_t[J'] \\ 
        &= \probst'[ \sigma_{AB'}] = \probst[\rho_{AB}] = \probstJ[J].
    \end{align}
    Moreover, as \(\Pi^B\) projects onto a subspace that is invariant under \(\Lambda\), it holds 
    \begin{equation}
        \Tr[\left(M^A \otimes \Pi^B \right) \chi_t(J)] = \Tr[\left(M^A \otimes \Pi^{B'} \right) \chi_t(J')].
    \end{equation}
    Including both observations in our previous lower bound on \(\hQKD\) lets us write
    \begin{equation}
    \begin{aligned}[t]
        \hQKD \geq &\inf_{\substack{\mbf{q} \in \Sacc, \\ J \in \mathrm{Pos}(A'B)}} \Bigg( \frac{\alpha D\rel{\mbf{q}}{\bsym{\nu}_{\CP}}}{\alpha - 1} + q(\gen) g_{\alpha}(J') \Bigg)\\
        \textrm{s.t. } &\Tr_{B}[J] = \idop_{A'}, \\
        &\bsym{\nu}_{\CP} = (\gamma \bsym{\nu}_{\CP|\test}, 1-\gamma)^T, \\
        &\bsym{\nu}_{\CP|\test} = \probstJ'[J'], \\
        &\sum_{\cP\in\mathcal{F}_t} \Psi'_{\cP}[J'] \geq \theta_1\left( \theta_2 - \Tr[\left(M^A \otimes \Pi^{B'} \right) \chi_t(J')] \right). \\
        %&\sum_{\cP\in\mathcal{F}_g} \Psi_{\cP}[J] \geq \theta_1\left( \theta_2 - \Tr[\left(M^A \otimes \Pi^B \right) \chi_g(J)] \right),
        \end{aligned}
    \end{equation}
    Finally, if we include the channel \(\Lambda\) in the optimization, the new optimization variable becomes \(J'\), which can only reduce the optimal value as one optimizes over a larger set. Therefore, the theorem statement for \(\hQKD\) follows. 

    If one repeats the same steps for the $\frenyiSandDown_\alpha$-normalization constant \(\kappa\), the equivalent statement for \(\kappa\) follows.
\end{proof}

\begin{rem}
    The constants \(\theta_1,\theta_2\) can be found exactly the same way as in \cite[App. B]{kamin_finite-size_2025}.
\end{rem}

\begin{rem}
    For notational simplicity, we did not include constraints from the states in the generation rounds, however one can include those straightforwardly.
    
    The construction would go as follows. One would define a similar subset \(\mathcal{F}_g \subseteq \mathcal{C} \) for constraints on generation rounds and a map acting similar to \(\Phi\) but with POVM elements conditioned on generation rounds. Then, the argument follows in the same manner as above.
\end{rem}

\section{Source Maps}\label{app:Source maps}
\newcommand{\Esource}{E_{\mathrm{sou}}}
\newcommand{\Vvir}{V_{\mathcal{E}_{\mathrm{vir}}}}
\newcommand{\Vreal}{V_{\mathcal{E}_{\mathrm{real}}}}
\newcommand{\Vsource}{V_{\Xi}}

For many imperfections, it is favorable to use so-called source maps \cite{nahar_imperfect_2023}. In \cite{nahar_postselection_2024} a proof-technique independent proof was given, however here we explicitly prove that source maps can indeed be incorporated into our framework.

Since the proof for the $\frenyiSandDown_\alpha$-normalization constant is simpler and already lays out the general idea, we start with it and then present the proof for \(\hQKD\).

\begin{theorem}[Source Maps in $f$-weighted Entropies]\label{thrm:Source maps QES}
    Let \(\sigma_{AA'}\) and \(\tau_{AA''}\) be the states emitted by the real and virtual source, respectively. Furthermore, let \(\Xi: A'' \rightarrow A' \) be a source map, i.e. a channel that satisfies \(\sigma_{AA'} = \Xi(\tau_{AA''})\), and \(\Vsource: A'' \rightarrow A' \Esource \) its Stinespring dilation. Then, for the QKD channel \(\EATchannQKD\) and the $\frenyiSandDown_\alpha$ entropy it holds
    \begin{equation}
    \begin{split}
        &\inf_{\Vvir} \frenyiSandDown_\alpha(S|\CP \tilde{T}E\Esource)_{\EATchannQKD(\Vvir \tau_{AA''} \Vvir^{\dagger})} \\ 
        &\leq \inf_{\Vreal} \frenyiSandDown_\alpha(S|\CP \tilde{T}E)_{\EATchannQKD(\Vreal \rho_{AA''} \Vreal^{\dagger})},
    \end{split}
    \end{equation}
    where \(\Vvir: A'' \rightarrow BE\Esource\) and \(\Vreal: A' \rightarrow BE\) are the Stinespring dilations of Eve's attack channels for the virtual and real source, respectively.

    The same statement holds upon replacing the $\frenyiSandDown_\alpha$ entropy with the $\frenyiSandUp_\alpha$ entropy.
\end{theorem}
\begin{proof}
    The proof is based on a simple calculation. 
    \begin{align}
        &\inf_{\Vvir} \frenyiSandDown_\alpha(S|\CP \tilde{T}E\Esource)_{\EATchannQKD(\Vvir \tau_{AA''} \Vvir^{\dagger})} \\ 
        &\leq \inf_{\Vreal} \frenyiSandDown_\alpha(S|\CP \tilde{T}E \Esource)_{\EATchannQKD(\Vreal \Vsource \tau_{AA''} \Vsource^{\dagger} \Vreal^{\dagger})} \\ 
        &\leq \inf_{\Vreal} \frenyiSandDown_\alpha(S|\CP \tilde{T}E)_{\EATchannQKD(\Vreal \Tr_{\Esource}[\Vsource \tau_{AA''} \Vsource^{\dagger}] \Vreal^{\dagger})} \\
        &= \inf_{\Vreal} \frenyiSandDown_\alpha(S|\CP \tilde{T}E)_{\EATchannQKD(\Vreal \sigma_{AA'} \Vreal^{\dagger})}.
    \end{align}
    The steps are as follows. The first inequality is due to optimizing over a smaller set, which can only increase the minimum. The second inequality is due to the data-processing inequality for $\frenyiSandDown_\alpha$ entropies \cite[Lemma 4.3]{arqand_generalized_2024}. The final equality holds because \(\Xi\) is a source map, i.e.
    \begin{equation}
        \Tr_{\Esource}[\Vsource \tau_{AA''} \Vsource^{\dagger}] = \Xi(\tau_{AA''}) = \sigma_{AA'}.
    \end{equation}
    The proof for $f$-weighted entropies follows by the same arguments upon replacing the data-processing inequality for $\frenyiSandDown_\alpha$ entropies with the data-processing inequality for $\frenyiSandUp_\alpha$ entropies \cite[Lemma 4.4]{arqand_marginal-constrained_2025}.
\end{proof}

This theorem, immediately allows us to include source maps in the $\frenyiSandDown_\alpha$-normalization constant \(\kappa\). Hence, next we will show that we can do the same for \(\hQKD\).

\begin{theorem}[Source maps in \(\hQKD\) and \(\hUpQKD\)]\label{thrm:Source mpas halpha}
    Let \(\sigma_{AA'}\) and \(\tau_{AA''}\) be the states emitted by the real and virtual source, respectively. Furthermore, let \(\Xi: A'' \rightarrow A' \) be a source map, i.e. a channel that satisfies \(\sigma_{AA'} = \Xi(\tau_{AA''})\), and \(\Vsource: A'' \rightarrow A' \Esource \) its Stinespring dilation. Then, \(\hQKD\) can be bounded from below as
    \begin{equation}
        \hQKD \geq \begin{aligned}[t]
        \inf_{\mbf{q} \in \Sacc} &\inf_{\substack{\Vvir, \\ \Vvir^\dagger \Vvir = \idop_{A''}} } \Bigg( \frac{\alpha}{\alpha - 1} D\rel{\mbf{q}}{\bsym{\nu}_{\CP}} \\
        &+ q(\gen) \renyiSandDown_{\alpha}(S|\tilde{T}E\Esource)_{\nu_{|\gen}} \Bigg),
        \end{aligned}
    \end{equation}
    where
    \begin{align}
        \nu_{S\tilde{T} \CP e \Esource|\gen} &= \EATchannQKD(\Vvir \tau_{AA''} \Vvir^{\dagger})_{|\gen}, \\
        \bsym{\nu}_{\CP} &= (\gamma \bsym{\nu}_{\CP|\test}, 1-\gamma)^T, \\
        \bsym{\nu}_{\CP|\test} &= \probst[\Tr_{E\Esource}[\Vvir \tau_{AA'}\Vvir^{\dagger}]].
    \end{align}
    Here we again defined \(\Vvir: A'' \rightarrow BE\Esource\) as the Stinespring dilation of Eve's attack channel for the virtual source.

    The same statement holds for \(\hUpQKD\), i.e. replacing \(\renyiSandDown_{\alpha}(S|\tilde{T}E\Esource)_{\nu_{|\gen}}\) with \(\renyiSandUp_{\alpha}(S|\tilde{T}E\Esource)_{\nu_{|\gen}}\).
\end{theorem}
\begin{proof}
    First, let us rewrite \(\hQKD\) from \cref{cor:Fixed-length security lower bnd halpha} (using the real source) as an optimization problem over Eve's attack channel characterized by its Stinespring dilation \(\Vreal: A' \rightarrow BE\). This can be written as
    \begin{equation}
        \hQKD = \begin{aligned}[t]
        \inf_{\mbf{q} \in \Sacc} &\inf_{\substack{\Vreal, \\ \Vreal^\dagger \Vreal = \idop_{A'}} } \Bigg( \frac{\alpha}{\alpha - 1} D\rel{\mbf{q}}{\bsym{\nu}_{\CP}} \\
        &+ q(\gen) \renyiSandDown_{\alpha}(S|\tilde{T}E)_{\nu_{|\gen}} \Bigg),
        \end{aligned}
    \end{equation}
    where
    \begin{align}
        \nu_{S\tilde{T} \CP E|\gen} &= \EATchannQKD(\Vreal \sigma_{AA''} \Vreal^{\dagger})_{|\gen}, \\
        \bsym{\nu}_{\CP} &= (\gamma \bsym{\nu}_{\CP|\test}, 1-\gamma)^T, \\
        \bsym{\nu}_{\CP|\test} &= \probst[\Tr_{E}[\Vreal \sigma_{AA'}\Vreal^{\dagger}]].
    \end{align}
    Since \(\Xi\) is a source map it satisfies
    \begin{equation}
        \Tr_{\Esource}[\Vsource \tau_{AA''} \Vsource^{\dagger}] = \Xi(\tau_{AA''}) = \sigma_{AA'},
    \end{equation}
    and we can rewrite
    \begin{align}
        \nu_{S\tilde{T}\CP E|\gen} &= \EATchannQKD(\Vreal \Tr_{\Esource}[\Vsource \tau_{AA''} \Vsource^{\dagger}] \Vreal^{\dagger})_{|\gen}, \\
        \bsym{\nu}_{\CP|\test} &= \probst[\Tr_{E\Esource}[\Vreal \Vsource \tau_{AA'} \Vsource^{\dagger} \Vreal^{\dagger}]].
    \end{align}
    Next, due to the data-processing inequality for \Renyi entropies \cite[Theorem 1]{frank_monotonicity_2013} (or \cite{muller-lennert_quantum_2013,beigi_sandwiched_2013,mosonyi_quantum_2015,tomamichel_quantum_2016}), we find
    \begin{equation}
        \begin{split}
        & \renyiSandDown_\alpha(S|\CP \tilde{T}E\Esource)_{\EATchannQKD(\Vreal \Vsource \tau_{AA''} \Vsource^{\dagger} \Vreal^{\dagger})_{|\gen}} \\ 
        &\leq \renyiSandDown_\alpha(S|\CP \tilde{T}E)_{\EATchannQKD(\Vreal \Tr_{\Esource}[\Vsource \tau_{AA''} \Vsource^{\dagger}] \Vreal^{\dagger})_{|\gen}}.
    \end{split}
    \end{equation}
    Combining these relations, we find a lower bound on \(\hQKD\) as
    \begin{equation}
        \hQKD \geq \begin{aligned}[t]
        \inf_{\mbf{q} \in \Sacc} &\inf_{\substack{\Vreal, \\ \Vreal^\dagger \Vreal = \idop_{A'}} } \Bigg( \frac{\alpha}{\alpha - 1} D\rel{\mbf{q}}{\bsym{\nu}_{\CP}} \\
        &+ q(\gen) \renyiSandDown_{\alpha}(S|\tilde{T}E\Esource)_{\nu_{|\gen}} \Bigg),
        \end{aligned}
    \end{equation}
    where
    \begin{align}
        \nu_{S\tilde{T}\CP E\Esource|\gen} &= \EATchannQKD(\Vreal \Vsource \tau_{AA''} \Vsource^{\dagger} \Vreal^{\dagger})_{|\gen}, \\
        \bsym{\nu}_{\CP} &= (\gamma \bsym{\nu}_{\CP|\test}, 1-\gamma)^T, \\
        \bsym{\nu}_{\CP|\test} &= \probst[\Tr_{E\Esource}[\Vreal \Vsource \tau_{AA'} \Vsource^{\dagger} \Vreal^{\dagger}]].
    \end{align}
    Finally, by optimizing over \(\Vvir \defvar \Vreal \Vsource \) instead of only \(\Vreal\), the theorem statement follows because the minimization runs over a bigger set that only decreases the infimum further.

    The proof for \(\hUpQKD\) follows exactly the same steps replacing \(\renyiSandDown_\alpha\) with \(\renyiSandUp_\alpha\) appropriately. The entropy \(\renyiSandUp_\alpha\) still satisfies the required data-processing inequality.
\end{proof}

\section{Proofs for Phase Imperfections}
\newcommand{\XiMeas}{\Xi_{\mathrm{meas}}}
To include phase imperfections in our framework, we require a statement allowing us to relate the approximately diagonalized states to exactly diagonalized states. The following lemma allows for exactly this application.

\begin{lemma}[Trace distance bounds under source-replacement scheme]\label{Lem:trace distance bounds source-replacement}
    Let \(\rho_{AA'A_s}, \sigma_{AA'A_s} \in \dop{=}(AA'A_s)\) be pure states of the form
    \begin{align}
        \rho_{AA'A_s} = \sum_{i,j} \sqrt{p_i p_j} \ketbra{i}{j}_A \ketbra{\psi_i}{\psi_j}_{A'A_s}, \\
        \sigma_{AA'A_s} = \sum_{i,j} \sqrt{p_i p_j} \ketbra{i}{j}_A \ketbra{\phi_i}{\phi_j}_{A'A_s},
    \end{align}
    defined by the source-replacement scheme. Furthermore, let \(\Tr_{A_s}[\ketbra{\psi_i}{\psi_i}] = \rho_{A'}^i\) and \(\Tr_{A_s}[\ketbra{\phi_i}{\phi_i}] = \sigma_{A'}^i\) satisfy
    \begin{equation}
        \norm{ \rho_{A'}^i - \sigma_{A'}^i }_1 \leq \epsilon_i.
    \end{equation}
    Then, it holds
    \begin{equation}
        \norm{ \Tr_{A_s} \circ \GMap \circ \left( \id_A \otimes \; \mathcal{E} \right) \left(\rho_{AA'A_s} - \sigma_{AA'A_s} \right) }_1 \leq \sum_i p_i \epsilon_i,
    \end{equation}
    where \(\mathcal{E}: A'\rightarrow B\) is Eve's channel.
\end{lemma}
\begin{proof}
    First, we make use of the special properties of Alice's measurements under the source-replacement scheme. Therefore, let \(\XiMeas\) be the channel measuring Alice's system in the computational basis defined by
    \begin{equation}
        \XiMeas(\rho_A) = \sum_i \ketbra{i}{i} \rho_A \ketbra{i}{i},
    \end{equation}
    for all \(\rho_A\). Under the source-replacement scheme \cite{bennett_quantum_1992, ferenczi_symmetries_2012}, we can always write \(\XiMeas\) this way and it satisfies 
    \begin{equation}
        \XiMeas \circ \XiMeas(\rho_A) = \XiMeas(\rho),
    \end{equation}
    for all \(\rho\). Then, based on the construction of the map \(\GMap\) in \cite{winick_reliable_2018}, it follows that  \(\GMap\) satisfies
    \begin{equation}
        \GMap = \GMap \circ \XiMeas.
    \end{equation}
    Now,we will exploit this property of the map \(\GMap\) to bound the trace distance. We find
    \begin{align}
        &\norm{ \Tr_{A_s} \circ \; \GMap \circ \; \id_A \otimes \mathcal{E} \left(\rho_{AA'A_s} - \sigma_{AA'A_s} \right) }_1 \\
        = &\begin{aligned}[t]
            \lVert &\GMap \circ \; \id_A \otimes \mathcal{E} \circ \Tr_{A_s} \circ \left( \XiMeas \otimes \id_{A'} \right) \\  &\left(\rho_{AA'A_s} - \sigma_{AA'A_s} \right) \rVert_1
        \end{aligned}.
    \end{align}
    Next, applying the data-processing inequality for the trace distance, see e.g. \cite[Theorem 9.2]{nielsen_quantum_2010}, we find
    \begin{align}
        &\norm{ \Tr_{A_s} \circ \; \GMap \circ \; \id_A \otimes \mathcal{E} \left(\rho_{AA'A_s} - \sigma_{AA'A_s} \right) }_1 \\
        &\leq \lVert \Tr_{A_s} \circ \left( \XiMeas \otimes \id_{A'} \right) \left(\rho_{AA'A_s} - \sigma_{AA'A_s} \right) \rVert_1.
    \end{align}
    We can simplify \(\left( \XiMeas \otimes \id_{A'} \right) \left(\rho_{AA'A_s} - \sigma_{AA'A_s} \right) \) further. For this, observe that
    \begin{align}
         \XiMeas \otimes \id_{A'} \left(\rho_{AA'A_s} \right) = \sum_i p_i \ketbra{i}{i} \otimes \ketbra{\psi_i}{\psi_i},
    \end{align}
    and similarly for \(\sigma_{AA'A_s}\). A quick calculation applying strong convexity \cite[Theorem 9.3]{nielsen_quantum_2010} and sub-additivity with respect to tensor products of the trace distance, yields
    \begin{align}
        &\lVert \Tr_{A_s} \circ \left( \XiMeas \otimes \id_{A'} \right) \left(\rho_{AA'A_s} - \sigma_{AA'A_s} \right) \rVert_1 \\
        &\leq \sum_i p_i \lVert \Tr_{A_s}[ \ketbra{\psi_i}{\psi_i} - \ketbra{\phi_i}{\phi_i}] \rVert_1 \\
        &= \sum_i p_i \lVert \rho_{A'}^i - \sigma_{A'}^i] \rVert_1.
    \end{align}
    After inserting the assumption \(\norm{ \rho_{A'}^i - \sigma_{A'}^i }_1 \leq \epsilon_i\) the claim follows.
\end{proof}

In order to incorporate the cutoff \(N_A\) into the map \(\probst\) or \(\probstJ\), we require the following lemma.

%Lemma for Bounds On the Statistics under Approximate diagonalization
\LemBoundsOnStatisticsApprox*
\begin{proof}\label{Proof:LemBoundsOnStatisticsApprox}
    First, note that under the source replacement scheme we can define \(M_{\alpha,x}^{A|\test,m} \) such that \((M_{\alpha,x}^{A|\test,m})^2 = M_{\alpha,x}^{A|\test,m}\). Then, it holds for every component \(\cP=\phi(\alpha,x,\beta,y) \) of \(\probst\)
    \begin{align}
        &\abs{\left(\Phi_{m} \circ \left(\id_A \otimes \mathcal{E}_m \right)\left[\ketbra{\xi^t_{|m}}{\xi^t_{|m}} - \ketbra{\widetilde{\xi}^t_{|m}}{\widetilde{\xi}^t_{|m}}\right] \right)_{\cP}} \\
        &= \abs{\Tr \Big[ M_{\beta,y}^{B} \left(\id_A \otimes \mathcal{E}_m \right) \left[Q \right] \Big]},
    \end{align}
    where we defined 
    \begin{equation}
        Q := M_{\alpha,x}^{A|\test,m} \left( \ketbra{\xi^t_{|m}}{\xi^t_{|m}} - \ketbra{\widetilde{\xi}^t_{|m}}{\widetilde{\xi}^t_{|m}} \right) M_{\alpha,x}^{A|\test,m}.
    \end{equation}
    Then, it follows by Hölder's inequality
    \begin{align}
        &\abs{\left(\Phi_{m} \circ \left(\id_A \otimes \mathcal{E}_m \right) \left[\ketbra{\xi^t_{|m}}{\xi^t_{|m}} - \ketbra{\widetilde{\xi}^t_{|m}}{\widetilde{\xi}^t_{|m}}\right] \right)_{\cP}} \\
        &\leq \Tr \Big[ \Big \vert M_{\beta,y}^{B} \left(\id_A \otimes \mathcal{E}_m \right) \left[Q \right] \Big \vert \Big] \\
        &\leq \norm{ \idop_A \otimes M_{\beta,y}^{B}}_{\infty} \cdot \norm{ \left(\id_A \otimes \mathcal{E}_m\right) \left[ Q\right]}_1.
    \end{align}
    Since \(M_{\beta,y}^{B}\) are POVM elements, they satisfy \(\norm{M_{\beta,y}^{B}}_{\infty} \leq 1\), and we can bound the first term by one. Additionally, we apply the data-processing inequality for the trace distance on the second term. Combining both, we find in summary
    \begin{align}
        &\abs{\left(\Phi_{m}\left[\ketbra{\xi^t_{|m}}{\xi^t_{|m}}\right]\right)_{\cP} - \left(\Phi_{m}\left[\ketbra{\widetilde{\xi}^t_{|m}}{\widetilde{\xi}^t_{|m}}\right] \right)_{\cP}} \\
        &\leq \norm{M_{\alpha,x}^{A|\test,m} \left( \ketbra{\xi^t_{|m}}{\xi^t_{|m}} - \ketbra{\widetilde{\xi}^t_{|m}}{\widetilde{\xi}^t_{|m}} \right) M_{\alpha,x}^{A|\test,m}}_1.
    \end{align}
    Hence, we continue by bounding the right-hand-side of this equation. Here, a quick calculation applying strong convexity \cite[Theorem 9.3]{nielsen_quantum_2010} and sub-additivity with respect to tensor products of the trace distance, yields
    \begin{align}
        &\norm{M_{\alpha,x}^{A|\test,m} \left( \ketbra{\xi^t_{|m}}{\xi^t_{|m}} - \ketbra{\widetilde{\xi}^t_{|m}}{\widetilde{\xi}^t_{|m}} \right) M_{\alpha,x}^{A|\test,m}}_1 \\
        \begin{split}
            &\leq \sum_{a,\mu} p(a,\mu|\test,m) \Big\lVert \ketbra{s_{m}(a,\mu)}{s_{m}(a,\mu)} \\
            &\qquad - \ketbra{w_{m}(a,\mu)}{w_{m}(a,\mu)} \Big\rVert_1.
        \end{split}
    \end{align}
    Now, by assumption, it holds
    \begin{equation}
        \begin{split}
            &\Big\lVert \ketbra{s_{m}(a,\mu)}{s_{m}(a,\mu)} \\
            &\qquad - \ketbra{w_{m}(a,\mu)}{w_{m}(a,\mu)} \Big \rVert_1 \leq \epsilon_{a,\mu}.
        \end{split}
    \end{equation}
    Therefore, we find 
    \begin{equation}\label{eq:proof of lemma Bounds Approx Diag}
        \abs{\left(\Phi_{m}\left[\ketbra{\xi^t_{|m}}{\xi^t_{|m}}\right]\right)_{\cP} - \left(\Phi_{m}\left[\ketbra{\widetilde{\xi}^t_{|m}}{\widetilde{\xi}^t_{|m}}\right] \right)_{\cP}} \leq \epsilon_m,
    \end{equation}
    for all components \(\cP=\phi(\alpha,x,\beta,y) \) of \(\probst\).

    Turning our attention to the probabilities \(\zeta_{m|\test}\) and \(\omega_{m|\test}\). We can bound the difference between \(\zeta_{m|\test}\) and \(\omega_{m|\test}\) by
    \begin{align}
        &\abs{ \zeta_{m|\test} - \omega_{m|\test} } \\
        &\leq \sum_{a,\mu} p(a,\mu|\test) \abs{\zeta_m(a,\mu) - \omega_(a,\mu)}\\
        &\leq \sum_{a,\mu} p(a,\mu|\test) \epsProj^{a,\mu} = \epsProb,
    \end{align}
    where the last inequality follows by assumption. Furthermore, since both \(\zeta_{m|\test}\) and \(\omega_{m|\test}\) are probabilities, we can rewrite their difference as upper and lower bounds on \(\zeta_{m|\test}\) such as
    \begin{align}
        &\zeta_{m|\test} \geq \max\{\omega_{m|\test} - \epsProb,0\} = \omega_{m|\test}^L, \\
        &\zeta_{m|\test} \leq \min\{\omega_{m|\test} + \epsProb,1\} = \omega_{m|\test}^U.
    \end{align}
    Next, by definition, we see that \(1-\mbf{p}_{\mathrm{tot}}(\widetilde{\xi})\) is bounded by
    \begin{align}    
        &1 - \mbf{p}_{\mathrm{tot}}(\widetilde{\xi}) \\ 
        &= 1-  \sum_{\substack{\cP \in \Ct \\ a \in \phi^{-1}(\cP)}} \sum_{m \leq \Ncons} \zeta_{m|\test} p(a|m,\test) \hat{e}_{\cP} \\
        &\leq \sum_{\substack{\cP \in \Ct \\ a \in \phi^{-1}(\cP)}} \sum_{m \leq \Ncons} \omega_{m|\test}^L p(a|m,\test) \hat{e}_{\cP} \\ 
        &= 1- \mbf{p}_{\mathrm{tot}}^L.
    \end{align}
    Finally, combining these results for the probabilities with \cref{eq:proof of lemma Bounds Approx Diag} yields the statement of the lemma.
\end{proof}

\ThrmPhaseImperfectRates*
\begin{proof}\label{Proof:ThrmPhaseImperfectRates}
     For simplicity we start the proof with \(\hUpQKD\), where we assume the special case of IID imperfections, i.e. \(\XiPhase^i \equiv \XiPhase\) for all \(i\).
     
     Furthermore, we note that since we only had continuity bounds available for \(\renyiSandUp_{\alpha}\), we are required to start from \(\hUpQKD\), which is defined in \cref{thrm:Fixed-length security generic bnd}.
    
    First, note that the block-tagged states \(\ket{\widetilde{\xi}}\) are related to the states without tagging through the source map \(\Tr_{\bar{A}}\). Furthermore, we need to include the source map \(\XiPhase\) from \cite[App. A]{nahar_imperfect_2023} connecting the model state with the phase imperfect laser pulses. Hence, we can apply \cref{thrm:Source mpas halpha} and bound the entropy by including the combined source map \(\XiPhase \circ \Tr_{\bar{A}}\) in Eve's attack. Thus, we find
    \begin{equation}
        \begin{aligned}[t]
            \hUpQKD \geq &\begin{aligned}[t]
                \inf_{\substack{\mbf{q} \in \Sacc, \\ J \in \mathrm{Pos}(A'\bar{A}B)}} \Bigg( &\frac{\alpha D\rel{\mbf{q}}{\bsym{\nu}_{\CP}}}{\alpha - 1} \\ &+ q(\gen) \renyiSandUp_{\alpha}(S|\tilde{T}E)_{\tilde{\nu}_{|\gen}(J)} \Bigg)
            \end{aligned}\\
            \textrm{s.t. } &\Tr_B[J] = \idop_{A'\bar{A}}, \\
            &\bsym{\nu}_{\CP} = (\gamma \bsym{\nu}_{\CP|\test}, 1-\gamma)^T, \\
            &\bsym{\nu}_{\CP|\test} = \probstJ_{\tilde{\xi}}[J],
            \end{aligned}
    \end{equation}
    where we made the dependence of \(\probstJ\) on the state \(\ket{\widetilde{\xi}^t}\) explicit. Next, we can insert \cref{eq:Continuity lower bnd key rate states} to bound the conditional \Renyi entropy and find
    \begin{equation}
        \begin{aligned}[t]
            \hUpQKD \geq &\inf_{\substack{\mbf{q} \in \Sacc, \\ J \in \mathrm{Pos}(A'\bar{A}B)}} \Bigg( \frac{\alpha D\rel{\mbf{q}}{\bsym{\nu}_{\CP}}}{\alpha - 1} - \Delta(\epsFull) \\ 
            &\quad + q(\gen) \renyiSandDown_{\alpha}(S|\tilde{T}E)_{\nu_{|\gen}(J)} \Bigg)\\
            \textrm{s.t. } &\Tr_B[J] = \idop_{A'\bar{A}}, \\
            &\bsym{\nu}_{\CP} = (\gamma \bsym{\nu}_{\CP|\test}, 1-\gamma)^T, \\
            &\bsym{\nu}_{\CP|\test} = \probstJ_{\tilde{\xi}}[J].
            \end{aligned}
    \end{equation} 
    As noted in \cref{subsec:Block-tagged states}, the block-tagged states satisfy assumptions \ref{Decoy assump block-diag}--\ref{Decoy assump QND Eve}. Hence, we can include the simplifications of \cref{sec:Block-Diag states} to reach
    \begin{equation}
    \begin{aligned}[t]
        \hUpQKD \geq &\inf_{\substack{\mbf{q} \in \Sacc, \bsym{\nu}_{\CP} \in \mathbb{P}(\alphCP)  \\ J_m \in \mathrm{Pos}(A'\bar{A}B)}} \Bigg( \frac{\alpha D\rel{\mbf{q}}{\bsym{\nu}_{\CP}}}{\alpha - 1} - \Delta(\epsFull) \\ 
        &\quad + q(\gen) g_{\alpha}^{\Nent}\left(J_0,\dots, J_{\Nent} \right) \Bigg)\\
        \textrm{s.t. } &\Tr_B[J_m] = \idop_{A'\bar{A}}, \\
        &\bsym{\nu}_{\CP} = (\gamma \bsym{\nu}_{\CP|\test}, 1-\gamma)^T, \\
        &\bsym{\nu}_{\CP|\test} \leq \sum_{m\leq \Ncons} \zeta_{m|\test} \probstJ_{\tilde{\xi}_m}[J_m] \\
        &\qquad+ 1-\mbf{p}_{\mathrm{tot}}(\widetilde{\xi}), \\
        &\bsym{\nu}_{\CP|\test} \geq \sum_{m\leq \Ncons} \zeta_{m|\test} \probstJ_{\tilde{\xi}_m}[J_m],
        \end{aligned}
    \end{equation}
    where we now explicitly highlighted that each \(\probstJ_{\tilde{\xi}_m}\) depends on the state \(\ket{\tilde{\xi}^t_{|m}}\). However, note that \(g_{\alpha}^{\Nent}\) is already defined in terms of the states \(\ket{\xi^g_{|m}}\) with a cutoff.

    Finally, replacing the states without a cutoff \(\ket{\widetilde{\xi}^t_{|m}}\) with \(\ket{\xi^t_{|m}}\) including the cutoff in the constraints follows by \cref{lem:Bounds on Statistics under Approx Diag}, which concludes the proof for \(\hUpQKD\).
    
    Next, we show the claim for the $\frenyiSandUp_\alpha$-normalization constants. Since here we assume independent imperfections the source maps \(\XiPhase^i\) are different for each \(j\). Again since we only had continuity bounds available for \(\renyiSandUp_{\alpha}\), we start from \(\kapup_i\) which includes the source map \(\XiPhase^i\) for each round \(i\).
    
    As before, we first relate the block-tagged states \(\ket{\widetilde{\xi}}\) to the states without tagging via the source map \(\Tr_{\bar{A}}\). Furthermore, the model state is related to the phase imperfect state via the source map \(\XiPhase^i\). Applying \cref{thrm:Source maps QES} we bound the $f$-weighted entropy by including the combined source map \(\XiPhase^i \circ \Tr_{\bar{A}}\) in Eve's attack, which lets us write
    \begin{align}
    \begin{aligned}
        \kapup_i \geq &\inf_{\substack{J \in \mathrm{Pos}(A'\bar{A}B), \\ \bsym{\nu} \in \mathbb{P}(\Ct) }} \Bigg( \frac{\alpha}{1-\alpha} \log \Big( \gamma \sum_{c\in \Ct} \bsym{\nu}(\cP)2^{\frac{\alpha-1}{\alpha}  f(\cP)} \\
        &\qquad + \left(1-\gamma\right)2^{-\frac{\alpha-1}{\alpha} \left( \renyiSandUp_{\alpha}(S|\tilde{T}E)_{\widetilde{\nu}_{|\gen}} - f(\gen)  \right)} \Big) \Bigg) \\
        \quad \textrm{s.t. } &\Tr_B[J] = \idop_{A'\bar{A}}, \\
        &\bsym{\nu} = \probstJ_{\tilde{\xi}}[J],
        \end{aligned}
    \end{align}
    Next, we make use of the monotonicity of the ``log-mean-exponential'' (\cref{lem:Monotonicity of Log-Mean-Exp} in \cref{app:Monotonicity of Log-Mean-Exp}) to obtain a lower bound by replacing the factors of \(\frac{\alpha}{1-\alpha}\) with \(\frac{1}{1-\alpha}\) (noting that since $\alpha>1$, we have $\frac{1-\alpha}{\alpha} \geq 1-\alpha$). This leaves us with
    \begin{align}
    \begin{aligned}
        \kapup_i \geq &\inf_{\substack{J \in \mathrm{Pos}(A'\bar{A}B), \\ \bsym{\nu} \in \mathbb{P}(\Ct) }} \Bigg( \frac{1}{1-\alpha} \log \Big( \gamma \sum_{c\in \Ct} \bsym{\nu}(\cP)2^{\left(\alpha-1\right) f(\cP)} \\
        &\qquad + \left(1-\gamma\right)2^{-\left(\alpha-1\right) \left( \renyiSandUp_{\alpha}(S|\tilde{T}E)_{\widetilde{\nu}_{|\gen}} - f(\gen)  \right)} \Big) \Bigg) \\
        \quad \textrm{s.t. } &\Tr_B[J] = \idop_{A'\bar{A}}, \\
        &\bsym{\nu} = \probstJ_{\tilde{\xi}}[J],
        \end{aligned}
    \end{align}
    Furthermore, we apply the same continuity bound as in \cref{eq:Continuity lower bnd key rate states} to bound \( \renyiSandUp_\alpha(S|\tilde{T}E)_{\widetilde{\nu}}\), which gives
    \begin{align}
    \begin{aligned}
        \kapup_i \geq &\inf_{\substack{J \in \mathrm{Pos}(A'\bar{A}B), \\ \bsym{\nu} \in \mathbb{P}(\Ct) }} \Bigg( \frac{1}{1-\alpha} \log \Big( \gamma \sum_{c\in \Ct} \bsym{\nu}(\cP)2^{(\alpha - 1) f(\cP)} \\
        &\left(1-\gamma\right)2^{(\alpha-1)\left(\Delta(\epsFull)+ f(\gen)\right)} \\ 
        &\qquad \cdot 2^{-(\alpha-1)\renyiSandDown_{\alpha}(S|\tilde{T}E)_{\nu_{|\gen}} } \Big) \Bigg) \\
        \quad \textrm{s.t. } &\Tr_B[J] = \idop_{A'\bar{A}}, \\
        &\bsym{\nu} = \probstJ_{\tilde{\xi}}[J].
        \end{aligned}
    \end{align}
    Again, the block-tagged states satisfy the assumptions \ref{Decoy assump block-diag}--\ref{Decoy assump QND Eve}. Hence, as for \(\hUpQKD\) one can proceed by applying the simplifications of \cref{sec:Block-Diag states} and with \cref{lem:Bounds on Statistics under Approx Diag} the claim follows.
\end{proof}

\section{Monotonicity of log-mean-exponential}\label{app:Monotonicity of Log-Mean-Exp}

\begin{lemma}[Monotonicity of log-mean-exponential]\label{lem:Monotonicity of Log-Mean-Exp}
Consider any (normalized) probability distribution $\mbf{p}$ on some finite alphabet $\alphCP$, and any tuple $\mbf{v}\in\mathbb{R}^{\abs{\alphCP}}$. Then the following ``log-mean-exponential'' quantity is nondecreasing with respect to $\zeta \in[-\infty,\infty]$:
\begin{align}\label{eq:lme}
\frac{1}{\zeta} \log \left(\sum_{\cP} p(\cP) 2^{\zeta v(\cP)} \right),
\end{align}
where the values at $\zeta \in \{-\infty,0,\infty \}$ are defined via the corresponding limits, which respectively have the following values:
\begin{align}\label{eq:lmelimits}
\min_{\cP \in \supp(\mbf{p})} v(\cP) , \qquad
\sum_{\cP} p(\cP) v(\cP), \qquad
\max_{\cP \in \supp(\mbf{p})} v(\cP), 
\end{align}
where $\supp(\mbf{p})$ denotes the support of the distribution $\mbf{p}$.
\end{lemma}
\begin{proof}
We first show the limits indeed exist and have the claimed values. 
For the $\zeta\to0$ limit, first we reparametrize $\zeta=\hat{\zeta}\log(e)$ so the logarithm and exponential can be expressed in base $e$ for ease of calculation, 
then apply l'H\^{o}pital's rule (noting that the conditions of that rule are satisfied) to find the limit to be
\begin{align}
\lim_{\hat{\zeta}\to0} \frac{\sum_{\cP} p(\cP) v(\cP) e^{\hat{\zeta} v(\cP)}}{\sum_{\cP} p(\cP) e^{\hat{\zeta} v(\cP)}}
=\sum_{\cP} p(\cP) v(\cP)
,
\end{align}
yielding the claimed result.
As for the other limits, observe that for any specific $\cP^\star \in \alphCP$, we can rewrite \cref{eq:lme} as
\begin{align}
\frac{1}{\zeta} \log \left(\sum_{\cP} p(\cP) 2^{\zeta (v(\cP) - v(\cP^\star))} \right) + v(\cP^\star) .
\end{align}
By choosing $\cP^\star$ to be any value attaining the minimum (resp.~maximum) in \cref{eq:lmelimits}, we obtain the claimed result for $\zeta\to-\infty$ (resp.~$\zeta\to\infty$), noting that the summation term in the above expression converges to a value that is finite and nonzero (the latter is ensured by noting that the $\cP=\cP^\star$ term in the sum has exponent zero and $p(\cP^\star)>0$).

It then suffices to show \cref{eq:lme} is nondecreasing on the intervals $\zeta\in(-\infty,0)$ and $\zeta\in(0,\infty)$, since the values at $\zeta \in \{-\infty,0,\infty \}$ are by definition equal to the respective (well-defined) limits, which quite straightforwardly implies the desired claim that the expression is nondecreasing over all $\zeta\in[-\infty,\infty]$.
To do so, we follow a technique developed in~\cite[Lemma~4.2]{arqand_generalized_2024} (based on earlier ideas from~\cite{dupuis_entropy_2020}), as follows. 

Consider any $M>0$ such that $M/2+v(\cP)>0$ for all $\cP$; note that this ensures $M+v(\cP)>M/2>0$. Let $A$ be a register with dimension $\max_{\cP}\ceil{2^{M+v(\cP)}}$, and let $\CP$ be a classical register with alphabet $\alphCP$. We construct a state $\rho_{A\CP}$ classical on $\CP$, as follows: the distribution on $\CP$ is $\mbf{p}$, and for each $\cP\in\alphCP$, the conditional state $\rho_{A|\cP}$ is a classical state corresponding to a uniform distribution with support size $\ceil{2^{M+v(\cP)}}$, sometimes called a \emph{flat state}. Note that for flat states, \emph{all} their \Renyi entropies are equal (to the logarithm of the support size). Hence we have, for all $\alpha\in[0,\infty]$,
\begin{align}
\renyiSandDown_\alpha(A)_{\rho_{|\cP}} = \renyiSandUp_\alpha(A)_{\rho_{|\cP}} = \log \ceil{2^{M+v(\cP)}}\nonumber\\
\in \left[M+v(\cP), M+v(\cP) + 2^{-M/2}\log e\right],\label{eq:flatstatebound}
\end{align}
where the upper bound in the interval holds because of the condition $M+v(\cP)>M/2>0$, as shown in the proof of~\cite[Lemma~4.2]{arqand_generalized_2024}.

With this, we consider any $\zeta,\zeta' \in (-\infty,0)$ such that $\zeta<\zeta'$, and set 
\begin{equation}
\alpha=1-\zeta, \quad \alpha'=1-\zeta', \label{eq:zetatoalpha}
\end{equation}
so we have $\alpha,\alpha' \in (1,\infty)$ with $\alpha>\alpha'$. Then we can write
\begin{align}
&\frac{1}{\zeta} \log \left(\sum_{\cP} p(\cP) 2^{\zeta v(\cP)} \right) \nonumber\\
\leq&\frac{1}{\zeta} \log \left(\sum_{\cP} p(\cP) 2^{\zeta \left(\renyiSandDown_\alpha(A)_{\rho_{|\cP}}-M\right)} \right) \nonumber\\
=&\frac{1}{1-\alpha} \log \left(\sum_{\cP} p(\cP) 2^{(1-\alpha) \renyiSandDown_\alpha(A)_{\rho_{|\cP}}} \right) -M \nonumber\\
=& \renyiSandDown_\alpha(A|\CP)_\rho - M \nonumber\\
\leq& \renyiSandDown_{\alpha'}(A|\CP)_\rho - M \nonumber\\
% =& \frac{1}{1-\alpha'} \log \left(\sum_{\cP} p(\cP) 2^{(1-\alpha') \renyiSandDown_{\alpha'}(A)_{\rho_{|\cP}}} \right) - M \nonumber\\
\leq& \frac{1}{\zeta'} \log \left(\sum_{\cP} p(\cP) 2^{\zeta' v(\cP)} \right) +2^{-M/2}\log e ,
\end{align}
where the second line holds by \cref{eq:flatstatebound} together with the observation that the expression is nondecreasing with respect to all the $v(\cP)$ terms (for any $\zeta\in\mathbb{R}$), the third line holds by substituting \cref{eq:zetatoalpha} and also extracting the $M$ term, the fourth line holds by \cite[Sec. III.B.2 and Prop. 9]{muller-lennert_quantum_2013} (or \cite[Prop. 5.1]{tomamichel_quantum_2016}), the fifth line is monotonicity of $\renyiSandDown_\alpha$ in $\alpha$, and the last line holds by repeating the reasoning in the previous steps in reverse order.

Crucially, recall that the start of this analysis, we picked \emph{any} arbitrary $M$ satisfying the initially stated criteria $M>0$ and $M/2+v(\cP)>0$ for all $\cP$; in particular, this means the above bound holds for arbitrarily large $M$. Thus by taking $M\to\infty$, the above bound implies 
\begin{align}
\frac{1}{\zeta} \log \left(\sum_{\cP} p(\cP) 2^{\zeta v(\cP)} \right) \leq \frac{1}{\zeta'} \log \left(\sum_{\cP} p(\cP) 2^{\zeta' v(\cP)} \right) ,
\end{align}
as desired. 

Next, we consider any $\zeta,\zeta' \in (0,\infty)$ such that $\zeta<\zeta'$, and instead set 
\begin{equation}
\alpha = \frac{1}{1+\zeta} , \quad \alpha' = \frac{1}{1+\zeta'} ,
\end{equation}
so we have $\alpha,\alpha' \in (1,\infty)$ with $\alpha>\alpha'$ (observing that the above expressions are nonincreasing in $\zeta,\zeta'$). Note that this gives $\zeta = \frac{1-\alpha}{\alpha}$ and $\zeta' = \frac{1-\alpha'}{\alpha'}$. Hence by similar reasoning to above, we can write
\begin{align}
&\frac{1}{\zeta} \log \left(\sum_{\cP} p(\cP) 2^{\zeta v(\cP)} \right) \nonumber\\
\leq&\frac{\alpha}{1-\alpha} \log \left(\sum_{\cP} p(\cP) 2^{\frac{1-\alpha}{\alpha} \renyiSandDown_\alpha(A)_{\rho_{|\cP}}} \right) -M \nonumber\\
=& \renyiSandUp_\alpha(A|\CP)_\rho - M \nonumber\\
\leq& \renyiSandUp_{\alpha'}(A|\CP)_\rho - M \nonumber\\
\leq& \frac{1}{\zeta'} \log \left(\sum_{\cP} p(\cP) 2^{\zeta' v(\cP)} \right) +2^{-M/2}\log e ,
\end{align}
which again yields the desired result by taking the $M\to\infty$ limit.
\end{proof}

The above proof essentially proceeded by relating the quantity to \Renyi entropies, allowing us to ``translate'' between monotonicity in $\zeta$ and monotonicity in $\alpha$. It seems plausible that a more ``direct'' proof of monotonicity in $\zeta$ might be possible by instead performing an analysis analogous to the proof that the \Renyi divergences (and thus conditional entropies) are monotone in $\alpha$, but we leave this possibility for future work. 

\section{Proofs of Convexity Statements}\label{app:Proof of Convexity}
In this appendix, we will prove the joint convexity of the optimization problems for both the single-round quantity \(\hQKD\) and the \(\frenyiSandDown_\alpha\)-normalization constant \(\kappa\). Since our numerical algorithm requires perturbations, we will construct a perturbed protocol map \(\GMapDelta\) and show the joint convexity for the perturbed optimization problems.

For clarity of notation, in all of this section, wherever necessary, we use the following identification of registers to the main text,
\begin{align}
    E \mapsto \tilde{T}E \quad \text{and} \quad Q \mapsto S_QIABXY.
\end{align}

Additionally, for completeness, similar to \cref{thrm:Petz Renyi simplifications}, we define a generic pinching map $\ZMap \in \CPTP(Q)$ and its Stinespring dilation.

\begin{definition}\label{def:zMapIsometry}
   Let $\{Z_i\}_{i=1\dots n} \subset \Pos(Q)$ be a set of projection operators on $Q$ satisfying $\sum_{i=1}^n Z_i = \idop_Q$ and $Z_iZ_j = \delta_{i,j}Z_i$. We define the isometry $V_{\ZMap} \in \Isom(Q,SQ)$ and the Z-map, $\ZMap \in \CPTP(Q)$, as
    \begin{gather}
        V_{\ZMap} = \sum_z \ket{i}_S \otimes Z_{i}, \\
        \mathcal{V}_{\ZMap}(\rho) = V_{\ZMap} \rho V_{\ZMap}^\dagger, \quad
        \ZMap(\rho) = \Tr_S \circ \mathcal{V}_{\ZMap} (\rho).
    \end{gather}
\end{definition}

We will prove the joint convexity of both optimization problems of the single-round quantity \(\hQKD\) and the \(\frenyiSandDown_\alpha\)-normalization constant \(\kappa\) by relating them to other convex functions through convexity-preserving transformations. One of our methods is to prepend convex (concave) functions with affine maps, which is well known to preserve convexity (concavity). Therefore, for reference, we restate the following lemma from Ref.~\cite{boyd_convex_2004} for the composition of convex functions and affine maps.

\begin{lemma} [Composition with affine functions preserves convexity; Section~3.2.2 in~\cite{boyd_convex_2004}]
\label{lem:ConvexCompAffine}
    Let $V_1$ and $V_2$ be vector spaces over the field $\R$, $C_1 \subseteq V_1$, $C_2 \subseteq V_2$ be convex sets, $f:C_1 \rightarrow \R$ a convex function on $C_1$, and $g: C_2 \rightarrow C_1$ an affine function on $C_2$, then the function $h = f \circ g: C_2 \rightarrow \R$ is convex on $C_2$.
\end{lemma}

\subsection{Convexity of Quantities related to Petz conditional \Renyi entropy}
As an initial step, we prove the convexity/concavity of several quantities related to the Petz conditional \Renyi entropy.

In \cref{thrm:Petz Renyi simplifications}, we used a dual relation to relate $\renyiSandDown_{\alpha}$ to $\renyiPetzUp_{\frac{1}{\alpha}}$ and the following expression to simplify $\renyiPetzUp_{\alpha}(S|Q)$,
\begin{equation}
    \renyiPetzUp_{\alpha}(S|Q)_\sigma = \frac{\alpha}{1-\alpha} \log \Tr\left[\ZMap\left(\rho_{Q}^{\alpha}\right)^{\frac{1}{\alpha}}\right],
\end{equation}
where the channel $\ZMap$ is again a pinching channel, as defined above, and \(\sigma_{SQE}\) is created from $\rho_{QE}$ as in \cref{thrm:Petz Renyi simplifications} by
\begin{equation}
    \sigma_{SQE} \defvar V_{\ZMap} \rho_{QE} V_{\ZMap}^{\dagger}.
\end{equation}
Thus, for simplicity, we define the quantity \(\Qopt_{\alpha}\) as
\begin{equation}\label{def:Qopt}
    \Qopt_{\alpha}(\rho_Q) \defvar \Tr\left[\ZMap\left(\rho_{Q}^{\alpha}\right)^{\frac{1}{\alpha}} \right], 
\end{equation}
for all $\rho \in \dop{=}(Q)$ and \(\alpha \in [0,2]\). With this definition, we can write
\begin{equation}
    \renyiPetzUp_{\alpha}(S|Q)_\sigma = \frac{\alpha}{1-\alpha} \log \left(\Qopt_\alpha(\rho_Q) \right).
\end{equation}

\begin{theorem}\label{thrm:Qopt concave}
    The function \(\rho \mapsto \Qopt_\alpha\left(\rho\right)\) is concave for all $\rho \in \dop{\leq}(Q)$ and $\alpha \in [0,1]$.
\end{theorem}
\begin{proof}
    From \cite[Lemma 5]{frank_monotonicity_2013} (originally shown by Epstein~\cite{epstein_remarks_1973} for $t \in [0,1]$), it follows that the function
    \begin{equation}
        \Tr[\left(K X^{t}K^{\dagger}\right)^\frac{1}{t}],
    \end{equation}
    is concave in $X \in \Pos(Q)$ for $t \in [-1,1], t\neq 0$ given any fixed operator $K$. Thus, we will define an operator $K$ such that we can rewrite $\Qopt_{\alpha}$ in the form above. 
    First, note that we can write 
    \begin{equation}
        \ZMap(\rho_Q) = \Tr_S[V_{\ZMap} \rho_Q V_{\ZMap}^{\dagger}],
    \end{equation}
    where $V_{\ZMap}$ is given as in \cref{def:zMapIsometry}.
    Next, let $S'$ be a Hilbert space of the same dimension as $S$ and let \(\ket{\Phi^+}_{SS'}\) be the unnormalized maximally entangled state on $SS'$. This allows us to rewrite the partial trace of any trace-class operator \(M_{SQ} \in \Pos(SQ)\) by
    \begin{equation}
        \Tr_S[M_{SQ}] = \bra{\Phi^+}_{SS'} \left(M_{SQ} \otimes \idop_{S'} \right) \ket{\Phi^+}_{SS'}.
    \end{equation}
    Specifically we find
    \begin{equation}
        \ZMap(\rho_Q) = \bra{\Phi^+}_{SS'} \left(V_{\ZMap} \rho_Q V_{\ZMap}^{\dagger} \otimes \idop_{S'} \right)\ket{\Phi^+}_{SS'}.
    \end{equation}
    Hence, let us define $K$ as
    \begin{equation}
        K\defvar \left(\bra{\Phi^+}_{SS'} \otimes \idop_Q\right) \left(V_{\ZMap} \otimes \idop_{S'} \right),
    \end{equation}
    which allows us to rewrite $\Qopt_{\alpha}(\rho_Q)$ as
    \begin{equation}
        \Qopt_{\alpha}(\rho_Q) =  \Tr\left[ \left(K \left( \rho_Q \otimes \idop_{S'} \right)^{\alpha} K^{\dagger}\right)^{\frac{1}{\alpha}}\right].
    \end{equation}
    Using this construction of $K$ the statement follows immediately from \cite[Lemma 5]{frank_monotonicity_2013}.
\end{proof}

\subsection{Perturbed objective function}
We conclude this preliminary part with the definition of the objective function \(g_{\alpha}\), similar to \cref{eq:defn galpha}, but in terms of the state \(\rho\) instead of the channel, which we will indicate by \(\renObj_{\alpha}\). 

\begin{definition}\label{def:renObj}
    Let $\ZMap$ be the pinching channel of \cref{cor:Fixed-length security lower bnd halpha} and $\GMap$ be the CPTP protocol map as in \cref{def:QKD channel}. We define $\renObj_\alpha: \dop{=}(AB) \rightarrow \R$ as
    \begin{equation}\label{eq:renObj}
    \begin{split}
        \renObj_\alpha (\rho_{AB}) &= \renyiSandDown_{\alpha}(S|\tilde{T}E)_{\EATchannQKD(\rho_{ABE})}, \\
        &= \frac{-1}{\alpha-1} \log \Tr\left[ \left(\ZMap\left(\GMap(\rho_{AB})^{\frac{1}{\alpha}}\right)\right)^{\alpha}\right],
    \end{split}
    \end{equation}
    where we made the negativity of the prefactor in front of the logarithm explicit.
\end{definition}
For simplicity of the later convexity proofs, we did not use the final simplification of the expression to CPTNI maps as in \cref{cor:Fixed-length security lower bnd halpha}. Importantly, this definition allows us to rewrite \(g_{\alpha}\) from \cref{eq:defn galpha} as 
\begin{equation}
    g_\alpha(J) = \bar{g}_\alpha \circ \chi_g(J).
\end{equation}

However, for the existence of gradients everywhere on the domain, we require a perturbed version of \(\renObj_\alpha\), which adds an affine perturbation to \(\GMap\). Thus, we first define the perturbed protocol map \(\GMapDelta\). The exact choice of perturbation we introduce here will be motivated in \cref{subsubsec:Choices for Affine Perturbation}.

\begin{definition}\label{def:renObj perturbed}
    Let $\GMap$ and $\hat{\GMap}$ be the CPTP and CPTNI protocol maps as in \cref{def:QKD channel}. We define the perturbed map \(\GMapDelta\) for \(\eps_1, \eps_2 \in [0,1)\) with \(\delta = \eps_1 +\eps_2 - \eps_1\eps_2\) ($\delta \in[0,1)$) as
    \begin{equation}
        \GMapDelta(\rho) \defvar \left(1-\delta\right) \GMap(\rho) + \delta \tau(\rho),
    \end{equation}
    where
    \begin{equation}\label{eq:tau perturbation}
    \begin{split}
        \tau(\rho) = &\left(\frac{p \idop_{\hat{S}_Q} }{\delta \dim(\hat{S}_Q)} + \left( \Tr[\rho] - \frac{p}{\delta} \right) \ketbra{\perp}{\perp}\right) \\ &\otimes  \frac{\idop_{ABXYI}}{\dim(ABXYI)},
    \end{split}
    \end{equation}
    and 
    \begin{equation}\label{eq:p in tau}
        p = \eps_1 \Tr[\hat{\GMap}(\rho)] \left(1-\eps_2\right) + \eps_2\Tr[\rho].
    \end{equation}
    Then, the perturbed objective \(\renObj_\alpha[\delta]\) in terms of the state is given by
    \begin{equation}\label{eq:renObj perturbed}
        \begin{split}
            \renObj_\alpha[\delta] (\rho_{AB}) \defvar \frac{-1}{\alpha-1} \log\left[ \Qopt_{\frac{1}{\alpha}} \circ \GMapDelta (\rho_{AB}) \right],
        \end{split}
    \end{equation}
    and the perturbed objective in terms of Eve's channel is 
    \begin{equation}
        \begin{split}
            g_\alpha[\delta] (J) \defvar \frac{-1}{\alpha-1} \log\left[ \Qopt_{\frac{1}{\alpha}} \circ \GMapDelta \circ \chi_g (J) \right].
        \end{split}
    \end{equation}
\end{definition}
We note that for \(\delta= 0\) in both cases we recover the unperturbed objective functions and most importantly the perturbed map \(\GMapDelta\) is still a linear function of \(\rho\).

Next, in preparation of the joint convexity proof of \(\hQKD\) in \cref{eq:halpha final} (and its perturbed version), we first show that the term inside the logarithm, i.e. $\Qopt_{\frac{1}{\alpha}} \circ \GMapDelta$ is still concave for \(\alpha >1\).

\begin{cor}\label{cor:perturbed Qopt with G concave}
    The function $\rho_{AB} \mapsto \Qopt_{\frac{1}{\alpha}} \circ \GMapDelta(\rho_{AB})$ is concave and $\renObj_{\alpha}[\delta]$ is convex for all $\rho \in \dop{=}(AB)$ and $\alpha \in [1,\infty)$. 
\end{cor}
\begin{proof}
    The perturbed protocol map $\GMapDelta$ applies an affine transformation to \(\rho\). Furthermore, by \cref{thrm:Qopt concave}, \(\Qbar_{\frac{1}{\alpha}}\) is concave for \(\alpha \in [1,\infty)\). Thus, since \(\GMapDelta\) is an affine map by \cref{lem:ConvexCompAffine}, $\rho_{AB} \mapsto \Qopt_{\frac{1}{\alpha}} \circ \GMapDelta(\rho_{AB})$ is concave.

    Additionally, since the negative logarithm, \(-\log(x)\), is convex and non-increasing, by \cite[Section 3.2.4]{boyd_convex_2004} the function $\renObj_{\alpha}[\delta]$ is convex.
\end{proof}

The perturbed objective is a lower bound on the unperturbed version and satisfies the following property.
\begin{lemma}\label{lem:perturbed g lower bound}
    For all $\rho_{AB} \in \dop{=}(AB)$, $\alpha >1$ and \(\delta \in [0,1)\), $\renObj_{\alpha}[\delta]$ satisfies
    \begin{equation}
        \frac{\renObj_{\alpha}[\delta](\rho_{AB})}{1-\delta} \leq \renObj_{\alpha}(\rho_{AB})
    \end{equation}
\end{lemma}
\begin{proof}
    First, we note that by \cref{lem:renObjNoG = 0}, \(\log[\Qopt_{\frac{1}{\alpha}}(\tau(\rho))] = 0\) for all \(\rho \in \dop{=}(AB)\).
    Next, using this and the fact that $\renObj_\alpha$ is a convex function, we find
    \begin{equation}
    \begin{split}
        \renObj_\alpha[\delta](\rho) & \leq (1-\delta) \renObj_\alpha(\rho) - \frac{ \delta}{\alpha -1} \log[\Qopt_{\frac{1}{\alpha}}(\tau(\rho))], \\
        & = (1-\delta) \renObj_\alpha(\rho).
    \end{split}
    \end{equation}
    Then, rearranging this equation yields the bound.
\end{proof}

Next, we show another property which is required for the joint convexity of the \(\frenyiSandDown_\alpha\)-normalization constant \(\kappa\).

\begin{cor}\label{cor:exp renyiObj is concave}
    The function \(\rho \mapsto 2^{-(\alpha-1)\frac{\renObj_\alpha[\delta](\rho)}{1-\delta}}\) is concave for all \(\rho \in \dop{=}(AB) \), $\alpha >1$ and $\delta \in [0,1)$.
\end{cor}
\begin{proof}
    We start by rewriting $2^{-(\alpha-1)\frac{\renObj_\alpha[\delta](\rho)}{1-\delta}}$ and a simple calculation yields,
    \begin{align}
        2^{-(\alpha-1)\frac{\renObj_\alpha[\delta](\rho)}{1-\delta}} &= \frac{\Tr[\left( \ZMap(\GMapDelta(\rho)^{\frac{1}{\alpha}}) \right)^{\alpha}]}{1-\delta} \\
        &= \frac{\Qopt_{\frac{1}{\alpha}} \circ \GMapDelta (\rho_{AB})}{1-\delta} .
    \end{align}
    Thus, using \cref{cor:perturbed Qopt with G concave} we can conclude that \(\rho \mapsto 2^{-(\alpha-1)\frac{\renObj_\alpha[\delta](\rho)}{1-\delta}}\) is concave for all \(\rho \in \dop{=}(AB) \), \(\alpha >1\) and $\delta \in [0,1)$.
\end{proof}

Now, we have all tools to prove the joint convexity of both the optimization problem of \(\hQKD\) and the \(\frenyiSandDown_\alpha\)-normalization constant \(\kappa\).

\subsection{Joint convexity of \texorpdfstring{$\hQKD$}{h}}
Before we state the theorem, let us introduce some more notation regarding the relative entropy for convenience. For two scalars $x,y\in \R_\geq0$, we denote $x\log(x/y)$ by
\begin{equation}\label{eq:def KL divergence scalars}
    D\rel{x}{y} \defvar x\log(x/y),
\end{equation}
such that for two vectors $\mbf{u},\mbf{v}\in \R^n_{\geq0}$ we can write
\begin{equation}
    D\rel{\mbf{u}}{\mbf{v}} = \sum_{i=1}^n D\rel{u_i}{v_i}.
\end{equation}

\begin{theorem}[Joint Convexity of perturbed $\hQKD$ (State)]\label{thrm:joint convexity perturbed halpha}
Let the feasible set of \(\hQKD\) of \cref{eq:defn_hQKD} be defined as
\begin{equation}
    \mathcal{D} :=\begin{aligned}[t]
        \{&\mbf{q} \in \Sacc, \rho \in \dop{=}(AB), \bsym{\nu}_{\CP} \in \mathbb{P}(\alphCP)| \\ &\Tr_B[\rho_{AB}] = \tau_A, \bsym{\nu}_{\CP} = (\gamma \bsym{\nu}_{\CP|\test}, 1-\gamma)^T, \\
        &\bsym{\nu}_{\CP|\test} = \probst[\rho_{|\test}]\},
    \end{aligned} 
\end{equation}
and for all $\alpha>1$ and $\delta \in [0,1)$ define the function
\begin{equation}\label{eq:halpha perturbed}
    \begin{aligned}[t]
        \bar{h}(\mbf{q}, \rho, \bsym{\nu}_{\CP}) \defvar &\frac{\alpha}{\alpha - 1}D\rel{\mbf{q}}{\bsym{\nu}_{\CP}} \\ 
        &+ q(\gen) \frac{\renObj_{\alpha}[\delta](\rho_{AB|\gen})}{1-\delta}.
        \end{aligned}
\end{equation}
Then, \(\bar{h}\) is jointly convex in $(\mbf{q}, \rho, \bsym{\nu}_{\CP})$ over \(\mathcal{D}\) for all $\alpha>\frac{1}{1-\delta}>1$ and $\delta \in[0,1)$. Furthermore, \(\bar{h}\) satisfies
\begin{align}
    \hQKD \geq \inf_{(\mbf{q}, \rho, \bsym{\nu}_{\CP}) \in \mathcal{D}} \bar{h}(\mbf{q}, \rho, \bsym{\nu}_{\CP}),
\end{align}
for all $\alpha>1$ and $\delta \in[0,1)$, with equality for \(\delta =0\).
\end{theorem}
\begin{proof}
    The fact that $\bar{h}$ is a lower bound on $\hQKD$ for all $\alpha>1$ and $\delta \in [0,1)$ follows from \cref{lem:perturbed g lower bound}.
    
    The proof for convexity will proceed by rewriting $\bar{h}$ into a sum of functions that are known to be jointly convex. We start by rephrasing $\rho_{|\gen}$ and following \cref{def:QKD channel}, we find
    \begin{align}
        \rho_{AB|\gen} &= \frac{\rho_{AB\wedge \gen}}{\Tr[\rho_{AB \wedge \gen}]}, \\
        \rho_{AB\wedge \gen} &= \sqrt{\Pi^{\gen}} \rho_{AB} \sqrt{\Pi^{\gen}}.
    \end{align}
    Furthermore, \(\bsym{\nu}_{\CP}\) satisfies in the generation component
    \begin{equation}
        \nu_{\CP}(\gen) = 1-\gamma = \Tr[\Pi^{\gen} \rho_{AB}].
    \end{equation}
    Inserting this into \(\bar{h}\) yields
    \begin{align}\label{eq:proof hbar inserted rhoGen}
        \bar{h}(\mbf{q}, \rho, \bsym{\nu}_{\CP}) &=
        \begin{aligned}[t]
            &\frac{\alpha}{\alpha - 1}D\rel{\mbf{q}}{\bsym{\nu}_{\CP}} \\ 
            &+ q(\gen) \frac{\renObj_{\alpha}[\delta]\left(\frac{\rho_{AB\wedge \gen}}{1-\gamma}\right)}{1-\delta}            
        \end{aligned}
    \end{align}
    The main step of this proof will be to rearrange the terms of the relative entropy. Thus, we expand the definition of \(\renObj_{\alpha}[\delta]\) to find
    \begin{align}\label{eq:proof hbar inserted gbar}
    \begin{aligned}[t]
        &\bar{h}(\mbf{q}, \rho, \bsym{\nu}_{\CP}) \\ = 
            & \frac{\alpha}{\alpha - 1}D\rel{\mbf{q}}{\bsym{\nu}_{\CP}} \\
            &- \frac{q(\gen)}{(\alpha-1)(1-\delta)} \log\left[ \Qopt_{\frac{1}{\alpha}} \circ \GMapDelta \left(\frac{\rho_{AB\wedge\gen}}{1-\gamma}\right) \right].
        \end{aligned}
    \end{align}
    
    Next, by using our definition of the relative entropy for scalars shown in \cref{eq:def KL divergence scalars}, we separate the test and generation components of the relative entropy between $\mbf{q}$ and $\bsym{\nu}_{\CP}$,
    \begin{equation}\label{eq:proof hbar expanded D}
        D\rel{\mbf{q}}{\bsym{\nu}_{\CP}} = \begin{aligned}[t]
            &D\rel{q(\gen)}{\nu_{\CP}(\gen)} \\ 
            &+ D\rel{\mbf{q}_{\wedge \test}}{\bsym{\nu}_{\CP\wedge\test}}.
        \end{aligned}
    \end{equation}
    Here $\mbf{q}_{\wedge\test}$ denotes the subnormalized distribution containing only the test components of \(\mbf{q}\) (and similarly for $\bsym{\nu}_{\CP}$), so we have e.g.~$\sum_{\cP} q_{\wedge\test}(\cP) + q(\gen) =1$.
    
    Finally, we note the following identities involving several combinations of \(\alpha\) and \(\delta\)
    \begin{align}
        &\frac{\alpha}{\alpha-1} = \frac{1}{\alpha-1}+1, \\
        &\frac{1}{\alpha-1} = \frac{1}{(\alpha-1)(1-\delta)} - \frac{\delta}{(\alpha-1)(1-\delta)}, \\
        &\frac{\alpha(1-\delta) -1}{(\alpha-1)(1-\delta)} = 1 - \frac{\delta}{(\alpha-1)(1-\delta)} , \\
        &-\frac{\alpha(1-\delta) -1}{(\alpha-1)(1-\delta)} = \frac1{(\alpha-1)(1-\delta)} -\frac{\alpha}{\alpha-1} .
    \end{align}

    A simple calculation exploiting the properties of the relative entropy while inserting the identities above together with \cref{eq:proof hbar expanded D}, lets us write
    \begin{equation}        
    \begin{aligned}[t]
        &\bar{h}(\mbf{q}, \rho, \bsym{\nu}_{\CP})\\ 
        = &\frac{D\rel{q(\gen)}{\Qopt_{\frac{1}{\alpha}} \circ \GMapDelta \left(\rho_{AB\wedge\gen}\right)}}{(\alpha - 1)(1-\delta)} \\ 
         &+ \frac{\alpha}{\alpha-1} D\rel{\mbf{q}_{\wedge \test}}{\bsym{\nu}_{\CP\wedge\test}} \\
         &+ \frac{\alpha(1-\delta) -1}{(\alpha-1)(1-\delta)} D\rel{q(\gen)}{\nu_{\CP}(\gen)}. 
    \end{aligned}
    \end{equation}
    This reformulation of $\bar{h}$ allows us to derive the claimed convexity.
    
    First, we note that the relative entropy is jointly convex in its arguments, see e.g. \cite[Example 3.19]{boyd_convex_2004}, and that a sum of jointly convex functions remains jointly convex. Thus, we will argue that each term is jointly convex. The second term is obviously jointly convex as it is a relative entropy between $\mbf{q}$ and $\bsym{\nu}_{\CP}$ omitting their \(\gen\) component.

    By the same reasoning the third term, the relative entropy between the two scalars $q(\gen), \nu_{\CP}(\gen)$ (in the sense of \cref{eq:def KL divergence scalars}) is jointly convex as long as 
    \begin{equation}
        0 \leq \alpha(1-\delta) -1 \Leftrightarrow \frac{1}{1-\delta} \leq \alpha.
    \end{equation}

    To see that the first term is jointly convex, consider 
    \begin{equation}
        \Qopt_{\frac{1}{\alpha}} \circ \GMapDelta \left(\rho_{AB\wedge\gen}\right) = \Qopt_{\frac{1}{\alpha}} \circ \GMapDelta \left( \sqrt{\Pi^{\gen}} \rho_{AB} \sqrt{\Pi^{\gen}} \right).
    \end{equation}
    This is concave due to \cref{cor:perturbed Qopt with G concave} and the composition with the affine transformation $\sqrt{\Pi^{\gen}} (\cdot) \sqrt{\Pi^{\gen}}$. Additionally, the relative entropy between two scalars (in the sense of \cref{eq:def KL divergence scalars}) is nonincreasing in the second argument.
    Thus, for any $ \mbf{q} \in \Sacc$, $\lambda \in [0,1]$ and $\rho_1, \rho_2 \in \dop{=}(AB)$, we find
    \begin{align}
        &D\rel{q(\gen)}{\Qopt_{\frac{1}{\alpha}} \circ \GMapDelta \left(\lambda \rho_{1\wedge\gen} + (1-\lambda) \rho_{2\wedge\gen} \right)} \\
        \leq & \begin{aligned}[t]
            D\Big(q(\gen) \Big\| &\lambda \Qopt_{\frac{1}{\alpha}} \circ \GMapDelta \left( \rho_{1\wedge\gen} \right) +\\ &(1-\lambda) \Qopt_{\frac{1}{\alpha}} \circ \GMapDelta \left(\rho_{2\wedge\gen}\right)\Big).
        \end{aligned}
    \end{align}
    Hence, using the joint convexity of the relative entropy, we find for all $ \mbf{q}_1, \mbf{q}_2  \in \Sacc$, $\lambda \in [0,1]$ and $\rho_1, \rho_2 \in \dop{=}(AB)$
    \begin{align}
    &\begin{aligned}[t]
        D\Big(&\lambda q_1(\gen) + (1-\lambda) q_2(\gen) \Big\| \\ &\Qopt_{\frac{1}{\alpha}} \circ \GMapDelta \left(\lambda \rho_{1\wedge\gen} + (1-\lambda) \rho_{2\wedge\gen} \right)\Big) 
    \end{aligned} \\
    \leq & \begin{aligned}[t]
            D\Big(&\lambda q_1(\gen) + (1-\lambda) q_2(\gen) \Big\| \\ &\lambda \Qopt_{\frac{1}{\alpha}} \circ \GMapDelta \left( \rho_{1\wedge\gen} \right) +\\ &(1-\lambda) \Qopt_{\frac{1}{\alpha}} \circ \GMapDelta \left(\rho_{2\wedge\gen}\right)\Big).
        \end{aligned}\\
    \leq &\begin{aligned}[t]
        &\lambda D\rel{q_1(\gen)}{\Qopt_{\frac{1}{\alpha}} \circ \GMapDelta \left( \rho_{1\wedge\gen} \right)} \\
        &+ (1-\lambda) D\rel{q_2(\gen)}{\Qopt_{\frac{1}{\alpha}} \circ \GMapDelta \left( \rho_{2\wedge\gen} \right)},
    \end{aligned}
    \end{align}
    and the first term is also jointly convex.
    
    Finally, the set \(\mathcal{D}\) is a convex set since it only contains linear trace constraints and we can write
    \begin{equation}
        \gamma\bsym{\nu}_{\CP|\test} = \gamma\probst[\rho_{|\test}] = \probst[\sqrt{\Pi^{\test}}\rho\sqrt{\Pi^{\test}}].
    \end{equation}
\end{proof}

From this theorem we can immediately conclude that the optimization problem for $\hQKD$ in \cref{eq:halpha final} is jointly convex.

\begin{cor}[Joint Convexity of perturbed $\hQKD$ (Channel)]\label{cor:Joint convexity halpha channel}
    Let the feasible set of $\hQKD$ of \cref{eq:halpha final} be defined as
    \begin{equation}
     \mathcal{D}' \defvar \begin{aligned}[t]
        \{&\mbf{q} \in \Sacc, J \in \Pos(A'B), \bsym{\nu}_{\CP} \in \mathbb{P}(\alphCP)| \\ &\Tr_B[J] = \idop_{A'}, \bsym{\nu}_{\CP} = (\gamma \bsym{\nu}_{\CP|\test}, 1-\gamma)^T, \\
        &\bsym{\nu}_{\CP|\test} = \probstJ[J] \},
    \end{aligned} 
    \end{equation}
    and for all $\alpha>1$ and $\delta \in [0,1)$ define the function
    \begin{equation}
        \begin{aligned}[t]
            h(\mbf{q}, J, \bsym{\nu}_{\CP}) \defvar &\frac{\alpha D\rel{\mbf{q}}{\bsym{\nu}_{\CP}}}{\alpha - 1}  + q(\gen) \frac{g_{\alpha}[\delta](J)}{1-\delta}.
            \end{aligned}
    \end{equation}
    Then, \(h\) is jointly convex in $(\mbf{q},J,\mbf{\nu}_{\CP})$ over \(\mathcal{D}'\) for all $\alpha>\frac{1}{1-\delta}>1$ and $\delta \in[0,1)$. Furthermore, \(h\) satisfies
    \begin{align}
        \hQKD \geq \inf_{(\mbf{q}, J, \bsym{\nu}_{\CP}) \in \mathcal{D}'} h(\mbf{q}, J, \bsym{\nu}_{\CP}),
    \end{align}
    for all $\alpha>1$ and $\delta \in[0,1)$, with equality for \(\delta =0\).
\end{cor}
\begin{proof}
    The map \(\chi_g\) is an affine map and its composition with $\renObj_{\alpha}[\delta]$ remains convex by \cref{lem:ConvexCompAffine}.
\end{proof}

\subsection{Joint convexity of normalization constant \texorpdfstring{$\kappa$}{k}}

Finally, we turn our attention to the \(\frenyiSandDown_\alpha\)-normalization constant \(\kappa\). With our previous lemmas at hand, we can immediately prove the convexity of the optimization in the \(\frenyiSandDown_\alpha\)-normalization constant. For brevity, in this section we shall refer to this optimization as the ``$\kappa$ computation''.

Although not stated in the main text, we first prove the joint convexity of the optimization in the $\kappa$-computation in terms of the state \(\rho\) and deduce the joint convexity in terms of Eve's channel from it.

\begin{theorem}[Joint Convexity $\kappa$ (State)]\label{thrm:kappa convex state}
    The optimization problem to find the $\frenyiSandDown_\alpha$-normalization constant \(\kappa\),
    \begin{align}
        \begin{aligned}
            \kappa = &\inf_{\substack{\rho \in \dop{=}(AB), \\ \bsym{\nu} \in \mathbb{P}(\Ct) }} \Bigg( \frac{1}{1-\alpha} \log \Big( \gamma \sum_{c\in \Ct} \nu(c)2^{(\alpha - 1) f(c)} \\
            &\qquad + \left(1-\gamma\right)2^{-(\alpha-1)\left(\frac{\renObj_{\alpha}[\delta](\rho)}{1-\delta} - f(\gen)  \right)} \Big) \Bigg) \\
            \quad \textrm{s.t. } &\Tr_B[\rho] = \tau_A, \\
            &\bsym{\nu} = \probst[\rho_{|\test}].
            \end{aligned}
        \end{align}
    is jointly convex in \((\rho,\bsym{{\nu}})\) for all \(\alpha > 1\) and \(\delta \in [0,1)\).
\end{theorem}
\begin{proof}
    Since \(-\log(x)\) is convex, by \cite[Section 3.2.4]{boyd_convex_2004} the $\frenyiSandDown_\alpha$-normalization constant \(\kappa\) will be convex in \((\rho, \bsym{\nu})\) if the sum within the logarithm is concave. We will show that this is indeed the case for all \(\alpha > 1\) and $\delta \in[0,1)$, yielding the claimed result. (The set over which the optimization is performed is clearly convex.)
    
    First, note that the function
    \begin{equation}
        \bsym{\nu} \mapsto \gamma \sum_{c\in \Ct} \bsym{\nu}(c)2^{(\alpha - 1) f(c)}
    \end{equation}
    is trivially concave because it is a linear transformation of \(\bsym{\nu}\).
    
    Next, the concavity with respect to \(\rho\) follows immediately from \cref{cor:exp renyiObj is concave}.

    Thus, it remains to show that the sum of both functions is jointly concave in \((\rho,\bsym{\nu})\). However, this also follows trivially, since any function \(f\) defined from concave functions \(f_1\) and \(f_2\) by
    \begin{equation}
        f(x,y) = f_1(x) + f_2(y),
    \end{equation}
    is concave. Therefore, the sum inside the logarithm is jointly concave, as claimed. 
    % 
    %Thus, by \cite[3.2.4]{boyd_convex_2004}, and since the set over which the optimization is performed is convex, the $\frenyiSandDown_\alpha$-normalization constant \(\kappa\) is jointly convex in \((\rho,\bsym{{\nu}})\) for all \(\alpha > 1\) and $\delta \in[0,1)$.
\end{proof}

Similarly, we prove the joint convexity of the optimization in the $\kappa$-computation characterized in terms of Eve's channel.

\begin{theorem}[Joint Convexity $\kappa$ (Channel)]\label{thrm:kappa convex}
    The optimization problem to find the $\frenyiSandDown_\alpha$-normalization constant \(\kappa\),
    \begin{align}
        \begin{aligned}
            \kappa = &\inf_{\substack{J \in \mathrm{Pos}(A'B), \\ \bsym{\nu} \in \mathbb{P}(\Ct) }} \Bigg( \frac{1}{1-\alpha} \log \Big( \gamma \sum_{c\in \Ct} \nu(c)2^{(\alpha - 1) f(c)} \\
            &\qquad + \left(1-\gamma\right)2^{-(\alpha-1)\left(\frac{g_{\alpha}[\delta](J)}{1-\delta} - f(\gen)  \right)} \Big) \Bigg) \\
            \quad \textrm{s.t. } &\Tr_B[J] = \idop_{A'}, \\
            &\bsym{\nu} = \probstJ[J].
            \end{aligned}
        \end{align}
    is jointly convex in \((J,\bsym{{\nu}})\) for all \(\alpha > 1\) and \(\delta \in [0,1)\) and equal to \cref{eq:kappa const} for $\delta = 0$.
\end{theorem}
\begin{proof}
    The proof has the same structure as \cref{thrm:kappa convex state} with one small change.

    The concavity of
    \begin{equation}
        2^{-(\alpha-1)\left(\frac{g_{\alpha}[\delta](J)}{1-\delta} - f(\gen)  \right)}
    \end{equation}
    follows from \cref{cor:exp renyiObj is concave} and \cref{lem:ConvexCompAffine}, because \(g_{\alpha}[\delta]\) is a composition of \(\renObj_{\alpha}[\delta]\) with the affine map \(\chi_g\). The remainder of the proof remains the same. 

    Equality to \cref{eq:kappa const} for $\delta = 0$ is also straight-forward since \(g_{\alpha}[0] = g_{\alpha} \).
\end{proof}

\section{Details of the Numerical Algorithm}\label{app:Details numerics}
\subsection{Introduction to the Frank-Wolfe Algorithm}

In the previous section we proved that a perturbed version of the term $g_{\alpha}$, $\renObj_\alpha$ is convex. We will use this to implement our main goal, numerically optimize~\cref{eq:halpha final} and \cref{eq:kappa const} from the main text. But how? Many convex optimization packages such as CVX and cvxpy follow the disciplined convex programming
(DCP) rule set~\cite{grant_disciplined_2006}. In this rule set, convex optimization problems are constructed from a base set of convex functions and sets (called atoms), and linked together by known operations that preserve convexity. 

However, many convex functions used in quantum information cannot easily be constructed from common atoms supported by DCP packages, and in cases where this is possible, it may result in very large problems~\cite{fawzi_semidefinite_2019} (though for the Umegaki divergence and ``QKD cone'', there has been work such as~\cite{hu_robust_2022,lorente_quantum_2025,he_exploiting_2024} aiming to resolve this difficulty). In particular, quantum \Renyi divergence is not supported by most packages for convex optimization, except in very recent work~\cite{he_operator_2025}.

To overcome this, we use the Frank-Wolfe optimization method to minimize~\cref{eq:halpha final} in a similar manor to~\cite{winick_reliable_2018}.

\begin{definition}
Let $V$ be a vector space over the field $\R$, and $S\subseteq V$ a convex subset. Let $f: S \rightarrow \R$ be a convex and differentiable function at all points in $S$. The convex optimization problem,
\begin{equation}
    \min_{\mbf{x} \in S} f(\mbf{x}),
\end{equation}
can be numerically solved by the following algorithm.
\begin{algorithm}[H]
\caption{Frank-Wolfe optimization \cite{frank_algorithm_1956}}\label{alg:FrankWolfe}
\begin{algorithmic}[1]
\State Let $\epsilon > 0$, $\mbf{x}_0 \in S$ and set $i = 0$.
\State Compute step direction $\Delta\mbf{x} \defvar \arg \min_{\Delta\mbf{x}}\inner{\grad f(\mbf{x}_i)}{\Delta \mbf{x}}$ subject to $\Delta\mbf{x}+\mbf{x}_i \in S$.
\State If $\inner{\grad f(\mbf{x}_i)}{\Delta \mbf{x}} < \epsilon f(\mbf{x}_i)$ then STOP.
\State Compute step size $\lambda \defvar \arg\min_{\lambda \in [0,1]} f(\mbf{x}_i+\lambda\Delta\mbf{x})$.
\State Set $\mbf{x}_{i+1} = f(\mbf{x}_i+\lambda\Delta\mbf{x})$ and $i \leftarrow i+1$. Go to step 2.
\end{algorithmic}
\end{algorithm}
\end{definition}

Different variations of this algorithm may be presented such as using an absolute stopping criteria instead of a relative one, or choosing the step-size in step 4 differently, but~\cref{alg:FrankWolfe} represents a ``vanilla" implementation of the Frank-Wolfe algorithm. Because $f$ is convex, the line search to minimize $\lambda$ is straightforward and can be done relatively easily with any one-dimensional optimization routine.

For differentiable functions, we can equivalently characterize their convexity by the so-called first-order conditions in the theorem below.

\begin{theorem}[First-Order conditions~\cite{boyd_convex_2004}] \label{thrm:FirstOrderConditions}
    Let $V$ be a vector space over the field $\R$ and $S\subseteq V$ a convex subset. Let $f:S\rightarrow \R$ be a differentiable function at all points in $S$. The function $f$ is convex if and only if for all points $\mbf{x},\mbf{y} \in S$,
    \begin{equation}
        f(\mbf{y}) \geq \inner{\grad{f}(\mbf{x})}{\mbf{y}-\mbf{x}}+ f(\mbf{x}).
    \end{equation}
\end{theorem}

The first-order conditions are very helpful for the case of QKD key rates due to the following issue. 

Let $\mbf{x}^*$ be the true minimum, and $\mbf{x}_n$ be the final point returned from the Frank-Wolfe algorithm. Then
\begin{equation}
    f(\mbf{x}_n) \geq f(\mbf{x}^*),
\end{equation}
but especially in the context of QKD, this is problematic as it is of utmost importance to determine a valid lower bound on the key rate. However, using~\cref{thrm:FirstOrderConditions} we lower bound the Frank-Wolfe solution \(\mbf{x}_n\) by~\cite{winick_reliable_2018},
\begin{equation}
    \begin{split}
        f(\mbf{x}^*) &\geq \inner{\grad{f}(\mbf{x}_n)}{\mbf{x}^*-\mbf{x}_n}+ f(\mbf{x}_n), \\
        & \geq \min_{\mbf{z}\in S}\inner{\grad{f}(\mbf{x}_n)}{\mbf{z}-\mbf{x}_n}+ f(\mbf{x}_n).
    \end{split}
\end{equation}
This is identical in form to computing the step direction in~\cref{alg:FrankWolfe}.

\subsection{Matrix Differentials}
The Frank-Wolfe algorithm above was defined over a generic vector space, which in our case amounts to Hermitian matrices. 
For this space, in order to compute the first-order derivatives required by the algorithm, it is convenient to use the concept of matrix differentials, as described in e.g. Refs.~\cite{harville_matrix_1997,hiai_introduction_2014,magnus_matrix_2002}\footnote{We note that many authors define derivatives for matrix calculus with different layout conventions. Therefore, care needs to be taken, when comparing between different sources.}. We present below the notation convention we use for such derivatives and differentials.

\begin{definition}
For an analytic function $f: \C^{n \times m} \rightarrow \C$, we define the differential and gradients by,
\begin{gather}
    \dd\rho \defvar \begin{bmatrix}
        \dd{\rho_{1,1}} & \cdots & \dd{\rho_{1,m}} \\
        \vdots & \ddots & \vdots \\
        \dd{\rho_{n,1}} & \cdots & \dd{\rho_{n,m}}
    \end{bmatrix}, \\
    (\grad f)^\dagger = \pdv{f}{\rho} \defvar \begin{bmatrix}
        \pdv{f}{\rho_{1,1}} & \cdots & \pdv{f}{\rho_{m,1}} \\
        \vdots & \ddots & \vdots \\
        \pdv{f}{\rho_{1,n}} & \cdots & \pdv{f}{\rho_{m,n}}
    \end{bmatrix}, \\
    df \defvar \Tr[\pdv{f}{\rho} \dd{\rho}] = \inner{\grad f}{\dd{\rho}}.
\end{gather}
\end{definition}

Note the swap of indices between $\dd\rho$ and $\pdv{f}{\rho}$. This is often referred to as the numerator layout. We are primarily concerned with functions $f: \Herm(Q) \rightarrow \R$, where,
\begin{gather}
    \dd{(\rho^\dagger)} = \dd{\rho}, \\
    (\grad{f})^\dagger = \grad f,
\end{gather}
and therefore, there is no distinction between $\grad f$ and $\pdv{f}{\rho}$.

Most importantly, in general $\rho$ and $\dd\rho $ do not commute, that is $\comm{\rho}{\dd\rho} \neq 0$, which complicates most derivatives. For example, the chain rule for analytic functions applied to matrices takes on the following form.

\begin{theorem}[Matrix derivatives (Theorem 3.25 \& 3.33 \cite{hiai_introduction_2014})] \label{thrm:matrixAnaylticDiff}
    Let $f: \C \rightarrow \C$ be an analytic function with derivative $f'$, and let $\sigma$ be a normal matrix with eigenvalues and (normalized) eigenvectors $\lambda_i$ and $\ket{\lambda_i}$ respectively. The differential of $f$ for operators is then given by
    \begin{equation}
        \dd{f}(\sigma) \defvar \sum_{i,j} \ketbra{\lambda_i} \dd{\sigma} \ketbra{\lambda_j} r_{i,j},
    \end{equation}
    where
    \begin{equation}
        r_{i,j} = \begin{cases}
            f'(\lambda_i), & \lambda_i = \lambda_j \\
            \frac{f(\lambda_i) -f(\lambda_j)}{\lambda_i - \lambda_j}, & \lambda_i \neq \lambda_j
        \end{cases}
    \end{equation}
\end{theorem}
Equivalently, the eigenvectors can be reorganized into the unitary matrix $U = \sum_i \ketbra{\lambda_i}{i}$, and $r_{i,j}$ compiled into $R = \sum_{i,j} r_{i,j} \ketbra{i,j}$ for the form
\begin{equation}
    \dd{f}(\sigma) = U \left( R \odot (U^\dagger\dd{\sigma} U) \right) U^\dagger,
\end{equation}
where $\odot$ is the Hadamard product.
\begin{definition} \label{def:matrixAnaylticDiffMap}
    Let $f$, $\ket{\lambda_i}$, $\sigma$, and $r_{i,j}$ be defined as in~\cref{thrm:matrixAnaylticDiff}. So long as all $f'(\lambda_i)$ exist, we define the linear map $f'[\sigma]$ of the tangent plane at the point $\sigma$ as,
    \begin{equation}
        f'[\sigma](\rho) \defvar \sum_{i,j} \ketbra{\lambda_i} \rho \ketbra{\lambda_j} r_{i,j}.
    \end{equation}
\end{definition}

We also note that $r_{i,j} = r_{j,i}$, and as a result, $f'[\sigma]$ is self-adjoint, that is $f'[\sigma] = f'^\dagger[\sigma]$. Furthermore, if $f(\R) \subseteq \R$ and $\sigma \in \Herm(Q)$, then $f'[\sigma]$ is hermitian preserving.

The usual intuition for the chain rule holds only under the trace.
\begin{lemma} \label{lem:matrixAnaylticDiffTrace}
    Let $f$ and $\sigma$ be defined as in~\cref{thrm:matrixAnaylticDiff}, then
    \begin{equation}
        \dd\left(\Tr[f(\sigma)] \right) = \Tr[f'(\sigma) \dd{\sigma}].
    \end{equation}
\end{lemma}
\begin{proof}
    The proof trivially follows from applying the cyclic property of the trace directly to $\dd{f(\sigma)}$ from~\cref{thrm:matrixAnaylticDiff}.
\end{proof}

\subsection{Application to \texorpdfstring{\Renyi}{Renyi }entropies}\label{subsec:Application to Renyi entropies}

Now we combine the statements for matrix differentials with the Frank-Wolfe algorithm. For this purpose, we need to carefully track the order in which the powers are applied, especially in the later part of the section where matrix differentials and gradients are required. As such we denote,
\begin{equation}
    \pow_{\alpha}(x) \defvar x^\alpha.
\end{equation}

In this section, we will omit the protocol map \(\GMap\) and assume its output as the input to \(\Qopt_{\frac{1}{\alpha}}\), see \cref{def:Qopt}. We will denote this function taking the output of the map \(\GMap\) as the input by $\renObjNoG_\alpha$. First, we construct the gradient of this map $\renObjNoG_\alpha$ based on the reduction to CPTNI maps shown in \cref{cor:Fixed-length security lower bnd halpha}. 

In the process, to compress our notation while still using the reduction to the CPTNI map \(\hat{\GMap}\) from \cref{cor:Fixed-length security lower bnd halpha}, we define the map $\theta: \dop{\leq}(Q) \rightarrow \R$
\begin{equation}
    \theta(\rho_Q) \defvar \Tr \circ \pow_\alpha \circ \hat{\ZMap} \circ \pow_{\frac{1}{\alpha}} (\rho_Q)  + 1- \Tr[\rho_Q],
\end{equation}
where $\hat{\ZMap}$ again indicates the map $\ZMap$ omitting the discard symbol like in \cref{cor:Fixed-length security lower bnd halpha}.
Then, we rewrite 
\begin{equation}
    \renObjNoG_\alpha(\rho_Q) = \frac{-1}{\alpha-1}\log[\theta (\rho_Q)].
\end{equation}

\begin{theorem} [Gradient of $\renObjNoG_\alpha$]
\label{thrm:diffBasic}
    Let $\alpha \in (1,\infty]$, $\rho_Q \in \dop{\leq}(Q)$, the gradient of $\renObjNoG_\alpha$ (when it exists) is given through the differential
    \begin{equation} \label{eq:diffGBasic}
        \dd{\renObjNoG_\alpha} = \frac{-1}{\alpha-1} \frac{\dd\theta}{\ln(2)\theta},
    \end{equation}
    with
    \begin{equation}
        \dd\theta = \inner{\grad\theta(\rho_Q)}{\dd\rho_Q}
    \end{equation}
    and
    \begin{equation}
        \grad\theta(\sigma) = \alpha \pow_{\frac{1}{\alpha}}'[\sigma] \circ \pow_{\alpha-1} \circ \hat{\ZMap} \circ \pow_{\frac{1}{\alpha}}(\sigma) -I.
    \end{equation}
\end{theorem}
\begin{widetext}
\begin{proof}
    \cref{eq:diffGBasic} is a simple application of the chain rule on $\log$. The term $\dd\theta$ requires more work.
    \begin{equation}
        \begin{split}
            \dd\theta &= \Tr \Big[ \left( \alpha\pow_{\alpha-1} \circ \hat{\ZMap} \circ \pow_{\frac{1}{\alpha}}(\rho) \right)
            \left( \hat{\ZMap} \circ \dd(\pow_{\frac{1}{\alpha}}(\rho)) \right) \Big] -\Tr[\dd{\rho}], \\
            &= \Tr \Big[ \left( \alpha\pow_{\alpha-1} \circ \hat{\ZMap} \circ \pow_{\frac{1}{\alpha}}(\rho) \right) 
            \left( \hat{\ZMap} \circ \pow'_{\frac{1}{\alpha}}[\rho](\dd\rho) \right) \Big] -\Tr[\dd{\rho}], \\
            &= \Tr\Big[\Big(\alpha\pow'^\dagger_{\frac{1}{\alpha}}[\rho] \circ \hat{\ZMap}^\dagger \circ \pow_{\alpha-1} \circ \hat{\ZMap} \circ \pow_{\frac{1}{\alpha}}(\rho) - I\Big) \dd\rho\Big], \\
            &= \Big\langle \alpha\pow'_{\frac{1}{\alpha}}[\rho] \circ \pow_{\alpha-1} \circ \hat{\ZMap} \circ \pow_{\frac{1}{\alpha}}(\rho) - I,\dd\rho \Big\rangle.
        \end{split}
    \end{equation}
    On line one we used~\cref{lem:matrixAnaylticDiffTrace} to take the derivative of $\pow_\alpha$. On line two we applied~\cref{thrm:matrixAnaylticDiff,def:matrixAnaylticDiffMap} to take the derivative of $\pow_{\frac{1}{\alpha}}$. On line three, we combined both traces into one (which defines the inner product), and applied the dual maps of $\hat{\ZMap}$ and $\pow'_{\frac{1}{\alpha}}$. Finally, for the last line, we used the facts that $\pow'_{\frac{1}{\alpha}}$ and $\hat{\ZMap}$ are self-dual, and that $\hat{\ZMap} \circ \pow_{\alpha-1} \circ \hat{\ZMap} = \pow_{\alpha-1} \circ \hat{\ZMap}$.
\end{proof}
\end{widetext}

Again, we did not include \(\GMap\) into \(\renObjNoG_\alpha\) like in the main text, see \cref{cor:Fixed-length security lower bnd halpha}. We also did not include the additional transformation to Choi states as in \cref{eq:defn galpha}. Both cases can be included by an affine map.

Most importantly, to ensure the existence of the derivative similar to \cite{winick_reliable_2018}, we need to include perturbations that can also be formulated as affine maps. This will result in the definition of \(\GMapDelta\) shown earlier in \cref{def:renObj perturbed} for our convexity proofs. Hence, we state the following theorem inserting arbitrary affine maps into \(\renObjNoG_{\alpha}\).

\begin{theorem}\label{thrm:generalPerturbationGradient}
     Let $\alpha \in (1,\infty]$, $\rho_Q \in \dop{\leq}(Q)$, and an affine map $\cal{A}: \dop{\leq}(Q) \rightarrow \dop{\leq}(Q)$, then
    \begin{equation}
        \dd{(\renObjNoG_\alpha \circ \mathcal{A})} = \frac{-1}{\alpha-1} \frac{\inner{\cal{A}'^\dagger \circ \grad \theta \circ \cal{A}(\rho_Q)}{\dd{\rho_Q}}}{\ln(2)\theta \circ \cal{A}(\rho_Q)}.
    \end{equation}
\end{theorem}
\begin{proof}
    The gradient follows the same steps as in~\cref{thrm:diffBasic} but with an extra chain rule for $\mathcal{A}$.
    \begin{equation}
        \begin{split}
            \dd(\theta \circ \mathcal{A}) &= \inner{\grad\theta \circ \mathcal{A}(\rho)}{\dd(\mathcal{A}(\rho))}, \\
            &= \inner{\grad\theta \circ \mathcal{A}(\rho)}{\mathcal{A}'(\dd\rho)}, \\
            &= \inner{\mathcal{A}'^\dagger \circ \grad\theta \circ \mathcal{A}(\rho)}{\dd\rho}.
        \end{split}
    \end{equation}
    Note that although in general the dual of an affine map is not defined, $\mathcal{A}'$ is linear so $\mathcal{A}'^\dagger$ is.
\end{proof}

In principle, the fact that $\renObjNoG(0) = 0$ means we can work with the semidefinite convex cone. However, both the Frank-Wolfe algorithm and~\cref{thrm:FirstOrderConditions} require the gradient to exist for all points in the convex set, which $\renObjNoG_\alpha$ fails whenever it has an eigenvalue of $0$.

\subsection{Perturbation and Stability}

As noted in the previous section, both the Frank-Wolfe algorithm and~\cref{thrm:FirstOrderConditions} require the objective function to be differentiable over the full convex set. For $\renObjNoG_\alpha$ with $\alpha \in (1,\infty]$ this reduces to two conditions:
\begin{enumerate}
    \item $\dd{(\log(\theta))}$ exists, therefore requiring $\theta >0$,
    \item $\pow'_{\frac{1}{\alpha}}[\rho](\sigma)$ exists, therefore requiring $\rho \succ 0$, which is easily violated.
\end{enumerate}
While the former condition is true for all $\rho_Q \in \dop{\leq}(Q)$, the later condition is not. In fact, the constraints and/or minimization may favor or even enforce an eigenvalue of zero. In Ref.~\cite{winick_reliable_2018}, this was overcome by applying an affine perturbation to push eigenvalues away from zero and using a continuity bound to ensure the result is still a valid lower bound on the key rate. Here, we instead exploit the convex properties of the function $\renObjNoG_\alpha$ to reduce the penalty.

In~\cref{thrm:generalPerturbationGradient} we prepended an affine map so we could add additional maps such as $\GMap$ or a transformation to Choi matrices. We can use the same structure to insert affine perturbations and maintain valid lower bounds. Therefore, we note the following lemma, which is a direct consequence of the definition of convex functions.
\begin{lemma}
\label{lem:convexDef0}
    Let $V$ be a vector space over the field $\R$, $C \subseteq V$ be a convex set, and $f:C \rightarrow \R$ a convex function on $C$. If there exists a point $x\in C$ such that $f(x) =0$, then for all $y \in C$ and $\theta \in [0,1]$
    \begin{equation}
        f((1-\theta) y + \theta x) \leq (1-\theta) f(y).
    \end{equation}
\end{lemma}
\begin{proof}
    By the definition of convexity, $f((1-\theta) y + \theta x) \leq (1-\theta) f(y) +\theta f(x)$, and we can replace $f(x)$ with 0.
\end{proof}

The gist of our approach is to pick affine perturbations $\mathcal{A}: \dop{\leq}(Q) \rightarrow \dop{\leq}(Q)$ that will resemble a convex combination of $\rho_Q$ with a state $\sigma$ where $\renObjNoG_{\alpha}(\sigma) =0$ (typically $\sigma$ is proportional to the identity). We tailor the choice of our affine map to the particular common problems later, and state the general case now.

\begin{theorem} \label{thrm:generalPerturbationBound}
    Let $\alpha \in (1,\infty]$, $\rho_Q \in \dop{\leq}(Q)$, and an affine map $\cal{A}: \dop{\leq}(Q) \rightarrow \dop{\leq}(Q)$. If there exists a number $\epsilon \in (0,1)$ such that for all $\tau_Q \in \dop{\leq}(Q)$ there exists an operator $\sigma_Q \in \dop{\leq}(Q)$ such that $\renObjNoG_\alpha(\sigma_Q) =0$ and $\mathcal{A}(\tau) = (1-\epsilon)\tau + \epsilon \sigma$, then
    \begin{equation}
        \frac{\renObjNoG_\alpha \circ \mathcal{A}(\rho_Q)}{1-\epsilon} \leq \renObjNoG_\alpha(\rho_Q).
    \end{equation}
    Furthermore, if $\sigma_Q \succ 0$, then $\grad[\renObjNoG_\alpha \circ \mathcal{A}](\rho_Q)$ exits.
\end{theorem}
\begin{proof}
    For the bound, let $\rho_Q$, $\sigma_Q$ and $\epsilon$ be defined as above. Using the fact that $\renObjNoG_\alpha$ is a convex function and \(\renObjNoG_{\alpha}(\sigma_Q) = 0\), we get
    \begin{equation}
    \begin{split}
        \renObjNoG_\alpha \circ \mathcal{A}(\rho_Q) & = \renObjNoG_\alpha((1-\epsilon)\rho_Q + \epsilon \sigma_Q), \\
        & \leq (1-\epsilon) \renObjNoG_\alpha(\rho_Q) + \epsilon \renObjNoG_\alpha(\sigma_Q), \\
        & = (1-\epsilon) \renObjNoG_\alpha(\rho_Q).
    \end{split}
    \end{equation}
    Then, rearranging this equation yields the bound. Note that for each $\rho_Q$ one can choose a different $\sigma_Q$. This subtlety means that $\mathcal{A}$ is not necessarily defined as $(1-\epsilon)\rho + \epsilon \sigma$, we only need a way of expressing each element of the image of \(\mathcal{A}\) in this form.

    Finally, we show that the map $\pow'_{\frac{1}{\alpha}}[\mathcal{A}(\rho_Q)]$ exists, which is equivalent to $\mathcal{A}(\rho_Q) \succ 0$. We know that $\mathcal{A}(\rho_Q)$ is a non trivial convex combination of positive semidefinite matrices. Hence, $\sigma_Q \succ 0$ implies $\mathcal{A}(\rho_Q) \succ 0$. Therefore, the gradient exists of $\renObjNoG_\alpha \circ \mathcal{A}(\rho_Q)$ for all $\rho_Q \in \dop{\leq}(Q)$.
\end{proof}

Here we note that \cref{thrm:generalPerturbationBound} is tighter than the data-processing inequality which only implies $\renObjNoG_\alpha \circ \mathcal{A}(\rho) \leq \renObjNoG_\alpha(\rho)$ if $\mathcal{A}$ is a quantum channel. This can be seen in the following way.

For $\epsilon \in [0,1)$, one can immediately deduce $\renObjNoG_\alpha \circ \mathcal{A}(\rho) \leq \frac{\renObjNoG_\alpha \circ \mathcal{A}(\rho)}{1-\epsilon}$. Therefore, \cref{thrm:generalPerturbationBound} improves upon the data-processing inequality used in \cite{winick_reliable_2018}. However, the improvement is quite small since any scenario where $\epsilon$ is not much smaller than one is already likely to result in a vanishing key rate.

\subsubsection{Choices for the Affine Perturbation}\label{subsubsec:Choices for Affine Perturbation}
In this section we present three different maps one can choose from for $\mathcal{A}$. For all three choices of maps, we make use of the following lemma.
\begin{lemma}\label{lem:renObjNoG = 0}
    Let $b\in [0,1]$ and $\alpha \in (1,\infty]$, and $d = \dim(Q)$. Then, $\renObjNoG_\alpha \left(\frac{b}{d}\idop_Q \right) =0$.
\end{lemma}
\begin{proof}
    Because $\renObjNoG_\alpha(\rho_Q) = \frac{-1}{\alpha-1}\log \circ \theta(\rho_Q)$, we only need to show that $\theta\left(\frac{b}{d}\idop_Q \right)=1$. Therefore,
    \begin{equation}
    \begin{split}
        \theta\left(\frac{b}{d}\idop_Q \right) &=
        \begin{multlined}[t]
            \Tr \circ \pow_\alpha \circ \hat{\ZMap} \\
            \circ \pow_{\frac{1}{\alpha}} \left(\frac{b}{d}\idop_Q \right) + 1- \Tr[\frac{b}{d}\idop_Q]
        \end{multlined} \\
        &= \Tr \circ \pow_\alpha \circ \hat{\ZMap} \left(\left(\frac{b}{d}\right)^{\frac{1}{\alpha}}\idop_Q\right) + 1 -b \\
        &= \Tr[\frac{b}{d} \idop_Q] +1 -b =1.
    \end{split}
    \end{equation}
    where we used that the map $\hat{\ZMap}$ is unital ($\hat{\ZMap}(\idop_Q) = \idop_Q$).
\end{proof}

The simplest choice for the perturbation map is the depolarizing channel $\pert_\epsilon$,
\begin{equation}\label{eq:depol channel perturbation}
    \pert_\epsilon (\rho_Q) = (1-\epsilon)\rho_Q + \epsilon \frac{\Tr[\rho_Q]\idop_Q}{\dim(Q)},
\end{equation}
and restricting $\epsilon \in (0,1)$. Since $\pert_\epsilon$ is trace preserving, the perturbation scales with $\Tr[\rho_Q]$, preventing it from dominating the bound. However, the depolarization channel does not suffice if $\rho_Q =0$, because the gradient only exists for \(\rho_Q \neq 0\). We summarize this statement for the depolarization channel in the next lemma.

\begin{lemma} \label{lem:gradDepolPerturbation}
    Let $\rho_Q \in \dop{\leq}(Q)$, and $\epsilon \in (0,1]$, then for \(\Delta_{\epsilon}\) as defined in \cref{eq:depol channel perturbation}, it holds
    \begin{equation}
        \grad[\renObjNoG_\alpha \circ \pert_\epsilon](\rho_Q)
    \end{equation}
    exists if and only if \(\rho_Q \neq 0\).
\end{lemma}
\begin{proof}
    Let $\rho_Q$ and $\epsilon$ be defined as above. Once again, we only need to focus on when the map $\pow'_{\frac{1}{\alpha}}[\Delta_\epsilon(\rho_Q)]$ is defined. Therefore, we can equivalently show that
    \begin{equation}
        \Delta_\epsilon(\rho_Q) \succ 0 \Leftrightarrow \rho_Q \neq 0.
    \end{equation}
    If $\rho_Q = 0$, then $\Delta_\epsilon(\rho_Q) =0$, and if $\rho_Q \neq 0$, then $\epsilon\frac{\Tr[\rho_Q]\idop_Q}{\dim(Q)} \succ 0$, and $\Delta_\epsilon(\rho_Q) \succ 0$.
\end{proof}

Therefore, if the constraints exclude $\rho_Q =0$ from the convex set, this map is already sufficient. Unfortunately, there exist scenarios were this can happen. In those cases we can use the affine perturbation defined by,
\begin{equation}
    \Psi_\epsilon(\rho_Q) = (1-\epsilon)\rho_Q + \epsilon \frac{\idop_Q}{\dim(Q)},
\end{equation}
where $\epsilon \in (0,1]$.
\begin{lemma}
     The gradient $\grad[\renObjNoG_\alpha \circ \Psi_\epsilon](\rho_Q)$ exist for all $\rho_Q \in \dop{\leq}(Q)$ and $\epsilon \in (0,1]$.
\end{lemma}
\begin{proof}
    Like in~\cref{lem:gradDepolPerturbation}, we only need to show that $\Psi_\epsilon(\rho_Q) \succ 0$, which trivially holds.
\end{proof}

Although $\Psi_\epsilon$ works for all $\rho_Q\in \dop{\leq}(Q)$, the affine perturbation will dominate the resulting value of \(\renObjNoG_{\alpha}\) if the eigenvalues of \((1-\epsilon) \rho_Q\) are comparable in magnitude to those of \(\epsilon \frac{\idop_Q}{\dim(Q)}\). Then, we observe
\begin{equation}
    \renObjNoG_{\alpha}\circ\Psi_{\epsilon}(\rho_Q) \approx \epsilon \renObjNoG_{\alpha}\left(\epsilon \frac{\idop}{\dim(Q)}\right) \approx 0.
\end{equation}
Physically, this is equivalent to having a high error rate or depolarization in the channel. Therefore, it is important to use the smallest possible value for $\epsilon$.

Since $\Delta_\epsilon$ is linear, it scales with $\Tr[\rho_Q]$ and does not cause effects as the one described above (or at least diminishes them). Hence, one would prefer to use $\Delta_\epsilon$ as much as possible and reserve $\Psi_\epsilon$ to handle the edge case of $\rho_Q \approx 0$. 

We achieve this by combining both maps into a single affine map $\Psi_{\epsilon_2} \circ \Delta_{\epsilon_1}$ where $\epsilon_2$ is set to a very small value to protect the stability of the algorithm. Hence, we define the map \(\Phi_{\epsilon_1,\epsilon_2}\) as
\begin{equation}\label{eq:perturbation map Phi}
    \Phi_{\epsilon_1,\epsilon_2}(\rho_Q) = \Psi_{\epsilon_2} \circ \pert_{\epsilon_1} (\rho_Q) = (1-\delta) \rho_Q + \delta \tau_Q
\end{equation}
where $\delta \defvar \epsilon_1 + \epsilon_2 - \epsilon_1\epsilon_2$ and
\begin{equation}
    \tau_Q \defvar \frac{\epsilon_1\Tr[\rho_Q] +\epsilon_2 -\epsilon_1\epsilon_2 \Tr[\rho_Q]}{\delta} \frac{\idop_Q}{\dim(Q)}.
\end{equation}

\begin{lemma}\label{lem:Existence of gradient for perturbation map Phi}
    Let \(\Phi_{\epsilon_1,\epsilon_2}\) be defined as in \cref{eq:perturbation map Phi}. The gradient $\grad[\renObjNoG_\alpha \circ \Phi_{\epsilon_1,\epsilon_2}](\rho_Q)$ exists for all $\rho_Q \in \dop{\leq}(Q)$ and $\epsilon_1,\epsilon_2 \in (0,1]$.
\end{lemma}
\begin{proof}
    First, we note that for all $\epsilon_1,\epsilon_2\in (0,1]$, the resulting \(\delta\) lies in $(0,1]$. Next, we need to show that $\tau_Q \in \dop{\leq}(Q)$ for all possible choices of $\epsilon_1,\epsilon_2\in (0,1]$. It is clear that $\frac{\idop_Q}{\dim(Q)} \in \dop{\leq}(Q)$, so we can focus on bounding the prefactor \begin{equation}
        \frac{\epsilon_1\Tr[\rho_Q] +\epsilon_2 -\epsilon_1\epsilon_2 \Tr[\rho_Q]}{\delta}.
    \end{equation}
    We note that $\epsilon_1(1-\epsilon_2) > 0$ and $\Tr[\rho_Q] \leq 1$. Therefore,
    \begin{equation}
    \begin{split}
        \delta  &=  \epsilon_1 + \epsilon_2 - \epsilon_1\epsilon_2 \\
        &=  \epsilon_2 + \epsilon_1(1 - \epsilon_2) \\
        &> \epsilon_2 + \Tr[\rho_Q]\epsilon_1(1 - \epsilon_2) \\
        &= \epsilon_1\Tr[\rho_Q] +\epsilon_2 -\epsilon_1\epsilon_2 \Tr[\rho_Q] >0.
    \end{split}
    \end{equation}
    Thus,
    \begin{equation}
        0 < \frac{\epsilon_1\Tr[\rho_Q] +\epsilon_2 -\epsilon_1\epsilon_2 \Tr[\rho_Q]}{\delta} < 1,
    \end{equation}
    and $\tau_Q \in \dop{\leq}(Q)$. Moreover, \(\tau_Q\) is positive definite. Hence, all conditions in~\cref{thrm:generalPerturbationBound} are satisfied and the gradient exists.
\end{proof}

Now, we can draw the connection to the definition of the perturbed objective function $\renObj_{\alpha}[\delta]$. We will simplify it using the CPTNI map $\hat{\GMap}$ following \cref{cor:Fixed-length security lower bnd halpha}, just like $\renObjNoG$ at the beginning of this section. Furthermore, let us define
\begin{equation}
    \hat{\tau} \defvar \frac{p}{\delta} \frac{\idop_{\hat{S}_QABXYI}}{\dim(\hat{S}_QABXYI)},
\end{equation}
where $p$ is as in \cref{eq:p in tau} given by
\begin{equation}
    p = \eps_1 \Tr[\hat{\GMap}(\rho_{AB})] \left(1-\eps_2\right) + \eps_2\Tr[\rho_{AB}].
\end{equation}
Using this definition of $p$ while assuming $\rho_{AB} \in \dop{=}(AB)$, we note that
\begin{equation}
    \left(1-\delta\right)\hat{\GMap}(\rho_{AB}) + \delta \hat{\tau} = \Phi_{\epsilon_1,\epsilon_2}(\hat{\GMap}(\rho_{AB})).
\end{equation}

Thus, for all $\rho_{AB} \in \dop{=}(AB)$ we find
\begin{align}\label{eq:relating renObj to renObjnoG with perturbations}
    &\renObj_\alpha[\delta] (\rho_{AB}) = \frac{-1}{\alpha-1} \log\left[ \Qopt_{\frac{1}{\alpha}} \circ \GMapDelta (\rho_{AB}) \right] \\
    &\begin{aligned}
        = \frac{-1}{\alpha-1} \log\Bigg[ &\Tr\left[ \hat{\ZMap}\left(\Phi_{\epsilon_1,\epsilon_2}(\hat{\GMap}(\rho_{AB}))^{\frac{1}{\alpha}} \right)^{\alpha} \right] \\
        &+ 1-\Tr\left[ \Phi_{\epsilon_1,\epsilon_2}(\hat{\GMap}(\rho_{AB}))\right] \Bigg]
    \end{aligned} \\
    &= \frac{-1}{\alpha-1}\log[\theta \circ \Phi_{\epsilon_1,\epsilon_2}(\hat{\GMap}(\rho_{AB}))].
\end{align}
Therefore, following the arguments presented in this section, the gradient of $\renObj_{\alpha}[\delta]$ exists for all $\rho_{AB} \in \dop{=}(AB)$. We summarize this in the following theorem.

\begin{theorem}
    Let \(\Phi_{\epsilon_1,\epsilon_2}\) be defined as in \cref{eq:perturbation map Phi}. The gradient $\grad[\renObj_{\alpha}[\delta]](\rho_{AB})$ exists for all $\rho_{AB} \in \dop{=}(AB)$ and $\epsilon_1,\epsilon_2 \in (0,1]$, such that $\delta \defvar \epsilon_1 + \epsilon_2 - \epsilon_1\epsilon_2$.
\end{theorem}
\begin{proof}
    The statement follows immediately from the calculation in \cref{eq:relating renObj to renObjnoG with perturbations} together with \cref{lem:Existence of gradient for perturbation map Phi}.
\end{proof}

\subsubsection{Stability in the Gradient}
In this section we outline the methods we chose to improve the numerical stability of the gradient. As such, all choices laid out here are our currently known best practices.

The numerical stability of the gradient is mainly influenced by:
\begin{enumerate}
    \item Instability when computing the finite difference term for $r_{i,j}$ in~\cref{def:matrixAnaylticDiffMap} which defines the map $\pow'_{\frac{1}{\alpha}} \circ [\mathcal{A}(\rho_Q)]$.
    \item Instability in computing small eigenvalues and its compound effect on $r_{i,j}$ and $\pow_{\alpha-1}$.
\end{enumerate}

First, let us consider the issue of finite differencing. In order to construct the map $\pow'_{\frac{1}{\alpha}}[\sigma_Q]$, we must evaluate the terms $r_{i,j}$ from~\cref{thrm:matrixAnaylticDiff} given by
\begin{equation}
    r_{i,j} = \begin{cases}
        \frac{1}{\alpha}\lambda_i^{\frac{1}{\alpha}-1}, & \lambda_i = \lambda_j \\
        \frac{ \lambda_i^{\frac{1}{\alpha}} -\lambda_j^{\frac{1}{\alpha}} }{\lambda_i - \lambda_j}, & \lambda_i \neq \lambda_j
    \end{cases}
\end{equation}
where $\{\lambda_i\}|_{i=1}^d$ are the eigenvalues (with multiplicity) of $\sigma_Q$. With regards to finite differencing, two clear problems emerge. First, numerical instability is likely to break any case where $\lambda_i =\lambda_j$. Second, if $\lambda_i \approx \lambda_j$ then the finite differencing case can easily cause numerical instability when dividing by $\lambda_i - \lambda_j$. This is especially problematic in conjunction with the first problem. We mitigate this issue by replacing $r_{i,j}$ with
\begin{equation}
    \hat{r}_{i,j} = \begin{cases}
        \frac{1}{2\alpha}\left( \lambda_i^{\frac{1}{\alpha}-1} + \lambda_j^{\frac{1}{\alpha}-1} \right), & \lambda_i \approx \lambda_j\\
        \frac{ \lambda_i^{\frac{1}{\alpha}} -\lambda_j^{\frac{1}{\alpha}} }{\lambda_i - \lambda_j}, & \text{otherwise}
    \end{cases}
\end{equation}
where $\lambda_i \approx \lambda_j$ is given by Matlab's ``ismembertol" function \cite{MATLAB}. For this function, $a,b \in \R$ are considered relatively equal, $a \approx b$, if they satisfy
\begin{equation}
    \abs{a-b} \leq 10^{-12}\max\{\abs{a},\abs{b}\}.
\end{equation}
In other words, we replaced the finite difference between close eigenvalues with the average of the derivatives. We justify this choice with the following lemma.

\begin{widetext}
\begin{lemma}[Taylor Series Convergence]
    Let $\lambda \geq 0$, $\epsilon > 0$, and $\alpha \in (1,\infty]$. Let $\lambda_i, \lambda_j$ be a pair of eigenvalues used to compute $r_{i,j}$ and $\hat{r}_{i,j}$ given by
    \begin{gather}
        \lambda_i = \lambda +\epsilon, \\
        \lambda_j = \lambda -\epsilon,
    \end{gather}
    If $\lambda_i \approx \lambda_j$, then the Taylor series expansions of the relative and absolute convergence near $\epsilon =0$ are given by
    \begin{gather}
        \frac{\hat{r}_{i,j}}{r_{i,j}} = 1 + \frac{(\alpha^{-1} -1)(\alpha^{-1}-2)}{3} \left( \frac{\epsilon}{\lambda} \right)^2 + \mathcal{O}\left( \left( \frac{\epsilon}{\lambda}\right)^4 \right), \\
        \hat{r}_{i,j} - r_{i,j} = \lambda^{\frac{1}{\alpha} -1}\left( 0 + \frac{\alpha^{-1}(\alpha^{-1} -1)(\alpha^{-1}-2)}{3} \left( \frac{\epsilon}{\lambda} \right)^2 + \mathcal{O}\left( \left( \frac{\epsilon}{\lambda}\right)^4 \right) \right).
    \end{gather}
\end{lemma}
\begin{proof}
    The proof follows by simply computing the Taylor series in $\epsilon$.
\end{proof}
\end{widetext}
Given the convergence properties in the above lemma, we expect sufficient stability and accuracy of $\hat{r}_{i,j}$.

The other concern with computing the gradient comes from numerical instability associated with handling small eigenvalues for $\alpha \gtrapprox 1$. Specifically, $\pow_{\alpha-1}(\sigma_Q)$ features a sharp transition, and $\hat{r}_{i,j}$ for $\pow'_{\frac{1}{\alpha}}[\sigma_Q]$ features vertical asymptotes for eigenvalues of zero. Because of this, instability in computing the eigenvalues can be magnified. We mitigated this issue through our choices of the affine maps in the previous section. Furthermore, we include safety checks ensuring that we do not accidentally use impossibly small eigenvalues.

The possible choices for the affine map $\mathcal{A}$ from our previous section \ref{subsubsec:Choices for Affine Perturbation} were specifically designed to remove all zero eigenvalues. Here we show that these affine maps act as non-decreasing functions on the minimum eigenvalue.

\begin{theorem}
    Let $\rho_Q \in \dop{\leq}(Q)$ and $\epsilon_1,\epsilon_2 \in [0,1]$, then
    \begin{gather}
        \eigmin \circ \Delta_{\epsilon_1}(\rho_Q) \geq \eigmin(\rho_Q), \\
        \eigmin \circ \Psi_{\epsilon_2}(\rho_Q) \geq \eigmin(\rho_Q), \\
        \eigmin \circ \Phi_{\epsilon_1,\epsilon_2}(\rho_Q) \geq \eigmin(\rho_Q).
    \end{gather}
\end{theorem}
\begin{proof}
    Let $\rho_Q \in \dop{\leq}(Q)$, $\epsilon_1,\epsilon_2 \in [0,1]$. A simple calculation yields
    \begin{equation}
    \begin{split}
        \eigmin \circ \Delta_\epsilon(\rho_Q)
        &= (1-\epsilon) \eigmin(\rho_Q) + \frac{\epsilon\Tr[\rho_Q]}{\dim(Q)} \\
        &= (1-\epsilon) \eigmin(\rho_Q) + \frac{\epsilon\sum_{i=1}^{\dim(Q)} \lambda_i}{\dim(Q)}  \\
        &\geq (1-\epsilon) \eigmin(\rho_Q) + \epsilon\eigmin(\rho_Q) \\
        & = \eigmin(\rho_Q).
    \end{split}
    \end{equation}
    Similarly for \(\Psi_\epsilon\),
    \begin{equation}
    \begin{split}
        \eigmin \circ \Psi_\epsilon(\rho_Q)
        & = (1-\epsilon)\eigmin(\rho_Q) + \frac{\epsilon}{\dim(Q)} \\
        & \geq (1-\epsilon) \eigmin(\rho_Q) + \frac{\epsilon\Tr[\rho_Q]}{\dim(Q)} \\
        & = \eigmin \circ \Delta_\epsilon(\rho_Q) \\
        &\geq \eigmin(\rho_Q).
    \end{split}
    \end{equation}
    Finally, combining the above statements, we have
    \begin{equation}
    \begin{split}
        \eigmin\circ\Phi_{\epsilon_1,\epsilon_2}(\rho_Q)
        &= \eigmin \circ \Psi_{\epsilon_2} \circ \Delta_{\epsilon_1} (\rho_Q) \\
        &\geq \eigmin \circ \Delta_{\epsilon_1} (\rho_Q) \\
        &\geq \eigmin (\rho_Q).
    \end{split}
    \end{equation}
\end{proof}

Because the affine maps lower \(\renObjNoG\), one typically tries to use the least amount of perturbation possible while maintaining a stable Frank-Wolfe optimization process. As an additional layer of safety, we track the smallest eigenvalues used for each matrix power and replace eigenvalues in the calculation that are deemed to be impossibly small. 

For example, if $\rho_Q \in \dop{\leq}(Q)$, any negative eigenvalue must be caused by numerical noise. We can then safely replace them with zero. We combine and propagate this through the affine maps for our best possible estimate of the minimum eigenvalue. For the perturbation $\Delta_\epsilon$ we gain a safe eigenvalue cutoff of,
\begin{equation}\label{eq:Min eig cutoff Delta_eps}
\begin{split}
    &\eigmin \circ \Delta_\epsilon(\rho_Q)\\
    &= (1-\epsilon) \max\left\{\eigmin(\rho_Q),0\right\}
    + \frac{\epsilon\Tr[\rho_Q]}{\dim(Q)},
    \end{split}
\end{equation}
and use this for computing $\hat{r}_{i,j}$ in $\pow'_{\frac{1}{\alpha}}[\Delta_\epsilon(\rho_Q)]$. 

Furthermore, for $\pow_{\alpha-1}$ we can track a bound on the lowest eigenvalue passing through
\begin{equation}
    \pow_{\alpha-1} \circ \hat{\ZMap} \circ \pow_{\frac{1}{\alpha}} \circ \mathcal{A}(\rho_Q).
\end{equation}
The minimum eigenvalue from $\pow_{\frac{1}{\alpha}} \circ \mathcal{A}(\rho_Q)$ can be computed similarly to \cref{eq:Min eig cutoff Delta_eps}. Additionally, we note that $\eigmin \circ \hat{\ZMap}(\sigma_Q) \geq \eigmin(\sigma_Q)$, which provides a safe eigenvalue cutoff for $\pow_{\alpha-1}$.

\section{Kraus Operators and States for Protocols}\label{app:Kraus ops protocols}
In this section, we summarize the states and Kraus operators required to calculate the key rates for each protocol. It is important to note that all Kraus operators are generated from Alice's POVM elements \emph{conditioned} on a generation round. Therefore, we will first state how those conditional POVM elements can be constructed in a generic way.

\subsection{Conditional POVM construction}\label{app:Conditional POVM construction}

\newcommand{\notproj}{\Lambda}

Here we especially focus on the case where a Schmidt decomposition was performed to reduce dimensions and show how the conditional POVM elements can be constructed in a generic way.

First, as in the main text, let \(\{M_k^A\}_{k=1\dots d_A}\) be Alice's POVM elements, however, here we do not set the ``test'' POVM elements via the relation \(M_k^{A|\test} = M^A_k\), and similarly for the ``generation'' POVM elements. This is because that approach is not entirely suitable for handling the POVM elements arising after a Schmidt decomposition.

Next, as in the main text, we partition the POVM elements into test and generation subsets. Furthermore, under a Schmidt decomposition the POVM elements \(M_k^A\) contain a factor proportional to the testing or generation fraction, depending on which subset they fall into. Therefore, for convenience of notation we indicate these POVM elements with a ``\(\wedge\)" similar to probabilities to indicate this factor, e.g. \(M_1^{A \wedge \test}\).
With this notation, the split into test and generation subsets can be written as
\begin{align}
    \{M^{A\wedge\test}_{\alpha,x} \}_{\alpha \in \mathcal{C}^{A,\test}, x \in \mathcal{X}^{\test}_{\alpha} } &= \{M^{A\wedge \test}_k\} , \\
    \{M^{A\wedge\gen}_{\alpha,x} \}_{\alpha \in \mathcal{C}^{A,\gen}, x \in \mathcal{X}^{\gen}_{\alpha} } &= \{M^{A\wedge \gen}_k\} .
\end{align}
Again, a ``\(\wedge\)" is supposed to indicate that
\begin{equation}
    \Tr[\rho M_k^{A \wedge \test} ] = p(k \wedge \test).
\end{equation}

Then, similar to the main text, we define the partition operators for test (generation) rounds as
\begin{align}
    \notproj^{\test} = \sum_{\substack{\alpha \in \mathcal{C}^{A,\test}, \\ x \in \mathcal{X}^{\test}_{\alpha}}} M_{\alpha,x}^{A\wedge\test}, \quad
    \notproj^{\gen} = \sum_{\substack{\alpha \in \mathcal{C}^{A,\gen}, \\ x \in \mathcal{X}^{\gen}_{\alpha}}} M_{\alpha,x}^{A\wedge\gen},
\end{align}
though now we allow for the possibility that these might not be projectors. Importantly however, note that summing positive semidefinite operators cannot result in an operator with smaller support, which implies we still have
\begin{align}
    \forall \alpha \in \mathcal{C}^{A,\test}, x \in \mathcal{X}^{\test}_{\alpha}, &\quad 
    \supp(M_{\alpha,x}^{A\wedge\test})\subseteq \supp(\notproj^{\test}) , \\ \forall \alpha \in \mathcal{C}^{A,\gen},  x \in \mathcal{X}^{\gen}_{\alpha}, &\quad
    \supp(M_{\alpha,x}^{A\wedge\gen}) \subseteq \supp(\notproj^{\gen}) ,
\end{align}
a property we will use later. Similarly we define states conditioned on test and generation as
\begin{align}
    \ket{\xi^t}_{AA'} \defvar \sqrt{\notproj^{\test}} \frac{\ket{\psi}_{AA'}}{\sqrt{\gamma}}, \\  
    \ket{\xi^g}_{AA'} \defvar \sqrt{\notproj^{\gen}} \frac{\ket{\psi}_{AA'}}{\sqrt{1 -\gamma}}.
\end{align}

We now aim to find the POVM elements conditioned on test and generation rounds, which we indicate as in the main text by
\begin{align}
    \{M^{A|\test}_{\alpha,x} \}_{\alpha \in \mathcal{C}^{A,\test}, x \in \mathcal{X}^{\test}_{\alpha} } &= \{M^{A|\test}_k\} , \\
    \{M^{A|\gen}_{\alpha,x} \}_{\alpha \in \mathcal{C}^{A,\gen}, x \in \mathcal{X}^{\gen}_{\alpha} } &= \{M^{A| \gen}_k\}.
\end{align}
We define the conditional POVM elements by
\begin{align}
    M^{A|\test}_{\alpha,x} &\defvar \sqrt{(\notproj^{\test})^{-1}}^{\dagger} M_{\alpha,x}^{A\wedge\test} \sqrt{(\notproj^{\test})^{-1}}, \\
    M^{A|\gen}_{\alpha,x} &\defvar \sqrt{(\notproj^{\gen})^{-1}}^{\dagger} M_{\alpha,x}^{A\wedge\gen} \sqrt{(\notproj^{\gen})^{-1}},
\end{align}
for all \(\alpha \in \mathcal{C}^{A,\test}, x \in \mathcal{X}^{\test}_{\alpha}\) and \(\alpha \in \mathcal{C}^{A,\gen}, x \in \mathcal{X}^{\gen}_{\alpha} \), respectively. Crucially, the $^{-1}$ notation in the above expression is to be understood as the Moore-Penrose pseudo-inverse rather than the usual inverse (which would be undefined if the operators are not full-support). By construction these operators are positive semidefinite and hermitian. 
Moreover, due to the properties of the pseudo-inverse, it holds
\begin{align}
    \sqrt{(\notproj^{\test})^{-1}} \sqrt{(\notproj^{\test})} = \Pi_{\supp(\notproj^{\test})}, \\
    \sqrt{(\notproj^{\gen})^{-1}} \sqrt{(\notproj^{\gen})} = \Pi_{\supp(\notproj^{\gen})},
\end{align}
where with \(\Pi_{\supp(A)}\) we indicate the projector onto the support of an operator \(A\). In addition, for each \(M_{\alpha,x}^{A\wedge\test}\) and \(M_{\alpha,x}^{A\wedge\gen}\) we have
\begin{align}
    \Pi_{\supp(\notproj^{\test})} M_{\alpha,x}^{A\wedge\test} \Pi_{\supp(\notproj^{\test})} = M_{\alpha,x}^{A\wedge\test}, \\
    \Pi_{\supp(\notproj^{\gen})} M_{\alpha,x}^{A\wedge\gen} \Pi_{\supp(\notproj^{\gen})} = M_{\alpha,x}^{A\wedge\gen}.
\end{align}
Furthermore, it holds for the POVM elements conditioned on test rounds
\begin{align}
    &\sum_{\substack{\alpha \in \mathcal{C}^{A,\test}, \\ x \in \mathcal{X}^{\test}_{\alpha}}} M^{A|\test}_{\alpha,x} \\ 
    &= \sum_{\substack{\alpha \in \mathcal{C}^{A,\test}, \\ x \in \mathcal{X}^{\test}_{\alpha}}} \sqrt{(\notproj^{\test})^{-1}}^{\dagger} M_{\alpha,x}^{A\wedge\test} \sqrt{(\notproj^{\test})^{-1}} \\
    &= \Pi_{\supp(\notproj^{\test})} ,
\end{align}
and similarly for generation rounds,
\begin{align}
    \sum_{\substack{\alpha \in \mathcal{C}^{A,\gen}, \\ x \in \mathcal{X}^{\gen}_{\alpha}}} M^{A|\gen}_{\alpha,x}  = \Pi_{\supp(\notproj^{\gen})}.
\end{align}
Therefore, the sets \(\{M_{\alpha,x}^{A|\test}\}\) and \(\{M_{\alpha,x}^{A|\gen}\}\) indeed form valid POVMs, but on the subspaces $\supp(\notproj^{\test})$ and $\supp(\notproj^{\gen})$ respectively rather than the entire Hilbert space of $A$.\footnote{In some cases though, such as the qubit BB84 protocol we consider below with the Schmidt-decomposition dimension reduction, these subspaces can turn out to be equal to the full Hilbert space. However, this fact does not affect any claims in this section.}

Next, inserting the above properties, we find (omitting identities on \(A'\)) that these operators indeed yield the correct conditional probabilities,
\begin{widetext}
\begin{align}
    &\Tr[\ketbra{\xi^t}{\xi^t}_{AA'} M^{A|\test}_{\alpha,x}] \\
    &= \frac{1}{\gamma} \Tr[\sqrt{\notproj^{\test}} \ketbra{\psi}{\psi}_{AA'} \sqrt{\notproj^{\test}}^{\dagger} \sqrt{(\notproj^{\test})^{-1}}^{\dagger} M_{\alpha,x}^{A\wedge\test} \sqrt{(\notproj^{\test})^{-1}}] \\
    &=\frac{1}{\gamma} \Tr[ \ketbra{\psi}{\psi}_{AA'} \sqrt{\notproj^{\test}}^{\dagger} \sqrt{(\notproj^{\test})^{-1}}^{\dagger} M_{\alpha,x}^{A\wedge\test} \sqrt{(\notproj^{\test})^{-1}} \sqrt{\notproj^{\test}} ] \\
    &= \frac{1}{\gamma} \Tr[ \ketbra{\psi}{\psi}_{AA'} \Pi_{\supp(\notproj^{\test})} M_{\alpha,x}^{A\wedge\test} \Pi_{\supp(\notproj^{\test})} ] \\
    &= \frac{1}{\gamma} \Tr[ \ketbra{\psi}{\psi}_{AA'} M_{\alpha,x}^{A\wedge\test}] = p(\alpha,x|\test),
\end{align}
\end{widetext}
and similarly for generation rounds.

In addition, they result in the correct state after the measurement. To see this, let us see what happens if we apply \(\sqrt{M^{A|\test}_{\alpha,x}}\) onto \(\ket{\xi^t}_{AA'}\). Since \(\supp(M_{\alpha,x}^{A\wedge\test}) \subseteq \supp(\notproj^{\test}) \), it follows
\begin{align}
    &\sqrt{M^{A|\test}_{\alpha,x}} \ket{\xi^t}_{AA'} \\ 
    &= \sqrt{M_{\alpha,x}^{A\wedge\test}} \sqrt{(\notproj^{\test})^{-1}} \sqrt{\notproj^{\test}} \frac{\ket{\psi}_{AA'}}{\sqrt{\gamma}} \\
    &= \sqrt{M_{\alpha,x}^{A\wedge\test}} \Pi_{\supp(\notproj^{\test})} \frac{\ket{\psi}_{AA'}}{\sqrt{\gamma}} \\
    &= \sqrt{M_{\alpha,x}^{A\wedge\test}} \frac{\ket{\psi}_{AA'}}{\sqrt{\gamma}}.
\end{align}
A similar calculation yields the equivalent result for the POVM elements conditioned on generation rounds.

Finally, we note that any operator and state unitarily equivalent to the ones above satisfies the same properties and could equally well be used. Now, in the following subsections, we will present the POVM elements constructed for each example shown in the main text.

\begin{widetext}

\subsection{Qubit BB84}\label{app:Kraus ops qubit BB84}
For this protocol, we reduced the dimensions using a Schmidt-decomposition for better numerical performance. The final dimensions of the systems involved, are
\begin{align}
    \dim_A = 2, \quad \dim_{A'} = 2, \quad \dim_B = 3.
\end{align}
Then, Bob's POVM elements are given by
\begin{equation}\label{eq:Bob POVMs qubit BB84}
	\begin{aligned}
		M^B_{(Z,0)} &= p_z^B \begin{pmatrix} 0 & 0 &0 \\ 0& 1 & 0\\ 0&0 & 0 \end{pmatrix}, \quad
		M^B_{(Z,1)} = p_z^B \begin{pmatrix} 0 & 0& 0\\ 0& 0 & 0\\ 0&0 & 1 \end{pmatrix}, \\
		M^B_{(X,0)} &= \frac{p_x^B}{2} \begin{pmatrix} 0&0 & 0 \\  0&1 & 1 \\ 0& 1 & 1 \end{pmatrix},
		\quad
		M^B_{(X,1)} = \frac{p_x^B}{2} \begin{pmatrix} 0&0 &0   \\  0&1 & -1  \\ 0 & -1 & 1  \end{pmatrix}, \\
		M^B_{\bot} &=\begin{pmatrix} 1 & 0 & 0 \\ 0 & 0 & 0 \\ 0 & 0 & 0 \end{pmatrix},
	\end{aligned}
\end{equation}
where \(p_z^B = \gamma\) and \(p_x^B = 1 - \gamma\). For Alice's side, the POVM elements are determined by the source replacement scheme \cite{bennett_quantum_1992, ferenczi_symmetries_2012} and after reducing the dimensions through the Schmidt-decomposition, they are given by 
\begin{equation}\label{eq:Alice POVMs qubit BB84}
	\begin{aligned}
		M^A_{(Z,0)} &= p_z^A \begin{pmatrix}
		    1 & \\
            & 0
		\end{pmatrix}, \quad M^A_{(Z,1)} = p_z^A \begin{pmatrix}
		    0 & \\
            & 1
		\end{pmatrix}, \\
		M^A_{(X,0)} &= \frac{p_x^A}{2} \begin{pmatrix}
		    1 & 1\\
            1 & 1
		\end{pmatrix}, \quad M^A_{(X,1)} = \frac{p_x^A}{2} \begin{pmatrix}
		    1 & -1\\
            -1 & 1
		\end{pmatrix},
	\end{aligned}
\end{equation}
where \(p_z^A = \gamma\) and \(p_x^A = 1 - \gamma\). The conditioning on \(\gen\) and \(\test\) follows immediately by dividing by \(p_z^A\) and \(p_x^A\), respectively, since the \(Z\)-basis is used exclusively for generation rounds and the \(X\)-basis is used exclusively for test rounds.

To define the signal states \(\ket{\xi^t}\) and \(\ket{\xi^g}\), for notational convenience, let \(\ket{\phi^+}_{AA'} \defvar \frac{1}{\sqrt{2}} \left( \ket{00}_{AA'} + \ket{11}_{AA'} \right)\), be a maximally entangled Bell-state on systems \(AA'\). Then, the signal states are given by
\begin{align}
    \ket{\xi^t}&= \ket{\phi^+}_{AA'}, \qquad
    \ket{\xi^g} = \ket{\phi^+}_{AA'}.
\end{align}

Next, we define the Kraus operators of the map \(\GMap\). Since in key generation rounds we only send the $Z$-basis, only one Kraus operator for the map \(\GMap\) is required, which is
\begin{equation}\label{eq:Kraus op qubit BB84}
	\begin{aligned}
		K_{Z} &= \sqrt{p_z^B} \left[
		\begin{pmatrix} 1 & \\  & 0 \end{pmatrix}_{A \rightarrow S_Q} +
		  \begin{pmatrix} 0 & \\  & 1 \end{pmatrix}_{A \rightarrow S_Q} \right] 
		\otimes 
		\begin{pmatrix} 1 & 0 & 0 \\ 0 & 1 & 0 \end{pmatrix}
		_B \otimes 1_I,
	\end{aligned}
\end{equation}
where \(p_z^B = \gamma\). Furthermore, \(1_I \) is just a scalar since there is only one announcement surviving sifting in generation rounds.

Finally, the Kraus operators of the map \(\ZMap\) are
\begin{equation}
	\begin{aligned}
		Z_1 &=
		\begin{pmatrix} 1 & \\ & 0 \end{pmatrix}
		\otimes \idop_{\dim_B-1}, \quad
		Z_2 &=
		\begin{pmatrix} 0 & \\ & 1 \end{pmatrix}
		\otimes \idop_{\dim_B-1}.
	\end{aligned}
\end{equation}

\subsection{Active Decoy BB84}\label{app:Kraus ops active decoy BB84}
This protocol uses the qubit squasher \cite{beaudry_squashing_2008,gittsovich_squashing_2014} for Bob's POVM elements, hence we recover the qubit POVM elements stated in \cref{eq:Bob POVMs qubit BB84}. Furthermore, we set \(\Nent =1\), i.e. we only use \(m=0\) and \(m=1\) photons for key generation.

We chose the dimension of \(A\) consistent over all photon numbers to be \(\dim_A=2\) as it simplifies the expressions of the states, POVM elements and Kraus operators. For \(m=0\) photons, this could be reduced to \(\dim_A=1\) via a another Schmidt-decomposition. 

Thus, given this setup, Alice's POVM elements are those of the qubit protocol \cref{eq:Alice POVMs qubit BB84} for both photon numbers. Defining the signal states with \(m=0\) and \(m=1\) photons, consistent with our choices above leads to
\begin{align}
    \ket{\xi^t_{|m=0}} &= \begin{pmatrix} 1 \\ 0 \end{pmatrix}_{AA'}, \qquad
    \ket{\xi^g_{|m=0}} = \frac{1}{\sqrt{2}} \begin{pmatrix} 1 \\ 1 \end{pmatrix}_{AA'} \\
    \ket{\xi^t_{|m=1}} &= \ket{\phi^+}_{AA'}, \qquad
    \ket{\xi^g_{|m=1}} = \ket{\phi^+}_{AA'}.
\end{align}

Next, we require the Kraus operators of the \(\GMap\) for \(m=0\) and \(m=1\) photons. Since in a key generation round we only send the $Z$-basis, there are only two Kraus operators for the \(\GMap\); one for \(m=0\) and one for \(m=1\) photons. However, with our current representation, both Kraus operators are equal, and both are given by the qubit Kraus operator shown in \cref{eq:Kraus op qubit BB84}. The same holds for the pinching map \(\mathcal{Z}\).

\subsection{Passive Decoy BB84}\label{app:Kraus ops passive decoy BB84}
This protocol uses the flag-state squasher \cite{zhang_security_2021}. Thus, we need to construct Bob's POVM elements including flags. First, Bob's POVM elements on the \(\leq N_B\)-photon subspace with \(N_B =1\) are
\begin{equation}
	\begin{aligned}
		\tilde{M}^{B}_{(Z,0)} &= p_z^B \begin{pmatrix} 0 & 0 &0 \\ 0& 1 & 0\\ 0&0 & 0 \end{pmatrix}, \;
		\tilde{M}^{B}_{(Z,1)} = p_z^B \begin{pmatrix} 0 & 0& 0\\ 0& 0 & 0\\ 0&0 & 1 \end{pmatrix}, \\
		\tilde{M}^{B}_{(X,0)} &= \frac{p_x^B}{2} \begin{pmatrix} 0&0 & 0 \\  0&1 & 1 \\ 0& 1 & 1 \end{pmatrix},
		\;
		\tilde{M}^{B}_{(X,1)} = \frac{p_x^B}{2} \begin{pmatrix} 0&0 &0   \\  0&1 & -1  \\ 0 & -1 & 1  \end{pmatrix}, \\
		\tilde{M}^{B}_{\bot} &=\begin{pmatrix} 1 & 0 & 0 \\ 0 & 0 & 0 \\ 0 & 0 & 0 \end{pmatrix},
	\end{aligned}
\end{equation}
where \(p_z^B = \gamma\) and \(p_x^B = 1 - \gamma\). Next, these POVM elements need to be padded with flags. Therefore, let \(E_i:=\diag(0,\dots,0,1,0, \dots ,0) \in \R^{5 \times 5}\), be a diagonal matrix with \(1\) at the \(i\)-th entry. Hence, Bob's full squashed POVM elements are
\begin{equation}
	\begin{aligned}
		M^B_{(Z,0)} &= \tilde{M}^{B}_{(Z,0)} \oplus E_1, \quad
		M^B_{(Z,1)} = \tilde{M}^{B}_{(Z,1)} \oplus E_2, \\
		M^B_{(X,0)} &= \tilde{M}^{B}_{(X,0)} \oplus E_3,
		\quad
		M^B_{(X,1)} = \tilde{M}^{B}_{(X,1)} \oplus E_4, \\
		M^B_{\text{mult}} &= \bar{0}_3 \oplus E_{5}, \quad  M^B_{\bot} = \tilde{M}^{B}_{\bot} \oplus \bar{0}_{5},
	\end{aligned}
\end{equation}
where \(\bar{0}_k\) indicates a matrix with only zeros of dimension \(k\).

Next, we construct the states and Alice's POVM elements. Again, we chose the dimension of \(A\) consistent over all photon numbers to be \(\dim_A=4\) and \(\Nent =1\), i.e. we only use \(m=0\) and \(m=1\) photons for key generation. Then, the signal states are given by
\begin{align}
    \ket{\xi^t_{|m=0}} &= \frac{1}{\sqrt{2}} \begin{pmatrix} 0 \\ 0 \\ 1 \\ 1 \end{pmatrix}_{AA'}, \qquad
    \ket{\xi^g_{|m=0}} = \frac{1}{\sqrt{2}} \begin{pmatrix} 1 \\ 1 \\ 0 \\ 0 \end{pmatrix}_{AA'} \\
    \ket{\xi^t_{|m=1}} &= \frac{1}{\sqrt{2}} \left( \begin{pmatrix}
        0 \\ 0 \\ 1 \\ 0
    \end{pmatrix}_{A} \otimes \frac{1}{\sqrt{2}}  \begin{pmatrix}
        1 \\ 1
    \end{pmatrix}_{A'} + \begin{pmatrix}
        0 \\ 0 \\ 0 \\ 1
    \end{pmatrix}_{A} \otimes \frac{1}{\sqrt{2}} \begin{pmatrix}
        1 \\ -1
    \end{pmatrix}_{A'} \right), \\
    \ket{\xi^g_{|m=1}} &= \frac{1}{\sqrt{2}} \left( \begin{pmatrix}
        1 \\ 0 \\ 0 \\ 0
    \end{pmatrix}_{A} \otimes \begin{pmatrix}
        1 \\ 0
    \end{pmatrix}_{A'} + \begin{pmatrix}
        0 \\ 1 \\ 0 \\ 0
    \end{pmatrix}_{A} \otimes \begin{pmatrix}
        0 \\ 1
    \end{pmatrix}_{A'} \right).
\end{align}

For all photon numbers Alice's POVM elements are,
\begin{equation}\label{eq:Alice POVMs passive decoy BB84}
	\begin{aligned}
		M^A_{(Z,0)} &= \begin{pmatrix}
		    1 & & & \\
            & 0 & & \\
            & & 0 & \\
            & & & 0
		\end{pmatrix}, \quad M^A_{(Z,1)} = \begin{pmatrix}
		    0 & & & \\
            & 1 & & \\
            & & 0 & \\
            & & & 0
		\end{pmatrix}, \\
		M^A_{(X,0)} &= \begin{pmatrix}
		    0 & & & \\
            & 0 & & \\
            & & 1 & \\
            & & & 0
		\end{pmatrix}, \quad M^A_{(X,1)} =  \begin{pmatrix}
		    0 & & & \\
            & 0 & & \\
            & & 0 & \\
            & & & 1
		\end{pmatrix}.
	\end{aligned}
\end{equation}
Here, no explicit conditioning on \(\test\) or \(\gen\) required, because these POVM elements are defined based on the generic source-replacement scheme, i.e. no Schmidt decomposition was applied.

Next, we define the Kraus operators of the \(\GMap\) for \(m=0\) and \(m=1\) photons. Since in key generation rounds we only send the $Z$-basis, there are only two Kraus operators, for the \(\GMap\). However, with our current representation, both Kraus operators are equal and given by
\begin{equation}
	\begin{aligned}
		K_Z &= \left[
		\begin{pmatrix} 1 \\ 0 \end{pmatrix}
		_S  \begin{pmatrix} 1 & 0 & 0 & 0 \end{pmatrix}_A +
		\begin{pmatrix} 0 \\ 1 \end{pmatrix}
		_S \begin{pmatrix} 0 & 1 & 0 & 0 \end{pmatrix}_A \right] \otimes \left[0 \oplus \sqrt{p_z^B} \idop_2 \oplus \left(E_1 + E_2 \right) \right]_B
	 	  \otimes 1_I
	\end{aligned},
\end{equation}
where \(1_I \) is just a scalar. Finally, the Kraus operators for the pinching map \(\mathcal{Z}\) are
\begin{equation}
	\begin{aligned}
		Z_1 &=
		\begin{pmatrix} 1 & \\ & 0 \end{pmatrix}
		\otimes \idop_{\dim_B}, \quad
		Z_2 &=
		\begin{pmatrix} 0 & \\ & 1 \end{pmatrix}
		\otimes \idop_{\dim_B}.
	\end{aligned}
\end{equation}

\subsection{Passive Decoy 4-6 Protocol}\label{app:Kraus ops passive decoy 46}

We again are required to use the flag-state squasher \cite{zhang_security_2021} for this protocol. Therefore, let us first define POVM elements on Bob's \(\leq N_B \)-photon subspace for \(N_B =1\). Those are
\begin{equation}
	\begin{aligned}
		\tilde{M}^{B}_{(Z,0)} &= p_z^B \begin{pmatrix} 0 & 0 &0 \\ 0& 1 & 0\\ 0&0 & 0 \end{pmatrix}, \;
		\tilde{M}^{B}_{(Z,1)} = p_z^B \begin{pmatrix} 0 & 0& 0\\ 0& 0 & 0\\ 0&0 & 1 \end{pmatrix}, \\
		\tilde{M}^{B}_{(X,0)} &= \frac{p_x^B}{2} \begin{pmatrix} 0&0 & 0 \\  0&1 & 1 \\ 0& 1 & 1 \end{pmatrix},
		\;
		\tilde{M}^{B}_{(X,1)} = \frac{p_x^B}{2} \begin{pmatrix} 0&0 &0   \\  0&1 & -1  \\ 0 & -1 & 1  \end{pmatrix}, \\
		\tilde{M}^{B}_{(Y,0)} &= \frac{p_y^B}{2} \begin{pmatrix} 0&0 & 0 \\  0&1 & -i \\ 0& i & 1 \end{pmatrix},
		\;
		\tilde{M}^{B}_{(Y,1)} = \frac{p_y^B}{2} \begin{pmatrix} 0&0 &0   \\  0&1 & i  \\ 0 & -i & 1  \end{pmatrix}, \\
		\tilde{M}^{B}_{\bot} &=\begin{pmatrix} 1 & 0 & 0 \\ 0 & 0 & 0 \\ 0 & 0 & 0 \end{pmatrix}.
	\end{aligned}
\end{equation}
As for the passive BB84 protocol, these POVM elements need to be padded with flags. Therefore, let \(E_i:=\diag(0,\dots,0,1,0, \dots ,0) \in \R^{7 \times 7}\), be a diagonal matrix with \(1\) at the \(i\)-th entry. Furthermore, one needs to include multi-clicks (mult) where multiple of Bob's detectors click, which for \(N_B=1\), are only part of the flags. Hence, Bob's full squashed POVM elements are
\begin{equation}
	\begin{aligned}
		M^B_{(Z,0)} &= \tilde{M}^{B}_{(Z,0)} \oplus E_1, \quad
		M^B_{(Z,1)} = \tilde{M}^{B}_{(Z,1)} \oplus E_2, \\
		M^B_{(X,0)} &= \tilde{M}^{B}_{(X,0)} \oplus E_3,
		\quad
		M^B_{(X,1)} = \tilde{M}^{B}_{(X,1)} \oplus E_4, \\
		M^B_{(Y,0)} &= \tilde{M}^{B}_{(Y,0)} \oplus E_5,
		\quad
		M^B_{(Y,1)} = \tilde{M}^{B}_{(Y,1)} \oplus E_6, \\
		M^B_{\text{mult}} &= \bar{0}_3 \oplus E_7, \quad
		M^B_{\bot} = \tilde{M}^{B}_{\bot} \oplus \bar{0}_{7},
	\end{aligned}
\end{equation}
where \(\bar{0}_k\) indicates a matrix with only zeros of dimension \(k\).

For this protocol we construct Alice's signal states and POVM elements numerically since we do the diagonalization of \cref{eq:phase imperfect states with projection} numerically. 

In the example presented in the main text Alice sends four states with two intensities. Therefore, let \(\{\ket{i}\}\) for \(i= 1 \dots 8\) label Alice's state and intensity choices.

Next, we calculate the states \(\ket{w_m(i)}\) in \cref{eq:phase imperfect states with projection}, where we relabeled \( i  = (a,\mu)\). For each \( i  = (a,\mu)\), we construct
\begin{equation}
    \rho_{A'}^{(a,\mu)} =  q \sum_{m=0}^{\infty} p(m|\mu) \ketbra{m}{m}_{a}  + (1-q) \ketbra{\sqrt{\mu}}{\sqrt{\mu}}_a,
\end{equation}
and diagonalize 
\begin{equation}
    \PiNA \rho_{A'}^{(a,\mu)} \PiNA.
\end{equation}
This gives us the following quantities
\begin{align}
    \{\omega_m(i)\}_{i = 1 \dots 8}, \quad \{\ket{w_m(i)}\}_{_{i = 1 \dots 8}},
\end{align}
from which, as described in the main text, we can construct the state
\begin{equation}
    \ket{\psi}_{AA_sA'\bar{A}} = \sum_{m=0}^{\infty}\sqrt{\omega_m} \ket{m}_{A_s} \ket{\psi_{|m}}_{AA'\bar{A}},
\end{equation}
where
\begin{equation}
    \ket{\psi_{|m}}_{AA'\bar{A}} = \sum_{i} \sqrt{p(i|m)} \ket{i}_{A} \ket{w_m(i)}_{A'} \ket{m}_{\bar{A}}.
\end{equation}

Next, we perform a Schmidt decomposition to reduce the dimensions of each block \(m\). For each \(m\) we can therefore write 
\begin{equation}
    \ket{\psi_{|m}}_{AA'\bar{A}} = \left( U_m \otimes V_m\right) \left(\diag(\bsym{\lambda}) \otimes \idop_{\tilde{A}'} \right) \ket{\bar{\phi}^+}_{\tilde{A}\tilde{A}'}\otimes \ket{m}_{\bar{A}},
\end{equation}
where \(\bsym{\lambda}\) is the vector containing the Schmidt coefficients, \(U\) and \(V\) isometries and \(\ket{\bar{\phi}}_{\tilde{A}\tilde{A}'}\) the unnormalized maximally entangled state.

Then, we define the states after the Schmidt-decomposition as
\begin{equation}
    \ket{\tilde{\psi}_{|m}}_{\tilde{A}\tilde{A}'\bar{A}} \defvar \left(\diag(\bsym{\lambda}) \otimes \idop_{\tilde{A}'} \right) \ket{\bar{\phi}^+}_{\tilde{A}\tilde{A}'}\otimes \ket{m}_{\bar{A}},
\end{equation}
and the POVM elements as
\begin{equation}
    M^{\tilde{A}}_i \defvar U_m^{\dagger} \ketbra{i}{i}_A U_m.
\end{equation}

Afterwards, to find the POVM elements conditioned on test and generation rounds for each block \(m\), we apply the procedure described in \cref{app:Conditional POVM construction}. Having calculated these conditional POVM elements, the Kraus operators of the map \(\GMap\) are also found numerically from the resulting POVM elements conditioned on generation rounds. For each block \(m\), they are given by
\begin{equation}
	\begin{aligned}
		K_{Z|m} &= \left[
		\begin{pmatrix} 1 \\ 0 \end{pmatrix}
		_S  \otimes \sqrt{M^{\tilde{A}|\gen,m}_{Z,0}} +
		\begin{pmatrix} 0 \\ 1 \end{pmatrix}
		_S \otimes \sqrt{M^{\tilde{A}|\gen,m}_{Z,1}} \right] \otimes \left[0 \oplus \sqrt{p_z^B} \idop_2 \oplus \left(E_1 + E_2 \right) \right]_B
	 	  \otimes 1_I
	\end{aligned},
\end{equation}
where \(1_I \) is again a scalar. Finally, the Kraus operators of the pinching channel \(\ZMap\) are for each block \(m\) given by
\begin{equation}
	\begin{aligned}
		Z_{1|m} &=
		\begin{pmatrix} 1 & \\ & 0 \end{pmatrix}
		\otimes \idop_{\dim_{\tilde{A}|m}\cdot\dim_B}, \quad
		Z_{2|m} &=
		\begin{pmatrix} 0 & \\ & 1 \end{pmatrix}
		\otimes \idop_{\dim_{\tilde{A}|m}\cdot\dim_B}.
	\end{aligned}
\end{equation}

In addition to the Schmidt decomposition, one can reduce the dimensions of the optimization variables, i.e. the Choi states of Eve's channels acting on each block \(m\), even further by noting the following.

Let \(J_m\) be the Choi state of Eve's channel \(\mathcal{E}_m:\tilde{A}'\bar{A} \rightarrow B\) acting on the \(m\)-th block. Then, for any state \(\rho\) in system \(\tilde{A}'\) it holds
\begin{align}
    \mathcal{E}_m[\rho_{\tilde{A}'} \otimes \ketbra{m}{m}_{\bar{A}}] = \Tr_{\tilde{A}'\bar{A}}\left[\left(\rho^T \otimes \ketbra{m}{m}^T \otimes \idop_B \right) J_m\right] = \Tr_{\tilde{A}'}\left[\left(\rho^T \otimes \idop_B \right) \bra{m}J_m\ket{m}\right].
\end{align}
If we define the operators \(\tilde{J}_m\) as 
\begin{equation}
    \tilde{J}_m \defvar \bra{m}J_m\ket{m},
\end{equation}
it is easy to verify that these are valid Choi states for a channel from \(\tilde{A}'\) to \(B\). Hence, we can reduce the dimensions of the Choi states and the signal states by optimizing over \(\tilde{J}_m\) instead.

\end{widetext}
\end{document}